\documentclass[12pt]{report}

\usepackage{graphicx} 
\usepackage{booktabs} 
\usepackage{multirow}
\usepackage[letterpaper, left=1.5in, right=1in, top=1in, bottom=1in]{geometry}
\usepackage{setspace}  
\usepackage[explicit]{titlesec}  
\usepackage[titles]{tocloft}  
\usepackage[page]{appendix} 

\usepackage[superscript]{cite}

\usepackage{rotating} 
\usepackage[normalem]{ulem}  

\usepackage{amsmath}
\usepackage{amsfonts}
\usepackage{amssymb}
\usepackage{siunitx} 
\usepackage{multirow}   
\usepackage{bm}
\usepackage{verbatim}

\usepackage[english]{babel}
\usepackage{blindtext}

\usepackage{caption}
\DeclareCaptionLabelFormat{bsc}{\textbf{\textsc{#1}\ #2}}
\cftsetindents{section}{6mm}{8mm}

\cftsetindents{subsection}{12mm}{11mm}
\cftpagenumbersoff{subsection} 

\cftsetindents{subsubsection}{18mm}{13mm}
\cftpagenumbersoff{subsubsection} 

\cftpagenumbersoff{subsection}

\titleformat{\chapter}[display]
{\Large\bfseries\filcenter}{\chaptertitlename\ \thechapter}{0pt}{#1}
\titleformat{\section}{\large\bfseries}{\thesection\hspace{2mm}}{0pt}{#1}
\titleformat{\subsection}{\bfseries}{\thesubsection\hspace{2.5mm}}{0pt}{#1}
\titleformat{\subsubsection}{\bfseries}{\thesubsubsection\hspace{3mm}}{0pt}{#1}

\usepackage[bottom,hang,flushmargin,perpage,symbol*]{footmisc} 
\DefineFNsymbols*{lamportnostar}[math]{\dagger\ddagger\S\P\|{\dagger\dagger}{\ddagger\ddagger}}
\setfnsymbol{lamportnostar}
\usepackage[lf]{Baskervaldx} 
\usepackage[bigdelims,vvarbb]{newtxmath} 
\usepackage[cal=cm]{mathalfa} 

\usepackage{nomencl}
\makenomenclature

\renewcommand{\nompreamble}{Effort has been made to use the blow symbols consistently. In the interest of clarity, symbols are occasionally used in isolation for other purposes, in which case their pertinent meaning will be stated.}
\usepackage{ifthen}
\renewcommand{\nomgroup}[1]{%
  \ifthenelse{\equal{#1}{A}}{ {\vskip 6mm} \item[\textbf{Roman Letters}]}{%
    \ifthenelse{\equal{#1}{G}}{ {\vskip 6mm} \item[\textbf{Greek Letters}]}{%
      \ifthenelse{\equal{#1}{M}}{ {\vskip 6mm} \item[\textbf{Mathematical Notations}]}{%
        \ifthenelse{\equal{#1}{S}}{ {\vskip 6mm} \item[\textbf{Subscripts and Superscripts}]}{%
        }
      }
    }
  }
}

\usepackage{algorithm}
\usepackage{algorithmicx}
\usepackage{algpseudocode}

\usepackage{booktabs}
\usepackage{tabularx}

\usepackage{amsmath, amssymb, amsfonts}
\usepackage{graphicx}
\usepackage{booktabs}
\usepackage{longtable}
\usepackage{pdflscape}
\usepackage{array}
\usepackage{tabularx} 
\usepackage{ltablex}  
\usepackage{microtype}

\usepackage[T1]{fontenc}
\usepackage[utf8]{inputenc}

\usepackage{fontawesome5}   

\usepackage{tikz}
\usetikzlibrary{calc}

\usepackage{pgfornament}

\usepackage{tabularx} 
\usepackage{booktabs} 
\usepackage{xcolor}   

\usepackage[bookmarksnumbered,pdfa]{hyperref}

\usepackage{cleveref}

\crefname{equation}{Eq.}{Eqs.}
\Crefname{equation}{Equation}{Equations}

\crefname{table}{Table}{Tables}
\Crefname{table}{Table}{Tables}

\crefname{figure}{Fig.}{Figs.}
\Crefname{figure}{Figure}{Figures}

\crefname{section}{Sec.}{Secs.}
\Crefname{section}{Section}{Sections}

\crefname{chapter}{Chap.}{Chaps.}
\Crefname{section}{Section}{Sections}

 \hypersetup{bookmarks=true,
         pdflang={en-US}, 
         pdfdisplaydoctitle=true, 
         pdfauthor={Pradosh Pritam Dash}, 
         pdftitle={Transcranial FUS Therapy and Monitoring using Nonlinear Acoustics}, 
         pdfsubject={Mechanical Engineering Dissertation},
         pdfkeywords={Focused Ultrasound (FUS), Transcranial Focusing, Acoustic Holograms, Topology Optimization,  Nonlinear Parametric Array Effect, Non-invasive Registration, Intracranial Pressure, Hydrocephalus, Neuromodulation},
         colorlinks=true,
         urlcolor=black,
         linkcolor=black,
         citecolor=black}
\usepackage[all]{hypcap}

\usepackage{siunitx}  

\newcommand{\videolink}[2]{
    \href{#1}{
        {\small\faVideo} \textcolor{black}{\underline{#2}}
    }
}

\begin{document}

\onehalfspacing
\frenchspacing


\newcommand{\thesisTitle}{ Transcranial FUS Therapy and Monitoring \\using Nonlinear Acoustics}
\newcommand{\yourName}{Pradosh Pritam Dash}
\newcommand{\yourSchool}{Mechanical Engineering}
\newcommand{\yourMonth}{May}
\newcommand{\yourYear}{2026}


\begin{titlepage}
\begin{center}

\begin{singlespacing}

{\Large\textbf{{\thesisTitle}}}\\
\vspace{10\baselineskip}
{\footnotesize {A Dissertation}}\\
{\footnotesize {Presented to}}\\
{\footnotesize {The Academic Faculty}}\\
\vspace{2\baselineskip}
{\footnotesize {By}}\\
\vspace{2\baselineskip}
\large\yourName\\
\vspace{2\baselineskip}
{\footnotesize {In Partial Fulfillment}}\\
{\footnotesize {of the Requirements for the Degree}}\\
{\footnotesize {Doctor of Philosophy in the}}\\
{\footnotesize {School of \yourSchool}}\\
\vspace{2\baselineskip}
{\large Georgia Institute of Technology}\\
\vspace{\baselineskip}
{\large \yourMonth{} \yourYear{}}
\vfill
\normalsize{{Copyright \textcopyright{} \yourName{} \yourYear{}}}
\vspace{\baselineskip}

\end{singlespacing}

\end{center}
\end{titlepage}









\currentpdfbookmark{Title Page}{titlePage}

\newcommand{\committeeMemberOne}{Prof.~Costas Arvanitis, Advisor}
\newcommand{\committeeMemberOneDepartment}{School of Mechanical Engineering \& \\ Dept. of Biomedical Engineering }
\newcommand{\committeeMemberOneAffiliation}{Georgia Tech \& Emory University}

\newcommand{\committeeMemberTwo}{Prof.~F.~Levent Degertekin}
\newcommand{\committeeMemberTwoDepartment}{School of Mechanical Engineering}
\newcommand{\committeeMemberTwoAffiliation}{Georgia Tech}

\newcommand{\committeeMemberThree}{Prof.~Karim Sabra}
\newcommand{\committeeMemberThreeDepartment}{School of Mechanical Engineering}
\newcommand{\committeeMemberThreeAffiliation}{Georgia Tech}

\newcommand{\committeeMemberFour}{Prof.~Levi Wood}
\newcommand{\committeeMemberFourDepartment}{School of Mechanical Engineering}
\newcommand{\committeeMemberFourAffiliation}{Georgia Tech}

\newcommand{\committeeMemberFive}{Prof.~Liang Han}
\newcommand{\committeeMemberFiveDepartment}{School of Biological Sciences}
\newcommand{\committeeMemberFiveAffiliation}{Georgia Tech}

\newcommand{\approvalDay}{17}
\newcommand{\approvalMonth}{March}
\newcommand{\approvalYear}{2026}

\begin{titlepage}
\begin{singlespacing}
\begin{center}

\textbf{\textsc{\thesisTitle}}\\
\vspace{10\baselineskip}

\end{center}
\vfill

\ifdefined\committeeMemberFour

Approved by:
\vspace{2\baselineskip}

\begin{minipage}[b]{0.4\textwidth}

    \committeeMemberOne\\
    \committeeMemberOneDepartment\\
    \textit{\committeeMemberOneAffiliation}\\

    \committeeMemberTwo\\
    \committeeMemberTwoDepartment\\
    \textit{\committeeMemberTwoAffiliation}\\

    \committeeMemberThree\\
    \committeeMemberThreeDepartment\\
    \textit{\committeeMemberThreeAffiliation}\\

\end{minipage}
\hspace{0.1\textwidth}
\begin{minipage}[b]{0.4\textwidth}

	\committeeMemberFour\\
	\committeeMemberFourDepartment\\
	\textit{\committeeMemberFourAffiliation}\\

	\ifdefined\committeeMemberSix

	  \committeeMemberFive\\
	  \committeeMemberFiveDepartment\\
	  \textit{\committeeMemberFiveAffiliation}\\

	  \committeeMemberSix\\
	  \committeeMemberSixDepartment\\
	  \textit{\committeeMemberSixAffiliation}\\

	\else

	  \committeeMemberFive\\
	  \committeeMemberFiveDepartment\\
	  \textit{\committeeMemberFiveAffiliation}\\

	\fi

\end{minipage}\\

\begin{minipage}[b]{0.4\textwidth}

  ~

\end{minipage}
\hspace{0.1\textwidth}
\begin{minipage}[b]{0.4\textwidth}
    Date Approved: \approvalMonth{} \approvalDay, \approvalYear
    \vspace{5\baselineskip}
\end{minipage}\\

\else

\hspace{0.6\textwidth}
\begin{minipage}[b]{0.4\textwidth}

	Approved by:
	\vspace{2\baselineskip}

	\committeeMemberOne\\
	\committeeMemberOneDepartment\\
	\textit{\committeeMemberOneAffiliation}\\

	\committeeMemberTwo\\
	\committeeMemberTwoDepartment\\
	\textit{\committeeMemberTwoAffiliation}\\

	\committeeMemberThree\\
	\committeeMemberThreeDepartment\\
	\textit{\committeeMemberThreeAffiliation}\\

	\vspace{2\baselineskip}

	Date Approved: \approvalMonth{} \approvalDay, \approvalYear
	\vspace{\baselineskip}

\end{minipage}

\fi

\end{singlespacing}
\end{titlepage}

\newcommand{\yourDedication}{
    \pgfornament[width=5cm]{84} \\
    \vspace{1.5cm}

{\Large \textit{To human endeavor\\  \\ which knows no bound.}} \\

    \vspace{1.5cm}
    \pgfornament[width=5cm]{88} \\
}

\begin{titlepage}
\begin{center}


\end{center}
\end{titlepage}

\pagenumbering{roman}
\addcontentsline{toc}{chapter}{Acknowledgments}

\setcounter{page}{5}

\clearpage

\begin{tikzpicture}[remember picture, overlay]

  \def\bindoffset{0.75cm}

  \def\topmargin{-2.5cm}
  \def\botmargin{3.0cm}
  \def\sidemargin{2.5cm}
  \def\ornsize{2cm}
  \colorlet{orncolor}{gray!30}

  \node[anchor=north west, text=orncolor] at
    ([xshift=\sidemargin+\bindoffset, yshift=\topmargin]current page.north west)
    {\pgfornament[width=\ornsize]{61}};

  \node[anchor=north east, text=orncolor] at
    ([xshift=-\sidemargin+\bindoffset, yshift=\topmargin]current page.north east)
    {\pgfornament[width=\ornsize, symmetry=v]{61}};

  \node[anchor=south west, text=orncolor] at
    ([xshift=\sidemargin+\bindoffset, yshift=\botmargin]current page.south west)
    {\pgfornament[width=\ornsize, symmetry=h]{61}};

  \node[anchor=south east, text=orncolor] at
    ([xshift=-\sidemargin+\bindoffset, yshift=\botmargin]current page.south east)
    {\pgfornament[width=\ornsize, symmetry=c]{61}};

\end{tikzpicture}

\vspace*{2cm}
\begin{center}
  {\LARGE Acknowledgments}\\
  \vspace{0.8cm}
\end{center}

My Ph.D. journey has rarely been a solitary endeavor, and I am deeply grateful and indebted to the many individuals who have been a blessing to me along the way. First and foremost, I would like to express my sincere gratitude to my advisor, Prof.\ Costas D.\ Arvanitis, for his invaluable guidance and support throughout my time in the Ultrasound Biophysics and Bioengineering Laboratory. His mentorship and vision have profoundly shaped my approach to research and life. I also express my gratitude to the members of my doctoral committee: Prof. Karim Sabra, Prof. Liang Han, Prof. Levi Wood, and Prof. F. Levent Degertekin. Their insightful feedback and rigorous evaluations shaped this work in ways I did not anticipate.

I owe a special debt of gratitude to Dr.\ Scott Schoen Jr., whose research I had the privilege to build upon and extend, and whose mentorship has been invaluable. I am also thankful to my colleagues and collaborators for their technical assistance and camaraderie. The countless discussions we shared were instrumental in refining my work, navigating the complexities of this research, and managing the stress of a PhD.

My appreciation also extends to my earlier mentors who laid the foundational stones for my academic journey: Profs.\ Ricardo Burdisso and Pablo Tarazaga at Virginia Tech, and Prof.\ Subrata Panda at NIT Rourkela, whose early guidance and belief in my potential set me on this path.

Finally, I want to thank my family and friends for their unwavering support and love. To spell out everything I owe to you will be pure hubris. Your presence has been a constant source of strength, and this milestone would not have been possible without you all.

\begin{center}
    \vspace{1.0cm}
    \pgfornament[width=5cm]{88} \\
\end{center}
\clearpage

{\singlespacing
\tableofcontents
}

\currentpdfbookmark{Table of Contents}{TOC}

\clearpage

\addcontentsline{toc}{chapter}{List of Tables}
\begin{singlespace}
	\setlength\cftbeforetabskip{0.5\baselineskip}
	\listoftables
  \label{sec:ListOfTables}
\end{singlespace}

\clearpage

\addcontentsline{toc}{chapter}{List of Figures}
\begin{singlespace}
\setlength\cftbeforefigskip{0.5\baselineskip}
\listoffigures
\end{singlespace}

\clearpage

\chapter*{List of Symbols}
\addcontentsline{toc}{chapter}{List of Symbols}

\renewcommand{\arraystretch}{0.9}

\noindent

\begin{longtable}{@{} p{0.11\textwidth} p{0.35\textwidth} @{\hspace{0.04\textwidth}} p{0.11\textwidth} p{0.35\textwidth} @{}}
\toprule
\textbf{Symbol} & \textbf{Definition} & \textbf{Symbol} & \textbf{Definition} \\
\midrule
\endfirsthead

\multicolumn{4}{@{}l}{{\bfseries \tablename\ \thetable{} -- continued from previous page}} \\
\toprule
\textbf{Symbol} & \textbf{Definition} & \textbf{Symbol} & \textbf{Definition} \\
\midrule
\endhead

\midrule
\multicolumn{4}{r}{{Continued on next page}} \\
\endfoot

\bottomrule
\endlastfoot

$AR$ &  pixel geometric aspect ratio ($L_{2\pi}/\Lambda$) &
$P_{sim}, P_{fwd}$ & Simulated high-res vs. forward-sampled fields \\

$d_r$ & Vector drag coefficient for dynamic ARF &
$p_{\Delta f}$ & Parametric array difference-frequency pressure \\

$E_{trapped}$ & Integrated trapped acoustic energy in skull &
$\mathcal{R}$ & Regularization penalty for topology smoothing \\

$F^{rad}$ & Time-averaged Acoustic Radiation Force (ARF) &
$S_{NL}$ & Virtual volumetric nonlinear source density \\

$f_1, f_2$ & Biosphere compressibility and density contrast &
$t^{(n)}, t_{max}$ & Computed hologram thickness and upper bound \\

$\mathcal{H}$ & HASA forward wave propagation operator &
$u^{(n)}$ & Unconstrained thickness optimization parameter \\

$\Delta I$ & Central pixel intensity drop (ARF tracking) &
$w_k$ & Kinetic energy density (elastodynamic shear) \\

$k_e$ & Effective stiffness constant of hydrogel matrix &
$w_p$ & Potential energy density (volumetric compression) \\

$L_{2\pi}$ & Modulation depth for a full $2\pi$ phase shift &
$\alpha_T$ & Amplitude transmission coefficient \\

$\mathcal{L}^{(n)}$ & HASA-ADAM topology optimization loss &
$\gamma$ & Speed of sound mismatch ratio \\

$m, \eta$ & Hydrophone linear and nonlinear sensitivities &
$\theta_c$ & Critical angle for elastodynamic shear conversion \\

$\mu$ & Squared refractive index ratio ($c_0^2/c^2$) &
$\Lambda$ & HASA spatial variation term \textit{or} Lateral pitch \\

$P_{primary}$ & Envelope of squared primary pressure &
$\phi_{gas}$ & Trapped gas volume fraction (pneumocephalus) \\

\end{longtable}

\clearpage

\clearpage
\begin{centering}
\textbf{SUMMARY}\\
\vspace{\baselineskip}
\end{centering}

Focused ultrasound (FUS) offers a promising, non-invasive method for modulating neural activity and delivering therapies deep within the brain with immense clinical potential. However, progress in developing transcranial ultrasound (TUS) for clinical applications has been hindered by several factors. The complexity of the human skull causes focal aberrations and attenuation, thereby presenting a major obstacle to the precise targeting of ultrasound waves. Although phased arrays can correct for these aberrations, their high cost and continuous reliance on magnetic resonance imaging (MRI) pose significant obstacles for widespread academic research and clinical translation. To address these challenges, this thesis proposes an innovative framework for the design, registration, and clinical application of acoustic holograms. First, we introduce a novel \textsl{frequency-domain topology optimization} method that overcomes the breakdown of traditional phase-only designs in the megahertz regime by accounting for volumetric wave-propagation effects, thereby achieving high-fidelity focusing. Second, we present a \textsl{non-invasive registration strategy that utilizes the nonlinear parametric array (PA) effect} to enable precise lens alignment without requiring any imaging modalities, such as MRI. Finally, we demonstrate the utility of this nonlinear parametric array (PA) effect as a tool for \textsl{monitoring ventricular dilation as a non-invasive proxy for intracranial pressure changes in hydrocephalus}. Collectively, these developments provide a path toward accessible, high-precision transcranial ultrasound systems for research and clinical use. In addition, we demonstrate a novel platform for \textsl{in vitro focused ultrasound neuromodulation} that leverages acoustics to advance therapeutic discovery.

\pagenumbering{gobble}

\clearpage
\pagenumbering{arabic}
\setcounter{page}{1}

\chapter{Introduction and Background}
\label{chap:Introduction}

\section{The Evolution of Therapeutic Ultrasound}

Ultrasound has evolved from a diagnostic novelty to a nearly ubiquitous tool in the medical ultrasound field \cite{szabo_diagnostic_2013} in the past five decades. Recent therapeutic advances extended the field into high-intensity regimes, with thermal ablation, cavitation, and shock wave-assisted therapies each entering clinical use \cite{miller_overview_2012,chaussy_extracorporeal_1984,skolarikos_extracorporeal_2006}.

Adoption in oncology and urology is now broad \cite{bachu2021high,chaussy2005technology,izadifar2020introduction}, yet CNS applications remained largely absent for decades. The obstacle is the complex geometry and heterogeneous microstructure of the human skull, which acts as an acoustic barrier. The skull attenuates energy and induces phase aberrations in the transmitted ultrasound fields, effects that worsen at the higher frequencies needed for precise spatial targeting.

Phased arrays, combined with Computed Tomography (CT) and magnetic resonance (MR) guidance \cite{cline1992mr,tanter1998focusing,hynynen1998demonstration,pernot2003high,aubry2003experimental}, have largely resolved this barrier. Phased arrays consist of hundreds of transducer elements, each driven with a pre-calculated phase and amplitude to correct for skull-related aberrations. Per-element control enabled successful pilot studies in thermal ablation \cite{mcdannold_transcranial_2010,jeanmonod_transcranial_2012,elias_pilot_2013,jung_bilateral_2015,fasano_mri-guided_2017} and contrast agent-enhanced drug delivery \cite{carpentier_clinical_2016,abrahao_first--human_2019,lipsman2018blood}.

Phased arrays carry drawbacks that limit access. They are prohibitively expensive, and element count is constrained by the packaging of the driving electronics, restricting the spatial resolution and wavefront complexity they can generate \cite{jolesz2014intraoperative,hertzberg2010ultrasound}. The reduced degrees of freedom narrow the treatable envelope: peripheral brain regions, the skull base, and irregularly shaped lesions are difficult to reach without depositing heat in adjacent healthy tissue or bone. The demand for larger transcranial treatment volumes and the need for conformal, high-resolution fields have outpaced what current phased-array-based solutions can deliver.

\section{Acoustic Holography}

Acoustic holography directly addresses the aforementioned limitations of phased arrays. Inspired by optical holography, this method encodes the desired acoustic wavefronts into a physical holographic lens or a metasurface \cite{melde2016holograms}.

A 3D-printed holographic lens modulates a single transducer's output to produce a 2D phase profile that generates targeted focusing \cite{shen2014anisotropic,maimbourg20183d,ferri2019enhanced} and complex volumetric wave fields with diffraction-limited resolution \cite{melde2016holograms}. The reconstruction degrees of freedom exceed those of commercial phased-array sources by two orders of magnitude, at a fraction of the cost.

\subsection{Historical Context and Fabrication}

Acoustic holography was first demonstrated in the 1990s. Recent gains in computational power and additive manufacturing have driven the revival of additive manufacturing in medical ultrasound. 3D printing now produces intricate, patient-specific holograms at low cost \cite{ferri2019enhanced}, which makes the hologram-based approaches more accessible. By calculating the required phase map using time-reversal or similar algorithms, a holographic lens can be tailored to an individual patient's skull and thereby correct specific aberrations without the need for complex active electronics associated with phased arrays \cite {brown2019}.

\subsection{Biomedical Applications}

Customizable acoustic fields have extended holography into several areas of noninvasive brain therapy. \textbf{Blood-Brain Barrier (BBB) Opening:} Holographic lenses achieved simultaneous bilateral BBB opening in non-human primates \cite{jimenez2023primate} and mouse models \cite{jimenez2021mouse} using 3D-printed holograms. \textbf{High-Intensity Focused Ultrasound (HIFU):} Holograms shape energy deposition to produce thermal holograms that match tumor geometry or target multiple locations simultaneously \cite{andres2022numerical,glickstein2024,he2022multitarget}. \textbf{Imaging:} Holograms tailor transmit fields to generate Bessel beams with extended depth of field \cite{jimenez2019generating} or enable compressive 3D imaging using single sensors \cite{kruizinga2017compressive}.

\section{Design Challenges in the High-Frequency Regime}
\label{sec:DesignChallenges}

Generating a target acoustic field requires careful specification of the hologram's phase and amplitude profiles. Available strategies include the \textbf{Iterative Angular Spectrum Approach (IASA)} \cite{melde2016holograms} and \textbf{Direct Search} methods \cite{Li2021Comparison}. Deep learning frameworks \cite{lee2022} and automatic differentiation (Diff-PAT) \cite{fushimi2021acoustic} have since accelerated the design of complex patterns.

High-fidelity holograms for transcranial use remain difficult to produce. Sub-millimeter accuracy for neurological interventions requires operation near $\SI{1}{MHz}$. Most rapid design methods rely on the \textbf{Thin-Element Approximation (TEA)} or \textbf{Thin-Film Approximation (TFA)}, which treats the lens as a pure phase screen, ignoring its physical thickness. At MHz frequencies and below, lens features become comparable to the wavelength and the lens is acoustically thick ($L \gg \lambda$), invalidating the phase-screen model. The result is \textbf{refractive walk-off}, where energy physically migrates to neighboring pixels within the lens, and volumetric diffraction effects that phase-only methods cannot capture. Full-wave time-domain simulations \cite{maimbourg20183d,ferri2019enhanced,jimenez2019holograms} model these effects correctly but are too slow for clinical use.

\textbf{Thesis Contribution (Chapter \ref{ch:hologram_design}):} \cref{ch:hologram_design} proposes HASA-ADAM, a frequency-domain lens topology optimization approach. The method accounts for volumetric wave propagation and medium heterogeneity. It scales to clinical dimensions without the computational cost of time-domain solvers.

\section{The Registration Bottleneck}
\label{sec:RegistrationChallenges}

The second major challenge is \textbf{registration} of the hologram with the patient's anatomy. A phase plate is a passive lens designed to accommodate the skull's geometry in a particular orientation. The heterogeneous speed-of-sound map for which the plate is optimized is tied directly to the skull's position and orientation relative to the transducer; shift one and the correction fails. Precise skull registration is, therefore, a prerequisite for correct operation. It has been shown that HASA fails to correct for skull aberrations when misregistration exceeds 10 wavelengths for point targets at $\SI{1}{MHz}$ \cite{schoen2021experimental}. Accurate registration also governs thalamic targeting \cite{o2016registration} and the spatial fidelity of real-time MR thermometry monitoring \cite{de2007mr}.

MR-guided tFUS (MRgFUS) \cite{kyriakou2014review} is the current gold standard: accurate, but expensive and slow. Neuronavigation (NaviFUS) \cite{chen2020neuronavigation} and augmented-reality holography \cite{vandoormaal2019clinical} use optical tracking and cost far less, but achieve only millimeter-level accuracy. That falls short of the sub-millimeter precision high-frequency holograms demand. As shown in Table~\ref{tab:comparison}, tFUS neuronavigation sits between these options in precision, cost, and clinical accessibility.

\begin{table}[h]
\centering
\caption{Comparison of FUS Registration Modalities}
\label{tab:comparison}
\small
\begin{tabularx}{\textwidth}{@{}l l l X@{}}
\toprule
\textbf{Modality} & \textbf{Est. Cost} & \textbf{Precision} & \textbf{Primary Limitations} \\
\midrule
\textbf{MRgFUS} \cite{elias2016thalamotomy, lipsman2018blood} & $\sim$\$2.37M & $< 1.0$ mm & Monopolizes MRI; high capital cost; rigid frame pinning. \\
\textbf{Optical (NaviFUS)} \cite{wei2013neuronavigation} & $\sim$\$80k & 1.5--3.5 mm & Blind to real-time acoustic distortion; line-of-sight limits. \\
\textbf{AR (HoloLens)} \cite{vandoormaal2019clinical} & $\sim$\$3.5k & 4.4--7.2 mm & Holographic drift; clinically unacceptable error margins. \\
\bottomrule
\end{tabularx}
\end{table}

\textbf{Thesis Contribution (Chapter \ref{ch:hologram_registration}):} The gap between MRgFUS accuracy and optical-tracker cost motivates a third approach. \cref{ch:hologram_registration} proposes a registration scheme based on the \textbf{nonlinear parametric array effect}. A misaligned lens amplifies nonlinear wave interactions within the skull, producing a measurable acoustic feedback signal. The method uses that signal to achieve sub-millimeter alignment without MR guidance.

\vspace{-10pt}
\section{Nonlinear Acoustics and Diagnostic Monitoring}
\label{sec:MonitoringChallenges}
\vspace{-10pt}

The nonlinear wave interactions exploited for registration have uses beyond skull-to-lens alignment. At therapeutic intensities, linear propagation assumptions break down, leading to harmonic generation \cite{sallam2023nonlinear} and parametric mixing. These nonlinear phenomena are sensitive to the acoustic nonlinearity parameter ($\beta$) of the medium. Brain tissue ($\beta \approx 6.6$) and cerebrospinal fluid (CSF, $\beta \approx 5.2$) have distinct nonlinear properties \cite{duck2013physical}. Because of this difference, the cumulative nonlinear signal should reflect changes in the intracranial environment, changes in the relative volumes of tissue and fluid alter the integrated $\beta$ along the propagation path. Hydrocephalus, defined by the accumulation of CSF, currently requires invasive intracranial pressure (ICP) monitoring, which carries risks of infection and hemorrhage \cite{jiang2022invention}.

\textbf{Thesis Contribution (Chapter \ref{chap:ICPMonitoring}):} \cref{chap:ICPMonitoring} applies nonlinear acoustic feedback to diagnostics. The central question is whether the parametric array effect can detect changes in ventricular volume. A positive result would allow hydrocephalus progression and shunt function to be tracked non-invasively, removing the need for surgical ICP probes.

\section{Challenges in Ultrasound Neuromodulation}
\label{sec:NeuromodulationChallenges}

Focused ultrasound also offers a route to noninvasive neuromodulation. Transcranial electric stimulation (tES) and transcranial magnetic stimulation (TMS) are established tools, but both are constrained by the diffuse spread of the fields they produce, which limits spatial selectivity \cite{nitsche2008transcranial,walsh2000transcranial,faria2011finite,deng2013electric}. Focused ultrasound propagates mechanical waves deep into neuronal tissue with precise spatial targeting \cite{haar2010ultrasound}.

The biophysical mechanisms governing ultrasound-neuron interactions remain poorly understood. A major obstacle is a hard physical trade-off: precision and penetration depth pull in opposite directions. Sub-millimeter, cellular-level targeting requires frequencies above 10~MHz, which confines useful penetration to superficial tissue \cite{cadoni2023ectopic}. \textsl{In-vivo} work is complicated further by thermal accumulation, bulk fluid streaming, and off-target auditory artifacts, which together make isolating specific mechanotransduction events very difficult \cite{sato_ultrasonic_2017,guo_ultrasound_2018}. Controlled \textsl{in-vitro} platforms are needed to hold those variables fixed to facilitate therapeutic discovery.

\textbf{Thesis Contribution (Chapter \ref{chap:inVitroNeuroModulation}):} \cref{chap:inVitroNeuroModulation} describes the development and validation of an \textsl{in-vitro} focused ultrasound neuromodulation platform. The apparatus isolates acoustic radiation force from confounders like thermal effects and fluid streaming, which can lead to false-positive neuronal activation. The chapter also introduces contrast-enhanced mechanostimulation using biocompatible metallic microspheres as local stress concentrators. These enable subcellular mechanical stimulation at clinically relevant low frequencies needed for deep tissue targeting.
\vspace{-10pt}

\section{Thesis Outline}
\vspace{-10pt}
This thesis addresses the design, registration, and monitoring challenges that may enable practical application of acoustic holography in clinical settings. It also introduces a platform for \textsl{in-vitro} focused ultrasound neuromodulation. In \textbf{chapter \ref{ch:hologram_design},}the HASA-ADAM topology optimization framework accounts for volumetric wave propagation at MHz frequencies, where the thin-element approximation is inadequate due to unaccounted refraction and diffraction within the acoustically thick lens. In \textbf{chapter \ref{ch:hologram_registration},}
a registration scheme built on the nonlinear parametric array effect achieves sub-millimeter lens alignment without any external imaging modality such as MRI. This makes the hologram-based focus ultrasound therapy portable and accessible. In \textbf{Chapter \ref{chap:ICPMonitoring}, }
nonlinear parametric array (PA) effect is applied to diagnostic monitoring: ventricular volume changes associated with hydrocephalus are monitored non-invasively by measuring the change in the PA signal measured outside the skull cavity. Lastly, in \noindent\textbf{Chapter \ref{chap:inVitroNeuroModulation},}
an \textsl{in-vitro} neuromodulation platform is developed which implements biocompatible metallic microspheres as stress concentrators to deliver subcellular mechanical stimulation at low frequencies, isolating radiation-force effects from thermal and streaming confounders.

\textit{Note: An earlier version of portions of the work (\textbf{chapter 2-4}) presented in this thesis is available as a preprint on arXiv (https://arxiv.org/abs/2508.07103) \cite{Dash2025}.}

\chapter{Hologram Topology Design}
\label{ch:hologram_design}
\footnotetext{An earlier version of the work presented in this chapter is available as a preprint on arXiv (https://arxiv.org/abs/2508.07103).\cite{Dash2025}}
\section{Introduction}
Recent advancements in acoustic holography based on holographic lenses offer a promising pathway for designing simpler, more economical, and more flexible ultrasound systems for a range of applications, including contactless manufacturing~\cite{melde2023compact}, consumer electronics~\cite{hirayama2019volumetric}, non-destructive testing~\cite{xie2016acoustic}, imaging~\cite{kruizinga2017compressive}, and targeted neuro-interventions~\cite{maimbourg20183d,maimbourg20183d,jimenez2019holograms}, among    others~\cite{melde2016holograms,hu2022airy,andres2023holographic}. These acoustic holograms, also known as phase plates, encode spatial phase patterns onto passive lenses, effectively transforming a single-element transducer into a device capable of generating complex volumetric pressure fields~\cite{jimenez2018adaptive,melde2016holograms}. This works because the effective pixel size sits well below the
acoustic wavelength~\cite{melde2016holograms}. Crucially, for large apertures (i.e., several cm$^2$) this is equivalent to dense phased arrays (e.g., $10^4 - 10^5$ elements) that are unrealizable, making possible field distributions that current phased-array technology cannot produce. Moreover, recent implementations that use {numerical methods for} wave propagation in heterogeneous media, such as the human skull, can account for skull-induced aberrations to produce focal spots~\cite{maimbourg20183d,daniel2024multifrequency} or pressure fields to concurrently target different brain regions~\cite{jimenez2019holograms,jimenez2024feasibility,he2022multitarget,kook2023multifocal,yao2025acoustic}.

Although this approach can achieve the desired phase distribution, converting the optimized phase pattern into a lens topology by scaling it with the relative wavenumber implies that the hologram is a thin element that alters only the phase, neglecting amplitude changes and wave propagation effects within the lens.

Frequencies ($\approx$\,1\, MHz) or below are needed for deep-tissue penetration. The lens feature size in this range approaches the acoustic wavelength ($\lambda < 1.5$\, mm), which invalidates the thin-element approximation and introduces significant thickness-dependent amplitude and phase errors. In this acoustically thick regime ($L \gg \lambda$), acoustic energy migrates laterally into neighboring pixels.

This causes volumetric crosstalk between the lens pixels, which phase-only models ignore. Lower frequencies penetrate the skull more easily; however, an operation near 1 MHz offers a workable compromise between spatial resolution and penetration depth. This frequency range is actively being explored for high-precision, non-thermal transcranial ultrasound (TUS) applications, such as targeted neuromodulation~\cite{meng2021applications} and blood-brain barrier opening~\cite{schoen2022towards}. Spatial precision is essential for these applications, but phase-based lens design methods fail. Off-target hotspots are an additional concern at these frequencies. Although time-domain simulations can accurately model heterogeneities, they often take hours to run~\cite{jimenez2019holograms}, which is too slow for iterating over the large apertures required in TUS~\cite{choi2024neuronavigation}. Frequency-domain methods are faster. The standard angular spectrum approach (ASA) is fast but ignores wave-speed heterogeneities~\cite{melde2016holograms}. Other frequency-domain methods, such as the Modified Mixed Domain Method (MMDM)~\cite{gu2020modified}, can handle strong heterogeneities and reflections, but folding them into fast iterative optimizers is non-trivial. Automatic differentiation has been used to accelerate hologram design~\cite{fushimi2021acoustic}, and gradient-based optimization has been applied to related problems~\cite{melde2023compact,sallam2024gradient}. A fast 3D-printed lens design that is simultaneously free of thickness-dependent errors and accounts for heterogeneous-medium aberrations, however, remains unsolved~\cite{jimenez2024feasibility}.

We introduce a unified framework for high-fidelity acoustic holography. This framework uses the heterogeneous angular spectrum approach (HASA), which is a fast spectral method for wave propagation in complex media~\cite{schoen2020heterogeneous}, and its ability to incorporate in-plane varying speed-of-sound maps and support differentiable optimization based on the ADAM iterative optimizer to design acoustic holograms with complex topologies. Unlike phase-only approaches, HASA-ADAM directly optimizes the lens topology and incorporates wave-propagation effects, as well as amplitude and phase changes within the lens material.

\section{Methods}

\subsection{Preparing the Transcranial Medium}
 We must have a realistic heterogeneous three-dimensional (3D) acoustic property map of the skull to perform patient-specific transcranial focused ultrasound (TUS) hologram optimization. The 3D geometry was acquired via a micro-CT scan of the skull secured in the fixture. This makes DICOM volumes co-registered directly with the transducer coordinate space. The raw 3D imaging data were interpolated onto a voxel resolution of 250 $ \ mu$m ($\sim\lambda/6$) as per the 1 MHz sampling requirement. The CT voxel intensities were normalized against the water and air baselines to obtain approximate Hounsfield units ($H$). These values were then used to calculate the volumetric porosity fraction of the bone, defined as $\Psi = 1 - H/1000$. A porosity-dependent semi-empirical mixture model~\cite{pichardo_multi-frequency_2011} was then used to convert these values to 3D acoustic properties. Specifically, the density ($\rho$) and longitudinal speed of sound ($c$) for each voxel were linearly interpolated between the baseline properties of pure water ($c = 1480$\,m/s, $\rho = 1000$\,kg/m$^3$) and the upper limits of the dense cortical bone ($c_{\max} = 2500$\,m/s, $\rho_{\max} = 2200$\,kg/m$^3$), proportional to the solid bone fraction ($1 - \Psi$).

\subsection{Holographic Lens Design}
\subsubsection{Heterogeneous Wave Propagation with ASA}
The first step in our design process was to model wave propagation through skull heterogeneity. For a time-harmonic pressure field $\tilde{\bm{p}}(\bm{r})e^{-i\omega t}$, where $\omega$ is the angular frequency, the angular spectrum $P$ is given by its 2D spatial Fourier transform
\begin{equation}
P(k_x, k_y, z) = \mathcal{F}_k[\tilde{p}(x, y, z)],
\end{equation}
For heterogeneous media where the spatial variation of sound speed $c(\bm{r})$ is less compared to the wavelength, the ordinary differential equation for the angular spectrum $P$ becomes
\begin{equation}
P_z + k_z^2P = \Lambda * P,
\end{equation}
Here, $\Lambda = \mathcal{F}_k[k_0^2(1 - \mu)]$, $k_0 = \omega/c_0$, $\mu = c_0^2/c^2$, $c_0$ is a reference (average) sound speed, and $*$ indicates a two-dimensional convolution over the component wavenumbers $k_x$ and $k_y$. For our design, the skull and tissue (speed-of-sound and density maps) were obtained from micro-CT scan data of a human skull~\cite{aubry2003experimental,arvanitis2014transcranial} (original resolution 95$\mu$m binned to 150$\mu$m, which amounts to 10 points per wavelength for $f_0 = 1$ MHz and considering equilibrium sound speed $c_0 = 1480$ m/s in water). {HASA is specifically designed to handle these spatially varying properties.} This wave propagation model captures the refraction and transmission of waves through the skull. In the current implementation, absorption was not considered in the calculations of the acoustic intensity.

An implicit solution of Eq. 2 may be obtained with a Green's function technique~\cite{morse1946methods}, and numerical approximation allows computation of $P$ at arbitrary $z$ via
\begin{equation}
P^{n+1} \approx P^n e^{ik_z\Delta z} + \frac{e^{ik_z\Delta z}}{2ik_z}(P^n * \Lambda) \times \Delta z,
\end{equation}
where $P^n = P(k_x, k_y, n\Delta z)$.
Provided that the marching step size $\Delta z$ is much smaller than the wavelength ($\lambda$), Eq. 3 enables the calculation of the field in the heterogeneous medium and ensures the computational stability. We chose $\Delta z = 150\mu$m. To prevent spatial aliasing and circular convolution errors inherent to the discrete Fourier transform during Angular Spectrum propagation, the computational grids were zero-padded by $N/2$ on all boundaries during the forward pass. Using the above wave propagator, the pressure distribution at the target plane $P_t$ can be efficiently computed from the initial pressure field $P_0$.
\textcolor{black}{
\subsubsection{Optimization Parameters}
To address the physics described above, we implemented a topology optimization framework (HASA-ADAM). To adapt the optimization for different frequencies for a fixed aperture ($60\,\text{mm}$) and focal depth ($45\,\text{mm}$), spatial sampling relative to the wavelength ($\lambda$) was kept constant: the voxel sizes ($ dx$ and $ dz$) were maintained at $\lambda/6$ to prevent numerical aliasing. The maximum lens thickness ($t_{max}$) determines its ability to induce a full $ 2\pi$ phase shift. Because the phase delay is proportional to the frequency, higher frequencies require thinner lenses to achieve the same phase-wrapping effect.}

\begin{table}[htbp]
\centering
\caption{Summary of Optimization Parameters}
\label{tab:params}
\resizebox{\columnwidth}{!}{
\begin{tabular}{lcccc}
\toprule
\textbf{Frequency} & \textbf{Wavelength ($\lambda$)} & \textbf{Resolution ($dx$)} & \textbf{Grid Size ($N$)} & \textbf{Max Thickness} \\
\midrule
0.5 MHz & 3.00 mm & 0.500 mm & 256 & 10.0 mm \\
1.0 MHz & 1.50 mm & 0.250 mm & 240 & 5.0 mm \\
2.0 MHz & 0.75 mm & 0.125 mm & 512 & 2.5 mm \\
\bottomrule
\end{tabular}
}
\end{table}

\textcolor{black}{
To address the stochastic nature of the ADAM optimizer and ensure exact reproducibility across studies, the hyperparameters and mathematical formulations utilized for both the topology and phase optimization pipelines to ensure convergence stability are detailed in Table~\ref{tab:hyperparams}.
}

\textcolor{black}{
\begin{table}[htbp]
    \centering
    \caption{\textcolor{black}{HASA-ADAM Hyperparameters for Topology Optimization}}
    \label{tab:hyperparams}
    \begin{tabular}{lc}
        \toprule
        \textbf{Parameter} & \textbf{Topology Optimization Value} \\
        \midrule
        \textbf{Learning Rate ($\eta$)} & 0.01 \\
        \textbf{Max Iterations ($N_{max}$)} & 500 \\
        \textbf{Optimizer} & ADAM \\
        \textbf{Initialization ($T_0$)} & 3.6 mm \\
        \textbf{Loss Function Metric} & L1 Intensity Loss\\
        \textbf{Variable Constraint} & $\min(\text{softplus}(u), 5\text{ mm})$ \\
        \textbf{Spatial Padding} & $N/2$ on all boundaries \\
        \bottomrule
    \end{tabular}
\end{table}
}
\subsubsection*{Approach 1 - Topology Optimization}
Topology optimization directly optimizes the lens thickness distribution to achieve a prescribed (known as reference) intensity profile at the target plane. This method employs the HASA algorithm to model skull heterogeneity and acoustic propagation through complex media. The algorithm is initialized with a uniform thickness distribution $\bm{u}^{(0)}$ and iteratively refined to minimize the loss function. At each iteration $n$, the thickness is computed as
\begin{equation}
t^{(n)} = \min\{\text{softplus}(u^{(n)}), t_{\max}\},
\end{equation}
where the softplus function ensures non-negative thickness values, and $t_{\max} = 5$mm (in the case of 1 MHz)imposes an upper bound on the hologram thickness. The complex pressure field $P^{(n)}$ is then calculated using the HASA forward model $\mathcal{H}(P_0, t^{(n)})$, where $P_0$ represents the initial pressure distribution of the transducer. The loss function quantifies the mismatch between the computed intensity $I^{(n)} = |P^{(n)}|^2$ and the target intensity $I_t$ at each spatial location $(x,y)$:
\begin{equation}
\mathcal{L}^{(n)} = \sum_{x,y} |I^{(n)}(x, y) - I_t(x, y)| + \lambda\mathcal{R}(t^{(n)}),
\end{equation}
where $\lambda\mathcal{R}(t^{(n)})$ is a regularization term that encourages smooth thickness variations, with $\lambda = 0.01$ controlling the strength of regularization. The optimization process utilizes the ADAM optimizer with a learning rate of $\eta = 0.01$ to update the thickness parameter as follows:\begin{equation}
u^{(n+1)} = u^{(n)} - \eta\nabla_u\mathcal{L}^{(n)},
\end{equation}
We used automatic differentiation with TensorFlow to calculate the gradients. The optimization iterations continue until the loss either converges below a threshold of $\epsilon = 0.001$ or the maximum number of iterations $N_{\max}$ is reached. We have kept $N_{\max}$ at 500 for single-point focusing and at about 2000 for complex 2D distributions. The steps for optimizing hologram thickness (500 iterations for single-point focusing) are detailed in Table~\ref{tab:algorithm}.

\subsubsection*{Approach 2 - Phase Optimization}
\begin{figure}
\centering
\includegraphics[width=0.8\linewidth]{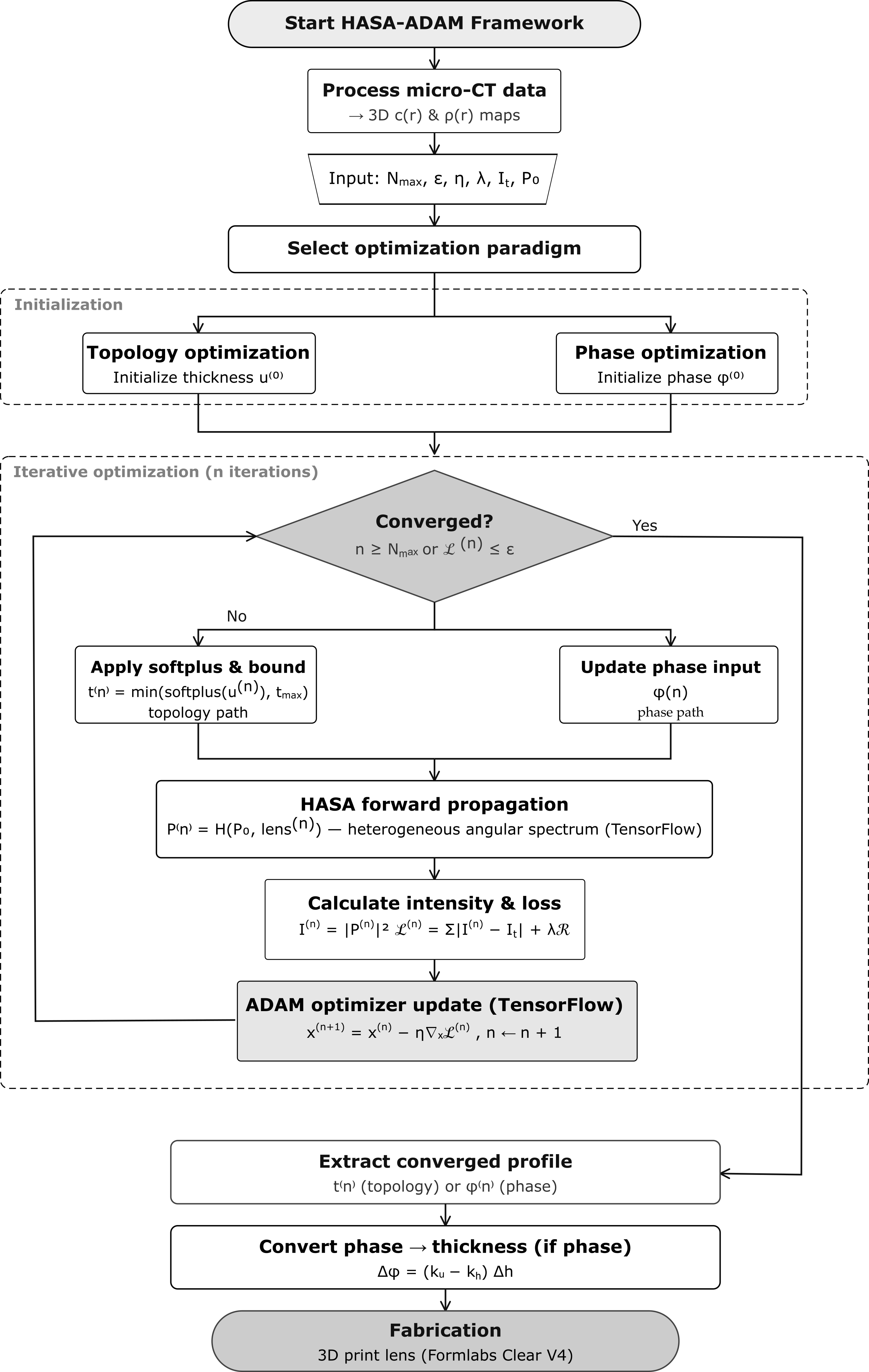}
\caption[Process Flow Chart for HASA-ADAM Phase and Topology Optimization]{Process Flow Chart for HASA-ADAM Phase and Topology Optimization. We start the process by extracting the acoustic properties from micro-CT scan data, which are then fed to the HASA-ADAM optimizer for phase or topology optimization. Upon convergence, a phase or topology map is generated. The topology map can be printed as is, whereas the phase map needs to be converted into a thickness map.}
\label{fig:HASA_ADAM_flowChart}
\end{figure}

The phase optimization approach \cite{Dash2023Heterogenous,arvanitis2025trans} iteratively refines the phase distribution $\phi$ at the hologram surface while maintaining a fixed amplitude profile. The algorithm begins with an initial uniform phase distribution $\phi^{(0)}$ and a zero-thickness assumption. At each iteration, the algorithm computes the pressure field using
\begin{equation}
P_t^{(n)} = \mathcal{H}(P_0, \phi^{(n)}),
\end{equation}
Here, $\mathcal{H}$ represents the HASA propagation operator. The intensity at the target plane is calculated as $I_t^{(n)} = |P_t^{(n)}|^2$. The loss function for phase optimization is expressed as follows:
\begin{equation}
\mathcal{L}^{(n)} = \sum_{x,y} |I_t^{(n)} - I_t| + \lambda\mathcal{R}(\phi^{(n)}).
\end{equation}
where $I_t$ is the target intensity distribution, and $\mathcal{R}(\phi^{(n)})$ regularizes the phase profile to encourage smoothness. The phase was then updated using the gradient descent with the ADAM optimizer. Following convergence, the optimized phase profile was converted into a thickness map. Furthermore, the optimization routine considers the impact of hologram thickness on amplitude transmission. This was achieved by converting the refined phase profile at the transducer surface into a thickness map of the adhesive layer. This conversion relies on the relationship between the phase change and thickness variation, represented by
\begin{equation}
\Delta\phi(x,y) = (k_w - k_h)\Delta h(x,y),
\end{equation}
Here, $k_w$ and $k_h$ denote the wave numbers of the water and hologram material, respectively. The thickness is used to compute the transmission coefficient $\alpha_T$ and the complex amplitude at the hologram plane, following established expressions~\cite{melde2016holograms}. The hologram phase optimization steps are summarized in Table~\ref{tab:algorithm} Additionally, Figure~\ref{fig:HASA_ADAM_flowChart}describes the process of designing a hologram via phase or topology optimization using a process flow chart.

\begin{table}[htbp]
\begin{center}
\caption{Table 1: HASA-ADAM algorithm for a) topology optimization, b) phase optimization}
\label{tab:algorithm}
\begin{tabular}{|l|l|}
\hline
a. HASA-ADAM: Topology Optimization &  HASA-ADAM: Phase Optimization \\
\hline
\( \begin{aligned} & \text { Input: } N_{\max }, \epsilon, \eta, \lambda, u^{(0)}, I_{t} \\ & \text { Output: } t, \mathcal{L} \\ & \text { while } n<N_{\max } \text { and } \mathcal{L}^{(n)}>\epsilon \text { do } \\ & t^{(n)}=\min \left\{\operatorname{softplus}\left(u^{(n)}\right), t_{\max }\right\} \\ & P^{(n)}=\mathcal{H}\left(P_{0}, t^{(n)}\right) \\ & I^{(n)}=\left|P^{(n)}\right|^{2} \\ & \mathcal{L}^{(n)}=\sum_{x, y}\left|I_{i}^{(n)}(x, y)-I_{t}(x, y)\right|+\lambda R\left(t^{(n)}\right) \\ & u^{(n+1)}=u^{(n)}-\eta \nabla_{u} \mathcal{L}^{(n)} \\ & n \leftarrow n+1 \\ & \text { end return } t^{(n)}, \mathcal{L}^{(n)} \end{aligned} \) & \( \begin{aligned} & \text { Input: } N_{\max }, \epsilon, \eta, \lambda, P_{0}, P_{t} \\ & \text { Output: } \phi, \mathcal{L} \\ & \text { while } n<N_{\max } \text { and } \mathcal{L}^{(n)}>\epsilon \text { do } \\ & P_{i}^{(n)} \leftarrow \mathcal{H}\left(P_{0}, \phi^{(n)}\right) ; I_{i}^{(n)}=\left|P_{i}^{(n)}\right|^{2} \\ & \mathcal{L}^{(n)} \leftarrow \sum_{x, y}\left|I_{i}^{(n)}-I_{t}\right|+\lambda \mathcal{R}\left(\phi^{(n)}\right) ; \\ & \phi^{(n+1)} \leftarrow \phi^{(n)}-\eta \nabla_{\phi} \mathcal{L}^{(n)} ; \\ & P_{0}^{(n+1)} \leftarrow \alpha_{h} P_{0}^{(n)} ; n \leftarrow n+1 ; \\ & \text { end } \\ & \text { return } \phi^{(n)}, \mathcal{L}^{(n)} \end{aligned} \) \\
\hline
\end{tabular}
\end{center}
\end{table}

\subsubsection{Fabrication}
The final step in the design process was the fabrication of the holograms. For topology optimization (i.e., Approach 1), we 3D-printed the thickness mask without any additional processing. To do this for phase opEq..e., approach 2), we utilize equation (9) to convert the optimized phase map to a thickness map and 3D-print it. In our design, we used a clear white v4 resin from Formlabs (Somerville, MA). {Lenses were printed using a Formlabs Form 3B printer.} This resin has low attenuation values across the frequency range of interest and a higher speed of sound (with group velocity $c_g = 2591$ m/s and attenuation $\alpha_0 = 2.922$ dBMHz$^{-1.044}$ cm$^{-1}$), making it suitable for 3D printing acoustic holograms among the materials characterized by Bakaric et al.~\cite{bakaric2021measurement}.

\subsection{Hologram Design Validation}
\subsubsection{Trans-Skull Simulations}
Holograms were tested using 3D linear acoustics simulations in the open-source k-wave toolbox with GPU acceleration~\cite{treeby2010k}. The elastic effects were ignored because the angle of incidence on the skull bone in our study was below the critical angle in most cases. All simulations used a spatial grid size of $\Delta x = \Delta y = 250\,\mu$m and a CFL number below 0.3. A porosity-dependent semi-empirical relationship converted Hounsfield units $H$ from the micro-CT scan to the speed of sound $c$, density $\rho$, and attenuation $\alpha$. Spatially varying maps for these properties were used throughout. The maximum speed and density of solid bone were taken as $c_{\text{bone}} = 2500$\,m/s and $\rho_{\text{bone}} = 2000$\,kg/m$^3$. Attenuation followed a power law $\alpha = \alpha_0 \cdot f^{\beta}$ with $\beta = 1.2$.

\textcolor{black}{
To capture shear mode conversion and propagation at the lens interfaces, we ran full-wave 3D elastic simulations using the \texttt{pstdElastic3D} solver in k-Wave~\cite{treeby2010k}. Skull bone was isolated by a density threshold ($\rho > 1100$\,kg/m$^3$), with a shear wave speed of $c_s = 1400$\,m/s and shear attenuation $\alpha_s = 20$\,dB/cm assigned to those voxels. A concern was whether the lower MSE in the elastic model reflected higher attenuation, which can artificially reduce the error amplitude. To rule this out, we zeroed the shear attenuation ($\alpha_s = 0$) while keeping the slow shear velocity intact (Table~\ref{tab:kwave_params}). Sharp fluid-solid boundaries can trigger numerical instability in elastic simulations. We addressed this by smoothing the property matrices and holding the CFL number to 0.1.
}

\begin{table}[htbp]
    \centering
    \caption{\textcolor{black}{Acoustic Parameters for k-Wave Fluid and Elastic Simulations}}
    \label{tab:kwave_params}
    \textcolor{black}{
    \begin{tabularx}{\textwidth}{Xcc}
        \toprule
        \textbf{Parameter} & \textbf{Fluid Model} & \textbf{Elastic Model} \\
        \midrule
        Background Density ($\rho_0$) & $1000\,\text{kg/m}^3$ & $1000\,\text{kg/m}^3$ \\
        Density Threshold ($\rho_{bone}$) & $> 1100\,\text{kg/m}^3$ & $> 1100\,\text{kg/m}^3$ \\
        Compressional Speed ($c_p$) & Derived from CT scan & Derived from CT scan \\
        Shear Speed ($c_s$) & N/A & $1400\,\text{m/s}$ \\
        Fluid Attenuation ($\alpha_0$) & $0.0022\,\text{dB/cm/MHz}^y$ & $0.0022\,\text{dB/cm/MHz}^y$ \\
        Compressional Atten. ($\alpha_p$) & Derived from CT scan & Derived from CT scan \\
        Shear Attenuation ($\alpha_s$) & N/A & $10-20\,\text{dB/cm/MHz}^y$ \\
        Courant Number (CFL) & $0.3$ & $0.1$ \\
        \bottomrule
    \end{tabularx}
    }
\end{table}

\subsubsection{Trans-Skull Experiments}
Ex vivo trans-skull experiments were performed using a degassed deionized water tank. A 60 mm diameter piston transducer (Precision Acoustics, Dorchester, UK) was coupled to a 3D printed hologram lens. {The transducer was driven by a function generator (HP Agilent Keysight, 33511B)) amplified by a power amplifier (E\&I, Rochester, NY, USA).} The transducer and lens assembly were then attached to the parietal region of an overnight degassed (approximately 12h) skull cap (Skulls Unlimited, Oklahoma City, OK, USA). The focal field of the lens was scanned using a calibrated 2 mm needle hydrophone (Precision Acoustics, Dorchester, UK) mounted on a three-axis positioning system (Velmex, Bloomfield, NY, USA) and recorded using a digital oscilloscope (Pico Technologies, St Neots, UK).
\subsection{Forward Modeling of Hydrophone Aperture and Sampling Effects}

Blurring and pixelation differences between simulations and measurements were reproduced by a forward model of the physical acquisition chain. The model has two steps, following the physics of the measurement setup:

\subsubsection*{Finite Aperture Spatial Averaging \& Discrete Subsampling}
The first step models the continuous physical interaction between the acoustic field and the sensor face. The simulation field $P_{sim}(x,y)$, generated on a high-resolution grid ($250\,\mu\text{m}$), was convolved with a spatial kernel representing the active area of the needle hydrophone (active element diameter $D = 2.0$\,mm):
\begin{equation}
P_{avg}(x,y) = P_{sim}(x,y) * H_{disk}(D)
\end{equation}
Performing this on the full-resolution grid captures the spatial averaging over the sensor face.

The second step models the digitisation of the stepper-motor scan. Despite the fine simulation grid, the experimental raster scan had a $500\,\mu$m step. We therefore downsampled $P_{avg}(x,y)$ to the experimental scan grid:
\begin{equation}
P_{fwd}(i,j) = P_{avg}(x_i, y_j) \quad \text{for} \quad x,y \in \text{Grid}_{exp}
\end{equation}
The result is a synthetic measurement that reproduces both the aperture blur and the step-limited pixelation seen in the data.

\section{Results}

\subsection{Hyperparameter Study of the HASA-ADAM Topology Optimization}

\begin{figure}[htbp]
\centering
\includegraphics[width=1.0\textwidth]{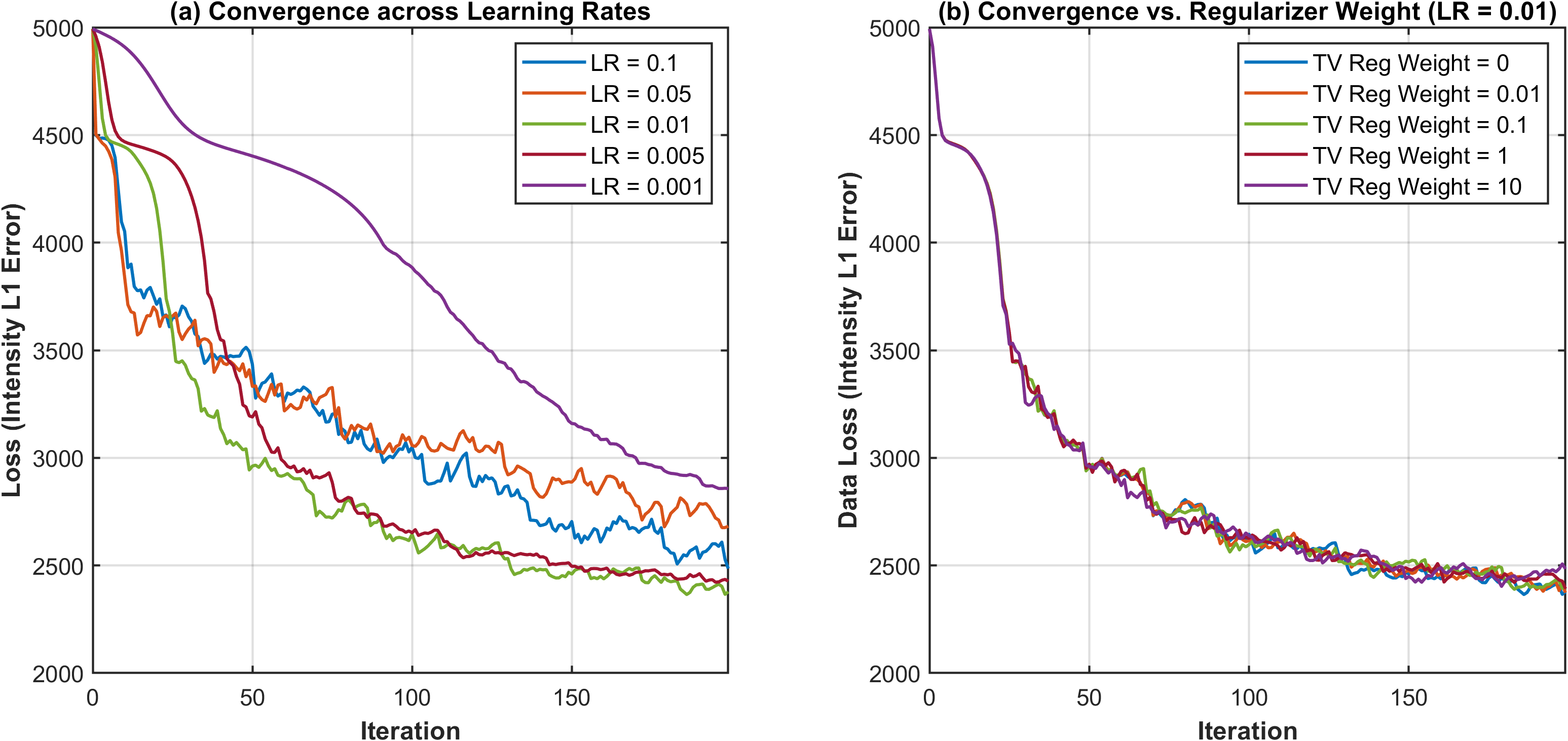}
\caption[HASA-ADAM Hyperparameter Study.]{\textcolor{black}{\textbf{HASA-ADAM Hyperparameter Study.} \textbf{(a)} Optimization convergence across varying learning rates. \textbf{(b)} Pure data loss across varying TV regularizer weights. The coincident plateaus confirm that the regularizer shapes the topology without reducing focal fidelity.}}
\label{fig:hyperparam_study}
\end{figure}

\textcolor{black}{
Stochastic optimizers can be sensitive to hyperparameter choices. We ran an empirical sweep to check convergence across configurations. We swept learning rates across $\eta \in \{0.001, 0.005, 0.01, 0.05, 0.1\}$ and observed stable, monotonic loss curves in every case (Figure~\ref{fig:hyperparam_study}a). High learning rates fell fast but oscillated near convergence; low rates were stable but slow. $\eta = 0.01$ was the best practical compromise. We also checked the Total Variation (TV) regularizer. TV smoothing is needed for manufacturability, but excessive weight could compromise acoustic accuracy. Sweeping $\lambda \in \{0.0, 0.01, 0.1, 1.0, 10.0\}$ while tracking the raw intensity L1 error over 200 iterations showed no separation between curves (Figure~\ref{fig:hyperparam_study}b). This confirmed that TV smooths the lens geometry without degrading the acoustic reconstruction in our optimizer.
}

\subsection{Canonical Validation of the HASA-ADAM Optimization Framework}
\begin{figure}[htbp]
\centering
\textcolor{black}{\includegraphics[width=0.8\textwidth]{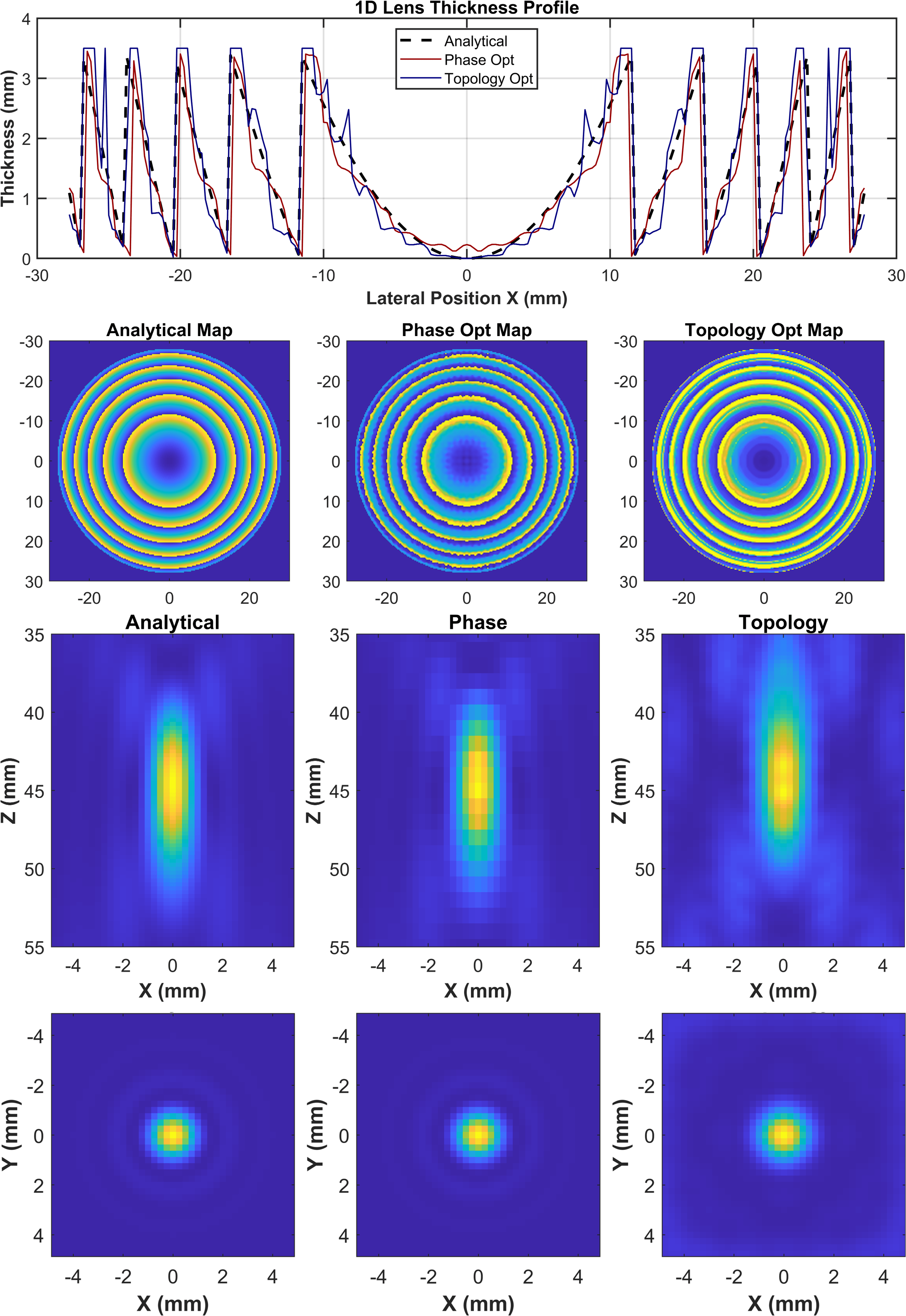}}
\caption[Canonical Validation of HASA-ADAM Optimization.]{\textcolor{black}{\textbf{Canonical Validation of HASA-ADAM Optimization.} \textbf{Top row:} 1D thickness profiles. \textbf{Second row:} 2D thickness maps. \textbf{Bottom rows:} Focal intensity in the axial (XZ) and lateral (XY) planes. Correlation drops from 0.892 (Phase Opt) to 0.70 (Topology Opt), but the topology optimizer finds an alternative valid topography that produces a well-confined focal spot.}}
\label{fig:canonical_validation}
\end{figure}

As a first test, we validated the optimizer on a canonical inverse problem---single-point focusing in a homogeneous free field---before introducing the skull.
We optimized a 1\, MHz Gaussian point focus at 45\, mm depth using a 60\, mm aperture in pure water ($c_0 = 1500$\,m/s), as per the steps in the computational graph of Section~2.2.1.2. A \textsl{softplus} activation was used to bound the lens to 5\,mm, and an L1 loss drove convergence. Grids were zero-padded to prevent aliasing.
The Rayleigh-Sommerfeld analytical solution for single-point focusing\cite{blackstock_fundamentals_2000} was our benchmark for this exercise. In this idealized environment, the required lens thickness $h_{analytical}(x,y)$ maps to a standard Fresnel zone plate:
\begin{equation}
h_{analytical}(x,y) = \left( \frac{\sqrt{x^2 + y^2 + F^2} - F}{1 - c_0/c_{lens}} \right) \pmod{L_{2\pi}}
\end{equation}
where $c_{lens} = 2591$\,m/s, and $L_{2\pi}$ is the material thickness for a full $2\pi$ phase shift. As shown in Figure~\ref{fig:canonical_validation}, both optimizers avoided local minima and produced tightly confined focal spots. Thickness profiles were compared with the analytical Fresnel geometry to assess structural fidelity. Phase optimization achieved a high structural correlation ($r = 0.892$), consistent with its planar phase-matching formulation. Topology optimization showed a lower morphological correlation ($r = 0.70$), with noticeable deviations from the analytical ideal. This divergence has a straightforward physical explanation. Topology optimization targets intensity at the focal plane rather than a prescribed surface phase, giving the optimizer far more freedom to distribute material. Since phase retrieval from intensity is ill-posed, many different thickness profiles produce nearly the same focal spot. The \textsl{softplus} activation also smooths the sharp $2\pi$ phase-wrap discontinuities required by the analytical lens. The result is a more continuous, printable lens that explores the volumetric wave-propagation space more fully, at the cost of a slight reduction in peak focal intensity. The optimizer still finds valid topographies that satisfy the focal objective, confirming that the HASA gradients are accurate.

\subsection{HASA combined with ADAM iterative optimizer can design holographic lens topologies for high fidelity acoustic holography}

\begin{figure}[htbp]
\centering
\includegraphics[width=0.75\textwidth]{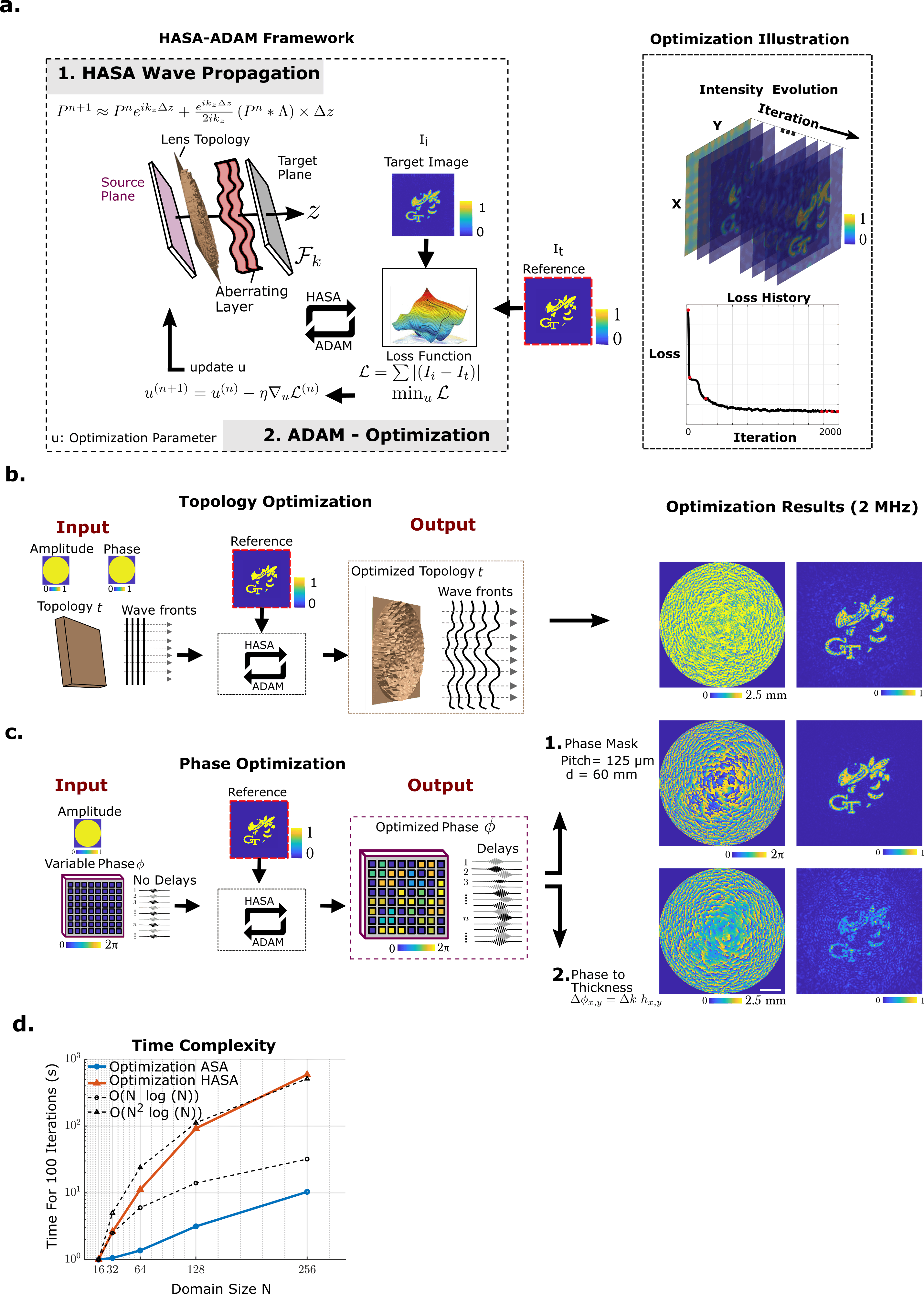}
\caption[HASA-ADAM provides a unified framework to design 3D high-fidelity, large aperture, acoustic holography.]{ (a) Schematic of the hologram design framework, comprising two key steps: 1) the Heterogeneous Angular Spectrum Approach (HASA) for wave propagation and 2) optimization using ADAM. The HASA algorithm for phase or thickness optimization via gradient descent is presented in Table~\ref{tab:algorithm}. (b) The HASA-ADAM framework was used to optimize the topology of a lens for a complex 2D hologram. At 2 MHz, this direct thickness optimization results in a high-fidelity target reconstruction. (c) HASA-ADAM phase optimization results for the complex 2D hologram at 2 MHz. Because the lens is acoustically thin at this higher frequency, the thin-element approximation holds; converting an ideal phase mask (top, utilizing a 125~$\mu$m pitch over a 60 mm aperture) into a physical thickness profile (bottom) produces good holographic results. (d) Complexity analysis of hologram optimization using HASA, showing that HASA optimization ($O(N^2\log N)$) requires more computational time than ASA ($O(N\log N)$).}
\label{fig:hasa_adam_framework}
\end{figure}

We introduce a framework that combines the Heterogeneous Angular Spectrum Approach (HASA), a fast spectral method for wave propagation in complex media~\cite{schoen2020heterogeneous} with the ADAM iterative optimizer to reduce a loss function (absolute difference in intensity between reference or target image and image plane)~\cite{fushimi2021acoustic}, for accelerated hologram optimization (Fig. ~\ref{fig:hasa_adam_framework}a and Table~\ref{tab:algorithm}). This approach takes advantage of HASA's ability to incorporate in-plane varying speed-of-sound maps and support a differentiable optimization of lens thickness profiles (that is, assuming an initial zero phase and constant amplitude at the source; Fig. ~\ref{fig:hasa_adam_framework}a and Table~\ref{tab:algorithm}). Consequently, the proposed framework allows for direct topology optimization and the design of holographic lens topologies that account for the physical effects of wave propagation within the lens, and consequently, generate holographic lenses that encode complex acoustic holograms in the megahertz frequency range (Fig. ~\ref{fig:hasa_adam_framework}b). This approach can also be used for acoustic holography based on the optimized phase (Fig. ~\ref{fig:hasa_adam_framework}c).

To test this concept and evaluate its performance, we first aimed to generate a complex holographic pattern at 2 MHz. We compared the performance of our direct topology optimization approach (Fig.~\ref{fig:hasa_adam_framework}b) with conventional acoustic holography methods, which rely on generating an optimized phase pattern and converting it to a physical thickness profile based on the thin-element approximation (Fig.~\ref{fig:hasa_adam_framework}c). {For the optimization and lens topology, a fine discretization of $\lambda/6$ (125~$\mu$m at 2~MHz) was used over a 60~mm aperture.} As shown in Fig.~\ref{fig:hasa_adam_framework}b and c, at 2 MHz, both direct topology optimization and the phase-to-thickness conversion successfully produce the target holographic pattern with high fidelity. Because the physical thickness required to achieve a full $2\pi$ phase shift at 2 MHz is relatively small, the lens remains acoustically thin, and the thin-element approximation holds. However, for many practical biomedical applications, such as transcranial ultrasound, operating at lower frequencies (e.g., 1 MHz or below) is critical to minimize acoustic attenuation and safely penetrate the skull. As the operating frequency drops, the acoustic wavelength increases, necessitating a proportionally thicker lens to achieve the same phase modulation.

The computational complexity ($\sim O(N^2\log N)$) of the proposed framework is higher than that of the homogeneous ASA ($\sim O(N\log N)$) (Fig. ~\ref{fig:hasa_adam_framework}d)~\cite{melde2016holograms}, the optimized hologram topology converged in approximately 20 min using a discretization of $\lambda/10$ at 1 MHz within a domain size of $N = 40$ mm (corresponding to a volume of 40 mm$^3$) on a system equipped with a 24 GB NVIDIA RTX 3090 GPU. Moreover, increasing the linear dimension by 50\% (i.e., 6 cm, which was the upper limit in the capacity of the GPU used) resulted in approximately a threefold increase in computational time (close to the 2.4-fold expected from the $O(N^2\log N)$ scaling and consistent with typical GPU-memory overhead). This highlights the method's scalability and efficiency for large aperture (i.e., clinical-scale) designs. A major reduction in optimization time ($\sim$15 min for 500 iterations) can be achieved by downsampling the domain to a discretization of $\lambda/6$ at 1 MHz, without any loss in performance. Thus, we adopted a $\lambda/6$ discretization for the subsequent studies. Together, HASA-ADAM constitutes a major advancement in acoustic holography, providing a unified framework for designing 3D-printed lenses for high-fidelity holography and enabling high-performance systems at a fraction of the cost.

\subsection{ Effect of Thin-Element Approximation on Hologram Optimization}

 Traditional hologram optimizations (such as Rayleigh-Sommerfeld diffraction \cite{blackstock_fundamentals_2000} or homogeneous Angular Spectrum approaches)use the Thin-Element Approximation (TEA), also known as the Thin-Film Approximation (TFA), to get the thickness map for 3D printing. Phase-only optimization may be insufficient for high-fidelity holography once the lens becomes acoustically thick (i.e., at sub-megahertz frequencies). The TEA treats phase shifts as purely longitudinal ( in other words, along the axis of wave propagation) phenomena and ignores lateral energy migration. Below roughly 1\, MHz, this assumption does not hold. Reducing the frequency shifts wave propagation from a locally planar regime to a volumetric diffraction regime. To motivate HASA-ADAM, we first examine where the TEA fails. The following subsections cover the theory and results across frequencies, which highlight this failure.

Resolving a lateral feature $\delta$ at depth $z$ requires:
\begin{equation}
\label{eq:master}
\boxed{ \underbrace{2 dx}_{\text{Sampling Limit}} \leq \underbrace{\frac{c}{f}}_{\lambda} \lesssim \underbrace{\delta}_{\text{Resolution}} \approx \underbrace{\frac{c \cdot z}{f \cdot D}}_{\text{Diffraction Geometry}} }
\end{equation}
Spatial resolution scales inversely with frequency: halving $f$ doubles the minimum resolvable feature. Applying our geometry ($f = 1$\,MHz, $D = 60$\,mm, $z = 45$\,mm) gives $\delta \approx 1.13$\,mm. The fine lines of the GT logo push directly against this boundary. The system is diffraction-limited.

\subsubsection{Numerical Aperture and Feature Scaling}
In acoustic holography, lens performance is characterized by its  numerical aperture (NA). For the geometry used here ($D = 60$\,mm, $F = 45$\,mm), the maximum steering angle is $\theta_{max} \approx 33.7^\circ$:
\begin{equation}
\text{NA} = \sin(\theta_{max}) = \sin(\arctan(D/2F)) \approx 0.55
\end{equation}
To steer a wavefront to angle $\theta$, the phase gradient on the lens surface must be $d\phi/dx = k_0 \sin(\theta)$. The sampling pitch $\Lambda$ must satisfy Nyquist to avoid grating lobes:
\begin{equation}
\Lambda \leq \frac{\lambda}{2 \cdot \text{NA}}
\end{equation}
For a fixed NA, holographic feature size $\Lambda$ must therefore scale linearly with $\lambda$.

\subsubsection{Acoustically Thick Regime}
The thin-film approximation fails because lens thickness $L$ shrinks far less favorably than feature size $\Lambda$ as the frequency increases. Unlike an electronic phased array, a passive lens generates phase delays by propagating through a material with a different sound speed $c_{lens}$. A full $2\pi$ wrap requires a modulation depth:
\begin{equation}
L_{2\pi} = \frac{\lambda}{|1 - c_0/c_{lens}|}
\end{equation}
For biocompatible polymers in water, the contrast is moderate ($\sim 0.6$), requiring 3--5\,mm at 1\,MHz. The total lens thickness $L$ is therefore much larger than $\lambda$ at these frequencies.

\subsubsection*{Refractive Walk-off and Grid Distortion}
Define the \textbf{ geometric aspect ratio } $\text{AR} = L_{2\pi}/\Lambda$. For high-NA lenses, $\text{AR} > 1$: the phase-modulating features resemble tall acoustic columns rather than a thin film. In this volumetric regime, the scalar TEA fails because the acoustic energy migrates laterally, commonly known as the refractive walk-off.

As illustrated by the ray-tracing analysis in Figure~\ref{fig:walkoff_merged}A, acoustic waves undergo pronounced refraction at the steep sawtooth features. Instead of propagating longitudinally, the energy migrates laterally inside the lens. The macroscopic result is a \textbf{Hologram Grid Distortion} (Figure~\ref{fig:walkoff_merged}B): the acoustic exit coordinates (solid blue grid) contract radially relative to the ideal design grid (dashed grey). Energy intended for one pixel leaks into neighbors, generating volumetric crosstalk that phase-only optimization does not consider.

\subsubsection*{Multi-Frequency Optimization Results}
\begin{figure}[htbp]
\centering
\includegraphics[width=0.95\textwidth]{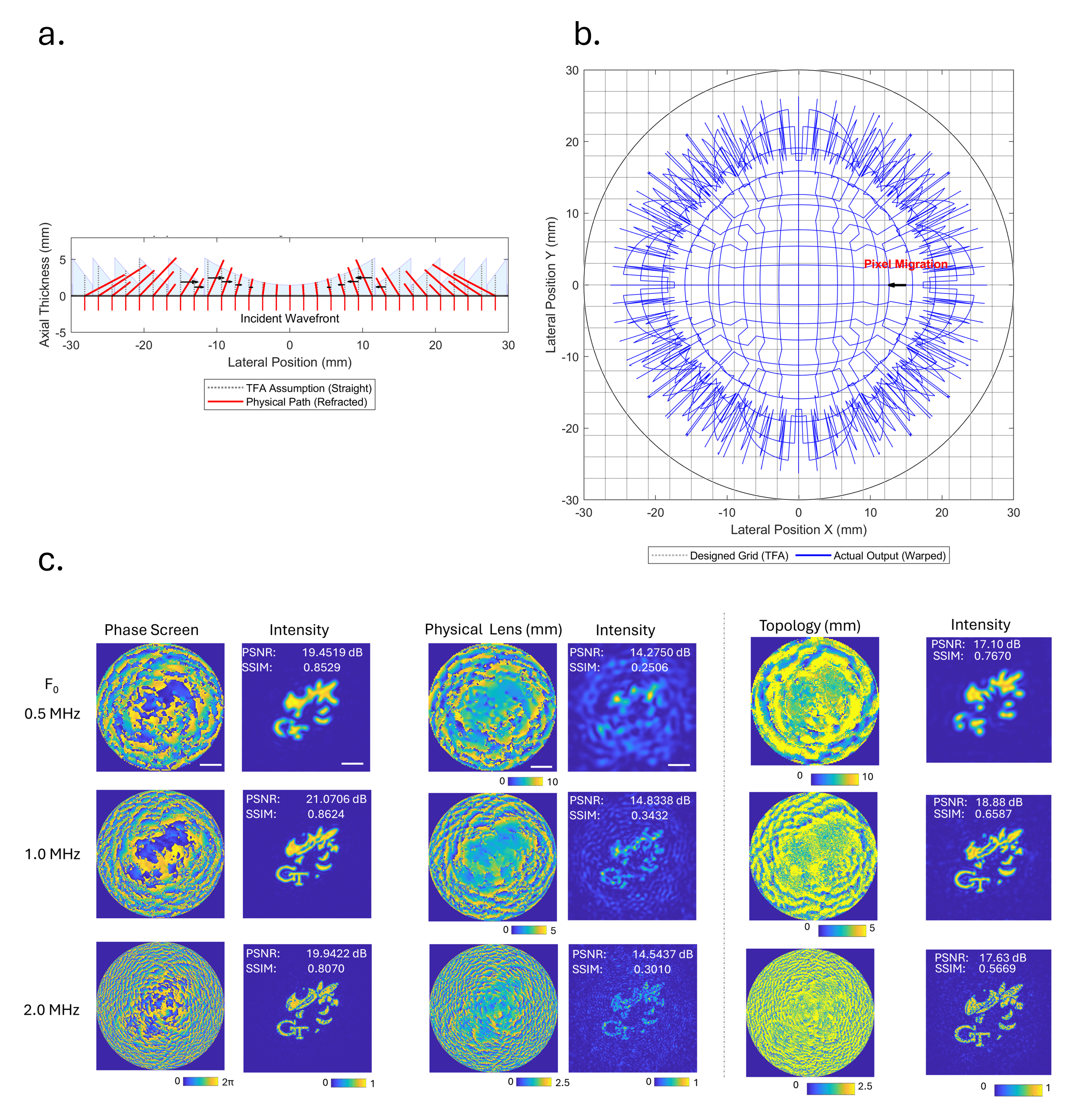}
\caption[Refractive Walk-off, Pixel Migration, and Reconstruction Analysis.]{
Simulations were performed for a high-NA acoustic lens ($f=1$MHz, $D=60$mm, $F=45$mm) using a sound speed of $c_{lens}=2500$m/s.
\textbf{(A) Mechanism:} Ray tracing through the lens cross-section. The Thin-Film Approximation (TFA) assumes that acoustic rays travel in straight lines (dashed gray), accumulating phase locally. In reality, the significant acoustic thickness causes the rays to refract according to Snell's law (solid red),\textbf{(B) Consequence:} Hologram Grid Distortion. A top-down view comparing the ideal pixel grid assumed by the TFA (dashed gray) with the acoustic exit locations (solid blue). The simulation reveals significant Pixel Migration. Energy intended for a specific spatial coordinate is displaced into neighboring pixels.\videolink{https://figshare.com/s/a0b266faa6d526e67c77}{Video S1 \& S2}
\textbf{(C) Reconstruction Results:} Comparison of holographic reconstructions across three frequencies ($F_0 = 0.5, 1.0, 2.0$~MHz). The columns display the idealized Phase Screen, the Lens model (incorporating walk-off effects), and the topology model, alongside their respective intensity fields. Quantitative metrics (PSNR, SSIM) demonstrate the severe degradation in image quality resulting from refractive walk-off when using conventional phase-to-thickness conversion.}
\label{fig:walkoff_merged}
\end{figure}

To assess volumetric cross-talk, we compared phase-based TEA optimization with HASA-ADAM topology optimization across operating frequencies (Figure~\ref{fig:walkoff_merged}C). Converting a 2D phase map to a 3D thickness map degrades performance, as expected from the walk-off theory. HASA-ADAM topology optimization compensates for this.

At \textbf{0.5\,MHz}, the required lens is $\sim$10\,mm thick. This maximizes the ray walk-off. The phase-converted lens degrades sharply (SSIM: 0.25). HASA-ADAM pre-compensates these effects, recovering the target fidelity (SSIM: 0.77) to a greater extent. \textbf{1.0\, MHz} offers the best geometry-acoustics tradeoff for the given aperture. The standard phase-to-thickness conversion gave a PSNR of 14.83\,dB. HASA-ADAM reached 18.88\,dB by capturing volumetric diffraction. Lastly, \textbf{2.0\, MHz}, being a higher frequency, should improve resolution. But the dense phase wrapping (which resets every multiple of the wavelength, i.e., $\sim$0.75\, mm) introduces edge diffraction and localized scattering. Even so, topology optimization (PSNR: 17.63\,dB) outperformed the TEA conversion using phase-only optimization (PSNR: 14.54\,dB).

Parametric sweeps also corroborate these trends. \videolink{https://figshare.com/s/a0b266faa6d526e67c77}{Video S1} shows a frequency sweep from 5.0 to 0.5\, MHz, illustrating how bulkier lens geometries at lower frequencies worsen walk-off. \videolink{https://figshare.com/s/a0b266faa6d526e67c77}{Video S2} isolates the effect of focusing strength. This shows that higher NA requires steeper phase gradients that worsen internal refraction.

\subsection{Comparison with the IASA phase optimization confirms degradation of performance with frequency scaling}
\begin{figure}[htbp]
\centering
\includegraphics[width=1.0\textwidth]{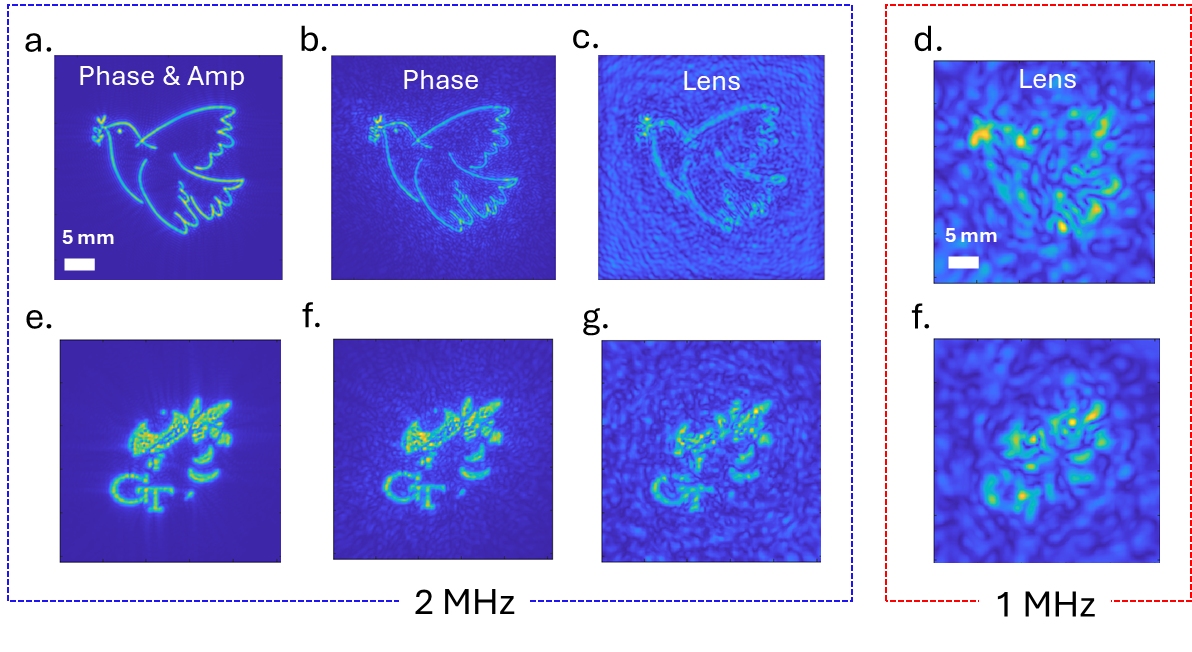}
\caption[Impact of Frequency Scaling on Hologram Fidelity.]{\textbf{Impact of Frequency Scaling on Hologram Fidelity.}
\textbf{(Blue Box) 2 MHz Regime:} Comparison of Dove (Top Row) and GT Logo (Bottom Row) reconstruction.
(a, e) Ideal Phase \& Amplitude.
(b, f) Phase only (Iterative Angular Spectrum Approach).
(c, g)  Lens. Note that at 2 MHz, the IASA phase map is coherent, but the Lens (c, g) shows distortion due to Refractive Walk-off.
\textbf{(Red Box) 1 MHz Regime:}
(d, f-right) Scaling the GT Logo design to 1 MHz results in a loss of fidelity. The diffraction limit $\delta$ doubles. This merges the fine details of the image. Thus, the IASA-designed lens fails to form a coherent image.}
\label{fig:comparison}
\end{figure}

To further understand this failure, we re-examined the benchmark holograms from Melde et al. (2016) \cite{melde2016holograms} and the GT logo to show the impact of frequency scaling (Figure. \ref{fig:comparison}). In this study, we used conventional Iterative Angular Spectrum Approach (IASA) phase optimization (similar to Melde et al. (2016)) to indicate where conventional phase optimization fails as we scale down the frequency.

At 2 MHz ($\lambda \approx 0.75$ mm), the diffraction limit $\delta$ is sufficiently fine for resolving the targets.
\textsl{IASA Performance (b, f):} The Phase-Only approximation produces coherent images for both the Dove (top) and GT Logo (bottom).
\textsl{Physical Lens (c, g):} However, the physical lens introduces distortion. This is the {Volumetric Failure} where the lens is acoustically thick, causing a refractive walk-off that IASA cannot predict. When we scale to 1 MHz (Bottom Right) while maintaining the same aperture ($D=50$ mm) and distance ($z=20$ mm), the system hits the diffraction limit.
\begin{itemize}
\item \textbf{Resolution Collapse:} Halving $f$ doubles the minimum feature size $\delta$. The fine lines of the GT logo are now smaller than the acoustic point-spread function.
\item \textbf{IASA Failure:} The standard IASA optimization fails in this condition. It attempts to create features that physics cannot support, resulting in noisy and unrecognizable distributions.
\end{itemize}

\begin{figure}[htbp] \centering \includegraphics[width=0.5\textwidth]{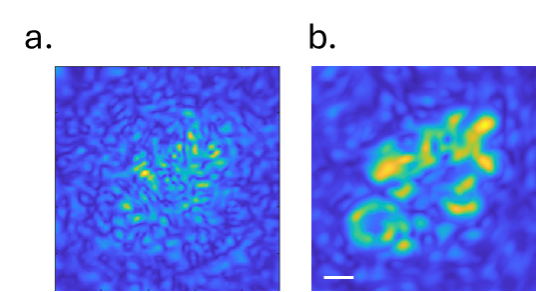} \caption[Phase-Only Optimization vs. Topology Optimization.]{\textbf{Phase-Only Optimization vs. Topology Optimization.} A comparison of acoustic pressure fields simulated using k-wave for the GT Logo target at $f=1$ MHz ($D=60$ mm, $z=45$ mm). \textbf{(a) IASA (Thin Film Approximation):} The standard phase-optimization approach fails to produce a recognizable image. The lens thickness introduces phase errors and lateral walk-off that the optimizer ignores, resulting in aberrations. \textbf{(b) Topology Optimization (This Work):} The proposed method, which models the volumetric wave propagation, restores the fidelity of the hologram. (Scale Bar 2 mm)} \label{fig:method_comparison}\end{figure}

We explicitly compare the performance of the proposed Topology Optimization (TO) against the standard IASA method using the specific geometric constraints of this study ($f=1$ MHz, $D=60$ mm, $z=45$ mm). {Figure \ref{fig:method_comparison}a} shows the pressure field resulting from a physical lens designed using the IASA. As predicted by the scaling analysis, the IASA failed to converge to a valid solution. The combination of the resolution limit ($\delta \approx 1.13$ mm) and the volumetric thickness (refractive walk-off) results in incoherent scattering.

In contrast, {Figure. \ref{fig:method_comparison}b} shows the result of the HASA-ADAM Topology Optimization. By solving the heterogeneous wave equation through the lens volume, the optimizer accounts for lateral energy transport and diffraction effects. It pre-compensated for the distortion and resulted in a coherent reconstruction of the GT logo.

The study confirms that converting phase maps to thickness is insufficient for high-fidelity acoustic holography (a $\sim5$-dB loss in PSNR due to refractive walk-off and volumetric diffraction). Failure here is governed by pixel migration and grid distortion in the thick-lens regime due to the Geometric Aspect Ratio of the lens features. Topology optimization compensates for these effects by accounting for volumetric wave propagation.

\subsection{Experimental Validation and the Impact of Shear Wave Mode Conversion at the Lens Interface}

We then experimentally validated the topology optimization framework at 1\, MHz. We 3D-printed an optimized lens for the GT Bee logo (at a 45 mm distance from the transducer with a 60 mm aperture), characterized its acoustic field at the focal plane using a hydrophone raster scan, and compared it with k-wave predictions.

Initially, we observed a significant discrepancy between the fluid (compressional-only) simulation and the experiments in terms of edge definition and diffuse background haze. The k-wave fluid simulation (Fig.~\ref{fig:shear_validation}a) predicts a tight, high-contrast spot, but the measured field (Fig.~\ref{fig:shear_validation}c) is broader and shows off-axis leakage that the fluid model misses. The fluid model MSE was 0.0230 when compared with the experimental scan.

We investigated elastic mode conversion at the lens-water interface as the likely cause. Although the lens polymer is homogeneous and isotropic, the optimized surface topology creates many locally varying oblique incidence angles. Such geometry introduces pronounced shear-wave interactions that are omitted in scalar fluid models. The Zoeppritz equations govern energy partitioning at fluid-solid boundaries. An incident P-wave at non-normal incidence splits into a transmitted P-wave and a mode-converted shear S-wave. For the photopolymer used here, $c_s \approx 1300$\,m/s---roughly half $c_p \approx 2590$\,m/s. The resulting S-waves degrade holographic fidelity via Phase Aberration and Refractive Steering.

\begin{enumerate}
\item \textbf{Phase Aberration:} The $S$-waves due to mode conversion propagate at a reduced velocity. They accumulate phase delays relative to the primary longitudinal wavefront, which propagates at a faster velocity.  This causes an uncorrelated phase, which in turn leads to destructive interference and a reduced peak focal intensity.

\item \textbf{Refractive Steering:} The trajectory of the mode-converted shear waves is governed by Snell's law for elastic media:
\begin{equation}
\frac{\sin \theta_s}{c_s} = \frac{\sin \theta_p}{c_p}.
\end{equation}
Owing to their lower sound speed ($c_s < c_p$), shear waves are refracted at steeper angles relative to the surface normal. This differential refraction actively steers the acoustic energy away from the designated target features, spatially dispersing it across the observation plane to form the diffuse background haze observed experimentally.
\end{enumerate}

\begin{figure}[htbp!]
\centering
\textcolor{black}{\includegraphics[width=1.0\linewidth]{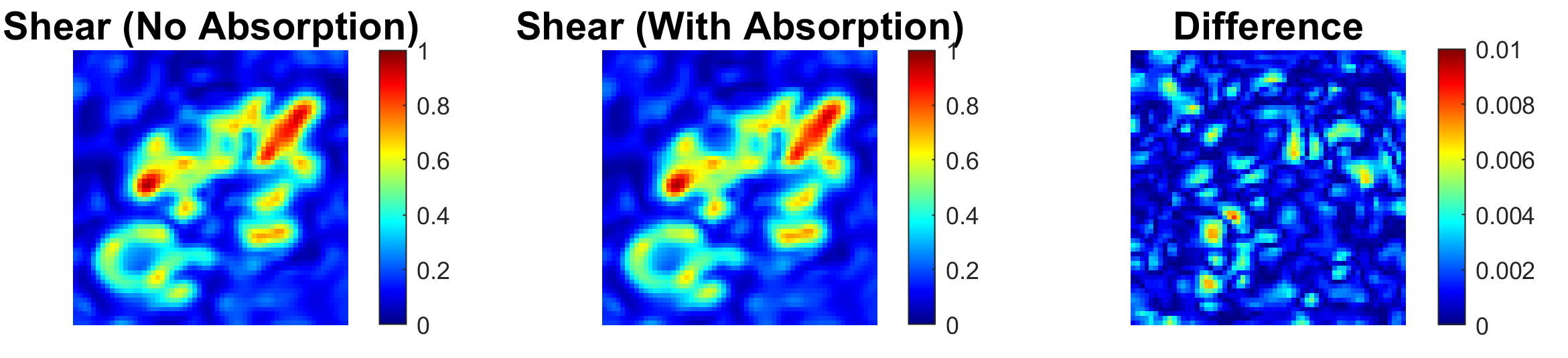}}
\caption[Verification of Phase Aberration vs. Shear Damping.]{\textcolor{black}{\textbf{Verification of Phase Aberration vs. Shear Damping.} Zeroing shear absorption (left) produces virtually the same field as the standard shear model (middle). The absolute difference (right) is negligible (MSE: 0.00, SSIM: 0.99, Correlation: 1.00), confirming that blurring originates from slow-shear-wave phase aberration and refraction, not attenuation. \videolink{https://figshare.com/s/57dc1ef8de85a04cc330}{Video S3 \& S4}.}}
\label{fig:shear_absorption_analysis}
\end{figure}

To confirm that the smearing results from slow-shear-wave aberration rather than simple attenuation, we ran an ablation study in the k-Wave elastic solver. We set $\alpha_s = 0$ while keeping $c_s = 1400$\,m/s. The resulting field is virtually identical to the fully attenuating case (Figure~\ref{fig:shear_absorption_analysis}). Correlation is 1.00, SSIM is 0.99, and the MSE is negligible. This shows that holographic degradation is driven by slow-shear-wave phase aberration and refractive steering at steep lens interfaces — not by energy absorption. Supplementary animations compare the fluid and elastic propagation fields (\videolink{https://figshare.com/s/57dc1ef8de85a04cc330}{Video S3\& S4}). Also, Figure~\ref{fig:lens_shear_simulation} in the appendix provides an elaborate description of the temporal evolution of shear waves relative to compressional waves at the lens interface and the effects of varying attenuation.

\begin{figure}[htbp!]
\centering
\includegraphics[width=1.0\linewidth,trim=10 10 10 10,clip]{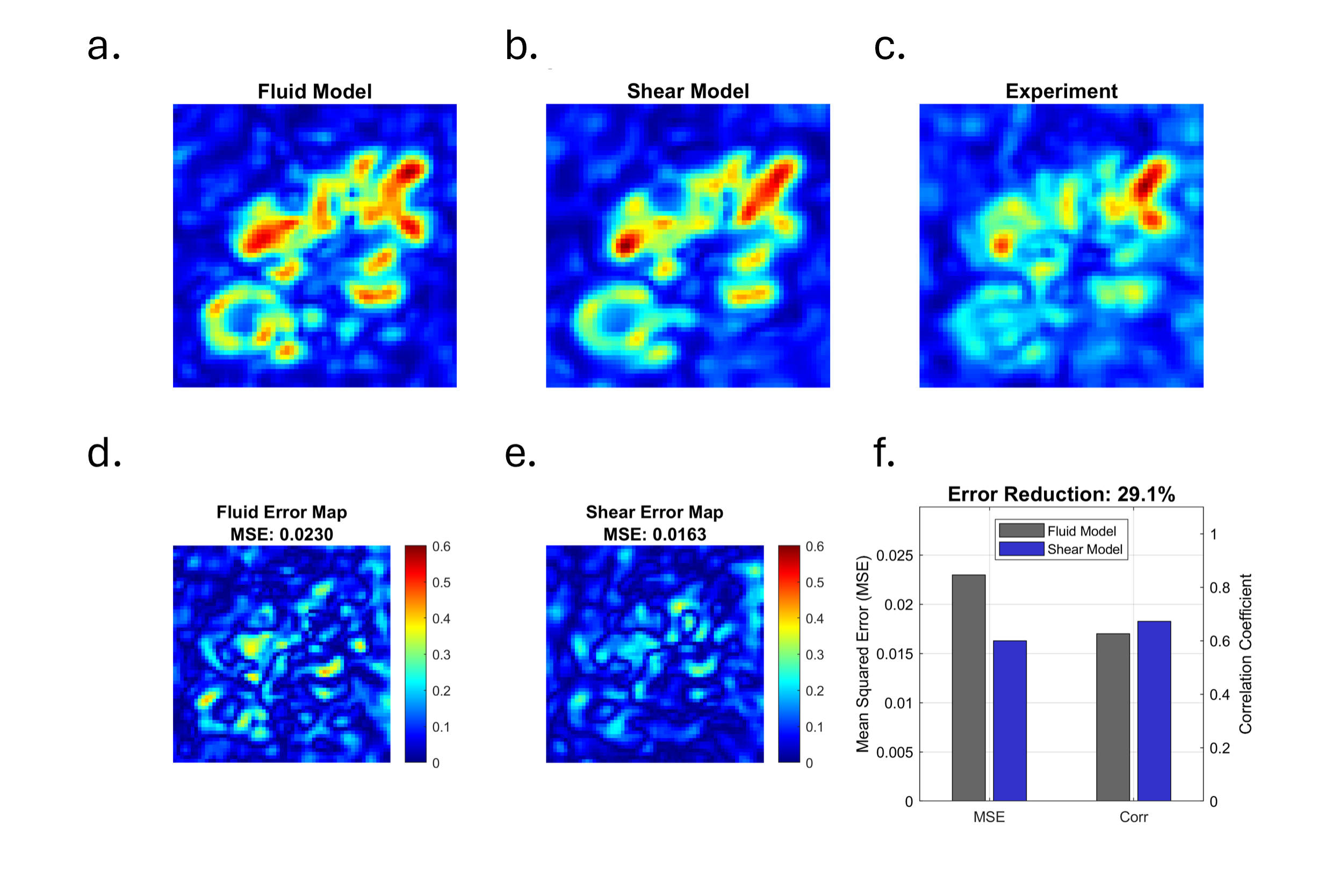}
\caption[Experimental validation of the holographic shear correction model.]{
Top row: Normalized intensity distributions of \textbf{(a)} the baseline Fluid Model, \textbf{(b)} the proposed Shear Model, and \textbf{(c)} the experimental ground truth.
Bottom row: Absolute error maps relative to the experimental data for \textbf{(d)} the Fluid Model and \textbf{(e)} the Shear Model. The Shear Model exhibited reduced residual artifacts and a lower Mean Squared Error (MSE: 0.0163) than the Fluid Model (MSE: 0.0230).
\textbf{(f)} Quantitative performance metrics showing the Mean Squared Error (left axis) and Correlation Coefficient (right axis). The Shear Model achieves a 29.1\% reduction in reconstruction error compared to the baseline.}
\label{fig:shear_validation}
\end{figure}

By incorporating these shear phenomena into a full-wave elastic simulation, the proposed Shear Model (Fig. ~\ref{fig:shear_validation}b) successfully reproduces the distortions and background scattering observed in the experimental measurements.

The corresponding error map for the Shear Model was substantially attenuated (decrease in the MSE to 0.0163)(Fig. ~\ref{fig:shear_validation}e). As per (Fig. ~\ref{fig:shear_validation}f), shear wave propagation accounts for a 29.1\% reduction in the reconstruction error and also improves the spatial correlation. These results suggest that the acoustic blurring and haze observed in high-frequency acoustic holography may be driven by elastic mode conversion at the lens's topographical interfaces.

\begin{figure}[htbp]
\centering
\includegraphics[width=0.9\textwidth]{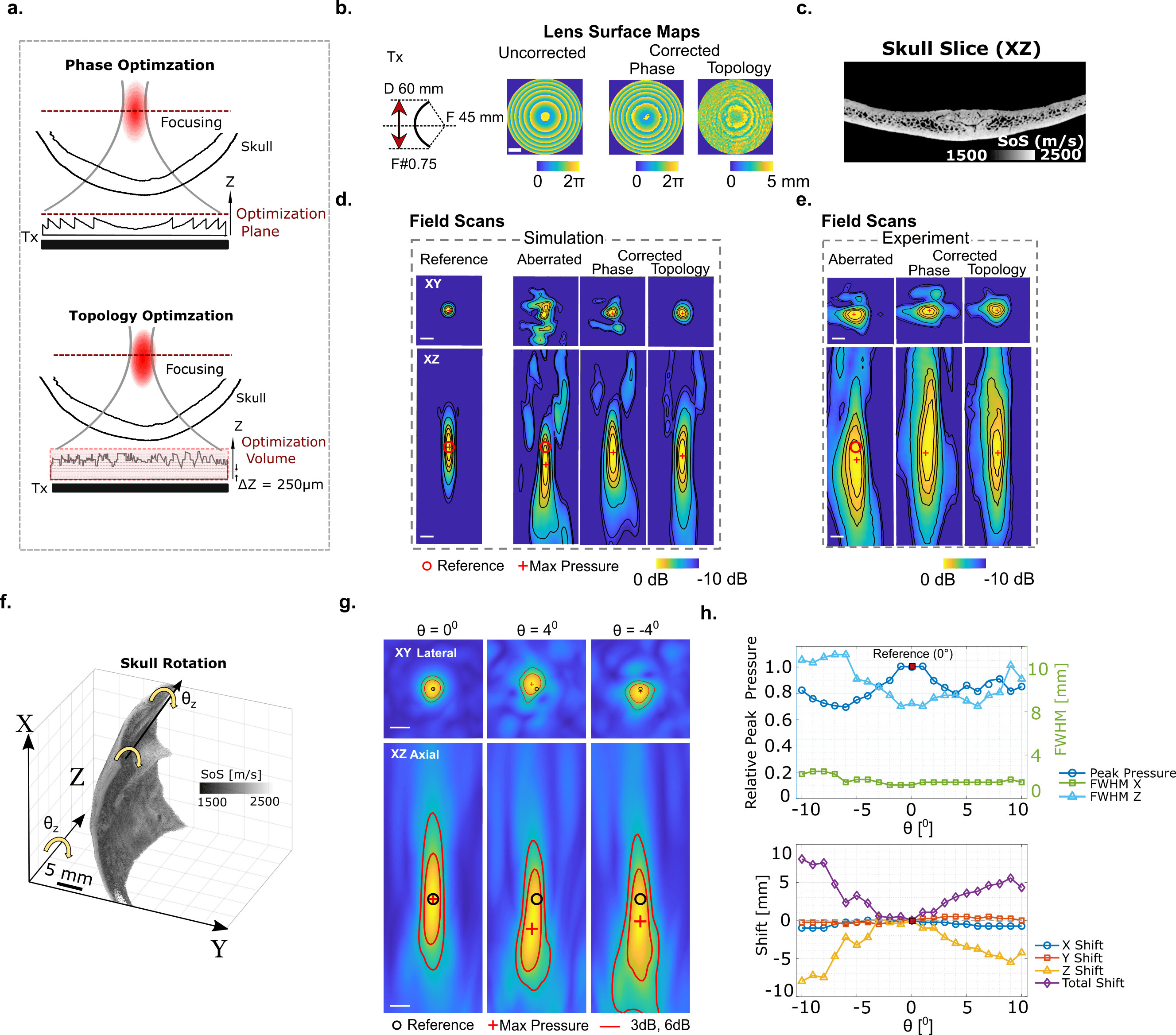}
\caption [Experimental validation of trans-skull hologram focusing and assessment of registration errors on focusing quality. ]{Experimental validation of trans-skull hologram focusing and assessment of registration errors on focusing quality. (a) Schematic of phase and topology optimization for single focusing. (b) Single-point focusing phase maps for a transducer with a diameter of 60 mm and an $F\#$ of 0.75, with and without skull corrections (scale bar: 1 cm) and equivalent topology map. (c) Skull speed of sound map obtained from micro-CT scan. (d) Lateral and axial focal profiles with contours obtained from simulations and (e) experiments, both with and without aberration correction for phase and topology optimized lenses (scale bar: 2 mm). (f) Simulation mask for skull rotation (scale bar: 5 mm). (g) Axial and lateral 2D surface maps demonstrating targeting and focusing errors due to a $\pm 4°$ skull rotation (scale bar: 2 mm). (h) Effect of skull rotation on peak amplitude and FWHM (Top) and focal shift and x, y and z directions (Bottom).}
\label{fig:trans_skull}
\end{figure}

\subsection{HASA-ADAM iterative optimizer for holographic lens topologies for TUS}
To demonstrate that HASA-ADAM thickness optimization can be used for transcranial ultrasound (TUS), we optimized topologies for a single focus (1 MHz with an F-number of 0.75) through the human skull (Fig. ~\ref{fig:trans_skull}a-b). {The optimization utilized spatially varying speed-of-sound (SoS) and density maps derived from micro-CT data (Fig.~\ref{fig:trans_skull}c).} We compared the thickness-optimized lens (Corrected Topology) with a lens design to generate a single focus in the free field (uncorrected/aberrated) and a lens optimized to account for aberration but designed with a standard phase-to-thickness conversion (Corrected Phase) (Fig. ~\ref{fig:trans_skull}a-c). The generated pressure was compared using both acoustic simulations and ex vivo trans-skull experiments (Fig. ~\ref{fig:trans_skull}d). We observed that the HASA-ADAM-based framework corrected for aberration and reduced the sidelobes in both the axial ($xz$) and lateral ($xy$) focal planes. These are indicated by both experimental and simulated data (Fig. ~\ref{fig:trans_skull}d).

The Corrected Topology optimization achieved a lower peak sidelobe level than the Corrected Phase approach.  Additionally, the optimized hologram that incorporates aberration correction (Corrected Topology) achieves a 24.5\% reduction in lateral 3 dB beam width and collimation of intended and actual focus(Fig. ~\ref{fig:trans_skull}d). Evidently, the focal pressure using the other two lenses was characterized by significant aberration (Fig. ~\ref{fig:trans_skull}d) and high side lobes, demonstrating suboptimal performance for high-frequency TUS.

The above data demonstrate the potential of the proposed framework to effectively correct skull-induced aberrations and lead to diffraction-limited performance; however, they also indicate that the focus attained with the experimental system is below the theoretical limits (Fig. ~\ref{fig:trans_skull}e). This discrepancy between the simulation and experiment is most likely due to registration errors or uncertainties in the skull and lens material properties or a combination of both. Past investigations have demonstrated that skull properties need to deviate by more than 20\% to lead to significant errors~\cite{schoen2021experimental}, which is unrealistic in many cases.

{To investigate this discrepancy, we performed a theoretical sensitivity analysis of the Speed of Sound (SoS) parameters. As shown in Table \ref{table:s1}, a $\pm 15\%$ mismatch in the skull or lens SoS can lead to a $\pm 30\%$ variation in the peak focal pressure and an axial focal shift of up to 2.0 mm. This axial shift is consistent with the broadening observed in the experimental data (Fig. ~\ref{fig:trans_skull}e). Additionally, the mechanical rotation of the skull fixture induces a coupled lateral translation (e.g., $\sim 10.6$ mm for a $6^{\circ}$ rotation). This misregistration likely enlarged the focal spot during the experimental scan.}

\begin{figure}[htbp]
\centering
\includegraphics[width=1.0\textwidth]{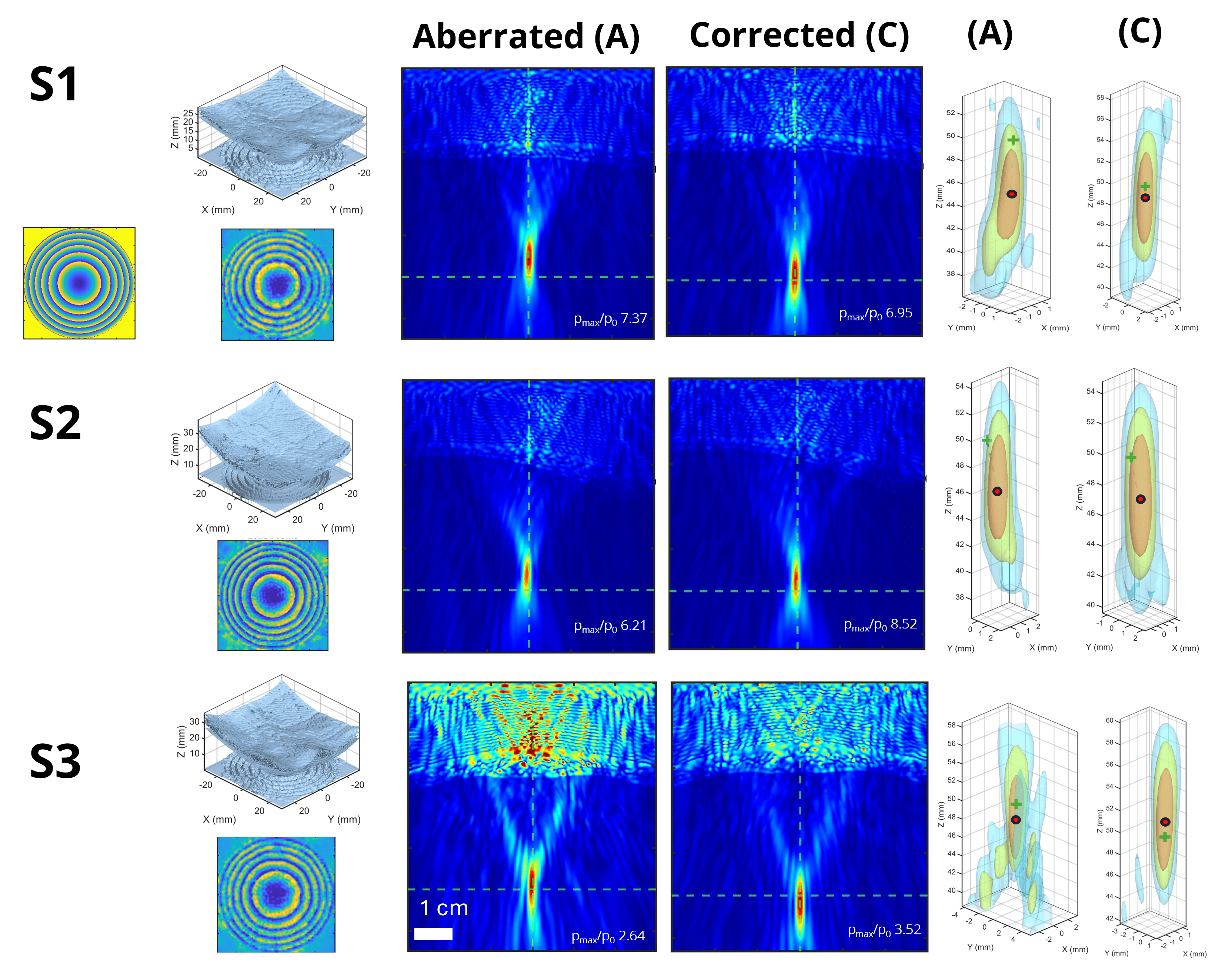}
\caption[Validating HASA-ADAM repeatability across different skull anatomies.]\textbf{Validating HASA-ADAM repeatability across different skull anatomies. 2D acoustic field maps and the 3D focal beam of the uncorrected beam are compared with our topologically corrected projections across three distinct human skulls (S1, S2, S3). Our results show three uncorrected failure modes due to skull aberration and the recovery after topology-based correction. Patient-specific topologies eliminate spatial targeting errors (Segment 1), recover lost acoustic pressure (Segment 2), and reconstruct defocused beams (Segment 3).}
\label{fig:multi_skull_correction}
\end{figure}

\begin{table}[htbp] 
\centering 
\caption{Focal Performance: Aberrated vs. Corrected (Segments S1, S2, S3)} 
\label{tab:performance_metrics} 
\renewcommand{\arraystretch}{1.2} 
\small 
\setlength{\tabcolsep}{3pt} 
\begin{tabularx}{\textwidth}{@{}Xcccccc@{}} 
\toprule 
\multirow{2}{*}{\textbf{Metric}} & \multicolumn{2}{c}{\textbf{Segment 1}} & \multicolumn{2}{c}{\textbf{Segment 2}} & \multicolumn{2}{c}{\textbf{Segment 3}} \\ 
\cmidrule(lr){2-3} \cmidrule(lr){4-5} \cmidrule(lr){6-7} 
& \textbf{Aberr.} & \textbf{Corr.} & \textbf{Aberr.} & \textbf{Corr.} & \textbf{Aberr.} & \textbf{Corr.} \\ 
\midrule 
\textbf{Peak Gain} & 7.37 & 6.95 & 6.21 & 8.52 & 2.64 & 3.52 \\ 
\textbf{Total Error (mm)} & 4.76 & 1.52 & 3.82 & 2.31 & 1.54 & 1.50 \\ 
\hspace{3mm}\textit{Lateral} & +0.00 & -0.25 & +0.50 & +0.25 & -0.25 & +0.00 \\ 
\hspace{3mm}\textit{Elevational} & -0.25 & -1.50 & +0.50 & +0.50 & +0.25 & -0.00 \\ 
\hspace{3mm}\textit{Axial} & -4.75 & +0.00 & -3.75 & -2.75 & -1.50 & +1.50 \\ 
\textbf{Axial FWHM (mm)} & 11.08 & 12.29 & 11.31 & 11.20 & 13.85 & 13.34 \\ 
\textbf{Lateral FWHM (mm)} & 2.08 & 2.14 & 1.99 & 1.99 & 2.34 & 2.14 \\ 
\textbf{Focal Vol. at -3 dB (mm$^3$)} & 9.9 & 11.1 & 8.6 & 8.8 & 875.4 & 59.5 \\ 
\bottomrule 
\multicolumn{7}{l}{\footnotesize \textit{Note: Aberr. and Corr. stands for Aberrated and Corrected cases, respectively.}} 
\end{tabularx} 
\end{table}

A single successful focusing trial is insufficient for clinical validation. Human skulls show large topographical variance. Accounting for patient-specific aberrations is needed to generalize our topology-optimization-based skull correction.  We expanded our 3D k-Wave simulations across three distinct human skull segments (S1, S2, and S3).

Without skull corrections, the acoustic beam suffers unpredictable failure modes (Table~\ref{tab:performance_metrics}). Each skull geometry introduces a somewhat different failure mode. Segment 1 causes spatial misalignment of the focus (4.76~mm off-target). Segment 2 drops the transmitted acoustic energy, and segment 3 results in defocusing of the focal volume into a large 875.4~mm$^3$ aberrating zone. Our topology optimization restores the focal quality for all the skull segments. For S1, it eliminates the axial shift and reduces the total spatial error by 68\%. For S2, it recovers transmission efficiency (a 37\% peak gain increase). For S3, the lens focuses the aberrated beam back into a confined focal spot of 59.5~mm$^3$. This reduces off-targeting errors. Thus, our trans-skull holograms account for the skull's geometry and ensure that acoustic energy goes to the targeted region as intended.(Fig.~\ref{fig:multi_skull_correction}).

\section{Analysis of Experimental Discrepancies}

\subsection{Speed of Sound (SOS) Estimation}
The longitudinal speed of sound in the ClearWhite v4 resin was determined using a through-transmission method. A reference signal was first acquired through a water path (baseline), followed by five measurements with the sample (thickness $d = \SI{15}{mm}$) inserted into the path at different positions between the transducer and hydrophone.

The arrival time was determined using a peak-threshold detection method of the envelope of the signal using the Hilbert transform, $E(t) = |\mathcal{H}(s(t))|$. The time shift ($\Delta t$) was calculated as the difference between the sample and baseline arrival times. The speed of sound in the sample ($c_{sample}$) was estimated using the relative ToF, given the sample thickness $d = \qty{15}{mm}$ and the speed of sound in water $c_{water} \approx \qty{1480}{m/s}$:
\begin{equation}
c_{sample} = \left( \frac{1}{c_{water}} - \frac{\Delta t}{d} \right)^{-1}
\end{equation}

\begin{figure}
\centering
\includegraphics[width=1.0\linewidth]{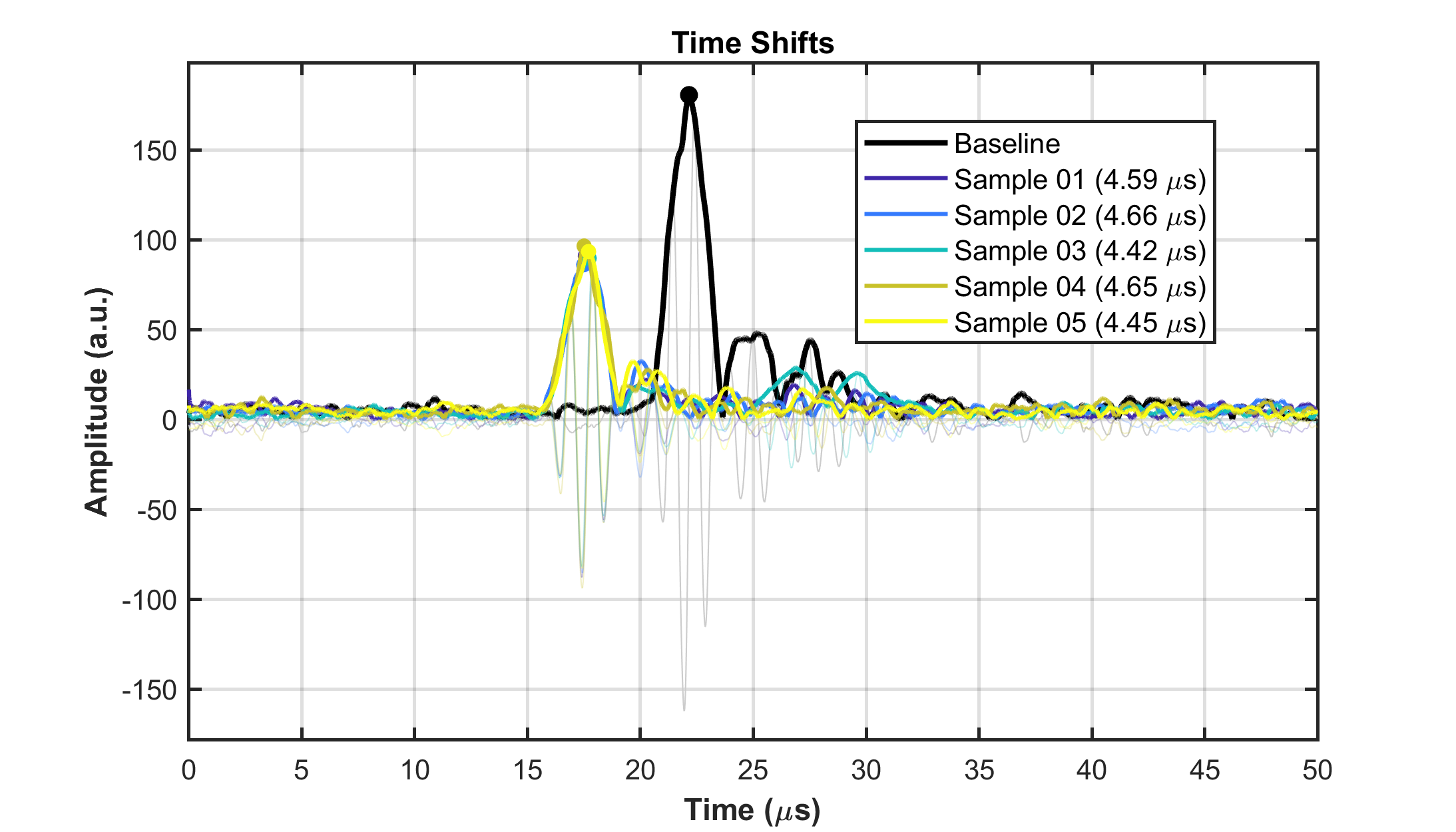}
\caption{Speed of Sound estimation using through transmit measurement}
\label{fig:placeholder}
\end{figure}
\begin{table}[H]
\centering
\caption{Measured Time Shifts and Speed of Sound Calculation}
\label{table:soc_meas_results}
\vspace{0.2cm}
\begin{tabular}{lc}
\toprule
\textbf{Measurement Pair} & \textbf{Time Shift ($\boldsymbol{\mu}$s)} \\
\midrule
Pair \#1 & 4.5920 \\
Pair \#2 & 4.6640 \\
Pair \#3 & 4.4240 \\
Pair \#4 & 4.6480 \\
Pair \#5 & 4.4480 \\
\midrule
\textbf{Average Shift ($\Delta t$)} & \textbf{4.56 $\pm$ 0.11 $\boldsymbol{\mu}$s} \\
\textbf{Est. Sound Speed ($c_{exp}$)} & \textbf{2689 $\pm$ 54 m/s} \\
\bottomrule
\end{tabular}
\end{table}

The average time shift is \textbf{4.56 $\pm$ 0.11 $\boldsymbol{\mu}$s} (mean $\pm$ standard deviation). Thus, the longitudinal speed of sound in the sample material is \textbf{2689 $\pm$ 54 m/s} which compares well with the standard literature value for the material ($c_{design} = \qty{2590}{m/s}$). The percentage error is:
\begin{equation}
\% \text{ Error} = \left| \frac{c_{exp} - c_{design}}{c_{design}} \right| \times 100 = \left| \frac{2689 - 2590}{2590} \right| \times 100 \approx 3.8\%
\end{equation}

The measured value deviated by approximately \textbf{3.8\%} from the reference design value, which is relatively minor and falls near the estimated error bounds.

\subsection{Lens Speed of Sound (SoS) Sensitivities: Longitudinal Shift and Wavefront Aberration}

{To investigate further discrepancies, we performed a sensitivity analysis on the Speed of Sound (SoS) parameters.}

Let $c_{des}$ be the speed of sound assumed in the optimization algorithm, and $c_{real}$ be the actual speed of sound of lens material. The phase accumulation $\phi$ through a thickness $h$ is governed by the refractive contrast with the background medium ($c_0$):
\begin{equation}
\phi_{real}(x,y) = \omega h(x,y) \left( \frac{1}{c_{0}} - \frac{1}{c_{real}} \right)
\label{eq:sos_phase_real}
\end{equation}
The optimization routine calculates the thickness $h(x,y) = \frac{\phi_{target}(x,y)}{\omega \left( \frac{1}{c_{0}} - \frac{1}{c_{des}} \right)}$ to achieve a target phase $\phi_{target}$ based on $c_{des}$.
Substituting this into Eq. \ref{eq:sos_phase_real}, the actual phase realized in the experiment is:
\begin{equation}
\phi_{real}(x,y) = \phi_{target}(x,y) \times \underbrace{\left[ \frac{\frac{1}{c_{0}} - \frac{1}{c_{real}}}{\frac{1}{c_{0}} - \frac{1}{c_{des}}} \right]}_{\gamma}
\label{eq:sos_mismatch}
\end{equation}
$\gamma$ is a scalar constant representing the mismatch ratio. Whether $c_{des}$ was chosen incorrectly or $c_{real}$ shifted due to curing, the result is identical: the output phase map is the target phase map scaled by $\gamma$.
However, because our sub-megahertz acoustic holograms operate in the acoustically thick regime (as established in Section 2.3.2), this paraxial assumption is incomplete. At the microscopic level, a change in the physical SoS alters the refractive index contrast at the lens-water interface. According to Snell's Law, this alters the internal angles of refraction for acoustic rays incident upon the steep topology of the lens. Consequently, the altered refraction angles simultaneously exacerbate the refractive walk-off effect.

\subsubsection{SoS Mismatch Primarily Causes a Z-Axis Shift for Single Point Focusing}
For point targetting however, the requirement for spatial coherence can be tempered. A lens SoS error shifts the hologram along the Z-direction instead of destroying it. In the paraxial approximation (Fresnel domain), a focusing element imparts a quadratic phase profile to the wavefront as follows:
\begin{equation}
\phi_{target}(r) \approx -\frac{k r^2}{2 F_{des}}
\end{equation}
where $k$ is the wavenumber, and $F_{des}$ is the design focal length.

Owing to the SoS mismatch derived from Eq. \ref{eq:sos_mismatch}, the physical phase profile becomes:
\begin{equation}
\phi_{real}(r) = \gamma \cdot \left( -\frac{k r^2}{2 F_{des}} \right) = -\frac{k r^2}{2 (F_{des}/\gamma)}
\end{equation}
This equation describes a perfect lens with a \textbf{new focal length} $F_{new}$:
\begin{equation}
F_{new} = \frac{F_{des}}{\gamma}
\end{equation}

Consequently, the internal phase relationships that create the hologram shapes are preserved because the {entire} phase map is scaled uniformly. The hologram is coherent and forms at $Z = F_{new}$. However, the coherence is not destroyed; instead, the plane of image formation is displaced.

\subsection{Skull Sound-speed (SOS) and Frequency variation}

Errors in the the estimation of the  of the speed of sound in the skull can lead to uncertainty in aberration correction. Because precise knowledge of skull acoustic properties is challenging to obtain under clinical conditions, it is important to understand how these uncertainties propagate through holographic reconstruction and  effect on focusing accuracy. Although our main analysis assumes operation at the design frequency, real transducers have finite bandwidths and may operate at frequencies that deviate from the nominal design value. We also assessed the effect of deviation from the design frequency on focusing performance.

\paragraph*{Speed of Sound (SOS) variation:} A phase-only hologram is designed for a skull speed $c_{\text{design}} = 2500$ m/s and average thickness $d_{\text{skull}} = 7$ mm. If the actual speed is $c = c_{\text{design}}(1 \pm 0.15)$, the one-way travel-time error is

\begin{equation}
\Delta t = \left(\frac{1}{c_0} - \frac{1}{c}\right)d_{\text{skull}} \approx \frac{\pm 0.15 d_{\text{skull}}}{c_0} = \pm 4.2 \times 10^{-7}\text{s},
\end{equation}

At the design frequency $f_0 = 1$ MHz this corresponds to a phase slip

\begin{equation}
|\Delta\phi| = 2\pi f_0|\Delta t| \approx 2.64\text{rad}(151^0),
\end{equation}

\begin{figure}[htbp]
\centering
\includegraphics[width=1.0\textwidth]{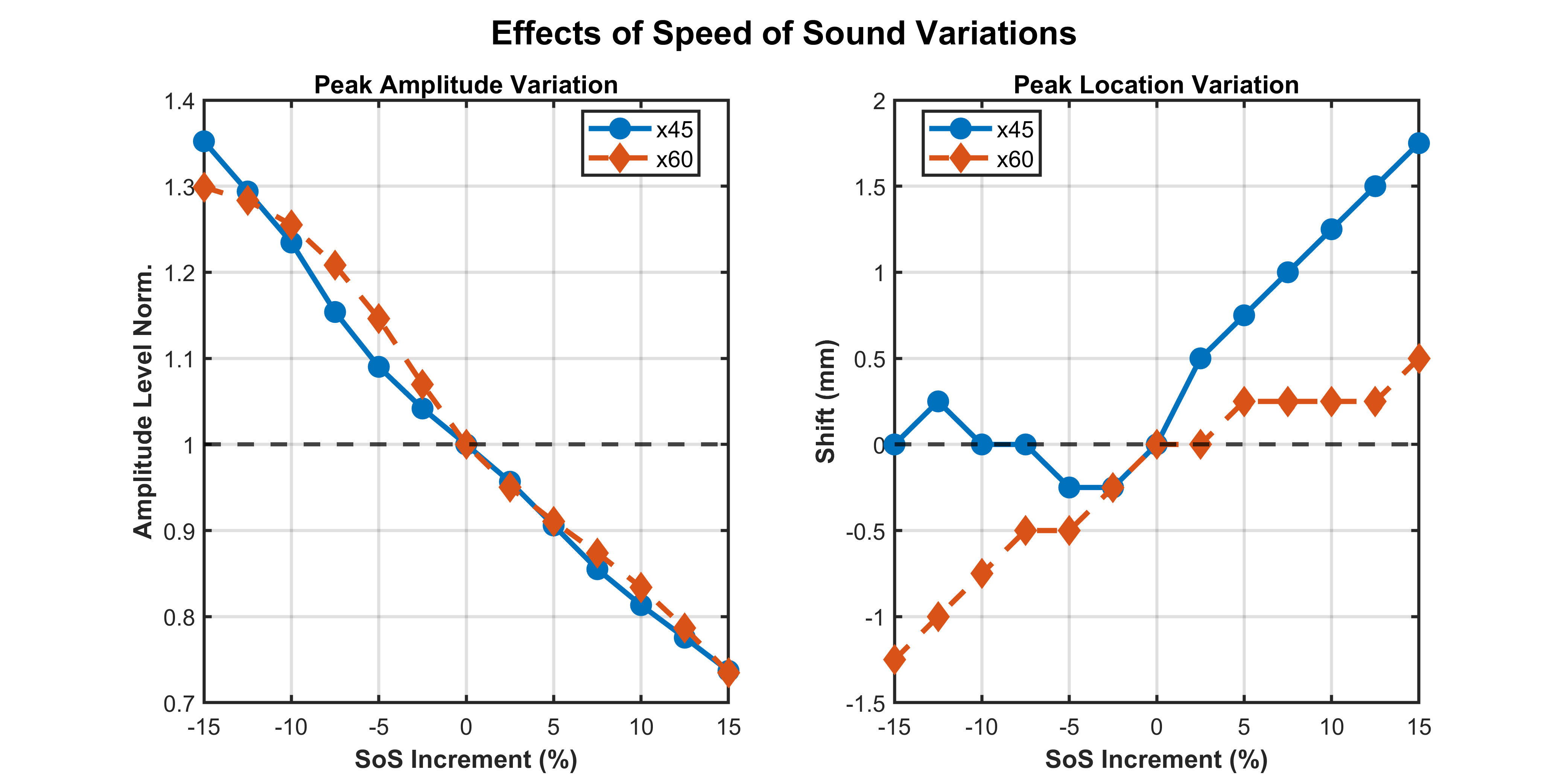}
\caption{Peak pressure and peak location versus SOS error ($\pm 15\%$) for both nominal foci.}
\label{fig:s1}
\end{figure}

Using simulations, we varied the speed of sound of the skull for two focusing configurations with phase-only lenses computed from the time of flight for a focus at 45 mm ($F\#0.75$) and 60 mm ($F\#1.0$) and observed the effect of peak amplitude and peak location variation. We observed approximately ($\pm 30\%$) variation in the peak focal pressure, whereas the axial drift ranges were relatively minor (Fig \ref{fig:s1} right). This is evident from the fact that SOS error ($\pm 15\%$) results in a maximum shift in focus (2.0 mm for x45 configuration, Table \ref{table:s1}), which slightly exceeds the wavelength in water at 1 MHz (1.5 mm). For applications that require high-precision targeting, this can still be a significant source of error, and methods to account for it should be considered.

\begin{table}[htbp]
\centering
\caption{Peak Amplitude and Location Statistics for Two Configurations}
\label{table:s1}
\begin{tabular}{lccccc}
\hline
Configuration & Min & Max & Range & Mean & Std Dev \\
\hline
\multicolumn{6}{c}{Peak Amplitude (Norm.)} \\
x45 & 0.737 & 1.352 & 0.615 & 1.016 & 0.200 \\
x60 & 0.735 & 1.299 & 0.564 & 1.027 & 0.197 \\
\multicolumn{6}{c}{Peak Location (mm)} \\
x45 & -0.250 & 1.750 & 2.000 & 0.500 & 0.685 \\
x60 & -1.250 & 0.500 & 1.750 & -0.212 & 0.548 \\
\hline
\end{tabular}
\end{table}

\paragraph*{Frequency variation:} To keep our analysis simple, we assume a homogenous medium. A planar aperture is driven with a static phase pattern

\begin{equation}
\phi_c(r) = -\frac{2\pi f_0}{c}\left(\sqrt{z_0^2 + r^2} - z_0\right),
\end{equation}

wrapped into the range $[0, 2\pi)$. Applying the same pattern at a new frequency, $f$ produces an effective time delay

\begin{equation}
\tau'(r) = \frac{\phi_c(r)}{2\pi f} = \frac{f_0}{f}\frac{\sqrt{z_0^2 + r^2} - z_0}{c},
\end{equation}

Therefore, the quadratic phase coefficient becomes $(f_0/f)$ times smaller. The new on-axis focus $z(f)$ satisfies

\begin{equation}
z(f) = \frac{f_0}{f}z_0,
\end{equation}

\begin{figure}[htbp]
\centering
\includegraphics[width=\textwidth]{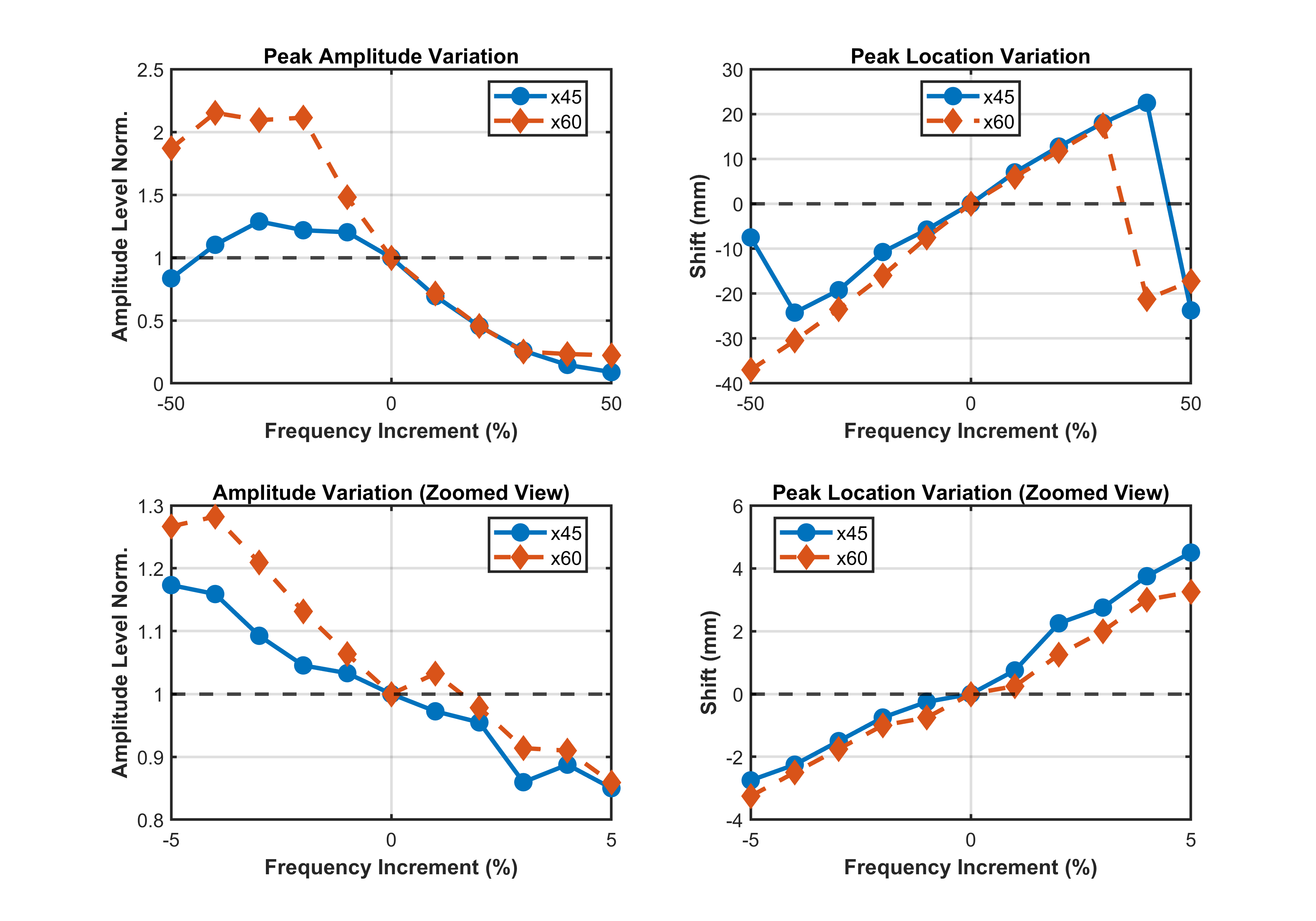}
\caption{Peak pressure and location versus frequency error ($\pm 50\%$) for both nominal foci.}
\label{fig:s2}
\end{figure}

Using simulations, we varied the frequency of excitation for two focusing configurations with phase-only lenses computed from the time of flight for a focus at 45 mm ($F\#0.75$) and 60 mm ($F\#1.0$) and observed the effect on focusing. The wavelength changes scale the beamwidth and depth-of-field inversely with $f$; a high frequency tightens and attenuates the beam, whereas a low frequency broadens and deepens it, moving the focus away from its intended position (Eqn. S5). The analysis suggests that for $\pm 5\%$ frequency shifts (which is the range of our registration trial with $f \pm \Delta f/2$ the max shift is comparable to 2-3 wavelengths in water at 1 MHz (Fig \ref{fig:s2} bottom right and Table \ref{table:s2}).

\begin{table}[htbp]
\centering
\caption{Peak Amplitude and Location Statistics}
\label{table:s2}
\begin{tabular}{lccccc}
\hline
Configuration & Min & Max & Range & Std Dev & Std Dev \\
\hline
\multicolumn{6}{c}{Peak Amplitude (Norm.)} \\
x45 & 0.850 & 1.173 & 0.323 & 1.003 & 0.112 \\
x60 & 0.859 & 1.282 & 0.423 & 1.059 & 0.147 \\
\multicolumn{6}{c}{Peak Location (mm)} \\
x45 & -2.750 & 4.500 & 7.250 & 0.591 & 2.432 \\
x60 & -3.250 & 3.250 & 6.500 & 0.045 & 2.159 \\
\hline
\end{tabular}
\end{table}

\subsection{Limitations and Failure Modes of the HASA-ADAM Framework}
\label{sec:hasa_limitations}
The HASA-ADAM framework relies on several physical and mathematical approximations. We first isolate its primary failure modes to define the operational bounds of this method: (1) compressional amplitude apodization and shear mode conversion, driven by steep topographical gradients in the free field, and (2) internal reflections and phase discontinuities induced by the highly heterogeneous human skull.

\subsubsection{Free-Field Limitations: Compressional Apodization and Shear Mode Conversion}
Projecting complex holograms (e.g., the GT Bee target) or tight focal spots requires steering acoustic energy at large off-axis angles. The principle of superposition requires that every point across the active transducer aperture contributes to the pressure field at the target plane. For an acoustic ray originating at a lateral aperture position $x$ and targeting a focal point at depth $Z$, the required steering angle relative to the optical axis is $\theta = \arctan(|x|/Z)$.

To steer via refraction from the solid lens ($c_{lens}$) into the water medium ($c_w$), the lens surface must possess a specific topographical slope. Let $\alpha$ be the angle of the surface normal relative to the incident longitudinal wave; as per Snell's law, the required slope ($|\nabla h| = \tan \alpha$) is governed by:
\begin{equation}
\tan \alpha = \frac{\sin \theta}{\cos \theta - c_w/c_{lens}}
\label{eq:lens_slope_exact}
\end{equation}

This relationship reveals two failure modes at high numerical apertures. Decreasing the focal depth $Z$ increases the required steering angle rapidly $\theta$. This forces the optimizer to generate steep slopes at the lens periphery.

\begin{figure}[htbp]
\centering
\includegraphics[width=\textwidth]{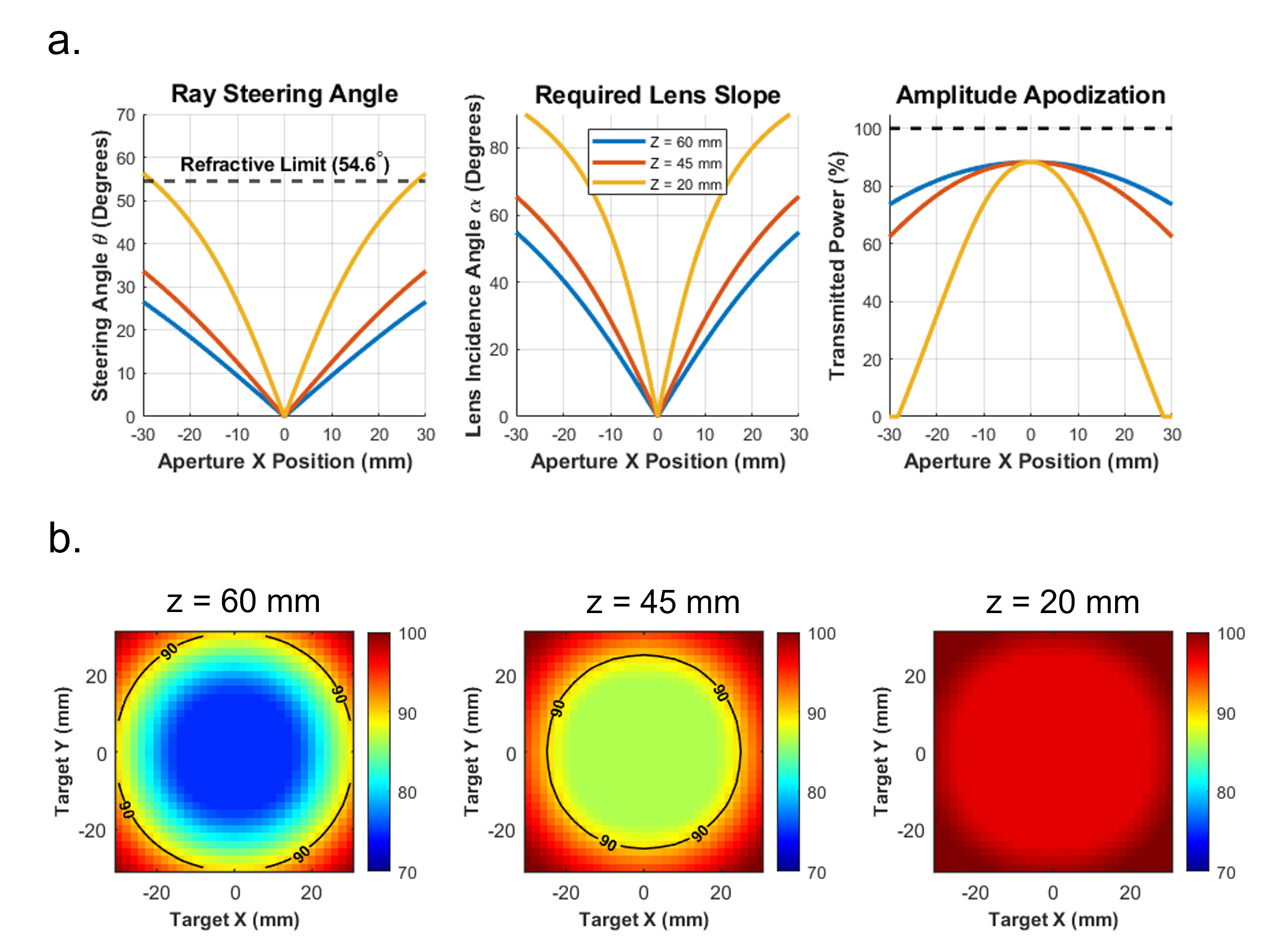}
\caption[Compressional and Shear Figure of Merit Analysis.]{Combined free-field failure modes at high numerical apertures. \textbf{(a) Compressional Limits (1D):} Focusing rays to a close focal plane ($Z=20$ mm) forces peripheral steering angles to approach the refractive limit ($\theta_{max} = 54.6^\circ$). The steeper lens slope causes a drop in transmitted power (Fresnel Apodization). The HASA algorithm erroneously models this as 100\% transmission. \textbf{(b) Shear Limits (2D FOM):} Geometric incidence analysis shows the percentage of the aperture exceeding the shear critical angle ($\theta_c \approx 30^\circ$) for every point on the target plane. Moving the target plane closer increases the shear risk at the periphery. This reduces the effective aperture.}
\label{fig:fom_analysis}
\end{figure}

\textbf{1. Compressional Fluid Limit (Fresnel Apodization):} As $\theta$ increases, the denominator in Eq.~\ref{eq:lens_slope_exact} approaches zero. This imposes an absolute refractive limit: $\theta_{max} = \arccos(c_w/c_{lens})$. For our 3D-printed lens ($c_{lens} \approx 2590$ m/s) in water ($c_w \approx 1500$ m/s), $\theta_{max} \approx 54.6^\circ$. Our 1D analysis (Fig.~\ref{fig:fom_analysis}a) shows that moving the focal plane to $Z=20$ mm pushes peripheral rays above this limit. This makes single-interface refraction impossible. Steep incidence angles trigger additional amplitude attenuation via Fresnel reflection. Because the one-way HASA forward propagator forces a reflection coefficient of zero ($R=0$), it overestimates the forward energy transmission through the aberrating layer(see appendix~\ref{app:parabolic_HASA_equation} for derivations). In other words, the portion of the lens at the periphery becomes inactive and does not contribute to the hologram at the focal plane. This reduces the effective numerical aperture, which is also known as apodization, leading to a loss of peak focal pressure.

\textbf{2.Shear Limit:}   Another mode of failure is the conversion of compressional waves to shear mode: High oblique incidence due to steeper slopes ($\alpha$) triggers substantial energy partitioning. Above the critical angle ($\theta_c \approx \arcsin(c_w / c_{s,lens}) \approx 30^\circ$), a large fraction of the incident acoustic energy converts into transverse shear waves (S-waves).  HASA, as a scalar fluid model ($\mu = 0$), cannot account for shear-mode conversion.  We define a geometric \textbf{Figure of Merit (FOM)} to quantify this limitation. It is defined as the percentage of the transducer aperture that requires a lens slope to exceed the critical shear-conversion threshold. As the target plane is moved to $Z=20$ mm, this FOM increases at the periphery (Fig.~\ref{fig:fom_analysis}b), mapping the spatial region where the fluid-based optimizer degrades. The result is that at extreme steering, the effective aperture that would constructively interfere is reduced, and unaccounted-for shear-mode energy leads to phase decorrelation.

\subsubsection{ Quantification of Free-Field Limits}
Let's focus on the fidelity of holographic reconstruction results from the free-field optimizations. We evaluated them using full-wave simulations (k-wave toolbox \cite{treeby2010k}) at multiple focal depths ($Z = 20, 45,$ and $60$ mm), spatial resolutions (point spacings from 2.5 to 40.0 mm), and target geometries (isolated 4-point targets and the continuous GT Bee pattern).

\begin{figure}[htbp]
\centering
\includegraphics[width=\textwidth]{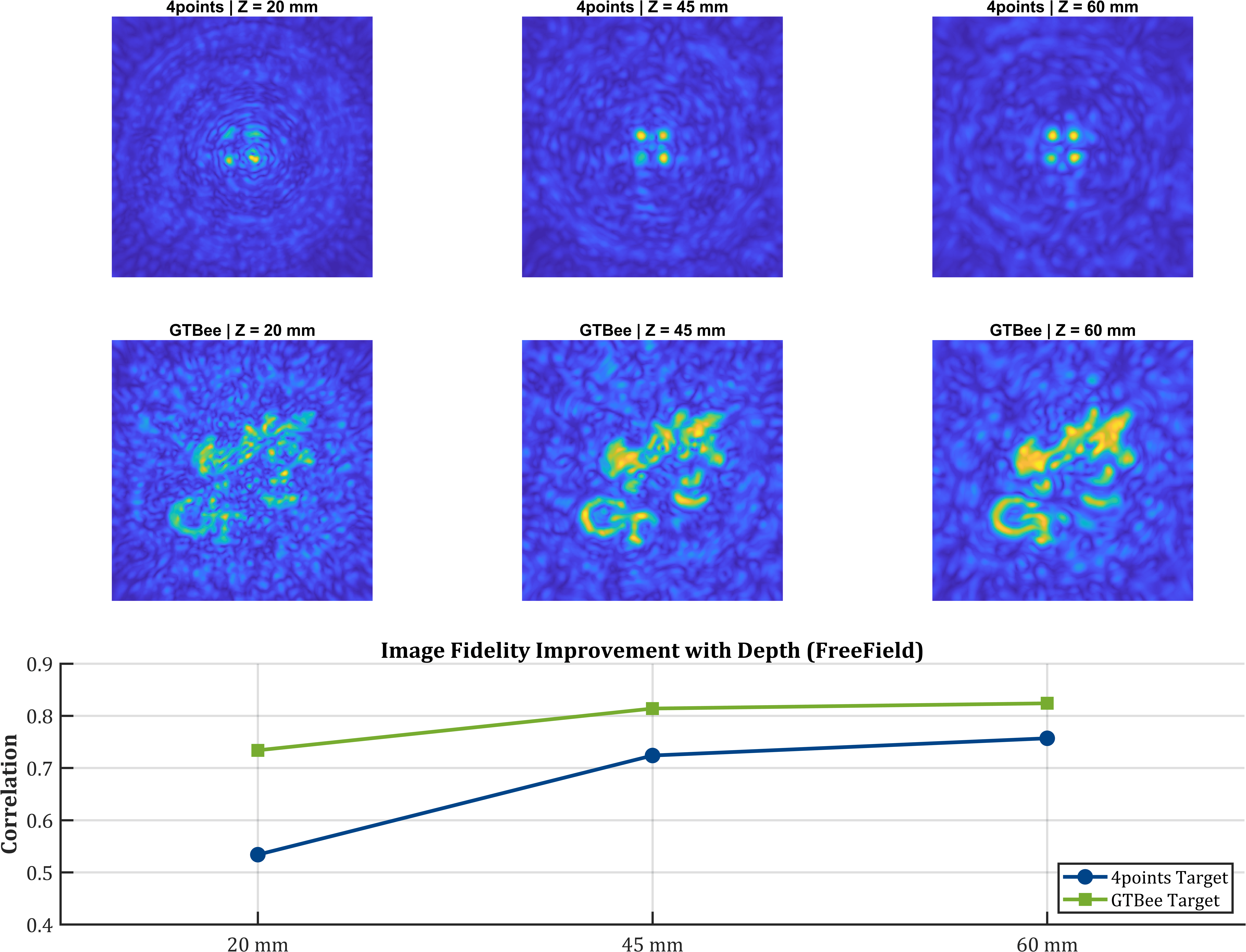}
\caption[Image Fidelity vs. Focal Depth in Free-Field Conditions.]{Empirical validation of compressional limits. Both the 4-point and GT Bee targets show improved correlation as the focal plane moves deeper (from $20$ mm to $60$ mm). The degradation at $Z=20$ mm aligns with the failure modes. The peripheral rays exceed the absolute refractive limit. They undergo significant Fresnel amplitude apodization and unmodeled shear-mode conversion.}
\label{fig:fidelity_vs_depth}
\end{figure}

The key insight from all the 2D optimization configurations is that near-field focusing ($Z=20$ mm)  forces the optimizer into more challenging steering regimes for our algorithm. This results in structural degradation (i.e., drop in correlation coefficient) and is consistent with refractive limits and Fresnel apodization. Quantitative analysis at this depth showed the highest overall error (MSE $> 0.34$) and the weakest structural definition (Correlation $\approx 0.534$ for the 4-point target). Moving the focal plane deeper to $45$ mm and $60$ mm relaxed the required steep steering angles. A deeper focusing plane reduces amplitude apodization and mode-conversion limits and results in a sharp recovery of image fidelity, increasing correlation by approximately $0.20$ for both target patterns. We also see a lower free-field MSE of $0.155$ at $Z=60$ mm (Fig.~\ref{fig:fidelity_vs_depth}). The deterioration we observed does not necessarily pose a strict limitation in a clinical setting; we can always steer the transducer so that the focal plane is in the far field, or use smaller-aperture transducers.

Another key factor is the portion of the aperture covered by the hologram in the focal plane. This decides whether the resulting pattern is a point target or a 2D image at two extremes.
Evaluating the effect of spacing between four-point targets across these depths revealed a sensitivity profile for effective aperture limits.

\begin{figure}[htbp]
\centering
\includegraphics[width=\textwidth]{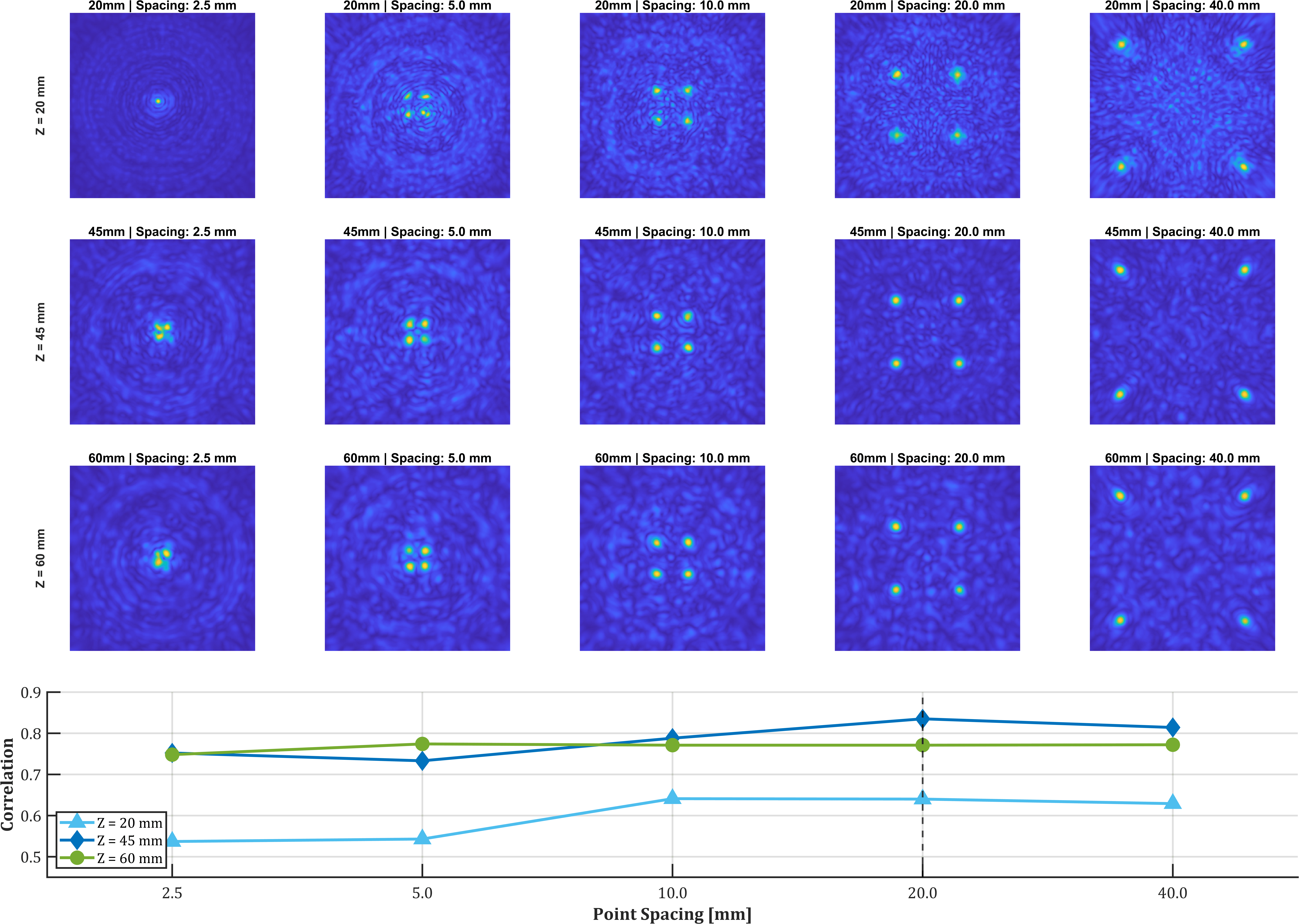}
\caption[ Sensitivity to Target Resolution Across Focal Depths.]{Spacing sensitivity across focal depths. The $45$ mm focal plane show a distinct performance curve, peaking optimally at a $20$ mm point spacing (Correlation $0.835$). The deeper $60$ mm plane shows high focal-plane tolerance, which has a flat performance profile across a wide range of target distributions. This confirms that the relaxed steering angles keep the rays within the high-transmission regime.}
\label{fig:spacing_sensitivity}
\end{figure}

At the optimal $Z=45$ mm focal plane, performance peaked at a $20.0$ mm point spacing (achieving an absolute peak system correlation of $0.835$), but degraded at tighter spacings ($<5.0$ mm). This is primarily due to acoustic diffraction. At wider spacings ($40.0$ mm) due to peripheral amplitude loss from extreme off-axis steering, the performance also dropped.(Fig.~\ref{fig:spacing_sensitivity}). The deeper $Z=60$ mm plane provided greater tolerance to spatial variation, although it did not achieve the absolute peak resolution of the $45$ mm plane. The reduced steering angles nonetheless allowed the optimizer to maintain a stable correlation profile ($\sim 0.771$ to $0.774$) across point spacings from $5.0$ mm to $40.0$ mm. Thus, it effectively avoided near-field failure modes.

\subsubsection{Transcranial Limitations in HASA-ADAM Optimization}
The HASA wave propagator is based on a parabolic (one-way) approximation of the heterogeneous Helmholtz equation. The second-order axial spatial derivative is neglected in the heterogeneous Helmholtz equation: $\frac{\partial^2 P}{\partial z^2} \approx 0$
This assumes that backscattering effects can be neglected. But, the  large difference in acoustic impedance between the water coupling medium ($Z_w \approx 1.5$ MRayl) and the dense cortical bone ($Z_b \approx 5.5 - 6.0$ MRayl) means that a reflection coefficient ($R$) is large:
\begin{equation}
R = \frac{Z_b - Z_w}{Z_b + Z_w} \approx 0.57
\end{equation}

This shows that about $R^2 \approx 32\%$ of the incident acoustic energy is reflected back. The multi-layered structure of the skull traps sound waves, producing complex reverberation within it. The one-way approximation in the HASA model imposes $R = 0$. It overestimates transcranial transmission and fails to account for phase decorrelation caused by internal standing waves. Additionally, the HASA convolution step ($\Lambda * P$) is derived under the slowly varying envelope approximation (also known as WKB approximation, see appendix~\ref{app:wkb_derivation} for details). \cite{schoen2019heterogeneous}. This requires the spatial variation of the medium's speed of sound to change slowly relative to the acoustic wavelength:
\begin{equation}
\frac{1}{k_0} \left| \frac{\nabla c}{c} \right| \ll 1
\end{equation}
 The speed of sound discontinuously jumps from $\sim 1480$ m/s to over $2500$ m/s within a sub-millimeter distance at the water-skull boundary. This spatial step-function violates the continuity assumption, introducing spectral leakage and artificial phase accumulation errors during HASA's spatial Fourier transforms. The strong heterogeneity of the skull may break the assumption on which the HASA model was built.

The skull's geometry may also cause shear conversion. Due to local anatomical curvature, approximately 50\% of the illuminated skull surface presents an incidence angle exceeding the critical threshold ($\theta_c \approx 30^\circ$) even under perfect registration (Chapter 3). Thus, the projection of high-spatial-frequency patterns through a strongly aberrating skull imposes a hard boundary condition. The fluid-based optimizer used here degrades due to multiple reflections, spectral leakage, phase discontinuities, and shear errors.

\begin{figure}[htbp]
\centering
\includegraphics[width=1.0\textwidth]{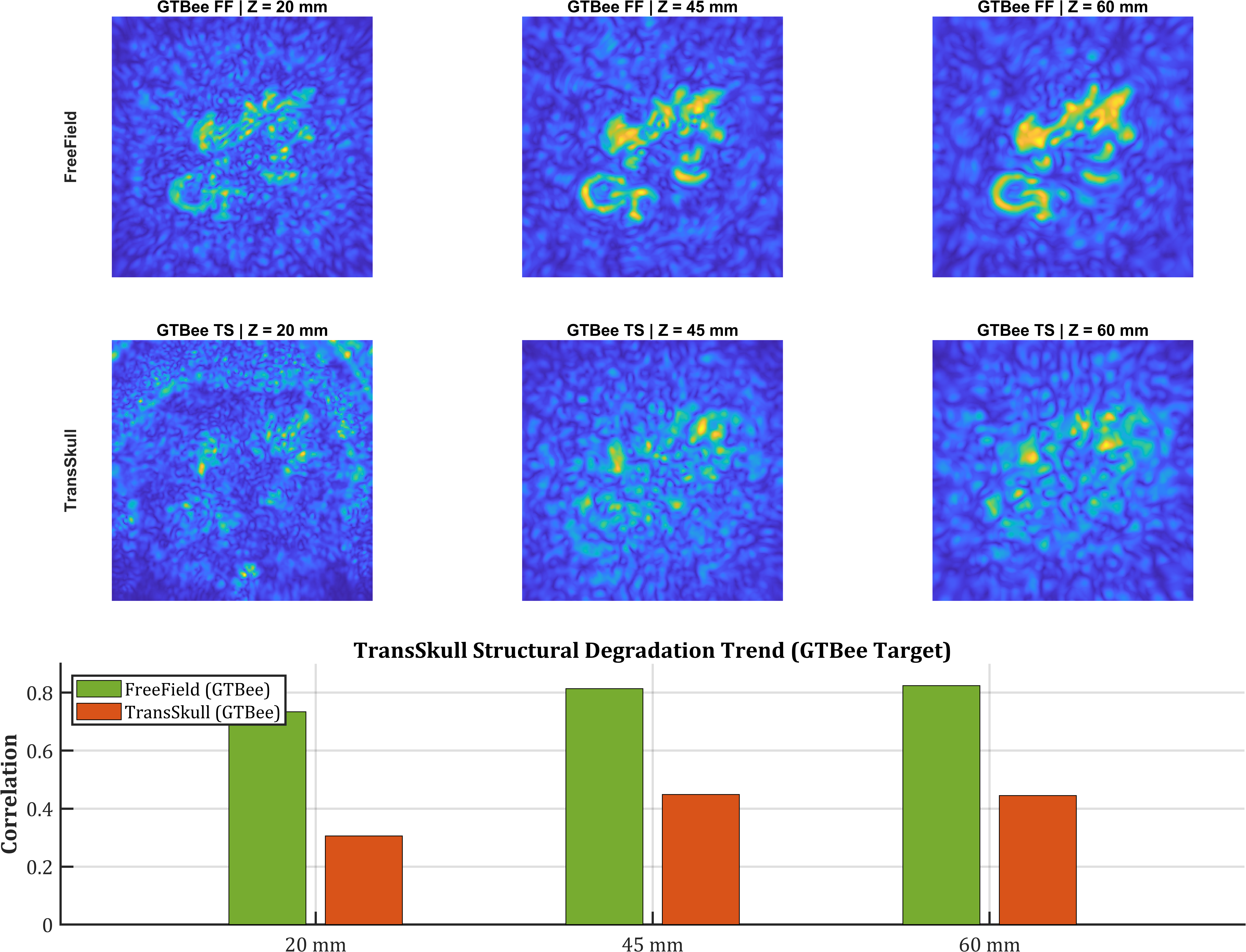}
\caption [Transcranial structural degradation of the high-spatial-frequency GT Bee target pattern.]{Transcranial correlation degradation of the high-spatial-frequency GT Bee target pattern. (Top) Simulated acoustic pressure fields demonstrating the target reconstruction in FreeField (FF) versus TransSkull (TS) conditions across varying axial depths ($Z = 20, 45,$ and $60$ mm). The Fluid-based HASA-ADAM optimizer accurately predicts the complex pattern in a homogeneous medium; focal blurring and distortion occur when it passes through the skull. (Bottom) correlation metrics highlighting the significant drop in fidelity. This degradation visualizes the failure of the HASA model's one-way parabolic and slowly varying envelope approximations when subjected to the heterogeneous skull for complex 2D targets.}
\label{fig:fail_transcranial_bee}
\end{figure}

\begin{figure}[htbp]
\centering
\includegraphics[width= 1.0\textwidth]{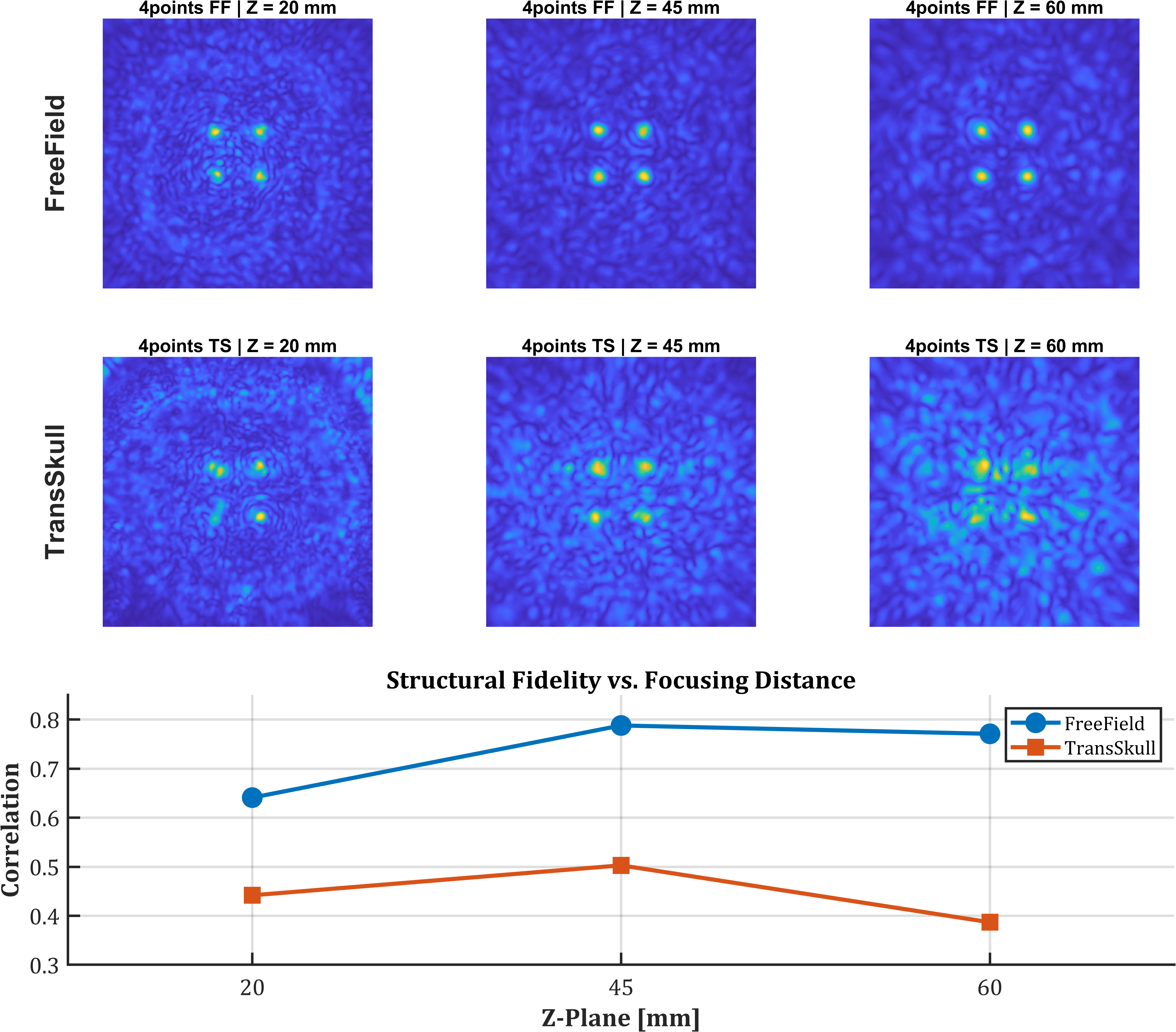}
\caption [Acoustic field reconstruction and structural fidelity for the 4-points target with a fixed 10.0 mm point spacing. ]{Acoustic field reconstruction and structural fidelity for the 4-points target with a fixed 10.0 mm point spacing. (Top and Middle) Simulated 2D maximum pressure planes ($p_{max}$) demonstrating target reconstruction in FreeField versus TransSkull conditions at axial depths of $Z = 20, 45,$ and $60$ mm. (Bottom) Quantitative trends comparing the correlation of the simulated fields against the ideal target. The Free Field condition demonstrates excellent focal fidelity that stabilizes at deeper planes (Correlation $> 0.77$); the introduction of the skull causes phase distortion and signal attenuation, particularly in the near-field ($Z = 20$ mm).}
\label{fig:4points_10mm_trend}
\end{figure}

Let's evaluate how focal depth and transcranial propagation interact using the previous example of four-point targets. Under Free-Field conditions, the optimizer resolves focal spots effectively at $Z = 45$ mm. The beam naturally widens at deeper planes. However, introducing the skull boundary degrades correlation across all depths. The skull blurs the focus. It also introduces additional sources of error. The optimal 45 mm depth focal plane case retains a fraction of its original shape (Correlation $\sim 0.50$).  At shallower focal planes(20 mm), the effects of near-field blurring and skull-induced phase aberrations reduce the correlations drastically (Fig.~\ref{fig:4points_10mm_trend}). This indicates that the phase distortions generated by the cranial barrier not only degrade resolution but also penalize targets positioned too close to the internal bone interface. These failure modes are worst when projecting complex targets. This leads to a major drop in correlation of the GT Bee pattern (Fig.~\ref{fig:fail_transcranial_bee}).

\section{Discussion}
We showed that the HASA-ADAM framework generates acoustically thick lens topologies that account for thickness-dependent effects, including amplitude errors, scattering, and edge diffraction. HASA-ADAM achieves this by implementing the physics of wave propagation within the lens topology. Such consideration is more important at sub-MHz frequencies, where the thin-element approximation (i.e., treating the lens as an ideal flat 2D phase screen) fails. Using HASA-ADAM, we achieve improved performance, including reduced sidelobes during trans-skull focusing (Fig.~\ref{fig:trans_skull}) compared to optimized phase-only methods. This capability is an improvement in acoustic holography, enabling the generation of holographic fields with fidelity that thin-element-lens-based systems and commercially viable phased arrays can not achieve (Fig. ~\ref{fig:hasa_adam_framework}). The framework, as such, can also be extended to use with dense or sparse phased arrays \cite{Kilinc2025Piezo} for transcranial applications.

We have used automated differentiation with the ADAM optimizer to reduce the risk of getting stuck in local minima and maintain robust, uniform focal accuracy. But the wave propagator has limitations that define the operational boundaries of our HASA-ADAM formulation. The underlying wave propagator for HASA-ADAM is built on the one-way wave equation. This means it does not account for multiple internal reflections (backscattering) or shear mode conversions within the cranial bone. For relatively simple targets, such as single-point focusing, the acoustic energy is steered at near-normal incidence, thereby minimizing shear-mode conversion at the water-lens interface. In these cases, the performance of HASA is only limited by diffraction. Projecting complex holographic patterns (e.g., the GT logo) through a highly aberrating skull, on the other hand, requires higher spatial frequencies and steeper angular steering gradients. As oblique acoustic rays incident on the skull near or beyond the critical angle ($\theta_c \approx 30^\circ$), a substantial percentage of energy is diverted into unmodeled shear waves. HASA-ADAM relies on a scalar fluid formulation and is blind to these solid-mechanics phenomena. It cannot pre-compensate for them. This represents a failure mode that bounds the fidelity of highly complex holographic projections through thick cortical bone.

In terms of computational speed, HASA-ADAM bridges the gap between computationally intensive methods such as time reversal ~\cite{sallam2024gradient,angla2023transcranial} and trained deep learning frameworks~\cite{bu2024deep,li2022acoustic,lee2022deep} that are opaque and reliant on extensive training datasets. {While methods such as MMDM~\cite{gu2020modified} offer high accuracy by including reflections (which HASA currently neglects), HASA's formulation is highly amenable to automatic differentiation, offering a significant computational advantage for iterative optimization. It provides a favorable balance of speed and accuracy for the forward optimization problem.} Although directly optimizing the thickness is more challenging than phase optimization and thus requires approximately twice as long to converge ($\sim$15 min at $\lambda/6$ discretization for 500 iterations), with improved tuning of the hyperparameter of optimization and initial conditions, the convergence can be further accelerated.
Likewise, hybrid strategies that merge the speed of deep learning methods with HASA's accuracy and interpretability of wave propagation of HASA with GPU acceleration can further mitigate the computational cost associated with the $O(N^2\log N)$ scaling, allowing real-time implementations, which can be desirable in some applications~\cite{jiang2022flexible,naor2012towards}.

Despite the remarkable performance of the proposed framework for designing holographic lens topologies, we observed discrepancies between the optimization results, k-wave simulations, and hydrophone scans, especially for high-fidelity holograms in the free field (Fig. ~\ref{fig:hasa_adam_framework}c and d). {As investigated in Section 5, these discrepancies primarily originated from \textsl{elastic mode conversion within the lens material} and uncertainties in the material properties of the 3D-printed lens.} First, the reported 2590 m/s group velocity~\cite{bakaric2021measurement} may not accurately represent the 3D-printed sample (using Clear White v4 resin) owing to curing-induced density variations and internal stresses. {Characterization of the specific sample used and subsequent updating of the simulations largely reconciled these differences.} Second, the manufacturing tolerances of $\pm$0.05-0.1 mm ($\sim$5\% of thickness)~\cite{lagerburg2025dimensional} may introduce additional timing errors. Another potential source of error is related to variations in the speed of sound (e.g., from the lens to the water), which, for the current HASA implementation, cannot be very high relative to the wavelength~\cite{schoen2020heterogeneous}.

Furthermore, the current HASA implementation does not account for absorption (see Methods), which may affect the optimization accuracy, particularly for highly attenuating media. We acknowledge this as a limitation that may be addressed in future iterations. Finally, understanding the impact of multiple reflections, which are neglected in the currently implemented HASA algorithm, and their role in optimization convergence~\cite{stanziola2023physics} may enable further improvements in hologram quality. Despite these potential sources of error, the HASA-ADAM framework delivers a balanced approach to scalable, rapid, and accurate hologram design. It also accommodates patient-specific skull variations, enabling transcranial targeting of specific brain regions during treatment planning.

\section{Conclusions}

This chapter resolves a major bottleneck in acoustic hologram design.  We established a framework for volumetric lens topology optimization that overcomes the limitations of the phase-based thin-film approximation to generate high-fidelity, patient-specific lenses in the sub-megahertz regime. This approach yields four primary contributions:\\ \\

\textbf{Failure Thin-Element Approximation in the Thick-Lens Regime:} We showed that at frequencies needed for transcranial applications ($\sim$1~MHz or lower), acoustic holograms operate in an acoustically thick regime where phase-to-thickness conversions fail. The resulting refractive walk-off and transverse energy migration degrade holographic fidelity. Our HASA-ADAM topology optimizer pre-compensates for these volumetric diffraction effects during the design.

\textbf{ Shear Mode Conversion:} We postulated the mechanisms driving the background acoustic haze observed in our 2D complex holographic experiments.

Full-wave elastic ablation studies indicated that this haze results directly from longitudinal-to-shear mode conversion at the lens's steep interfaces. It causes phase aberration and refractive steering.  Actively modeling this shear wave propagation dropped the experimental reconstruction error by 29.1\%.

\textbf{Validation of Patient-Specific Transcranial Aberration Correction:} We showed the robustness and repeatability of the HASA-ADAM framework across highly variable human skull segments (S1, S2, and S3). In every case, the patient-specific topologies improved the dominant acoustic failure modes. The lenses corrected severe spatial misalignments (reducing targeting error by 68\% in S1), recovered lost acoustic pressure (driving a 37\% increase in peak gain in S2), and forced diffused energy back into tightly confined focal spots.

\textbf{Operational Optimization Boundaries:} We mapped the failure modes of our HASA-ADAM holographic optimization. We conclude that forcing the system to generate high numerical apertures (e.g., $Z=20$~mm) triggers Fresnel apodization and mode conversion. Finally, we bounded the framework's transcranial limitations to unmodelled multiple reflections ($R \approx 0.57$) and violations of the slowly varying envelope approximation.

\chapter{Hologram Registration}
\label{ch:hologram_registration}
\footnotetext{An earlier version of the work presented in this chapter is available as a preprint on arXiv (https://arxiv.org/abs/2508.07103).\cite{Dash2025}}

\section{Introduction}

Our ability to reconfigure the potentially disruptive technology of acoustic holography for biomedical applications such as transcranial ultrasound also depends on our ability to accurately register the holographic lens to the patient's anatomy, orientation, and position relative to the transducer/lens plane~\cite{maimbourg2018printed,jimenez2024feasibility}. For instance, complex hologram designs and pressure field topologies, which require a higher operation frequency ($\geq 0.7$ MHz) and skull-compensating lens topologies, are very sensitive to lens-skull misalignment~\cite{andres2022numerical}. Hence, high-quality registration (i.e., sub-wavelength accuracy, $<$1.5 mm at 1 MHz) is required to preserve the fidelity and targeting accuracy. Unfortunately, {current non-MRI-based registration methods (e.g., standard neuronavigation) typically achieve an accuracy of only $\sim$2 mm, which} can lead to targeting errors of a few millimeters (i.e., 1–2 wavelengths) and reduced performance~\cite{choi2024neuronavigation,chen2021neuronavigation,wei2013neuronavigation,pouliopoulos2020clinical}. Therefore, robust and accurate registration strategies that can accurately align the lens to the patient's skull anatomy are critical for designing cost-effective and portable transcranial ultrasound (TUS) systems for high-precision (i.e., subwavelength) neurointerventions.

To address this critical challenge of registration, we hypothesize that nonlinear acoustic effects can be leveraged for noninvasive lens–skull registration. \cite{Dash2025Leveraging} The hypothesis was tested through modelling and experiment. A parametric array (PA) signal, the low-frequency tone produced when two high-frequency beams mix nonlinearly~\cite{westervelt1963parametric, Dash2021Non}, serves as the registration metric. We investigations suggest that a misaligned (aberrated) lens augments the finite-amplitude wave propagation effects within a highly nonlinear skull, giving rise to a strong PA signal that can penetrate the skull with minimal losses. Thus, minimizing the PA signal leads to an effective acoustic feedback mechanism for noninvasively aligning the holographic lens with the skull.

In the following sections, we summarize the rationale and theory behind hologram registration using nonlinear PA feedback, followed by simulation and experimental results, as well as a sensitivity analysis. They help us explore the limitations of this method and draw key insights for clinical applications.

\section{Methods}

\subsection{Mechanism of Lens Registration}

To establish the theoretical basis for our registration strategy, we modeled the interaction of finite-amplitude ultrasound waves within a medium of spatially varying nonlinearity. We utilize the Westervelt equation to model the generation of the difference frequency component, where the finite-amplitude primary waves act as a volumetric driving source.

\paragraph*{The Nonlinear Source Term}
Under the quasilinear approximation, the secondary difference frequency field $p_{\Delta f}(\mathbf{r}, t)$ is driven by a {virtual volumetric source density}, $S_{NL}$:
\begin{equation}
\nabla^2 p_{\Delta f} - \frac{1}{c_0^2}\frac{\partial^2 p_{\Delta f}}{\partial t^2} = -\underbrace{\frac{\beta(\mathbf{r})}{\rho_0 c_0^4} \frac{\partial^2}{\partial t^2} \langle p_{primary}^2 \rangle}_{S_{NL}(\mathbf{r}, t)}
\label{eq:source_term}
\end{equation}
Where:
\begin{itemize}
\item $\beta(\mathbf{r})$ is the spatially dependent coefficient of nonlinearity.
\item $\langle p_{primary}^2 \rangle$ is the envelope of the squared primary pressure field.
\item $S_{NL}$ represents the local strength of nonlinear generation.
\end{itemize}

The amplitude of the parametric signal is proportional to the volume integral of the source term over the interaction domain $V$:
\begin{equation}
P_{\Delta f}(\mathbf{r}_{obs}) \propto \int_V \frac{\beta(\mathbf{r})}{\rho_0 c_0^4} \left| p_{primary}(\mathbf{r}) \right|^2 G(\mathbf{r}, \mathbf{r}_{obs}) \, dV
\label{eq:integral}
\end{equation}

\paragraph*{The Material Contrast Mechanism}
To determine the sensitivity of this method, we analyzed the significant contrast in the nonlinearity parameter $\beta$ between the skull bone and the surrounding soft tissue.
\begin{itemize}
\item \textbf{Soft Tissue / Water:} $\beta \approx 3.5 - 4.5$.
\item \textbf{Skull Bone:} $\beta \approx 188$ (derived from $B/A \approx 374$).
\end{itemize}
Because $\beta_{skull} \gg \beta_{tissue}$, the integral in Eq. \ref{eq:integral} is primarily dominated by the volume of the skull illuminated by high-intensity ultrasound.

Two limiting cases bound the expected PA emission:
\begin{enumerate}
\item \textbf{Misaligned:} In this state, the lens no longer corrects the skull's aberration profile. Primary-beam energy scatters and reverberates inside the porous bone (high $\beta$). The product $\beta(\mathbf{r})\cdot|p_{\text{primary}}|^{2}$ is large throughout the bone volume, so the skull radiates a strong PA signal.

\item \textbf{Aligned:} Phase aberration due to the skull are compensated; constructive interference occurs \textit{beyond} the skull. Primary energy propagates through the bone quickly and focuses in low-$\beta$ brain tissue. The overlap integral (Eq.~\ref{eq:integral}) is minimized under this condition.
\end{enumerate}

\subsection{Practical Implementation: Double-Layer Propagation}

\paragraph*{Frequency-Dependent Transmission}
For external detection, the PA signal must cross the distal skull layer. The primary beam ($f_0 \approx 1$~MHz) after the focal plane attenuates heavily on this second pass ($\alpha \propto f^{b}$), but the 100~kHz difference frequency crosses the distal skull with negligible loss.
\begin{itemize}
\item At $f_0 = 1$ MHz, attenuation is high ($\sim 15$ dB/cm). The primary beam was effectively filtered out by the exit layer.
\item At $\Delta f = 100$ kHz, attenuation is negligible ($\sim 0.15$ dB/cm).
\end{itemize}
Therefore, the PA signal generated at the entry layer propagates through the brain and exits the skull layer with minimal energy loss.

\subsection{Volumetric Analysis of Energy Trapping}
To rigorously confirm that the registration signal arose from the acoustic energy trapped within the skull, we performed a 3D volumetric analysis of the primary pressure fields obtained from the k-Wave simulations.

The nonlinear source term $S_{NL}$ is proportional to the square of the primary pressure inside the bone ($S_{NL} \propto \beta_{skull} |p_{primary}|^2$). We defined the skull interaction volume via the acoustic impedance map ($c > 1700$ m/s) and calculated the \textbf{Integrated Source Potential} ($E_{trapped}$) specifically within the bone matrix as follows:
\begin{equation}
E_{trapped} = \int_{V_{skull}} |p_{primary}(\mathbf{r})|^2 \, dV
\end{equation}

\subsection{{Calculation of Elastodynamic Energy Partitioning and Parametric Amplification}}

To assess whether mode conversion played a significant role in the angular sensitivity of the acoustic feedback, we calculated the elastodynamic energy partitioning and parametric amplification within the solid skull matrix.

\textbf{Geometric Incidence and Critical Angle.} The threshold for longitudinal wave transmission at the fluid-bone interface is analytically quantified using Snell's law for elastic media. Assuming that the speed of sound in the water coupling medium is $c_w \approx 1480$~m/s, and the longitudinal speed of sound in the cortical bone is $c_b \approx 2800$~m/s, the critical angle $\theta_c$ is
\begin{equation}
    \theta_c = \arcsin\left(\frac{c_w}{c_b}\right) \approx \arcsin(0.528) \approx 31.9^\circ
\end{equation}
For computational thresholding in the Shear Mode Risk analysis, this value was approximated as $30^\circ$.

\textbf{Stress Tensor and Energy Densities.} In the k-wave elastic solver, the macroscopic scalar pressure $P$ is derived exclusively from the trace of the stress tensor (normal stresses):
\begin{equation}
    P = -\frac{1}{2}(\sigma_{xx} + \sigma_{yy})
\end{equation}
To rigorously quantify the elastodynamic partitioning of acoustic energy within a solid skull matrix, the total energy density was separated into kinetic and potential components. The \textit{Kinetic Energy Density} ($w_k$), which captures the total particle motion (including both compressional and shear contributions), is defined via the peak particle velocity vector $\mathbf{u} = (u_x, u_y)$:
\begin{equation}
    w_k(\mathbf{r}) = \frac{1}{2} \rho(\mathbf{r}) \left( |u_x(\mathbf{r})|^2 + |u_y(\mathbf{r})|^2 \right)
\end{equation}
where $\rho(\mathbf{r})$ denotes the local mass density. The \textit{Potential Energy Density} ($w_p$), representing the energy stored purely in volumetric compression, is approximated using the peak scalar pressure $P$:
\begin{equation}
    w_p(\mathbf{r}) = \frac{1}{2} \frac{P(\mathbf{r})^2}{\rho(\mathbf{r}) c_L(\mathbf{r})^2}
\end{equation}
where $c_L$ is the speed of longitudinal sound. Because pure shear waves are isochoric, their energy is captured almost exclusively by the kinetic term $w_k$. The total trapped energy, $E_{total}$, is obtained by volume-integrating these densities over the spatial domain of the skull mask, $\Omega_{skull}$:
\begin{equation}
    E_{total} = \sum_{\mathbf{r} \in \Omega_{skull}} \left( w_k(\mathbf{r}) + w_p(\mathbf{r}) \right) \Delta x \Delta y
\end{equation}

\textcolor{black}{\textbf{Derivation of the $c^{-3}$ Scaling in Bone.} The difference-frequency pressure $p_{df}$ is derived from the Westervelt equation source term $S$~\cite{westervelt1963parametric,hamilton_nonlinear_2008}:
\begin{equation}
    \nabla^2 p_{df} - \frac{1}{c^2}\frac{\partial^2 p_{df}}{\partial t^2} = -\underbrace{\frac{\beta}{\rho c^4} \frac{\partial^2 p_p^2}{\partial t^2}}_{S}
\end{equation}
The local source strength scales as:
\begin{equation}
    S \propto \frac{\beta \cdot \Delta\omega^2}{\rho c^4}
\end{equation}
In the absorption-limited case (common in cortical bone), the total accumulated pressure $P_{df}$ is the integral of the source over the effective interaction length $L_{eff} = 1/\alpha$:
\begin{equation}
    P_{df} \approx \int_0^\infty S e^{-\alpha z}dz = \frac{S}{\alpha}
\end{equation}
Given that the absorption coefficient $\alpha$ in bone roughly scales with $1/c$ for a fixed frequency~\cite{clement_enhanced_2004}, the total efficiency $\eta$ scales as:
\begin{equation}
    \eta \propto \frac{1}{c^4} \cdot c = \frac{1}{c^3}
\end{equation}
Therefore, hypothetically, for shear waves ($c_s \approx 1400$ m/s) vs. longitudinal waves ($c_l \approx 2800$ m/s):
\begin{equation}
    \frac{\eta_s}{\eta_l} = \left(\frac{c_l}{c_s}\right)^3 = 2^3 = 8
\end{equation}
This derivation suggests that if fast longitudinal waves are mode-converted into slow shear waves, an approximately 8-fold generation efficiency boost could \textit{potentially} occur. However, we present this strictly as an exploratory hypothesis to help explain possible experimental discrepancies, rather than a definitively validated physical mechanism.}

\subsection{Robustness and Artifact Analysis Methodology}

\textbf{Robustness to Hydrophone Placement:} To determine whether the registration metric is robust to the positioning of the receiving hydrophone, a requirement for clinical translation, we evaluated its spatial invariance. We simulated a finite aperture receiver (e.g., piston hydrophone or ultrasound transducer) scanned across a region of interest $40-100$ mm axially and $\pm 25$ mm laterally behind the skull. To quantify the benefits of using larger detectors, we performed a parameter sweep by varying the receiver aperture from 2 mm to 30 mm.

\textbf{Pseudo-Sound Quantification:} To verify that the measured low-frequency signals originated from true parametric generation within the medium rather than from the nonlinear transfer function of the hydrophone, we quantified the contribution of pseudo-sound. Pseudo-sound arises from the nonlinearity of the hydrophone ($e \approx mp + \eta p^2$). We took advantage of near-field measurements taken at the focus, 60 mm from the focus, and 120 mm from the focus to quantify and isolate pseudo-sounds based on how true difference-frequency waves scale differently with range.

\textbf{Intrinsic Skull Nonlinearity vs. Microbubble Artifacts:} To investigate whether the observed Difference Frequency (DF) stemmed from the intrinsic classical cumulative nonlinearity of the bone matrix rather than artifactual resonant bubble nonlinearity from trapped gas, we analyzed the spectral content predicted by two competing models of nonlinearity. Degassed human skull samples were submerged in a water tank. A bichromatic excitation pulse ($f_1, f_2 \approx 1$ MHz, $\Delta f \approx 100$ kHz) was transmitted through the skull, and the resulting acoustic field was measured using a hydrophone along the propagation axis ($z$). A system governed by classical modular nonlinearity produces only integer linear combinations of input frequencies, whereas microbubble artifacts are characterized by bifurcation and the emission of subharmonics ($f_0/2$).

\textbf{Simulation of Intracranial Trapped Gas:} We simulated the effect of trapped gas within the porous structure of the skull on PA generation, to evaluate the robustness of the proposed acoustic lens system against realistic postoperative conditions (e.g., pneumocephalus). We used the k-wave toolbox to solve nonlinear coupled wave equations. A morphological void-filling algorithm assigned the acoustic properties of air to a specified volume fraction ($\phi_{gas}$) of the voids. The source was driven by bi-frequency excitation to generate a difference frequency ($\Delta f = 100$ kHz) at the target depth.

\section{Simulation and Experimental Results}

\subsection{Skull-compensating lens misalignment augments nonlinear wave propagation and parametric array signal}
We observed in Figure~\cref{fig:trans_skull} that transcranial focusing at 1 MHz is highly sensitive to registration errors.  We evaluated whether non-linear wave propagation through the human skull could serve as an active acoustic feedback mechanism to indicate the status of skull-lens alignment. Briefly, the parametric array effect is a nonlinear wave propagation effect~\cite{westervelt1963parametric}, where two (primary) high-frequency sound beams of finite amplitude interact to produce (secondary) sum- and difference-frequency beams. The strength of the difference frequency $|p_{\Delta f}|$, here termed the parametric array signal (PA signal), which has several unique properties, including high directionality and penetration through the skull, is proportional to the medium's nonlinearity parameter $\beta$, the square of the primary beam amplitude $p_{f_{1,2}}$, and their propagation length~\cite{hamilton1997nonlinear}. Considering the characteristics of the PA signal, we hypothesized that skull-compensating lens misregistration (i.e., suboptimal aberration correction) can affect the amplitude of $|p_{\Delta f}|$ which once detected and quantified (e.g., using a hydrophone) can provide a real-time feedback mechanism to noninvasively align the holographic lens to the patient's skull (Fig. ~\ref{fig:nonlinear_concept}a). Registration is performed before sonication by adjusting the lens pose until the PA signal is minimised.

\begin{figure}[!htbp]
\centering
\includegraphics[width=0.7\linewidth]{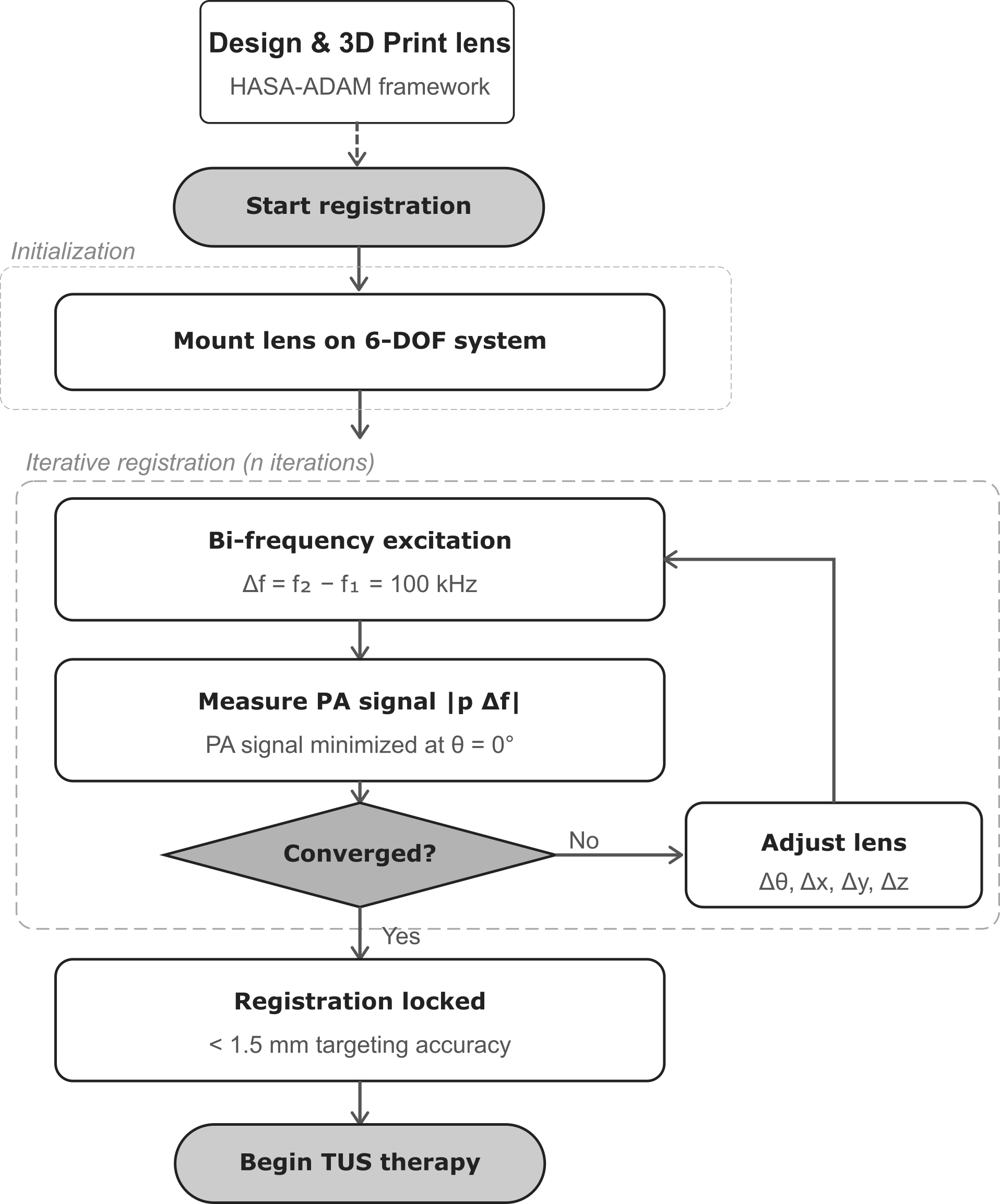}
    \caption[Process flow chart for trans-skull hologram registration in a clinical setting.]{\textbf{Process flow chart for trans-skull hologram registration in a clinical setting.} We mount a 3D-printed acoustic lens (fabricated via HASA-ADAM) on a 6-DOF stage and align it for transcranial ultrasound (TUS) therapy using an iterative process. At each iteration, we drive the transducer with bi-frequency excitation ($\Delta f = f_2 - f_1 = 100\,\text{kHz}$) and record the PA signal magnitude $|p_{\Delta f}|$. We seek a minimum at $\theta = 0^\circ$. If the loop has not converged, we apply positional and angular corrections ($\Delta\theta,\,\Delta x,\,\Delta y,\,\Delta z$) and repeat. Once converged, we lock the lens position and initiate TUS therapy, achieving a targeting accuracy of $ < 1.5\,\text{mm}$.}
\label{fig:flowchart_regd}
\end{figure}

In our fixture the pivot axis ($\mathbf{P}_{\text{pivot}} = [443,\,644,\,23]$) is offset from the transducer centre. A pure $z$-rotation therefore couples into lateral and axial translation. Tracking a surface voxel ($\mathbf{T}_{\text{initial}} = [1120,\,620,\,590]$) through $\pm6^\circ$ of rotation gives a 10.6~mm total displacement (Table~\ref{tab:voxel_shift}; Figure~\ref{fig:combined_mechanics}a). Even small angular errors produce large linear misregistration.

\begin{table}[!htbp]
\centering
\small
\caption{Geometric displacement of central voxel relative to pivot over the range $\pm 6^{\circ}$.}
\label{tab:voxel_shift}
\vspace{0.1cm}
\renewcommand{\arraystretch}{1.1}
\begin{tabular}{@{}c r r r r@{}}
\toprule
\textbf{Angle ($^{\circ}$)} & \textbf{$\Delta X$ (mm)} & \textbf{$\Delta Y$ (mm)} & \textbf{$\Delta Z$ (mm)} & \textbf{Total Euclidean Shift (mm)} \\
\midrule
-6 & -0.1800 & +10.6346 & 0.0000 & 10.6361 \\
-5 & -0.0727 & +8.8644& 0.0000 & 8.8647 \\
-4 & +0.0038 & +7.0925& 0.0000 & 7.0925 \\
-3 & +0.0492 & +5.3197& 0.0000 & 5.3199 \\
-2 & +0.0638 & +3.5462& 0.0000 & 3.5467 \\
-1 & +0.0474 & +1.7728& 0.0000 & 1.7734 \\
\textbf{0} & \textbf{0.0000} & \textbf{0.0000} & \textbf{0.0000} & \textbf{0.0000} \\
+1 & -0.0783 & -1.7717& 0.0000 & 1.7734 \\
+2 & -0.1875 & -3.5419& 0.0000 & 3.5467 \\
+3 & -0.3276 & -5.3098& 0.0000 & 5.3199 \\
+4 & -0.4985 & -7.0750& 0.0000 & 7.0925 \\
+5 & -0.7002 & -8.8370& 0.0000 & 8.8647 \\
+6 & -0.9326 & -10.5951 & 0.0000 & 10.6361 \\
\bottomrule
\end{tabular}
\end{table}

\begin{figure}[htbp]
\centering
\includegraphics[width=\textwidth]{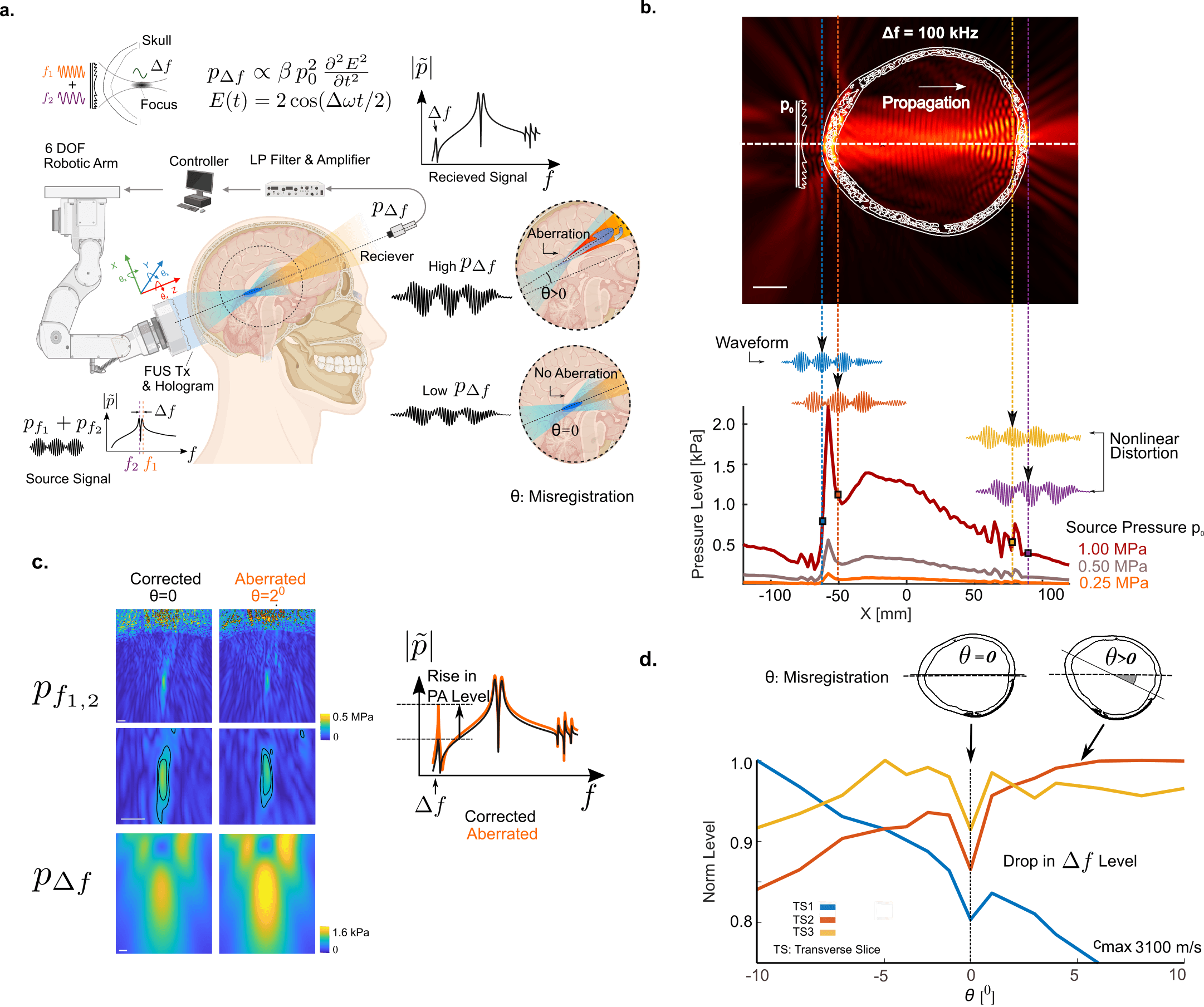}
\caption[Conceptual framework of skull-compensating lens registration using nonlinear acoustic feedback. ]{(a) Top: Nonlinear mixing of two high-frequency waves generated by a flat ultrasound (US) transducer attached to a hologram lens. At higher intensities, a low-frequency difference frequency arises from nonlinear steepening. This is known as the parametric array (PA) effect. Bottom: Schematic illustration of the hologram-assisted FUS therapy device integrated with a six-degree-of-freedom (6 DOF) robotic arm. The system utilizes standard acoustic coupling (e.g., a gel or water bolus). The receiver was placed at a fixed position relative to the head. The acoustic feedback from nonlinear parametric array signals is then used for accurate registration. (b) Top: 2D acoustic simulation demonstrating the generation of a 100 kHz parametric field within the skull cavity and its subsequent transmission through the skull cavity. Bottom: Waveforms showing increasing nonlinear distortion as primary waves attenuate due to skull-induced losses, and absolute pressure traces along the axial direction for varying source pressures at the transducer. (c) Top: (Row 1), Simulation of the primary ultrasound field with 3dB (in red) and 6 dB (in black) contour maps in the skull region showing local pressure maxima in the skull for skull rotation $\theta=0^\circ$ and $\theta=2^\circ$, (Row 2) with zoomed view of the focal region as ROI with 3dB and 6dB contours (both in black); Bottom: The parametric field with ($\theta=2^\circ$) and without ($\theta=0^\circ$) skull rotation indicating aberration leads to higher PA signal (plot on the right). (d) 2D simulation results illustrate a decrease in parametric pressure corresponding to zero skull rotation ($\theta=0^\circ$) for various transverse skull slices.}
\label{fig:nonlinear_concept}
\end{figure}

Nonlinear k-Wave simulations (Westervelt/PSTD) confirmed that the effect is detectable in a realistic skull geometry~\cite{treeby2010k}.

These models were run along with bone and tissue nonlinearity parameters from the literature~\cite{renaud2008exploration,pinton2011effects}. Using primary frequencies of 0.95 MHz and 1.05 MHz that resulted in 100 kHz difference frequency and pressures ranging from 0.25 to 1 MPa (safe exposure), we found that the parametric signal was immediately evident after the primary beams passed through the skull (Fig. ~\ref{fig:nonlinear_concept}b). Despite the formation of a weak standing wave ($\leq$1.5 kPa), which is expected owing to the difference in frequency used~\cite{baron2009simulation}, the parametric signal outside the skull was also evident and well within the detection limits of many piezoelectric detectors (Fig. ~\ref{fig:nonlinear_concept} (b), bottom). While the observed temporal profile is atypical of nonlinear propagation (i.e., the peak positive pressure tends to be higher), this is due to the accumulation of nonlinearities over extended propagation distances, combined with the substantially higher attenuation of the primary and secondary MHz-range fields. In a clinical geometry the signal must cross two skull layers. The primary MHz tones are heavily attenuated on the return pass; the 100~kHz PA signal passes through with minimal loss ($\alpha \propto f^{b}$). The skull itself acts as a low-pass filter that isolates the diagnostic tone.

Replacing the intracranial medium with water ($\beta_{\text{water}}=3.5$) or brain ($\beta_{\text{brain}}=4.45$) changed the primary field negligibly but suppressed the PA signal when skull nonlinearity was absent. Bone ($\beta \approx 188$) overwhelms the volume integral in Eq.~\ref{eq:integral}; the skull region under the primary beam dominates PA generation.

Building on these observations, we tested the impact of lens misregistration on $|p_{\Delta f}|$. We found that in the presence of misregistration, the PA signal increases substantially (Fig. ~\ref{fig:nonlinear_concept}c, plot on the right). Further analysis revealed that the aberrated beams, in addition to distorting the pressure field (i.e., defocusing), also lead to a higher pressure buildup in the highly nonlinear skull. The quantitative integration of the volumetric data revealed an accumulation of acoustic energy in the misaligned state. We established a baseline potential of $7.729 \times 10^6$ Pa$^2$ m$^3$ in the aligned state ($\theta = 0^\circ$). Integration of the squared primary pressure within the skull bone volume revealed a +9.3\% surge in trapped acoustic energy during a mere $2^\circ$ misalignment, increasing to $8.446 \times 10^6$ Pa$^2$ m$^3$ (Table \ref{tab:energy_quant}, Fig. ~\ref{fig:combined_mechanics}b). This acts as a volumetric pump, significantly amplifying the nonlinear source term ($S_{NL} \propto \beta |p_{primary}|^2$) and supporting the physical mechanism of the alignment feedback.

\begin{figure}[!htbp]
    \centering
    \includegraphics[width=1.0\textwidth]{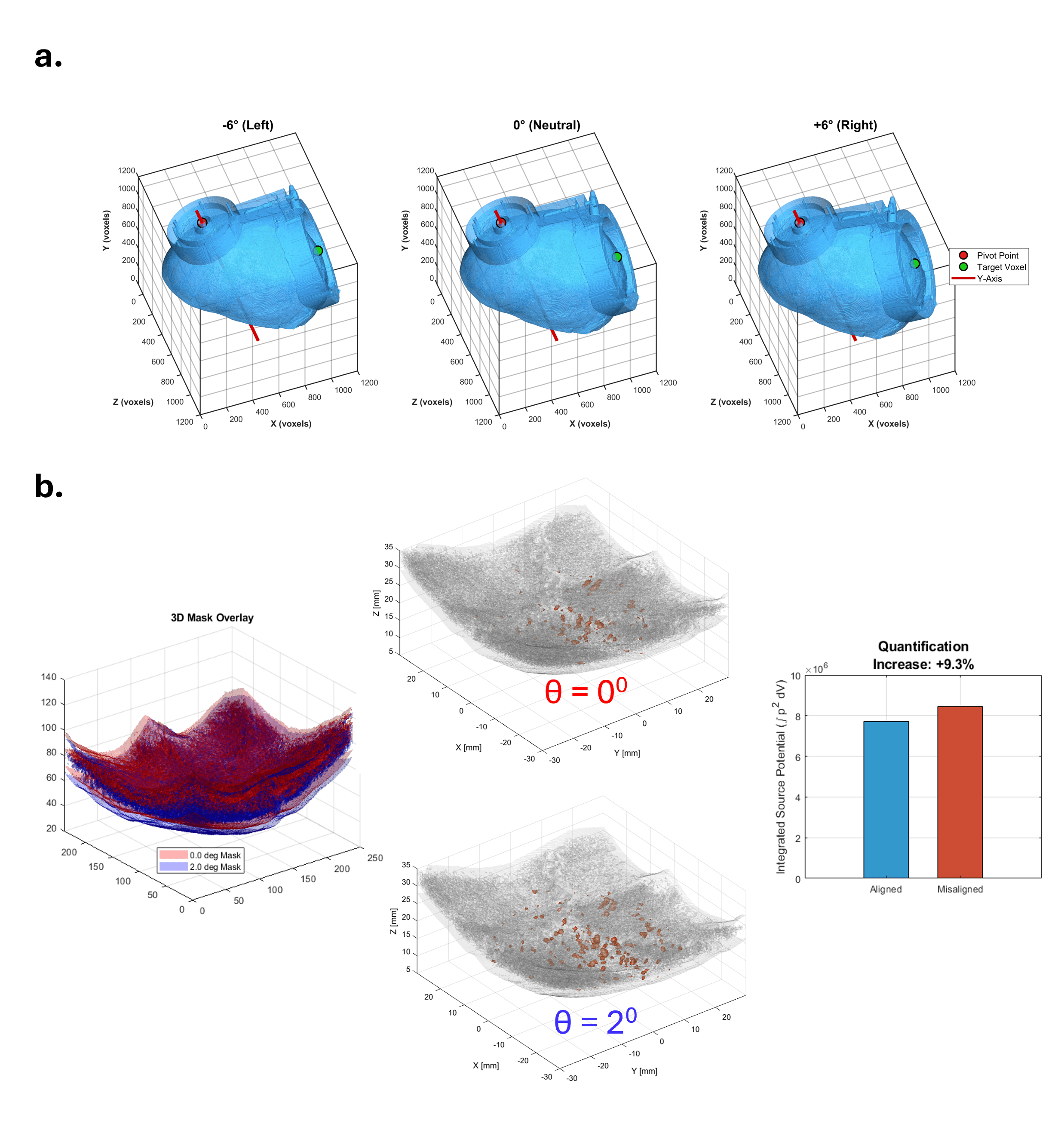}
    \caption[Misregistration and Volumetric Energy Trapping.] {
    \textbf{(a)} Visualization of skull fixture displacement. The skull is shown at \textbf{-6$^{\circ}$ (left)}, \textbf{0$^{\circ}$ (neutral)}, and \textbf{+6$^{\circ}$ (right)}. The \textbf{Red Line} represents the fixed pivot axis. The \textbf{Green Marker} tracks the target voxel at the center of the skull segment facing the transducer. This shows a translation in the X-Y plane due to the pivot offset. \videolink{https://figshare.com/s/ea80df73bdca5153a23d}{Fixture Rotation Animation}
    \textbf{(b)}Analysis of Energy Trapped between the skull layers. \textit{Left:} 3D Mask Overlay verifying spatial coherence between the skull geometry and the pressure field grid. \textit{Middle:} Isosurfaces of acoustic pressure hotspots trapped within the skull bone, showing a denser distribution of scattering nodes in the misaligned state ($\theta=2^\circ$) compared to the aligned state ($\theta=0^\circ$). \textit{Right:} integration confirms a \textbf{+9.3\%} increase in trapped energy during misalignment.}
    \label{fig:combined_mechanics}
\end{figure}

\begin{table}[!htbp]
    \centering
    \caption{Quantification of acoustic energy trapped within the high-nonlinearity skull volume.}
    \vspace{0.2cm}
    \begin{tabular}{lc}
        \toprule
        \textbf{Registration State} & \textbf{Integrated Source Potential} [Pa$^2$ m$^3$] \\
        \midrule
        Aligned ($\theta = 0^\circ$) & $7.729 \times 10^6$ \\
        Misaligned ($\theta = 2^\circ$) & $8.446 \times 10^6$ \\
        \midrule
        \textbf{Relative Increase} & \textbf{+9.3\%} \\
        \bottomrule
    \end{tabular}
    \label{tab:energy_quant}
\end{table}

This effectively extends the nonlinear interaction region, which is critical for the development of finite-amplitude effects~\cite{hamilton1997nonlinear}. Finally, we assessed the influence of misregistration on the PA signal by rotating the skull (Fig.~\ref{fig:nonlinear_concept}d). Interestingly, we found a steep increase in the PA signal for very small angles (i.e., small misregistration errors). We also verified the robustness of this metric to translational misalignments. 3D nonlinear simulations confirm that any error degrading aberration correction increases energy trapping, making the method highly sensitive to both rotational and translational shifts driven by the coupled kinematic moments discussed earlier.

Crucially, these observations persisted for different skull slices (Fig. ~\ref{fig:nonlinear_concept}d), indicating that the PA signal drop is persistent and sensitive to the skull-compensating lens alignment. Together, these findings supported the notion that the low frequency acoustic signal generated by the nonlinear mixing of high frequency beams can be leveraged to attain accurate skull-compensating lens alignment.

\subsection{Sensitivity analysis reveals that parametric array signal is a robust and sensitive surrogate to skull-compensating lens alignment}
\begin{figure}[htbp]
\centering
\includegraphics[width=\textwidth]{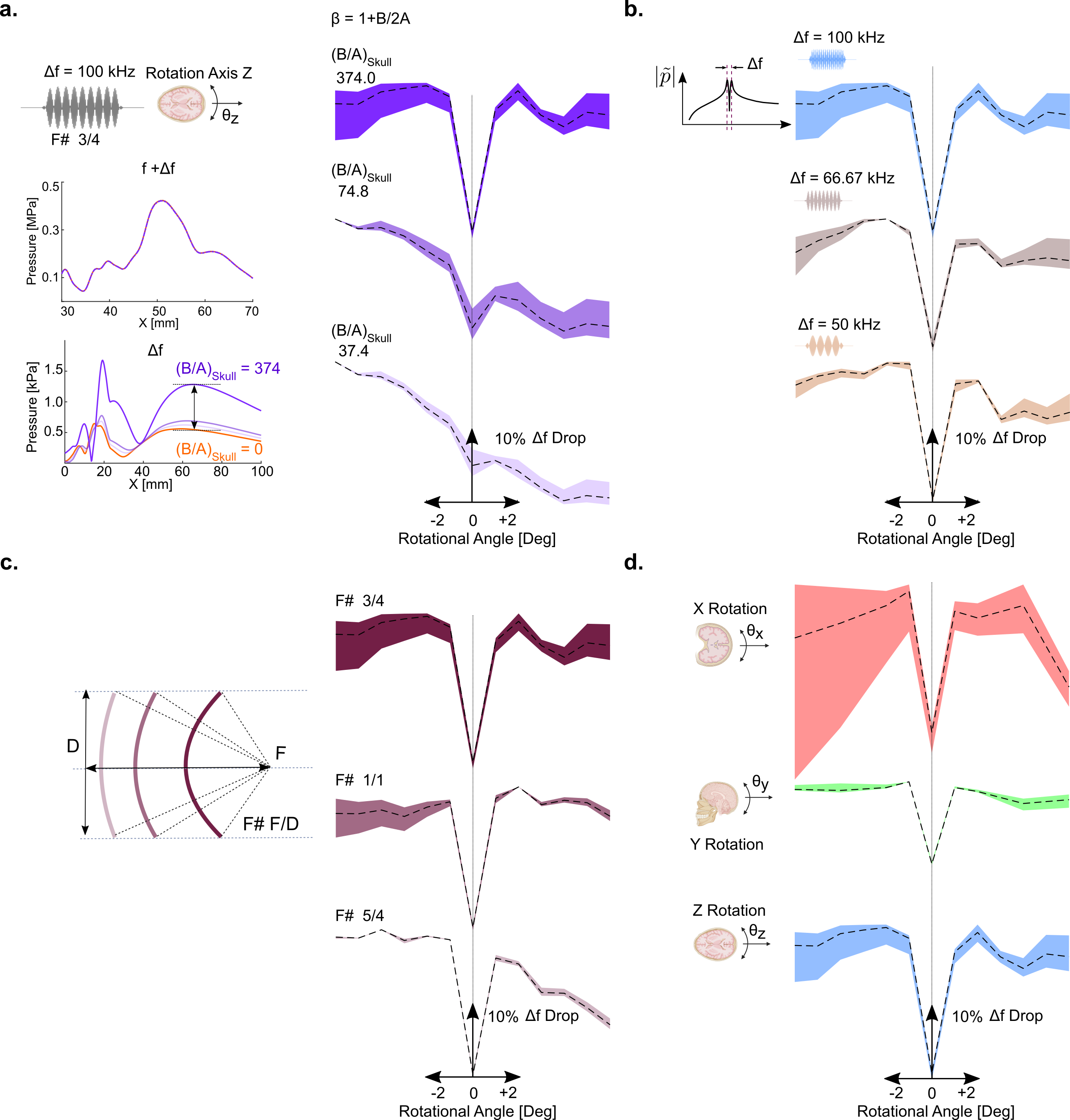}
\caption[Multiparametric sensitivity analysis reveals a robust and persistent drop in parametric pressure during optimal skull-compensating lens alignment.] {(a) Left: Primary and parametric array pressures under conditions of high and low skull nonlinearity. Right: Effect of varying skull nonlinearity on parametric pressure drop. The observed drop in parametric pressure indicates optimal registration of the hologram lens with the skull. (b) Influence of difference frequency on the parametric pressure. (c) Variation in parametric pressure drop with different depths of focusing (or F\#) across varying levels of nonlinearity, difference frequencies, and focusing parameters. (d) Impact of skull rotation about the x, y, and z axes on parametric pressure.}
\label{fig:sensitivity}
\end{figure}

{Motivated by these initial observations of volumetric energy trapping and augmented PA signals, we systematically evaluated the robustness of this metric using comprehensive three-dimensional (3D) nonlinear simulations.} First, we investigated the impact of skull nonlinearity $(B/A)_{\text{skull}}$ on parametric generation. As expected, the primary field (1.05 MHz) remained unaltered across different levels of skull nonlinearity; however, the parametric field (100 kHz) decreased markedly when skull nonlinearity was absent (Fig. ~\ref{fig:sensitivity}a). To further clarify this observation, we varied the skull nonlinearity parameter ($(B/A)_{\text{skull}} = 374, 74.8$, and 37.24; these are equivalent to Goldberg numbers of 3.0, 0.62, and 0.32, respectively) and performed multiple registration iterations by rotating the 3D skull in the transverse plane. Evidently, the parametric pressure drop closely followed skull nonlinearity (Fig. ~\ref{fig:sensitivity} (a), right). Crucially, when skull misalignment is minimized (i.e., $\theta=0^\circ$), the drop in the PA signal becomes even more pronounced in the more realistic 3D simulations, as compared to 2D, for the same B/A parameters.

We also explored the effect of varying the difference frequency $\Delta f$ (Fig. ~\ref{fig:sensitivity}b) and observed that the PA signal drop appears to be insensitive to $\Delta f$ when it ranges from 50 to 100 kHz. Notably, the relationship between the PA signal amplitude and downshift ratio ($f/\Delta f$) follows established parametric array theory~\cite{westervelt1963parametric}, where larger downshift ratios yield smaller PA signals owing to lower nonlinear interaction efficiency. Conversely, smaller downshift ratios, while potentially producing stronger signals, require transducers with broader bandwidths and are hindered by higher frequency-dependent attenuation through the propagation medium. This complex interplay of contributing factors determines the sensitivity of the PA signal changes to downshift-ratio variations. Next, we assessed the influence of focal depth by reducing the f-number while maintaining a constant aperture, producing a progressively weakly focused beam. Although we did not observe any major differences, lower f-numbers appeared to have higher variation. Finally, the drop in the PA signal during optimal alignment is robust to different axes of rotation, although rotations about the z-axis resulted in more substantial decreases in parametric pressure (Fig. ~\ref{fig:sensitivity}d).

{To assess clinical feasibility, we investigated the robustness to receiver placement. Because the parametric source ($\lambda_{\Delta f} \approx 15$ mm) is larger than the skull thickness, it acts as a subwavelength source radiating quasi-omnidirectionally. This ensures the feedback signal is detectable even with ipsilateral receiver placement.}

Across all tested parameters the PA signal dropped $\geq 20$\% at optimal alignment, provided the skull nonlinearity was high.

\subsection{Parametric array signal provides a real-time feedback mechanism to noninvasively align skull-compensating holographic lens to human skull}
\begin{figure}[htbp]
\centering
\includegraphics[width=0.7\textwidth]{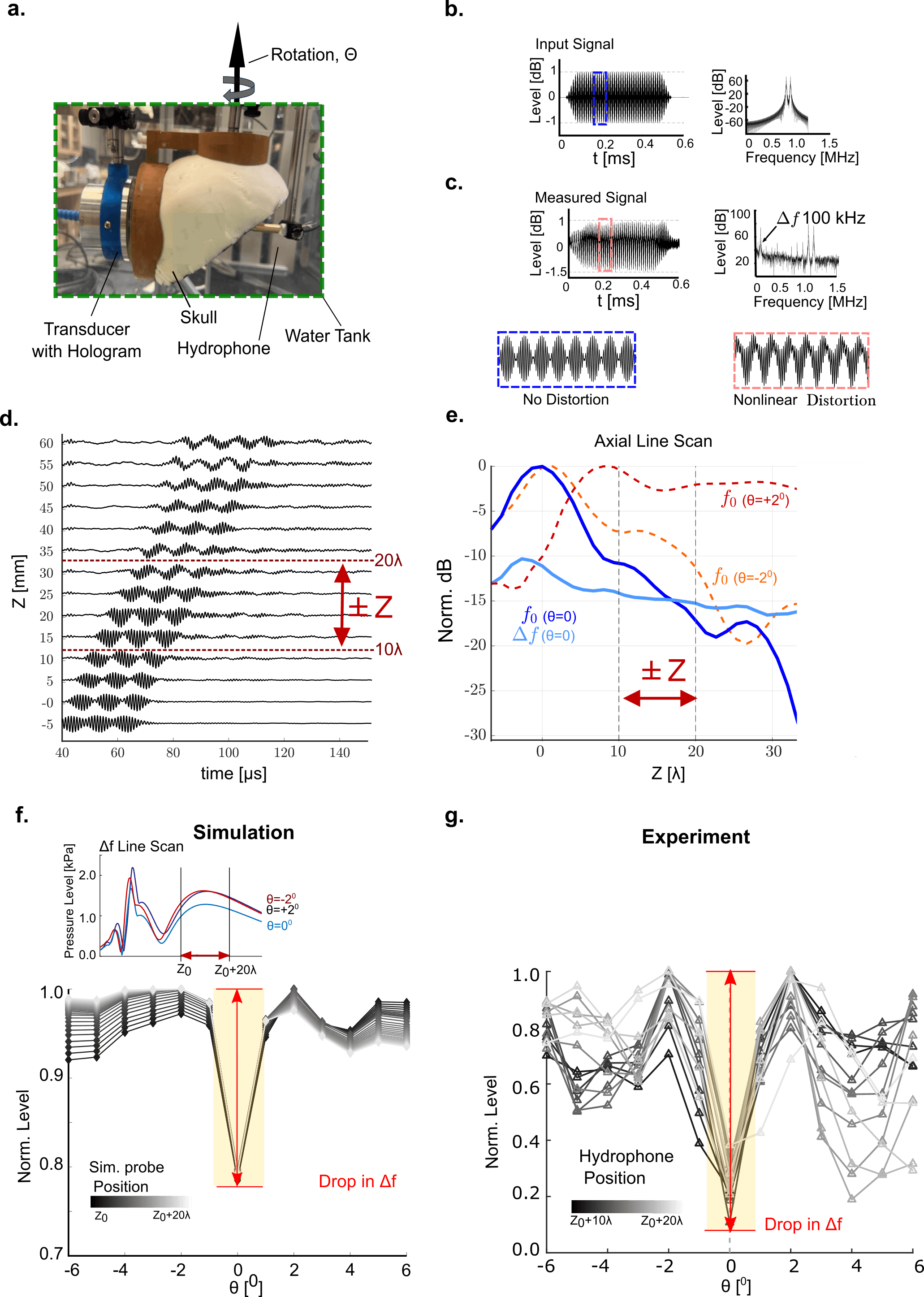}
\caption[Low frequency acoustic feedback generated by nonlinear mixing of high frequency waves enables accurate skull-compensating lens alignment ex vivo.]{ (a) Experimental setup with the skull mounted on a rotating fixture{, with the skull cavity filled with degassed water}. (b) Sample bi-frequency normalized input signal in both time and frequency domains; and Zoomed-in sections highlighting the absence of nonlinear distortion. (c) Sample measured signals using hydrophone and after 600kHz low pass filtering with 40 dB gain in both time and frequency domains; and Zoomed-in sections highlighting the presence of nonlinear distortion. (d) Stacked waterfall plot indicating progressive nonlinear distortion of the measured signal (for $\theta=0^\circ$). (e) Hydrophone line scans demonstrating Primary and parametric signals. (f) 3D simulation mimicking the experimental setup, illustrating the variation of parametric pressure with skull rotation in the transverse plane. (g) Experimental variation of parametric pressure, showing a decrease corresponding to zero registration error. {In (f) and (g), normalization is performed across the different rotational angles ($\theta$) for measurements taken at specific axial positions (e.g., $Z_0+20\lambda$).}}
\label{fig:experimental_validation}
\end{figure}

{To experimentally validate the theoretical sensitivity of this nonlinear acoustic feedback mechanism, we designed a holographic lens} using the HASA-ADAM framework (Fig.~\ref{fig:hasa_adam_framework}) and conducted experiments with a 1 MHz transducer with an active aperture D = 60 mm, coupled with a single focusing lens (F\# 0.75). The transducer was excited with a bi-frequency input signal (containing 0.95 MHz and 1.05 MHz) at 0.2 MPa peak-to-peak pressure to produce a 100 kHz nonlinear difference frequency. {The excitation consisted of short tone bursts (30 cycles) to avoid standing wave artifacts.} This signal was then recorded with a needle hydrophone after 40 dB of low-pass filtering and compared with 3D simulations using the same geometry and skull segment (Fig. ~\ref{fig:experimental_validation}). To perform axial scans and characterize the PA signal at different distances from the skull using a hydrophone, we removed part of the skull (Fig. ~\ref{fig:experimental_validation}a). The skull cavity was immersed in degassed water to provide a standardized acoustic environment. This choice allowed us to isolate the effect of skull-induced aberration and nonlinearity on the PA signal, as water has low nonlinearity, and helped minimize the risk of pseudosound artifacts due to hydrophone nonlinearity~\cite{song2021experimental}.

Axial line scans confirmed that a 0.2 MPa (peak-to-peak) primary field (peak pressure at the focus) produced increasing nonlinear distortion along the axis and beyond the focal position. This was evident in the waterfall plot showing progressive self-demodulation (Fig. ~\ref{fig:experimental_validation}d).

To minimize pseudo-sound effects that can appear in the measurements when the hydrophone is subject to strong primary pressure fields, the measurement window was extended several wavelengths away from the focus (Fig.~\ref{fig:experimental_validation}d). We established the presence of a measurable PA signal for clinically relevant primary pressures (M.I.= 0.1), we rotated the skull at $1^\circ$ increments around the z-axis and measured its amplitude, as in the simulations above (Figs. ~\ref{fig:nonlinear_concept} and \ref{fig:sensitivity}). The line scans for the primary frequency revealed broadening of the axial beamwidth for both positive ($\theta>0^\circ$) and negative ($\theta<0^\circ$) registration errors (Fig. ~\ref{fig:experimental_validation}e). These measurements, aggregated across z-axis positions, not only closely aligned with the simulation predictions, but also confirmed the pronounced drop in parametric pressure when the skull rotation was zero (Fig. ~\ref{fig:experimental_validation}f-g). {The normalization in Fig.~\ref{fig:experimental_validation}f-g is performed across the different rotational angles ($\theta$) for measurements taken at specific axial positions (e.g., $Z_0+20\lambda$), not by spatial averaging.} This procedure is accessible in a clinical setting; the operator monitors the signal at a fixed location while adjusting the lens orientation to find the minimum PA signal.

Taken together, these findings (Figs. ~\ref{fig:nonlinear_concept}–\ref{fig:experimental_validation}) support our hypothesis that the feedback from parametric acoustic array effect is sensitive to skull-aberrations caused by misregistration. We thus demonstrated its potential to provide real-time feedback to align the skull-compensating holographic lens to the patient's skull.

\begin{figure}[!htbp]
\centering
\includegraphics[width=1.0\textwidth]{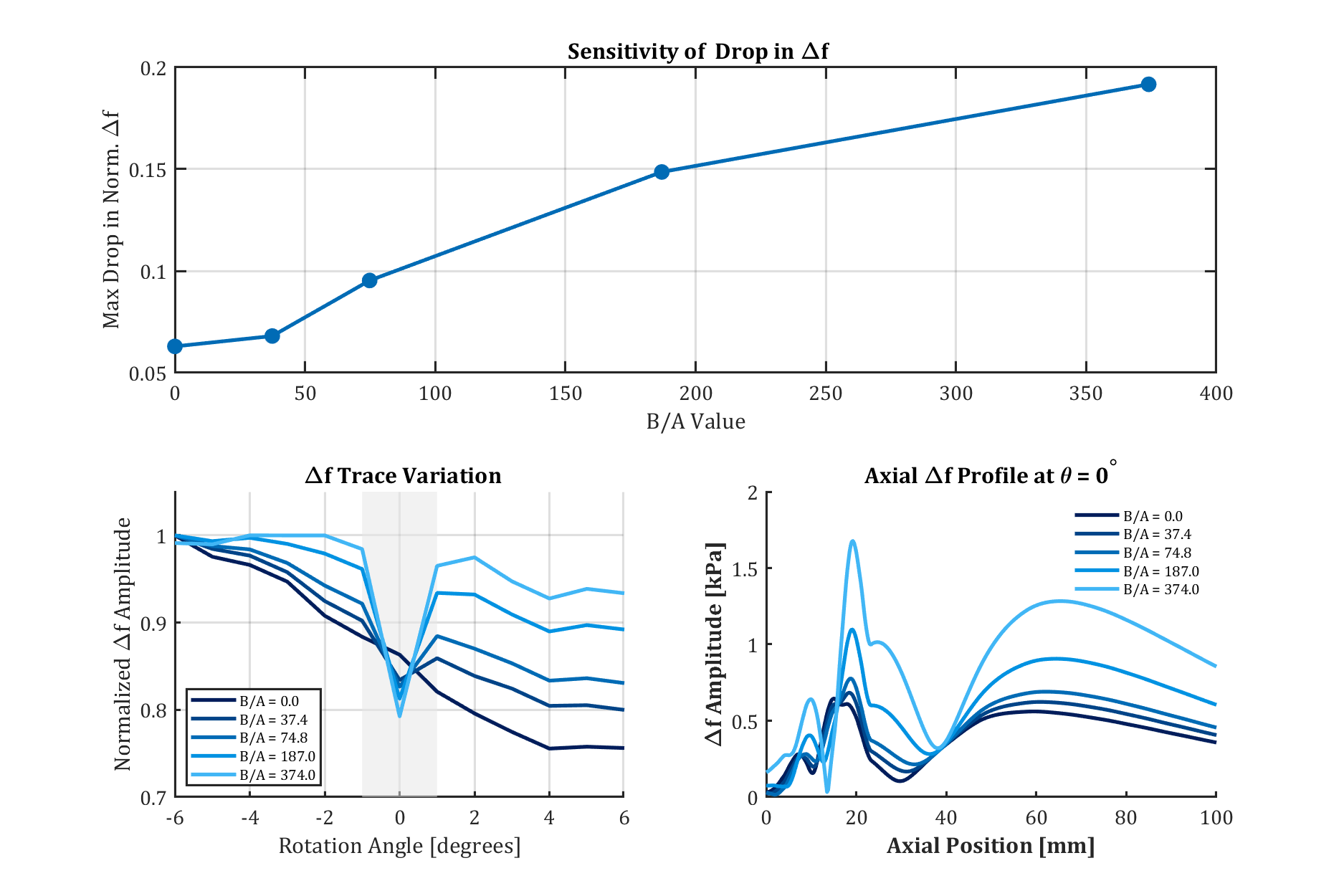}
\caption[Effect of Skull Nonlinearity on Angular Sensitivity.]{\textcolor{black}{\textbf{Effect of Skull Nonlinearity on Angular Sensitivity.} \textit{Top:} The magnitude of the relative drop in the normalized difference frequency ($\Delta f$) amplitude increases monotonically with the skull's nonlinearity parameter ($B/A$), ranging from a 5.8\% drop at $B/A=0.0$ to a 19.1\% drop at $B/A=374.0$. \textit{Bottom Left:} Normalized $\Delta f$ amplitude as a function of skull rotation angle for varying $B/A$ values, demonstrating that higher nonlinearity yields a steeper and more pronounced registration dip. \textit{Bottom Right:} Axial profiles of the $\Delta f$ amplitude at optimal alignment ($\theta=0^\circ$), illustrating the baseline enhancement of parametric generation with increasing $B/A$.}
\label{fig:boa_sensitivity}
}
\end{figure}

\subsection{Mechanisms Governing the Angular Sensitivity of PA Acoustic Feedback}
In the preceding sections, the key result that a consistent parametric array signal drop at $\theta = 0^\circ$ is observed in both simulations and the experiment was shown, which proved our primary hypothesis. We investigate a secondary issue here: experiments showed a steeper angular roll-off than the fluid model predicted. Three candidate explanations are mechanical fixture inaccuracy, an underestimated skull $B/A$, and unmodelled shear-wave generation. To evaluate the impact of the skull's nonlinearity on the registration metric, we quantified the normalized drop in the difference frequency ($\Delta f$) across a range of $B/A$ values (Fig.~\ref{fig:boa_sensitivity}). The analysis reveals a direct, positive correlation: as the assumed nonlinearity of the skull increases, the magnitude of the signal drop at optimal alignment ($\theta = 0^\circ$) increases substantially. For instance, while a purely linear skull approximation ($B/A=0.0$) yields a relative drop of roughly 5.8\% (calculated between $\pm 1^\circ$ misalignment and $0^\circ$ alignment), increasing $B/A$ to upper physiological estimates ($B/A=374.0$) yields a pronounced, steep signal reduction of 19.1\%. This parametric relationship provides strong evidence that the steep angular drop-off observed in our \textit{ex vivo} experiments could be largely attributed to a higher true $B/A$ value of the skull segment than the conservative estimates we used in our simulations. Consequently, underestimating the skull's nonlinearity is a highly probable, simple explanation for the discrepancy between the PA drop simulation and the experiment. Nevertheless, to examine all possibilities, we also explored mode conversion at the fluid-bone interface.

\textbf{Geometric incidence and shear mode risk.} Continuous parametric sweeps of the rotation angle ($\pm 10^\circ$) revealed a piecewise linear sensitivity to spatial misalignment (Fig. ~\ref{fig:composite_shear_analysis}a). We observed a threshold-like abrupt change in sensitivity at ($\pm 4^\circ$) angular rotation. The mode-conversion risk is minimized at zero fixture rotation ($\theta \approx 0^\circ$), which indicates optimal longitudinal transmission. However, at the rotation angle exceeding $\pm 4^\circ$, the primary beam interacts with the steeper skull curvature that pushes the acoustic aperture above the critical angle($\theta_c \approx 30^\circ$). We can see these abrupt jumps in the animation for skull fixture rotation (Fig. ~\ref{fig:composite_shear_analysis}a). Even small positioning errors can result in a rapid loss of the effective transmitting aperture. This increases the absolute shear risk by nearly 10\% at a $6^\circ$ misalignment. Such geometric dependence of mode conversion also explains why projecting complex holographic patterns is more difficult than projecting a simple single-focus pattern (as discussed in Chapter 2). Simpler point-focusing tasks typically involve near-normal incidence  (i.e., low mode conversion), complex holographic patterns require steep phase gradients and highly oblique incidence angles. This increases their susceptibility to such scattering and mode conversion.

\textbf{The shear energy trap.} While fluid-based models partially explain the baseline signal increase during misalignment, capturing the spatial redistribution of acoustic hotspots from the weakly nonlinear brain tissue back into the highly nonlinear cranial bone ($\beta_{eff} \approx 40$), they fail to capture the magnitude of the experimental PA signal spike. This is because fluid solvers inherently neglect solid mechanics. By comparing the elastic simulations against fluid solvers (Fig. ~\ref{fig:composite_shear_analysis}b), we show that geometrically triggered mode conversion could possibly act as an angle-dependent acoustic energy trap.

In the elastic regime, angular misalignment ($\theta_i > 30^\circ$) causes the incident longitudinal energy to be mode-converted into transverse shear waves ($S$-waves). Shear waves are volume-preserving. This means their kinetic energy does not appear in scalar pressure measurements. They only show up instead as a sharp drop in forward-transmitted compressional energy (Fig. ~\ref{fig:composite_shear_analysis}c).

\textbf{Parametric amplification via velocity mismatch.} \textcolor{black}{As derived in Section 3.2.4, the efficiency of nonlinear parametric interaction is theoretically proportional to $c^{-3}$. Assuming this scaling can be extrapolated from fluids to elastodynamic modes in a solid, the slower velocity of shear waves in bone ($c_S \approx 1400$~m/s) compared to longitudinal waves ($c_L \approx 2800$~m/s) implies that a unit of shear energy could be approximately eight times more efficient at generating nonlinear byproducts. By potentially converting fast-propagating waves into slow-propagating ones, the misaligned skull could effectively force acoustic energy to linger inside the highly nonlinear diploe layer for twice the duration, acting as a highly efficient volumetric pump.}

This shear-amplification explanation needs dedicated future studies for verification. No existing solver, unfortunately, couples 3D elastodynamics with dual-frequency nonlinear parametric mixing. Greater shear attenuation in bone may also reduce the effect before it accumulates. An underestimated $B/A$ or fixture error can be a simpler and more plausible explanation. The conclusion remains unaffected. The PA signal drop we observe at accurate registration is replicable in both simulation and experiment.

\begin{figure}[!htbp]
\centering
\includegraphics[width=1.0\textwidth]{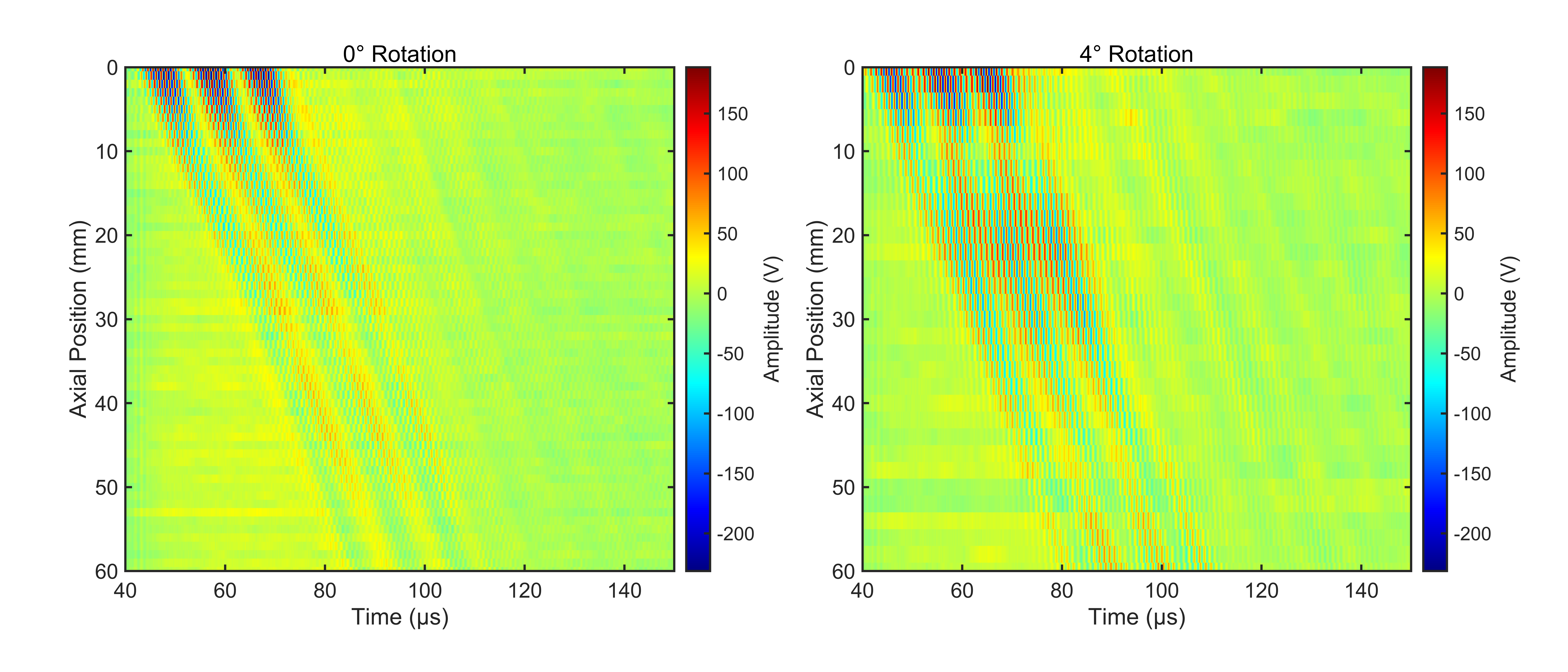}
\caption[Spatiotemporal evidence of Shear Mode Conversion ]{\textbf{Spatiotemporal evidence of Shear Mode Conversion}
\textbf{(a)} At optimal alignment ($\theta = 0^\circ$), the received signal envelope is compact in the time domain. This is consistent with Longitudinal propagation ($c_L \approx 2900$ m/s), where energy goes through the skull quickly.
\textbf{(b)} At $4^\circ$ rotation, the wave packet exhibits distinct \textbf{temporal elongation} (a delayed energy tail).
\textbf{ Mechanism:} This delayed energy possibly corresponds to Shear Modes ($c_S \approx 1400$ m/s) generated by mode conversion at the oblique interface. We know that shear waves propagate at approximately half the speed of longitudinal waves; so they could be effectively trapped within the high-nonlinearity diploe layer for a longer duration. This lingering energy density within the skull bone acts as a potential source for nonlinear mixing ($S_{PA} \propto P^2$). This is one possible mechanism to explain the steeper drop-off in sensitivity observed in experiments.}
\label{fig:shear_evidence}
\end{figure}

\begin{figure}[!htbp]
    \centering
    \includegraphics[width=0.75\linewidth]{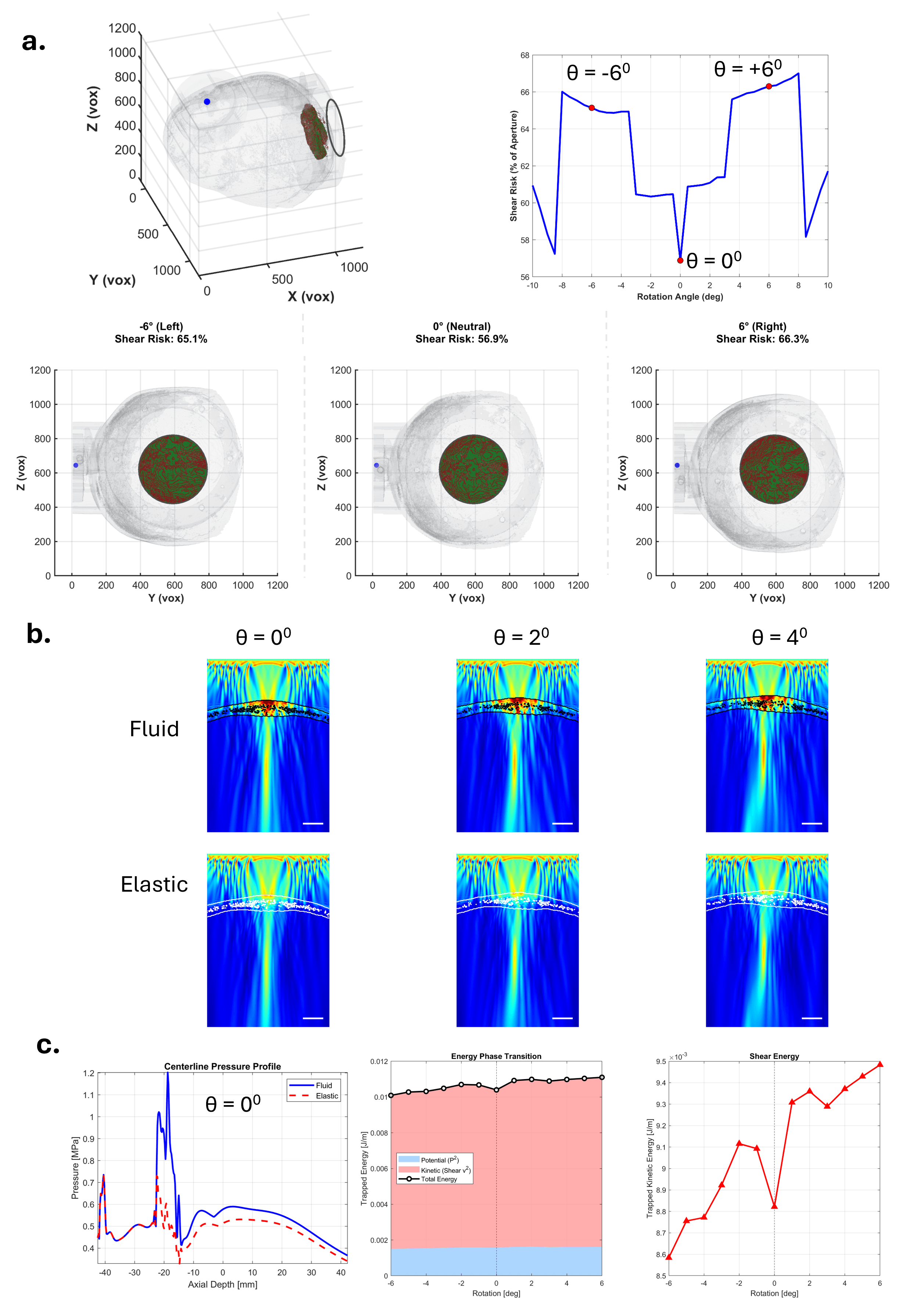}
    \caption[Shear Mode Conversion during Skull Rotation] {
    \textbf{(a) Geometric Incidence and Shear Risk:} \textit{Top:} 3D beam-skull mapping and Shear Risk profile ($\theta_c = 30^\circ$) as a function of misalignment. \textit{Bottom:} Incidence maps for $-6^\circ, 0^\circ, +6^\circ$; \textcolor{green!60!black}{green} indicates safer transmission for longitudinal or compressioanl waves. \textcolor{red!80!black}{Red} highlights the portion on the skull cap prone to shear mode conversion due to angle of incidence exceeding the critical angle for fluid-bone interface(\videolink{https://figshare.com/s/e8941dace20252512d4a}{Animation}).
    \textbf{(b) Full-Wave Propagation:} Comparison of Fluid (top) and Elastic (bottom) steady-state pressure fields across $0^\circ, 2^\circ, 4^\circ$ misalignment, showing energy dampening due to shear scattering.
    \textbf{(c) Energy Dynamics and Amplification:} \textit{Left:} Centerline profiles showing fluid model overestimation of internal standing waves. \textit{Middle:} Stacked area plot of energy transfer from compressional (blue) to kinetic shear modes (red) with increasing rotation. \textit{Right:} Isolated trend of trapped kinetic shear energy \textcolor{black}{exploring its potential as a driver for} parametric amplification. ( \videolink{https://figshare.com/s/71fa3fe423dc778a835a}{Animation}).}
    \label{fig:composite_shear_analysis}
\end{figure}

\textbf{Spatiotemporal waveform validation.} We also empirically probed this hypothesis via waveform analysis of \textit{ex vivo} transmissions.  Any parametric signal generated via shear-mode mixing must arrive at the detector at a distinct time lag due to slower shear wave speeds (roughly 0.5 times the compresional wave speed). As shown in Fig.~\ref{fig:shear_evidence}, the experimental waveform at the optimal normal incidence ($\theta=0^\circ$) is temporally compact and consistent with longitudinal propagation. However, at an oblique incidence of $4^\circ$, the wave packet exhibits temporal elongation. But this is absent in the aligned case. This delayed signature offers a possible mechanism in which a slower-propagating shear mode contributes to nonlinear generation. This lends circumstantial support to the shear hypothesis as a contributing factor. Our primary finding, regardless, is that the robust drop in the PA signal indicating alignment remains validated independently of this effect.

\subsection{Registration metric is robust to hydrophone placement and aperture size}

\begin{figure*}[t!]
\centering
\includegraphics[width=1.0\textwidth]{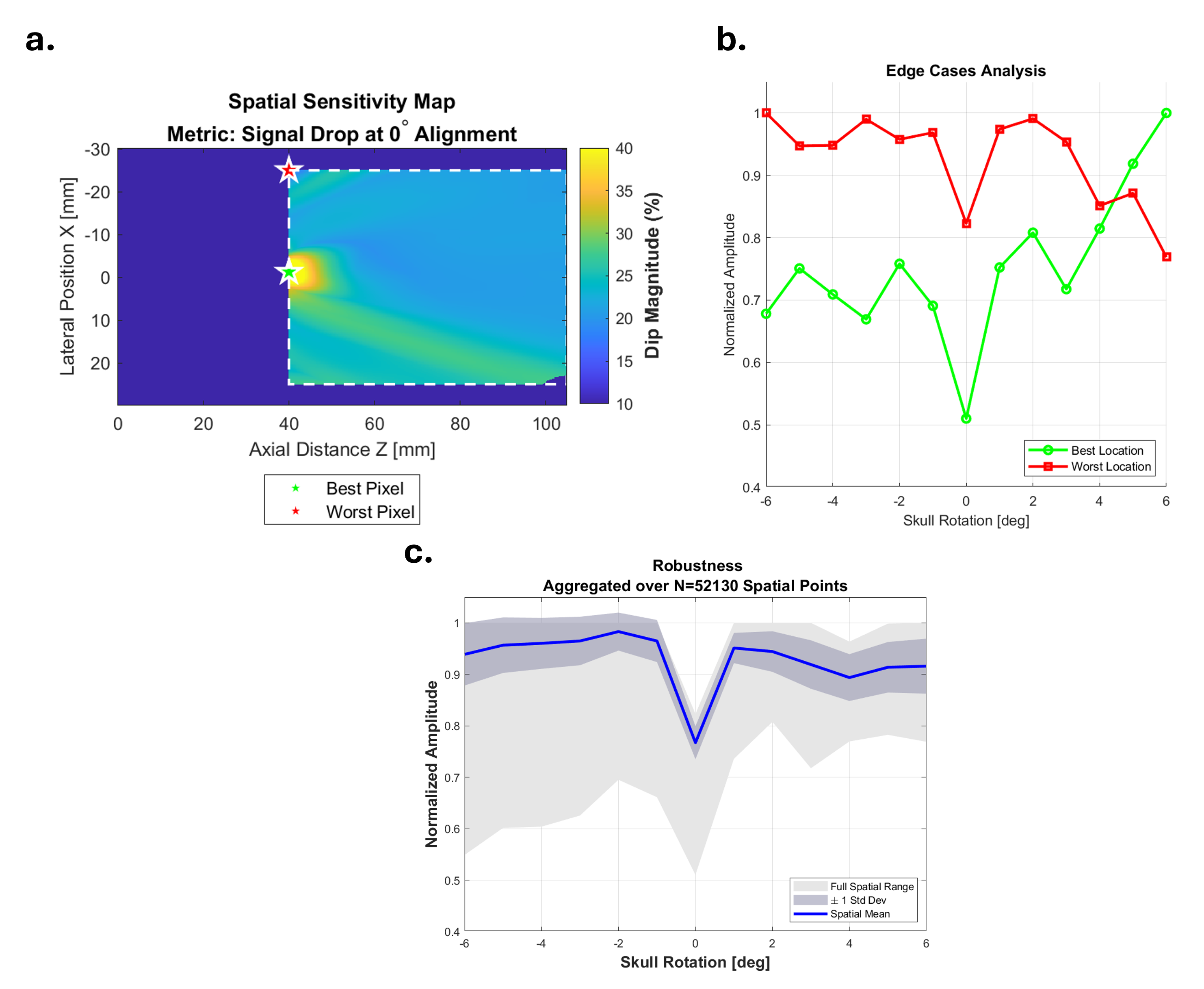}
\caption[Spatial Robustness of Nonlinear Acoustic Feedback.]{\textbf{Spatial Robustness of Nonlinear Acoustic Feedback.} \textbf{(a)} Spatial Sensitivity Map showing the magnitude of the registration dip (signal drop at $0^\circ$) across a $60 \times 60$ mm region behind the skull. The signal was detectable across most of the fields.
\textbf{(b)} Edge-case analysis contrasting the signal trace at the most sensitive spatial pixel (Green) versus a diffraction node (Red).
\textbf{(c)} Robustness analysis aggregated over a large number of spatial points ($N=52,130$). While individual point measurements vary (Gray envelope), the spatially averaged response (Blue line), which can also be observed with a larger-aperture receiver, exhibits a global minimum at accurate registration.}
\label{fig:spatial_robustness}
\end{figure*}

\begin{figure*}[t!]
\centering
\includegraphics[width=1.0\textwidth,trim= 12 0 0 12,clip]{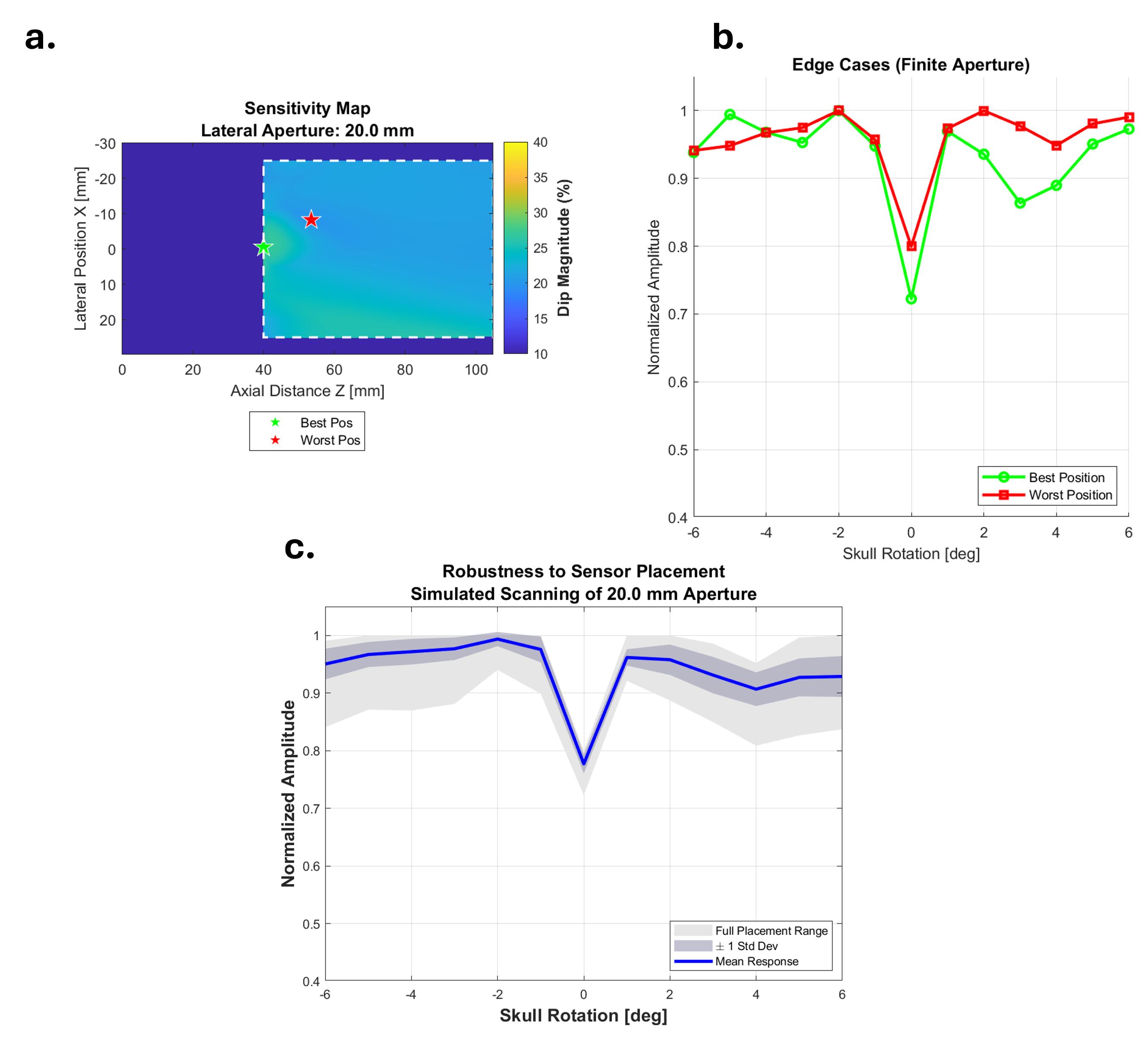}
\caption[Effect of Sensor Aperture on Signal Stability.]{\textbf{Effect of Sensor Aperture on Signal Stability.}
\textbf{(a)} Spatial Sensitivity Map showing the magnitude of the registration dip across a $60 \times 60$ mm region.
\textbf{(b)} Comparing the signal trace at the best versus worst sensor positions. The difference between the Best and Worst detection points decreases as the aperture increases. A 20 mm aperture averages sub-wavelength interference and ensures consistent detection.
\textbf{(c)} The larger 20 mm aperture acts as a spatial filter, which lowers uncertainty limits (Gray region) and gives a reliable mean response (Blue line).}
\label{fig:aperture_sweep}
\end{figure*}

Figure \ref{fig:spatial_robustness} shows the sensitivity map of the registration dip magnitude across a $60 \times 60$ mm region behind the skull. The signal was detectable across most of the fields. The difference between the Best and Worst detection points decreases significantly as the aperture increases (Figure \ref{fig:aperture_sweep}). A 20 mm aperture averages sub-wavelength interference to minimize dead spots and ensure detection in a robust manner.

The radiation pattern of the parametric source within the skull is governed by the diffraction limit $kD$, where $D$ denotes the skull thickness.
The \textbf{primary beam ($\lambda \approx 1.5$ mm)} is highly directional; it requires precise targeting. On the other hand, the \textbf{parametric signal ($\lambda_{\Delta f} \approx 15$ mm)} has a wavelength larger than the skull thickness ($L < \lambda_{\Delta f}$, $L \approx 7$ mm). Thus, the interaction volume acts as a sub-wavelength acoustic source.

A source smaller than $\lambda_{\Delta f}$ radiates as a monopole. The receiver, therefore, needs no phase alignment with the transmitter; any acoustically coupled position on the head captures the alignment-induced energy dip.

\subsection{Near-field measurements suggest parametric generation over pseudo-sound}

\begin{figure}[!htbp]
\centering
\includegraphics[width=0.8\textwidth]{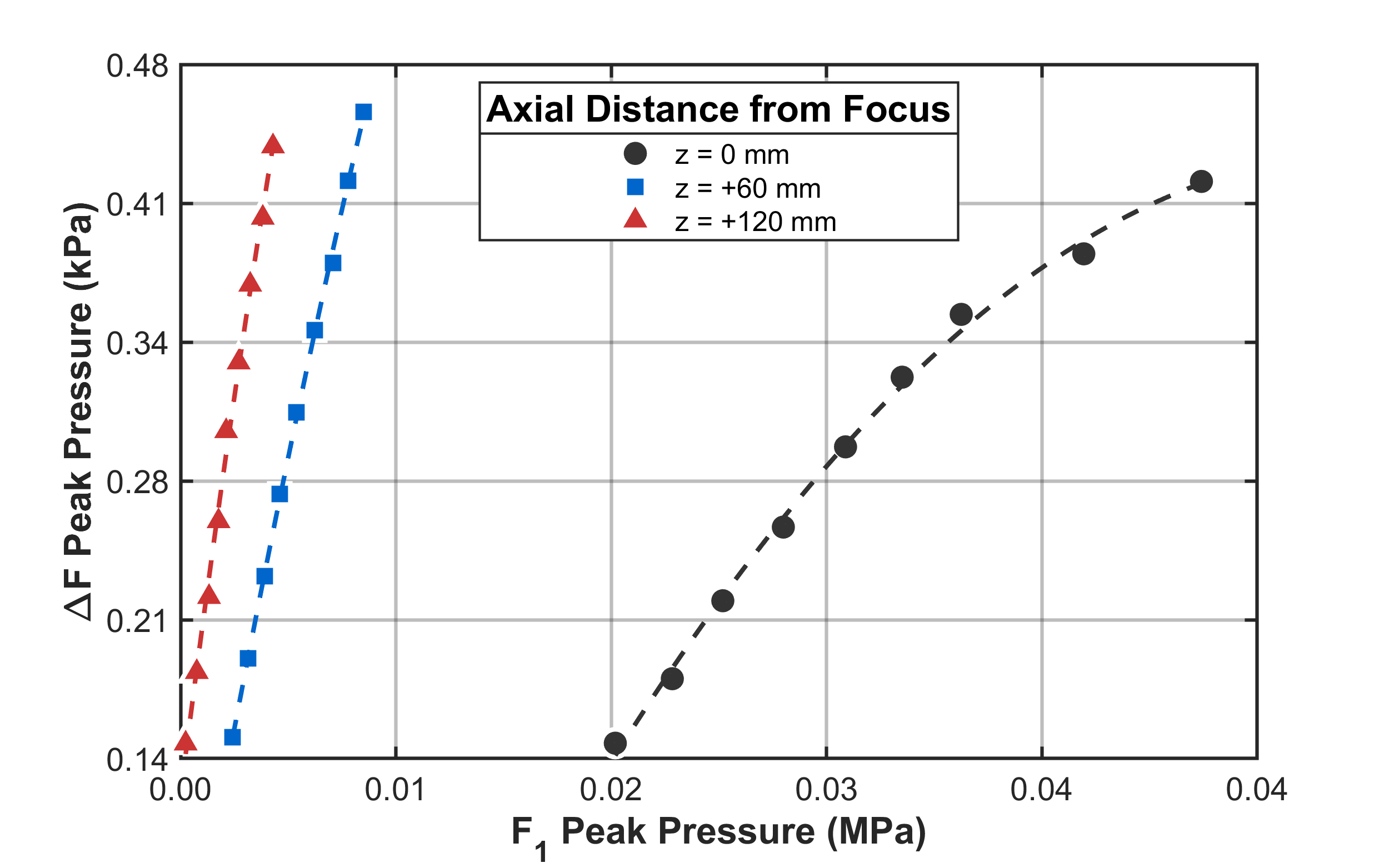}
\caption[Pseudo-sound quantification using near-field measurements.]{Pseudo-sound quantification using near-field measurements. Axial Distance from Focus: z = 0 mm (black circles), z = +60 mm (blue squares), z = +120 mm (red triangles).}
\label{fig:s3}
\end{figure}

Figure \ref{fig:s3} shows the difference in frequency at $\Delta f = 100$ kHz measured using a needle hydrophone as a function of the primary pressure $\Delta f_1 = 1.05$ MHz for three separate conditions: at focus, 60 mm from the focus, and 120 mm from the focus. Evidently, as the hydrophone moves away from the focus, the primary pressure decreases (indicated by the rightward shift in the curves); however, the difference in frequency pressure at 60 mm and 120 mm from the focus remains the same and has an almost linear relationship with the primary pressure. Conversely, the difference in frequency pressure at the focus has a quadratic relationship with the focal pressure (i.e., the quadratic term associated with the pseudo-sound is significant).

The primary pressure decreases with increasing distance, following the inverse-square law. Pseudo-sound depends on this primary pressure amplitude and is expected to drop following a similar trend. However, the PA signal we observe remains relatively flat. We can conclude that the measured signal at those distances is not due to hydrophone nonlinearity and may be due to true parametric generation that accumulates over a longer propagation distance.

\subsection{Spectral analysis supports intrinsic skull nonlinearity over microbubble artifacts}

\begin{figure}[!htbp]
\centering
\includegraphics[width=1.0\textwidth]{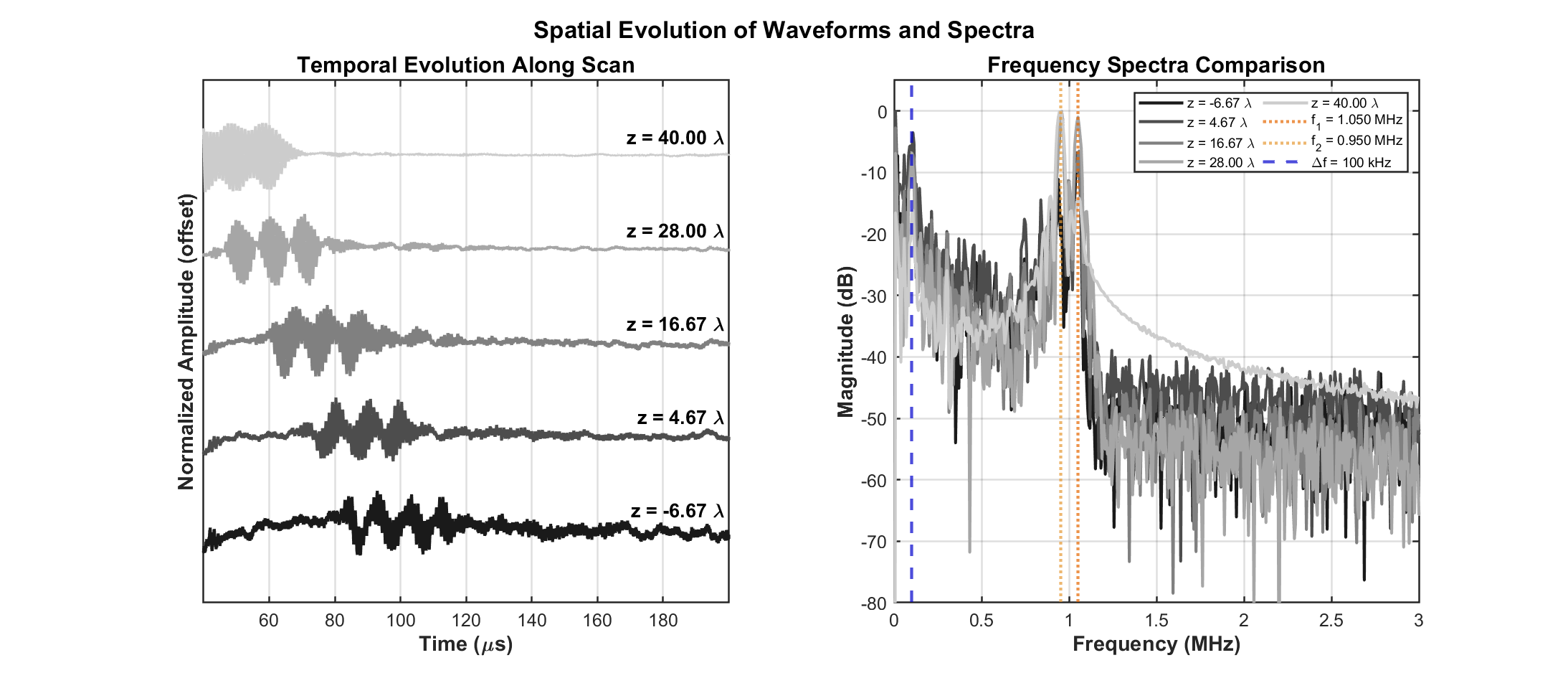}
\caption[Time-domain waveforms and Frequency Spectra along the propagation axis.]{\textbf{Time-domain waveforms and Frequency Spectra along the propagation axis.} The left column shows the raw voltage recorded by the hydrophone at increasing depths (Top to Bottom). The column on the right shows the corresponding FFT results. Dashed lines indicate the Difference Frequency ($\Delta f$, blue), the Primary Frequency ($f_1$ and $f_2$, orange and yellow)}
\label{fig:fft_analysis}
\end{figure}

Figure \ref{fig:fft_analysis} shows the time-domain waveforms (left column) and their corresponding frequency spectra (FFT) (right column) at various depths $z$. The spectral data provide evidence regarding the physical origin of nonlinearity:

\begin{enumerate}
\item \textbf{Confirmation of Classical Mixing:}
As seen in the FFT plots (positions 18 through 71), there are distinct, high-SNR peaks corresponding to the mixing terms predicted by the Westervelt model: The Primary inputs ($f_1, f_2$) around 1 MHz and the target \textbf{Difference Frequency ($\Delta f$)} at $\approx 100$ kHz.
\item \textbf{Absence of Bubble Signatures:}
We examined the spectral region corresponding to the subharmonic frequency ($f_{sub} \approx 500$ kHz). As shown in the FFT plots, the noise floor in the $0.2 - 0.8$ MHz range remained flat. There is \textbf{no detectable energy} at $f_0/2$.
\end{enumerate}

 If incident pressures typically exceed the threshold for inertial cavitation and period-doubling bifurcations, any trapped gas bodies would be driven into a strong-scattering regime. This is characterized by the emission of subharmonics ($f_0/2$) and broadband noise. Therefore, the observation of a robust Difference Frequency signal and absence of subharmonic content suggests that microbubbles are not the source of the nonlinearity. Thus, the signal is attributable to the intrinsic cumulative nonlinearity of the bone matrix.

We further analyzed the signal integrity relative to thresholds for bubbly media~\cite{karpov2003}. It has been demonstrated that gas-saturated layers exhibit softening nonlinearity; incident pressures of only 50 kPa are sufficient to distort the carrier wave into a steep sawtooth, resulting in a large number of high-frequency harmonics. In contrast, in our experiments, we used focal pressures of approximately 0.1 MPa. If trapped gas microbubbles were present in the diplo\"{e} layer, this pressure would force the system into inertial cavitation. And we would see significant signal degradation, along with the presence of sub-harmonics and ultra-harmonics. Our spectral analysis (Fig. \ref{fig:fft_analysis}) reveals transmission of the primary frequencies and a clean Difference Frequency ($\Delta f$) peak. The medium's ability to support 0.1 MPa propagation without degrading into the shock regime provides further evidence that the propagation path is free of resonant bubbles.

\subsection{Impact of Intracranial Trapped Gas on Parametric Array Generation}
\begin{figure}[!htbp]
\centering
\includegraphics[width=0.75\textwidth]{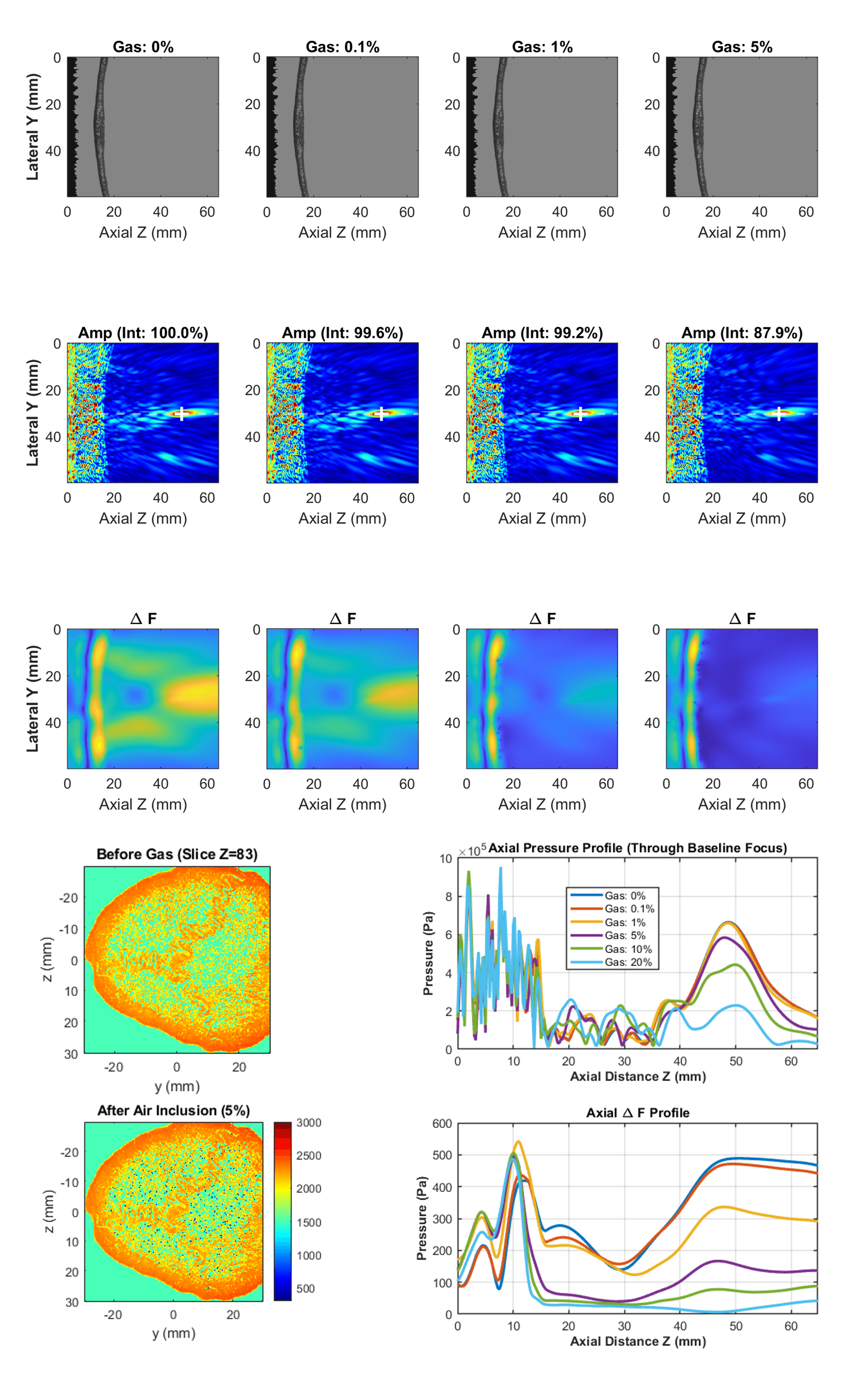}
\caption[Effect of trapped gas on acoustic propagation.]{ \textit{(Top Panel)} Longitudinal field maps showing skull geometry (row 1), fundamental frequency amplitude (row 2), and difference frequency amplitude (row 3). \textit{(Bottom Left Panel)} Skull slice of speed of sound without (Row 1) and with (Row 2) gas inclusions in the microstructure \textit{(Bottom Right Panel)} Axial pressure profiles extracted through the geometric focus. The top plot shows the fundamental frequency ($f_1$), whereas the bottom plot shows the difference in frequency ($\Delta f$). A critical transition occurs at $\phi_{gas} \ge 1\%$, where the difference frequency signal is effectively extinguished.}
\label{fig:gas_analysis}
\end{figure}

The results illustrated in Figure \ref{fig:gas_analysis} demonstrate a nonlinear threshold response regarding the survival of the parametric array compared to that of the fundamental beam under gas exposure. As shown in the axial pressure profiles (Figure \ref{fig:gas_analysis}, Right Panel), two distinct regimes were observed. In the \textbf{Weakening Regime (orange curve) ($\phi_{gas} = 0.1\%$)}, the primary beam undergoes scattering but retains a focal structure. The difference frequency was also similarly attenuated, but remained coherent and visible. In the \textbf{Elimination Regime (yellow curve) ($\phi_{gas} \ge 1\%$)}, a transition occurs at 1\% gas inclusion. The scattered and broadened primary beam still transmits through the skull, but the difference frequency (PA) signal drops to the noise floor immediately after the skull. This suppression of the difference frequency ($\Delta f$) could be due to the disruption of the nonlinear mixing zone inside the skull bone. The amplitude of the parametrically generated wave, $P_{\Delta f}$, scales with the coefficient of nonlinearity ($\beta$) of the medium as follows:
\begin{equation}
P_{\Delta f} \propto \beta \cdot P_{f1} \cdot P_{f2}^*
\label{eq:gas_mixing}
\end{equation}
The skull volume acts as a local amplifier for the generation of different frequencies. Trapped gas bubbles disrupt this mechanism by randomizing the phases of the primary waves, which destroys the phase coherence required for cumulative generation. Additionally, because difference frequency generation scales with the \textit{product} of the primary pressures (Eq. \ref{eq:gas_mixing}), a linear reduction in primary amplitude due to scattering results in a quadratic reduction in the secondary source strength. Therefore, we conclude that gas inclusions act as passive scatterers. The elimination of the parametric array is caused by the high acoustic impedance mismatch, which scatters the primary energy and disrupts the nonlinear coherence.

\section{Discussion}
To comply with the targeting requirements in the brain, where mistargeting can pose safety risks, or when accurately targeting specific brain regions or neuronal circuits is essential, ~\cite{meng2021applications,airan2017neuromodulation} a millimeter (i.e., subwavelength) targeting accuracy is required. To achieve this level of accuracy, our investigations addressed the long-standing challenge of skull-compensating lens registration by uncovering the close relationship between skull nonlinearity and aberrations caused by misregistration. The PA minimum provides sub-wavelength alignment---below the ${\sim}2$~mm floor of current neuronavigation~\cite{choi2024neuronavigation,chen2021neuronavigation,wei2013neuronavigation,pouliopoulos2020clinical}. {The proposed PA feedback method complements neuronavigation by providing the sub-wavelength accuracy required for high-frequency TUS, and our simulations show it is robust to both rotational (Fig. 4) and translational misalignments.} \textcolor{black}{Regarding the clinically acceptable margin of error, our experimental data demonstrated that a registration tolerance of $\pm 1^\circ$ was acceptable for the specific skull segment tested---maintaining targeting accuracy and focal pressure within safe therapeutic margins. However, human skulls show high variability in thickness, geometry, porosity, and internal composition. Thus, determining a universal clinical tolerance for registration will require future studies across a large dataset of varying skull geometries. The core mechanism of tracking the PA signal minimum should remain robust (i.e., a drop of more than 10-20\% at accurate registration) across these variations.}

Beyond its immediate application to skull-compensating lens registration, one natural extension is that PA acoustic feedback, combined with the HASA-ADAM framework, can also be utilized for noninvasive aberration correction of phased arrays, where it can be used as an objective function for noninvasive in vivo phase and amplitude optimization of each element. Together, these conceptual contributions and advancements support the design of simple, economical, and high-performance ultrasound systems for high-precision neurointerventions. Such systems may also support daily/weekly treatments, possibly in outpatient and/or limited resource settings, without compromising performance, thereby supporting the effective translation and broad dissemination (i.e., similar to US imaging) of this technology~\cite{schoen2022towards,rincon2022biomarkers}. This may also alleviate the need for repeated use of intraoperative MRI during TUS interventions such as targeted drug delivery or liquid biopsy, which can complicate or even prevent their implementation (e.g., the average time to obtain an MRI appointment can be several months~\cite{hofmann2023variations}).

Although our experimental and numerical results indicate a substantial drop in the PA signal during good skull-compensating lens alignment, we noticed some discrepancies that can be attributed to several interrelated factors. \textcolor{black}{As discussed, the steeper signal drop in experiments compared to baseline fluid simulations may stem from unmodeled shear wave parametric generation, physical fixture errors, or an underestimation of the skull's nonlinearity. For instance,} the nonlinearity parameter $\beta$, which fundamentally governs PA signal generation, exhibits frequency-dependent behavior that is often oversimplified in simulations~\cite{panfilova2021review,zhang2001experimental}. Additionally, $\beta$ for the skull, which is a highly porous structure, has not been characterized in the literature, suggesting that the current values used in theoretical investigations may not be optimal. The experimental uncertainties may also have contributed to this. Most notably, microscopic air bubbles trapped in skull pores may persist~\cite{tang2011effect} despite the extended degassing we performed (see Methods). These microbubbles, which are resonant in the MHz range and exhibit extreme nonlinearity even at very low void fractions, may contribute to skull non-linearity~\cite{cavaro2011microbubble,overvelde2010nonlinear}. Additionally, the skull bone follows complex frequency-dependent attenuation and has high interindividual variability that is often underestimated in simulations~\cite{pinton2012attenuation,pinton2011effects}. Together, these sources of uncertainty can lead to a higher Goldberg number (i.e., nonlinearities) and PA signal under the experimental conditions. Accurate, frequency-resolved measurements of skull $B/A$ and shear parameters are now a priority. The principal validation stands: the PA signal drops at optimal alignment in both computation and experiment, confirming its value as a registration metric.

Collectively, the proposed research, by accelerating hologram design and introducing robust registration strategies to support the design of high-fidelity transcranial holography, may support the effective translation and broad dissemination of this technology in the clinic. Although our work is primarily focused on biomedical applications, the implications of high-fidelity acoustic holography are much broader and will invite researchers to explore this new capability across a range of applications~\cite{melde2023compact,hirayama2019volumetric,xie2016acoustic,kruizinga2017compressive,maimbourg20183d,jimenez2019holograms,melde2016holograms}. Our research also lays the groundwork for future studies exploring low-frequency nonlinear acoustic feedback for the diagnosis, monitoring, and treatment of brain diseases and highlights the importance of relatively thin and highly nonlinear media to augment finite-amplitude effects.

\section{Conclusions}
To conclude, nonlinear wave propagation through the skull provides a feedback signal for registering holographic lenses to the skull bone. The key findings of this chapter are:

\noindent\textbf{PA feedback as a registration metric.}
The difference-frequency signal generated inside the skull drops to a minimum when the lens is correctly aligned. This dip was observed in both fluid simulations and \textit{ex vivo} experiment.

\noindent\textbf{Energy-trapping mechanism and shear hypothesis.}
Misalignment traps primary-beam energy in the high-$\beta$ bone, increasing the PA source term by ${\sim}9$\%. Our secondary hypothesis that mode-converted shear waves amplify the effect via a $c^{-3}$ scaling is consistent with temporal waveform data. It requires rigorous future studies for validation. Under-estimated $B/A$ and fixture error are equally plausible explanations.

\noindent\textbf{Intrinsic skull nonlinearity isolated.}
Near-field pseudo-sound measurements and the absence of subharmonic spectral lines ruled out microbubble cavitation. The registration signal originates from cumulative quadratic nonlinearity of the bone matrix.

\noindent\textbf{Spatial robustness confirmed.}
The PA interaction volume ($\lambda_{\Delta f}\approx 15$~mm $>$ skull thickness) acts as a monopole source. A receiver anywhere on the acoustically coupled head surface detects the alignment dip; a 20~mm aperture virtually eliminates dead spots.

\chapter{Monitoring Intracranial Pressure in Hydrocephalus}
\label{chap:ICPMonitoring}
\footnotetext{An earlier version of the work presented in this chapter is available as a preprint on arXiv (https://arxiv.org/abs/2508.07103).\cite{Dash2025}}

\section{Introduction}

In this chapter, we explore whether the parametric acoustic (PA) array can be used to non-invasively monitor ventricular expansion as a proxy for changes in intracranial pressure (ICP). We believe this is possible by detecting shifts in the brain's effective acoustic nonlinearity in response to relative changes in ventricular size.  We have simulated transcranial bi-frequency nonlinear ultrasound transmission to assess the diagnostic feasibility of this approach.

Hydrocephalus is a disturbance in cerebrospinal fluid (CSF) dynamics that results in enlarged ventricles and elevated intracranial pressure (ICP). The management of hydrocephalus requires frequent monitoring of ICP, which is a key biomarker for tracking disease progression to help guide treatment by removing excess CSF (e.g., via shunt-based treatments). Invasive ICP monitoring using external ventricular drains (EVDs) or parenchymal microsensors is the gold standard~\cite{zhang2017invasive} in clinical practice. This approach, though, carries several associated risks such as infection, hemorrhage, and mechanical failure~\cite{jiang2022invention, fischer2020non}. Such invasive monitoring is also episodic. Clinicians often lack continuous insight into intracranial dynamics in the outpatient setting after catheter removal. Non-invasive surrogates such as Transcranial Doppler (TCD) ultrasonography or Optic Nerve Sheath Diameter (ONSD) measurements can be used to mitigate this. But, they often lack a direct physical correlation with ventricular volume or result in significant operator variability~\cite{qiu2025noninvasive, jiang2025advancements}.  Therefore, non-invasive techniques that accurately detect ventricular volume changes and reliably assess shunt function can reduce complications by closing the monitoring gap.

Because clinical investigations show that the removal of excess CSF leads to a decrease in ICP, which in turn results in reduced ventricular space (or increased space occupied by brain tissue)~\cite{czosnyka2004monitoring}, we used the distinct physical contrast between the acoustic nonlinearity parameters of brain parenchyma ($B/A \approx 7.4 \rightarrow \beta \approx 4.7$) and the protein-poor, water-like CSF ($B/A \approx 5.2 \rightarrow \beta \approx 3.6$)~\cite{duck2013physical, blomqvist2021sulfatide}. As ventricles expand, they displace higher-nonlinearity brain tissue with lower-nonlinearity CSF. We hypothesize that this volumetric substitution suppresses the cumulative generation of difference-frequency ultrasound along the transcranial path.

Linear pulse-echo ultrasound (e.g., at 500 kHz) could be used to track the geometric expansion of the ventricles by measuring the distance to the brain-water interface. But such techniques rely heavily on precise registration to capture specular reflections from the ventricular walls. This makes them highly operator-dependent and prone to signal loss due to skull-induced scattering or misalignment. In contrast, the Parametric Array-based method may serve as a continuous, bulk proxy measurement to track the drop in the effective nonlinearity parameter ($\beta$) as CSF replaces brain tissue. Because this volumetric signal is more tolerant to lateral misalignment, it may provide a reliable basis for a simple, operator-independent wearable sensor.

\section{Methods}

\subsection{Nonlinear Acoustic Sensing}
The origin of acoustic nonlinearity can provide a basis for detecting ventricular expansion. The propagation of finite-amplitude ultrasound in thermoviscous tissue is modeled by the Westervelt equation~\cite{hamilton_nonlinear_2008}, where the local nonlinearity parameter $\beta$ acts as a virtual source density. When the medium is excited by two primary frequencies ($\omega_1$ and $\omega_2$), the nonlinear interaction generates a low-frequency wave at the difference frequency ($\omega_d = |\omega_1 - \omega_2|$). The amplitude of this difference frequency, $P_{\Delta f}$, grows cumulatively over an interaction length $L$ and under quasilinear approximation can be written as :
\begin{equation}
P_{\Delta f}(L) \propto \omega_d^2 \int_0^L \frac{\beta(z)}{\rho_0 c_0^5} P_1(z) P_2(z) e^{-\alpha_d z} \, dz
\label{eq:parametric}
\end{equation}
This relationship indicates that the received signal acts as a path integral of the nonlinearity $\beta(z)$ weighted by the primary pressure fields. Consequently, if expanding ventricles displace brain tissue with CSF in the focal region, the local value of $\beta(z)$ drops, thereby reducing the integrated signal. Thus, $P_{\Delta f}$ can serve as a non-invasive volumetric indicator of tissue composition.

\subsection{Computational Modeling of Ventricular Expansion}
High-resolution two-dimensional (2D) pseudo-spectral time-domain (PSTD) simulations examined the relationship between ventricular expansion and the PA signal using the k-Wave MATLAB toolbox~\cite{kwave_toolbox}.

We developed a realistic 2D head model derived from $\mu$CT data, using the Evans Index (EI)(the ratio of frontal horn width to maximum skull diameter) to define pathological states~\cite{jaraj2017estimated}. The ventricular mask was dynamically resized to simulate expansion from a baseline cross-sectional area (CSA) of 20 cm$^2$ (representing moderate ventriculomegaly) to 40 cm$^2$ (representing severe hydrocephalus). Primary frequencies of 0.95 MHz and 1.05 MHz were emitted by a focused transducer to generate a 100 kHz difference frequency.

The pulse length for this bi-frequency excitation sequence can be constrained to avoid auditory neuromodulation or other adverse artifacts caused by long, low-frequency pulses. Our simulated sonications used very short bursts consisting of 30 to 40 cycles of the primary 1 MHz pulse (resulting in a total pulse duration of 30--40 $\mu$s). Also, these pulses can be delivered at a very low Pulse Repetition Frequency (PRF) of 1-2 Hz as per the desired sensor refresh rate. These pulse parameters are sufficient to generate parametric array signals without crossing thermal and mechanical safety thresholds. Our simulation domain included the entire realistic 2D skull-brain slice to capture the nonlinear propagation effects as the beam passes through the proximal skull layer, the brain parenchyma/ventricles (interaction zone), and the distal skull layer.  We eventually detect the external PA signal in transmission mode outside the skull cavity.

\subsection{ Lateral Misalignment Robustness}
We performed a robustness analysis to evaluate performance under realistic conditions with sensor placement errors. We modeled a fixed hardware setup that incorporates a physical acoustic plano-concave lens to focus the 1 MHz beam at a depth of 60 mm. We introduced lateral positioning errors by shifting the patient anatomy relative to the fixed probe in 1 mm increments over a $\pm 5$ mm range. The diagnostic metric was defined as the maximum amplitude of the signal envelope captured by a broad receiver window (Region of Interest, ROI) of 20 mm positioned in the far-field. This broad ROI was selected to determine if spatial integration could maintain diagnostic contrast despite localized geometric beam steering caused by the skull.

\subsection{Frequency Dependence}
We also probed the frequency dependence of our sensing modality's performance.
We characterized the sensitivity across a range of different frequencies ($\Delta f$) from 20 kHz to 200 kHz. The goal of this study was to understand the optimal operating window. We considered the trade-off between nonlinear conversion efficiency (which favors higher frequencies) and the practical limitations of transducer bandwidth.

\section{Results}

\subsection{Parametric acoustic (PA) feedback detects ventricular expansion}
Our in silico models showed expected propagation behaviors when propagated across the skull-brain layers. The primary 1 MHz field suffered $\sim$20 dB attenuation and scattering at the entry skull interface; the parametric array signal, or the difference-frequency field generated at 100 kHz, indicated better penetration due to favorable inverse-frequency attenuation. Also, the difference frequency is generated cumulatively along the propagation path within the tissue. This allows it to maintain a coherent beam profile that effectively tunnels through the distal skull layer.

\begin{figure}[htbp]
\centering
\includegraphics[width=1.0\textwidth,trim=8 5 10 10,clip]{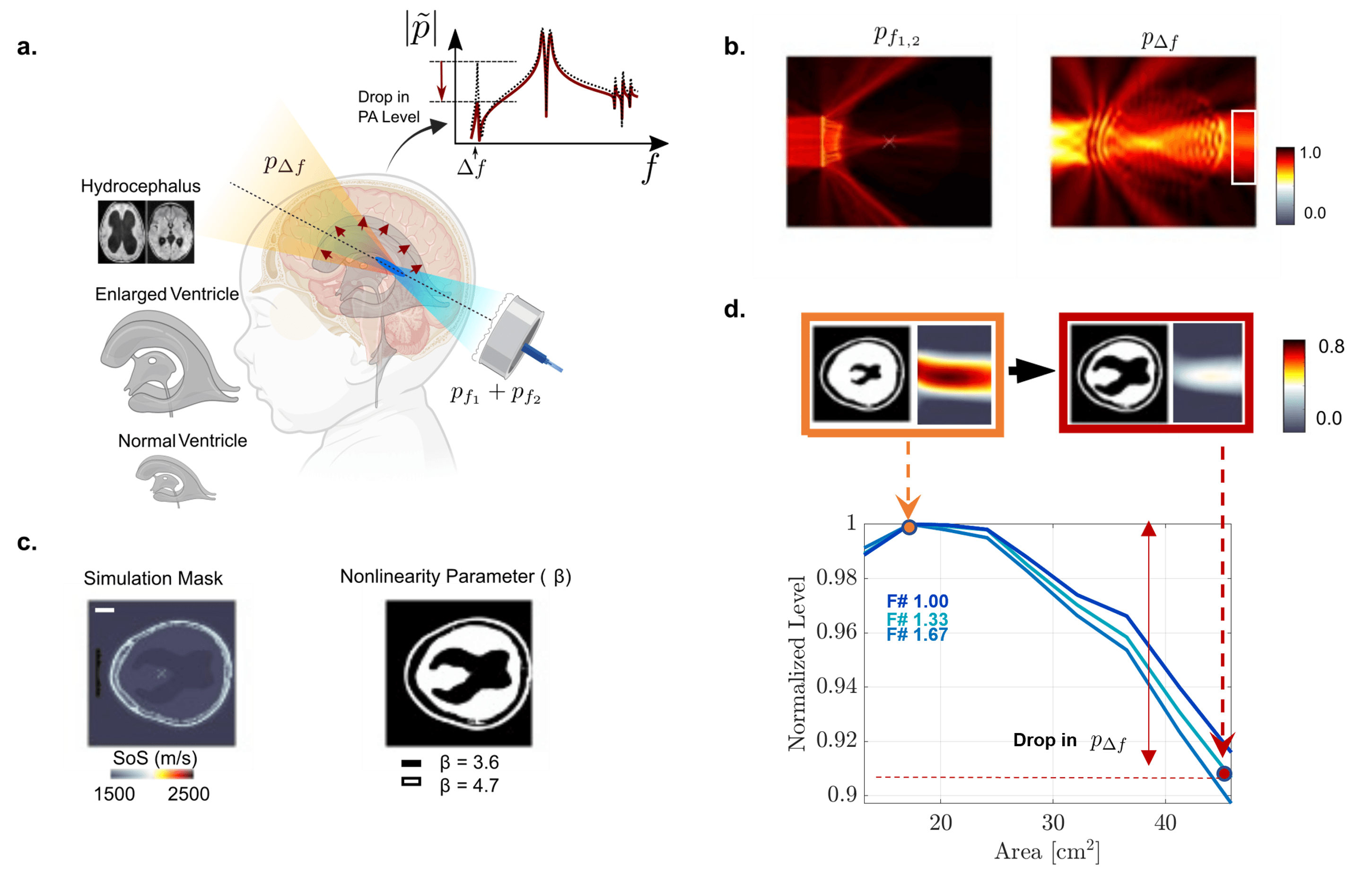}
\caption[Nonlinear acoustic feedback can detect relative changes in ventricular size.] {Nonlinear acoustic feedback can detect relative changes in ventricular size. (a) Schematic showing hydrocephalus monitoring using the parametric array effect: increasing hydrocephalus leads to a drop in the PA signal. (b) The simulation mask and the nonlinearity parameter map correspond to enlarged hydrocephalus. (c) Primary and parametric field obtained from simulation. (d) Peak parametric signal pressure outside the skull cavity decreases with increasing ventricular size.}
\label{fig:hydrocephalus}
\end{figure}

Next, we simulated ventricular expansion using a clinically relevant scenario for untreated hydrocephalus~\cite{bendella2024brain}. Our results show that as the ventricular size doubles (progressively expanding from 20 cm$^2$ to 40 cm$^2$ ventricular cross-sectional area), the PA signal measured outside the skull cavity drops by approximately 10\% (Fig.~\ref{fig:hydrocephalus}d).  This drop in amplitude suggests a drop in nonlinear generation efficiency, as expected. As the interaction zone is increasingly occupied by low-nonlinearity CSF, the generated PA signal drops as per Equation~\ref{eq:parametric}. This decrease is also sensitive to the transducer F-number. Tightly focused beams ($F\#1.00$) leading to a larger relative drop in parametric pressure.

\subsection{Robustness analysis shows a consistent diagnostic margin}
\begin{figure}[htbp]
\centering
\includegraphics[width=0.7\linewidth]{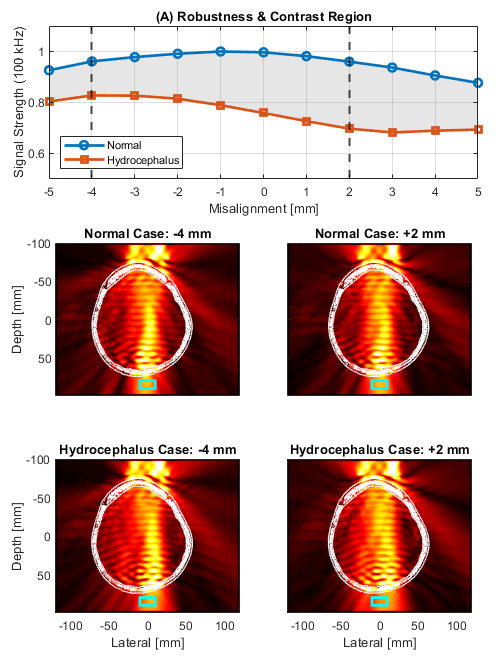}
\caption[{Robustness Analysis and Diagnostic Margin in Hydrocephalus Monitoring.} ]{Robustness Analysis and Diagnostic Margin in Hydrocephalus Monitoring. (A) Normalized parametric signal strength (100 kHz) as a function of lateral probe misalignment ($\pm 5$ mm). The trend lines for Normal (blue circles) and Hydrocephalus (orange squares) do not intersect. This creates a distinct \textbf{Diagnostic Contrast Region} (shaded gray). The vertical dashed lines correspond to the misalignment scenarios shown in the lower panels.
(B) 2D acoustic field maps showing the spatial distribution of the difference frequency pressure. The top row displays the Normal case, and the bottom row displays the Hydrocephalus case with lateral shifts of -4 mm and +2 mm (indicated by the cyan receiver box). The Hydrocephalus cases demonstrate a steady reduction in signal intensity due to the volumetric replacement of high-nonlinearity brain tissue with low-nonlinearity CSF. }
\label{fig:robustness}
\end{figure}

Lateral probe misalignment ($\pm 5$ mm) simulations were used to determine system reliability under typical operational conditions. The results( Fig.~\ref{fig:robustness}) show that the PA signal levels outside the skull cavity for the Normal state (ventricle size 20 cm$^2$) and the Hydrocephalus state (ventricle size 40 cm$^2$)  do not intersect.

In linear acoustic models, complex skull scattering can lead to overlapping signal distributions for normal and disease states. Our nonlinear analysis showed that the Normal signal amplitude remains higher than the Hydrocephalus signal across the entire sweep. This points to the robustness of the method. The detector acts as a spatial integrator due to a broad receiver window in the far field. This broad ROI captures the bulk of the forward-propagating energy flux and averages out local geometric steering artifacts caused by skull curvature. The hydrocephalic brain produces less nonlinear signal due to CSF replacement, which leads to lower integrated signal at the difference frequency. This creates a persistent diagnostic margin (Fig.~\ref{fig:robustness}A, shaded region) for reliable differentiation between the two states.

\begin{figure}[!htbp]
\centering
\includegraphics[width=0.7\linewidth]{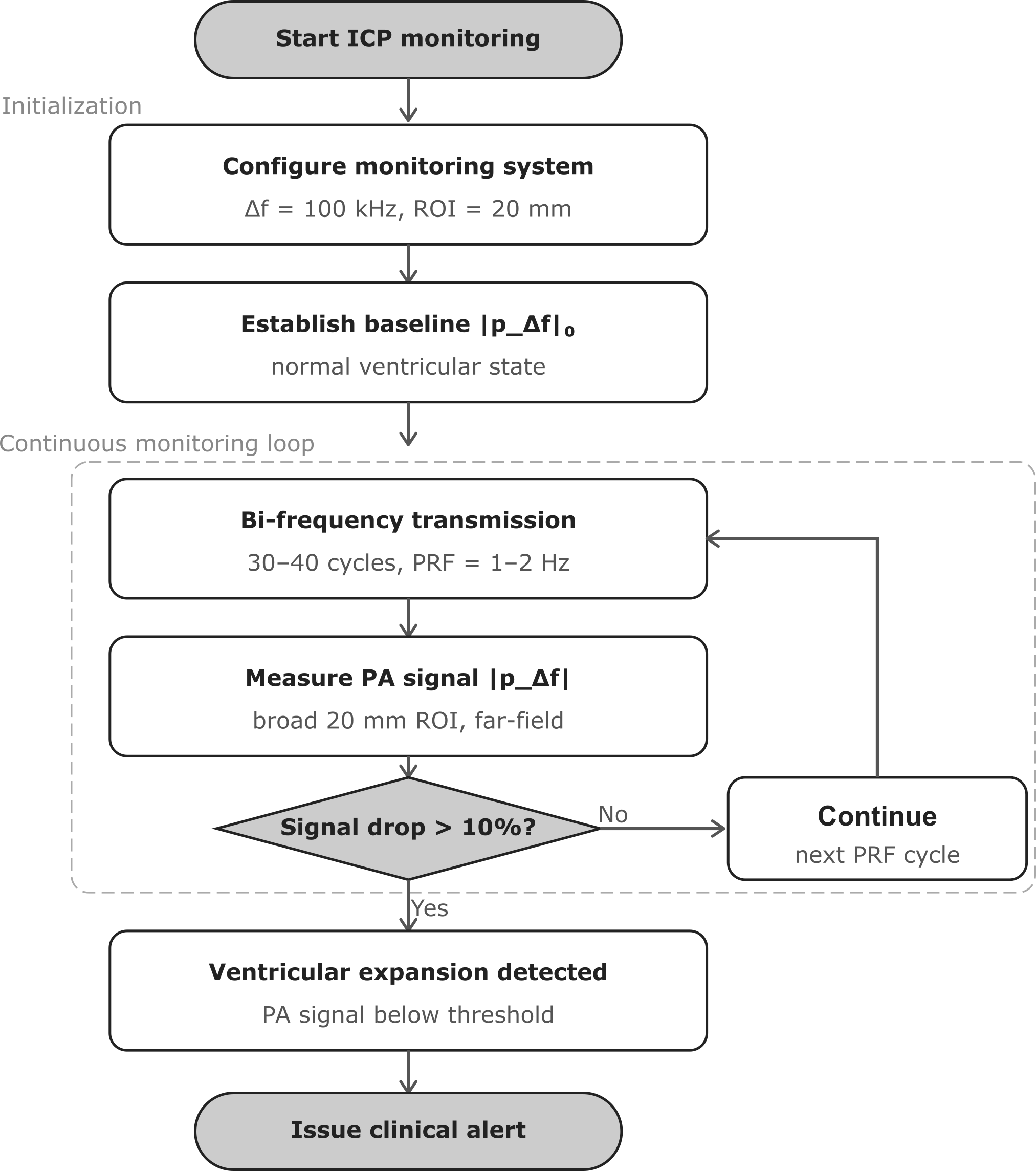}
    \caption[Process Flow Chart for ICP Monitoring in a Clinical Setting using PA signal]{\textbf{Process Flow Chart for ICP Monitoring in a Clinical Setting using PA signal} The protocol has an initialization phase and a continuous monitoring loop.  The system is configured with a difference frequency ($\Delta f$) of 100 kHz and a region of interest (ROI) of 20 mm, establishing a baseline PA signal magnitude ($|p_{\Delta f}|_0$) representing baseline ventricular state.  The system then uses bi-frequency transmission (30--40 cycles, PRF = 1--2 Hz) to measure the variation in PA signal ($|p_{\Delta f}|$) in a loop. If the measured signal drops by more than 10\% relative to the baseline, the system detects ventricular expansion and issues a clinical alert.}
    \label{fig:icp_monitoring_flowchart}
\end{figure}

\subsection{Frequency dependence shows an optimal operational window}
\begin{figure}[htbp]
\centering
\includegraphics[width=0.8\linewidth,trim= 2 0 5 0,clip]{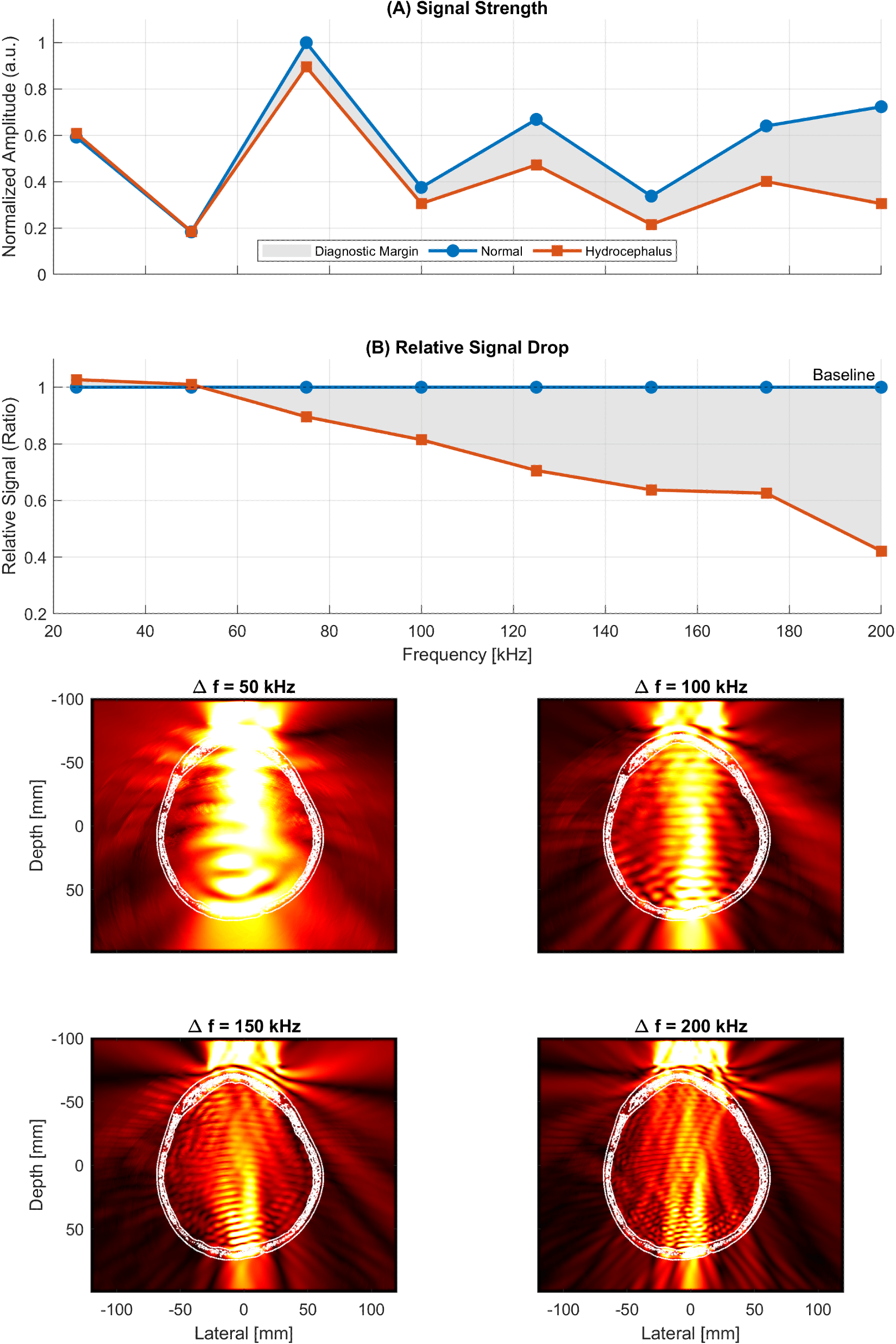}
\caption[Frequency Robustness Analysis in Hydrocephalus Monitoring.] {Frequency Robustness Analysis in Hydrocephalus Monitoring. (A) Normalized signal strength across various difference frequencies ($\Delta f$). (B) Relative signal drop (Hydrocephalus/Normal). A stable diagnostic contrast (Ratio $< 1.0$) appears above 75 kHz. (Bottom Panels) Acoustic field maps show beam collimation. At low frequencies ($\Delta f = 50$ kHz), the beam is diffuse due to diffraction, which reduces spatial sensitivity. At higher frequencies ($\Delta f \geq 100$ kHz), the parametric array forms a collimated beam. This maximizes the interaction with the ventricular volume.}
\label{fig:freq_robustness}
\end{figure}

The analysis of difference frequencies ($\Delta f$) revealed a trade-off between system performance and diagnostic contrast (Fig.~\ref{fig:freq_robustness}). At low difference frequencies ($\Delta f \le 50$ kHz), the nonlinear conversion efficiency dropped as expected from the theory of parametric acoustic arrays~\cite{westervelt_parametric_1963}. Additionally, the larger acoustic wavelengths ($\lambda \approx 30$ mm) led to diffraction, resulting in a diffuse beam that failed to interact with the ventricular volume, thereby lowering the diagnostic contrast.

On the other hand, generating higher difference frequencies (e.g., 200 kHz) requires a wider separation between the primary frequencies; this separation is limited by the bandwidth of piezoelectric transducers (typically high-Q). A larger separation reduces primary pressure amplitudes. Our simulation identified the 75--125 kHz range as the optimal operational window. In this regime, the parametric conversion efficiency is sufficient to generate a measurable signal. Also, the beam maintains effective collimation to interact with the ventricles. Thus, the diagnostic contrast remains stable at $\sim$10--15\%.

\section{Discussion}

Monitoring relative changes in the parametric pressure outside the skull cavity can help us gauge ventricular expansion during hydrocephalus progression. It can also be used to detect ventricular shrinkage during successful shunt treatment. The separation between normal and hydrocephalus states (termed as diagnostic contrast) shown during robustness testing (Fig.~\ref{fig:robustness}) indicates that a decrease in nonlinear gain is the dominant mechanism behind the sensitivity; the contrast relies on changes in the medium's effective nonlinearity parameter ($\beta$). Expanding ventricles replace higher-nonlinearity tissue with lower-nonlinearity CSF in the interaction zone. Such displacement results in lower total nonlinear acoustic energy at the difference frequency and in a stable signal offset. Additionally, the non-intersecting nature of these robustness curves for normal and hydrocephalus states enables threshold-based classification. A PA signal below a baseline threshold can indicate pathological ventricular expansion. This does not need complex image reconstruction or expert operator interpretation. The diagnostic margin of the signal across $\pm 5$ mm of lateral misalignment implies that precise stereotactic placement is not required. This misalignment tolerance highlights the key advantage of the PA method over conventional 500 kHz linear pulse-echo ultrasound. The latter would rapidly lose the required specular reflection from the ventricular wall under similar probe translation. In other words, for meaningful interpretation, we need to put the probe in the exact configuration that it was during baseline calibration. This makes the linear pulse-echo method quite restrictive.
Using our PA-based approach, on the other hand, a patient or caregiver applying a wearable sensor based on general anatomical landmarks would have a high probability of obtaining a valid diagnosis. This built-in tolerance to user error provides a safety margin against false negatives (missed shunt failures). This matters most when monitoring elevated ICP via the proxy of ventricular expansion) in an outpatient setting. ICU patients with External Ventricular Drains (EVDs) have direct ICP measurement. We target the monitoring gap that opens once the patient leaves the hospital.  Our PA-based monitoring approach provides a reliable, non-invasive surrogate for tracking ventricular volume after the EVD has been removed. This addresses a major limitation in long-term management of hydrocephalus.

 Our results argue for safer, non-invasive diagnostics to reduce infection and hemorrhage risks~\cite{geraldini2022transcranial, caricato2014echography} as compared to gold standard invasive ICP monitoring ~\cite{zhang2017invasive}. By using tissue acoustic properties alone, this modality also offers a favorable safety profile compared to contrast-enhanced ultrasound~\cite{zhang2021wearable}. Future efforts will require extending our models to complex 3D human skull geometries and validating them experimentally in large-animal models (e.g., non-human primates). Additionally, characterizing the nonlinearity parameter of CSF and the brain in vivo will improve experimental correlation.

Apart from its immediate implications for monitoring ICP post-shunt treatment, this approach could also be used for early detection and disease progression monitoring in resource-limited settings. Alternatively, it could also be integrated into existing ultrasound imaging modalities for diagnosing hydrocephalus (e.g., in infants) to improve their accuracy, as our method does not rely on
 operator skill and  favorable acoustic windows~\cite{jiang2022invention, moskowitz2010cumulative, filippou2018recent}.

\section{Conclusions}

This chapter showed the computational feasibility of using the parametric acoustic array effect to non-invasively monitor variations in ventricular volume (a proxy for ICP changes) associated with hydrocephalus. The acoustic nonlinearity contrast between brain parenchyma and cerebrospinal fluid produces a steady, detectable signal attenuation in transmission mode. PA suppression as a diagnostic contrast for hydrocephalus remains stable across a wide operational frequency band (75--125 kHz) and is tolerant to lateral probe misalignment ($\pm 5$ mm). It thus addresses the main reliability barriers in wearable sensor design.

 To conclude, our monitoring mechanism supports the development of operator-independent, continuous, portable monitoring tools that rely on tissue acoustic properties.  Hopefully, this could reduce the clinical risks and healthcare burdens associated with current invasive hydrocephalus management methods.

\chapter{\textsl{In-vitro} High-throughput Ultrasound Neuromodulation: Platform Development and Proof-of-Concept}
\label{chap:inVitroNeuroModulation}

\section{Introduction}
Noninvasive neuronal stimulation allows researchers to modulate brain activity without surgical intervention. Traditional modalities like transcranial electric stimulation (tES) \cite{nitsche2008transcranial} and transcranial magnetic stimulation (TMS) \cite{walsh2000transcranial} are widely used in basic and translational neuroscience, but they suffer from low spatial selectivity because their applied electric and magnetic fields diffuse easily \cite{faria2011finite,deng2013electric}.

In contrast, Focused Ultrasound (FUS) offers an alternative by facilitating the propagation of mechanical waves deep within the neuronal tissue without compromising spatial targeting precision. These mechanical waves, based on the pulsing regime, can elicit several thermal and non-thermal bioeffects (such as acoustic cavitation, fluid streaming, and radiation pressure) \cite{haar2010ultrasound}. Consequently, ultrasound as a tool for noninvasive neuromodulation has garnered increasing interest in recent years. Supporting investigations have revealed the role of polymodal US-neuron interaction, including acoustic cavitation \cite{krasovitski_intramembrane_2011,king_effective_2013,plaksin2014intramembrane}, shear stress \cite{prieto_activation_2018,liao_optimal_2021}, and radiation pressure \cite{hoffman_focused_2022,kubanek_ultrasound_2018}, across diverse neuronal populations \cite{hoffman_focused_2022,yoo_focused_2022,cotero_noninvasive_2019}, as well as across different species (such as worms, rodents, and non-human primates) \cite{kubanek_ultrasound_2018,folloni_manipulation_2019,tufail_transcranial_2010,kubanek_remote_2020}. However, despite these encouraging findings, the complex nature of US-neuron interactions remains elusive. US exposure settings that may lead to a desired robust activation or suppression of certain neuronal populations are poorly understood \cite{blackmore_ultrasound_2019,darmani_non-invasive_2022}. Furthermore, the specific genes and molecular pathways responsible for neuromodulation within neurons are not fully characterized. This underscores the need for further controlled \textsl{in-vitro} investigations to methodically disentangle the cellular and molecular basis of US-mediated neuromodulation.

A critical challenge in advancing ultrasound neuromodulation is the fundamental trade-off between spatial precision and penetration depth. Mechanical stimuli elicit diverse neuronal responses, ranging from transient to sustained, depending on their magnitude and location \cite{gaub2020neurons}. To apply mechanical stimuli at cellular and subcellular levels, patch clamp tips or atomic force microscopy (AFM) cantilevers have been used in the literature at nanometer spatial resolution and piconewton force sensitivity \cite{alsteens2017atomic,krieg2019atomic}. However, therapeutic ultrasound (1-10 MHz) operates on much larger scales (0.15-1.5 mm). While the acoustic radiation force is an established biophysical mechanism for ultrasound neuromodulation \cite{rabut2020ultrasound,mohammadjavadi2022transcranial}, both the targeting precision and magnitude of mechanically induced stress show a positive dependence on frequency. This creates a challenging scenario: higher spatial targeting (sub-millimeter) necessitates the application of very high frequencies (i.e., exceeding 10 MHz) \cite{cadoni2023ectopic}. However, increasing the frequency restricts the penetration depth, thereby limiting the proposed method to superficial applications.

We address these limitations by developing an in-vitro apparatus \cite{Arvanitis2023Ultrasound} that isolates the specific mechanical effects of ultrasound (such as acoustic radiation force (ARF)) from thermal and streaming confounders. We also show that using contrast-enhanced ARF in combination with biospheres can enable targeted, subcellular mechanical stimulation at lower, clinically relevant frequencies. Our efforts in this chapter focus on the preliminary work required to design and validate this in vitro platform. We first outline the design of the apparatus, which is optimized to enhance acoustic radiation force while suppressing thermal effects and fluid shear forces \cite{sato_ultrasonic_2017,guo_ultrasound_2018,rabut2020ultrasound}. We then introduce the theoretical framework for using micron-scale, biocompatible metallic spheres to locally amplify mechanical stress via their high acoustic impedance. Finally, we present preliminary physical and biological experiments utilizing Dorsal Root Ganglion (DRG) neurons and surrogate 4T1 cells to validate the platform's capabilities.

\section{Methods}

\subsection{Experimental Setup Design}
\begin{figure*}[!htb]
    \centering
    \includegraphics[width=1\textwidth,trim= 8 8 8 8,clip]{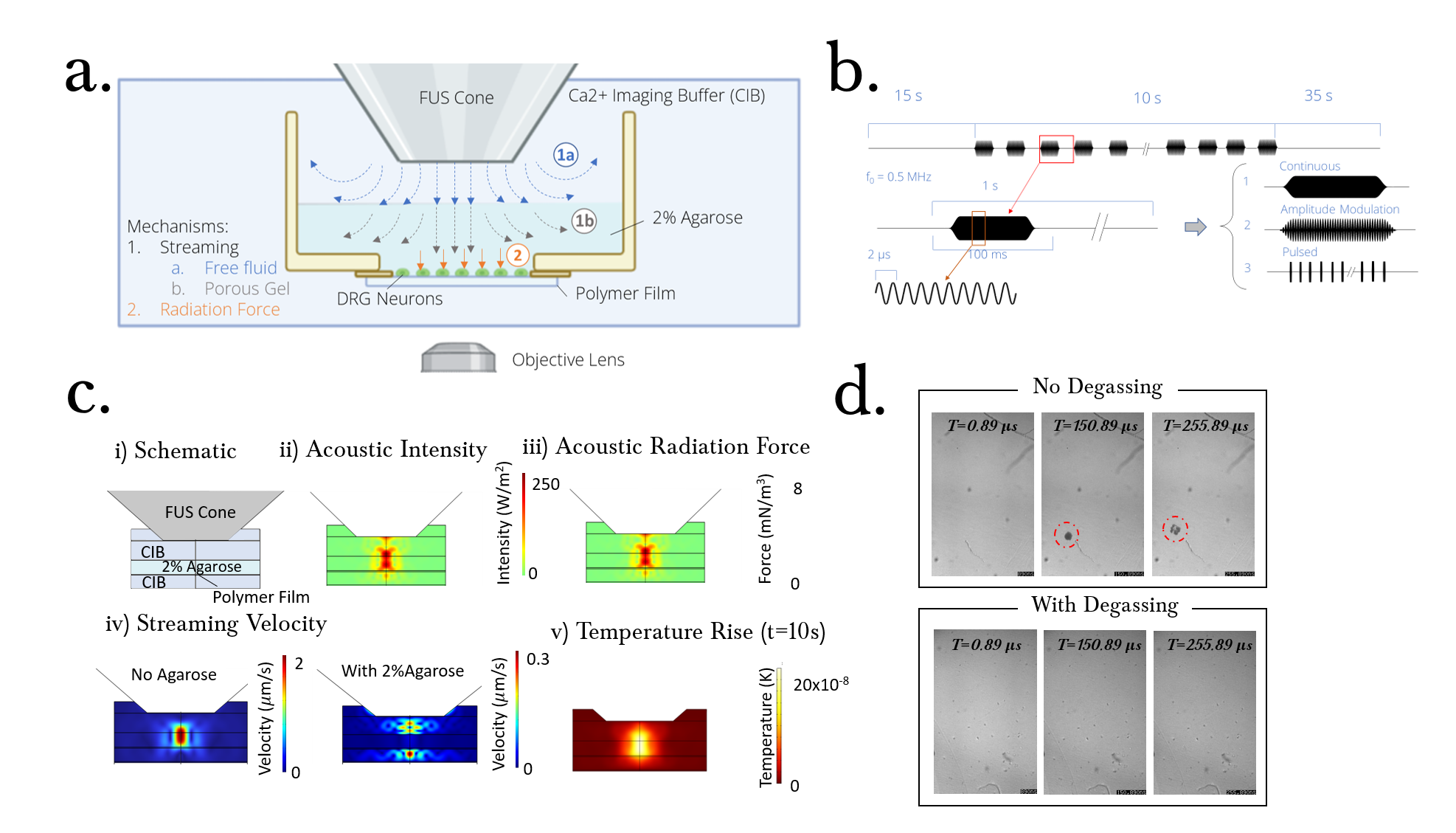}
    \caption[High Throughput US Apparatus for in-vitro Neuromodulation]{ High Throughput US Apparatus for in-vitro Neuromodulation a) Schematic showing in-vitro aparatus with poly-modal US interaction with DRG neurons, b) US pulse sequence used in the in-vitro characterization, c) Finite elmement simulation for acoustic characterization (i-ii), Prevention of fluid streaming using agarose layer(iv), and Thermal effects (v) d) Ultrasonic degassing and qualitative estimation of cavitation with high frame rate optical microscopy  }
    \label{fig:DRGNsetup}
\end{figure*}

To provide a controlled environment to promote the acoustic radiation force while minimizing thermal effects and fluid-streaming-related shear forces, our experimental setup was designed for simultaneous US exposure and high-throughput calcium imaging (Figure~\ref{fig:DRGNsetup}a). The setup was based on an inverted fluorescent microscope (Nikon Ti, Japan). US waves were transmitted from above using a custom-made \SI{40}{mm} diameter 0.5 MHz focused transducer (F\# 0.75) in a direction perpendicular to the bottom of the experimental chamber. The transducer was coupled to the experimental chamber using a 3D-printed cone filled with degassed deionized water and an acoustically transparent film at the interface. The outer experimental chamber was a 3D-printed cylindrical well fitted with a glass bottom and filled with $Ca^{2+}$ imaging buffer solution, in which the cell culture dish was placed at a \SI{2}{mm} elevated position. This ensured that the pressure release surface due to the water-air interface at the bottom was away from the area where the cells were cultured (Figure~\ref{fig:DRGNsetup}b). Furthermore, to avoid the formation of local minima caused by standing waves, we established a consistent distance of \SI{3}{mm} between the substrate and the tip of the FUS cone or collimator. This parameter was determined through numerical simulations and was validated experimentally through pulse and echo measurements.

To optimize acoustic transmission and cell viability, our design entailed a comprehensive investigation of different materials for cell culture substrates (such as glass coverslips, mylar, and polymers) and several types of coatings on top (poly-d-lysine plus laminin, matrigel, and collagen). We found that culturing cells directly on a polymer bottom dish (ibidi, Germany) with poly-D-lysine plus laminin pre-coating provided optimal results. The selection of a thick polymer substrate ($\SI{150}{\mu m}$) allowed maximal US transmission and minimal heating owing to acoustic absorption. To avoid fluid streaming (Figure~\ref{fig:DRGNsetup}b) due to the presence of free fluid between the substrate and FUS cone, a \SI{2}{\%} wt/vol agarose layer was added over the cells. This demonstrably reduced the streaming without affecting the acoustic fields.

Finally, to minimize the unpredictable effects of cavitation, a rigorous degassing protocol was established. We degassed all relevant fluids that came in contact with the cells, such as the calcium imaging buffer and agarose solution, using an ultrasonic degasser (Branson CPX2800H) at a temperature below the agarose gelation point (~50°C). We verified the absence of cavitation activity optically by observing high frame time recordings of sonication (\SI{10}{million} frames per second) (Figure~\ref{fig:DRGNsetup}d), confirming the efficacy of our degassing protocol.

\subsection{US-Pulse Parameters}
Our ultrasound pulsing sequence for varying acoustic stimuli involves a 3-layered structure(Figure~\ref{fig:DRGNsetup}c). Careful design of pulse parameters controls the degree to which thermal and streaming effects are amplified relative to mechanical stimulation. In the top layer, the total sonication duration (SD) was kept at \SI{10}{sec}. This is interleaved with \SI{15}{sec} and \SI{25}{sec} of the idle period for background measurements. Each SD is, in turn, divided into a set of burst durations (BD) lasting \SI{100}{msec} with a repetition frequency of \SI{1}{Hz} or \SI{2}{Hz}. Each burst can have various pulse distributions (e.g., continuous, modulated, or pulsed). Lastly, each pulse contains a waveform at the carrier frequency of \SI{0.5}{MHz}.

\subsection{Theoretical Formulation of Acoustic Radiation Force}
To establish the physical basis for applying localized mechanical stress to individual cells using contrast agents, we evaluated the Acoustic Radiation Force (ARF). ARF is a time-averaged net force on an object (e.g. a sphere) placed in an acoustic field due to the interaction between incident and scattered acoustic field from the object.\cite{sarvazyan2010biomedical}

For objects of arbitrary size, $\boldsymbol{F}^{\mathrm{rad}}$ can be calculated by integrating the second-order pressure (also known as radiation pressure) over the surface of the object.\cite{sapozhnikov2013radiation,gor2014forces,yosioka1955acoustic}. Based on this definition, a relatively easier and versatile approach \cite{glynne2013efficient} to quantify the ARF is to compute the scattered acoustic field owing to the presence of the object and employ a numerical approach to compute the change in momentum flux.\cite{sapozhnikov2006radiation}. Assuming the fluid medium to be inviscid the second-order radiation pressure can be expressed in terms of first-order linear quantities as \cite{bruus2007theoretical}:

\begin{equation}
\left\langle p_2\right\rangle=\frac{1}{2 \rho_0 c_0{ }^2}\left\langle p_1{ }^2\right\rangle-\frac{1}{2} \rho_0\left\langle v_1{ }^2\right\rangle
\end{equation}

where $\rho_0$ and $c_0$ are the equilibrium fluid density and speed of sound, and $p_1$ and $v_1$ are the time-harmonic linear acoustic pressure and particle velocities, respectively. Here $\langle.\rangle$ denotes time averaging. By integrating the normal component of $p_2$ over the surface of the particle ($S_0$) acoustic radiation pressure can be computed as \cite{yosioka1955acoustic}:
\begin{equation}
\boldsymbol{F}^{\mathrm{rad}}=\int_{S_0} p_2 \mathbf{n} d \mathbf{a}-\int_{S_0} \rho\langle(\mathbf{v_1 n}) \cdot \mathbf{v_1}\rangle d \mathbf{a}
\end{equation}

Here, the second term on the right side is a compensating term that accounts for the convective momentum flux due to the time-dependent movement of the surface enclosing the particle $s(t)$ which is considered a fixed $S_0$ in the boundary integration.

For a small spherical object of radius $a$ in a progressive ultrasound field of wavelength $\lambda$ such that $a\ll\lambda$, ARF can be expressed as \cite{settnes2012forces}:

\begin{equation}
\boldsymbol{F}^{\mathrm{rad}}=\frac{4 \pi}{3} a^3\left[\operatorname{Im}\left[f_1\right] \frac{\kappa_0}{2}\left\langle p_{\text {in }}^2\right\rangle+\operatorname{Im}\left[f_2\right] \frac{3 \rho_0}{4}\left\langle v_{\text {in }}^2\right\rangle\right] \boldsymbol{k}
\end{equation}
Where $\kappa_0$ and $\rho_0$ is the isentropic compressibility and density of the background medium, $p_{\text {in }}$ and $v_{\text {in }}$ are incident first-order pressure and velocity fields. $f_1$ and $f_2$ are the compressibilities and density contrast factors, respectively, on which the magnitude of the radiation force depends. This strong dependence on acoustic contrast indicates that introducing a bio-sphere with a significantly higher acoustic impedance than the surrounding tissue can generate highly localized mechanical stress.

\subsection{Formulation of Dynamic Acoustic Radiation Force}
To explore the theoretical potential of simulating transient tactile stimuli, we evaluated Dynamic Radiation Force (DRF). Several studies have demonstrated that tactile perception arises primarily from the leading and trailing edges of transient mechanical pulses, highlighting heightened neural sensitivity to the temporal derivative of the applied stimulation \cite{gavrilov1984use,gavrilov2012focused}. This indicates that neural structures might be more sensitive to the gradient of applied stimulation. Thus, exploring the effect of time-varying radiation force or Dynamic Radiation Force (DRF), widely used in vibroacoustography \cite{fatemi2000probing}, on the activation of bio-sphere-labeled neurons is worth exploring in future iterations. DRF on an area $S$ subjected to an acoustic wave with energy density $\langle E \rangle$ can be described as:
\begin{equation}
F = d_rS\langle E \rangle
 \label{eqn:DARF}
\end{equation}
where $d_r$ is the vector drag coefficient in the wave propagation direction. To produce an oscillating radiation force, incident ultrasound ($f_0 = \omega_0/2\pi$) is typically amplitude modulated at a desired low frequency ($\Delta f = \Delta \omega/2\pi$), typically in the kHz range \cite{fatemi1999vibro}. Notably, nonlinearities within the medium can induce parametric amplification of the DRF, leading to even finer subwavelength localization of forces within the kilohertz range \cite{silva2006parametric}.

\subsection{Quantification of ARF-induced Bio-sphere Displacement}
To empirically quantify the displacement of biospheres owing to the ARF without biological confounders, we utilized an \textsl{in-vitro} phantom setup similar to that shown in Figure~\ref{fig:DRGNsetup}a. This setup consists of metallic bio-spheres embedded in a layer of an agarose gel matrix (\SI{1}{\%} wt/vol ). Under the assumption of small linear deformation,  $F^{rad} = -k_{e}x$ (Hooke's law), where  $k_e$ is the effective stiffness constant of the hydrogel. Consequently, measuring the displacement allows for the estimation of the ARF.

Consider a sphere placed in the focal plane; the scattered field $u(r,z)$ from the object interferes with the undiffracted reference field to generate an interference pattern characterized by a central bright spot and surrounding rings. A change in this interference pattern was observed upon the application of ultrasound, which displaced the sphere downwards, effectively shifting the focal plane of the microscope upwards. Thus, by tracking the central pixel intensity as a function of $z$, we can obtain a set of calibration images that serve as a look-up table. We then compared them with \textsl{in-vitro} recordings of sphere movements due to ultrasound and quantified the ARF-induced displacement with precision.

\subsection{Preparation of Cellular Models}
To evaluate the operational capacity of the apparatus, we utilized Dorsal Root Ganglion (DRG) sensory neurons isolated from $\text{Pirt}^\text{GCaMP6f}$ mice, in which the pan-sensory neuronal Pirt promoter drives the expression of GCaMP6f in more than 95\% of sensory neurons. These neurons were selected due to their baseline sensitivity to mechanical stimuli.

Furthermore, to demonstrate the practical feasibility of physically attaching micro-particles to living cells within a 3D matrix, we conducted a proof-of-concept labeling assay. Because robust protocol optimization was required to refine the avidin-biotin attachment chemistry prior to utilizing sensitive primary DRG neurons, we pragmatically employed GFP-expressing 4T1 cells (chosen for their accessibility, robust adherence, and ease of manipulation) as a technical surrogate, and labeled them with avidin-coated iron oxide biospheres (Banglabs, USA). The protocol involved the creation of a 3D cell culture environment: a base layer of agarose gel (\SI{1}{\%} wt/vol) was prepared, followed by the addition of a collagen (TeloCol 6, Advanced Biomatrix, USA) layer. After growing the 4T1 cells to confluency, they were incubated with anti-GFP biotin, and then the biospheres were added.  These biosphere-labeled cells were then plated onto a collagen matrix, with a layer of agarose added on top.

\section{Preliminary  Results}

\subsection{ DRG Activation and Acoustic Parameter Study}
 We conducted preliminary experiments using the isolated DRG sensory neurons to test the apparatus. We must emphasize that these findings are preliminary and currently lack comprehensive controls. Future trials need to include proper controls such as sham sonications, temperature monitoring, and pharmacological blockers. However, they serve as a proof of concept for the apparatus design.

\begin{figure*}[!htb]
    \centering
    \includegraphics[width=1.0\textwidth,trim= 8 8 8 8,clip]{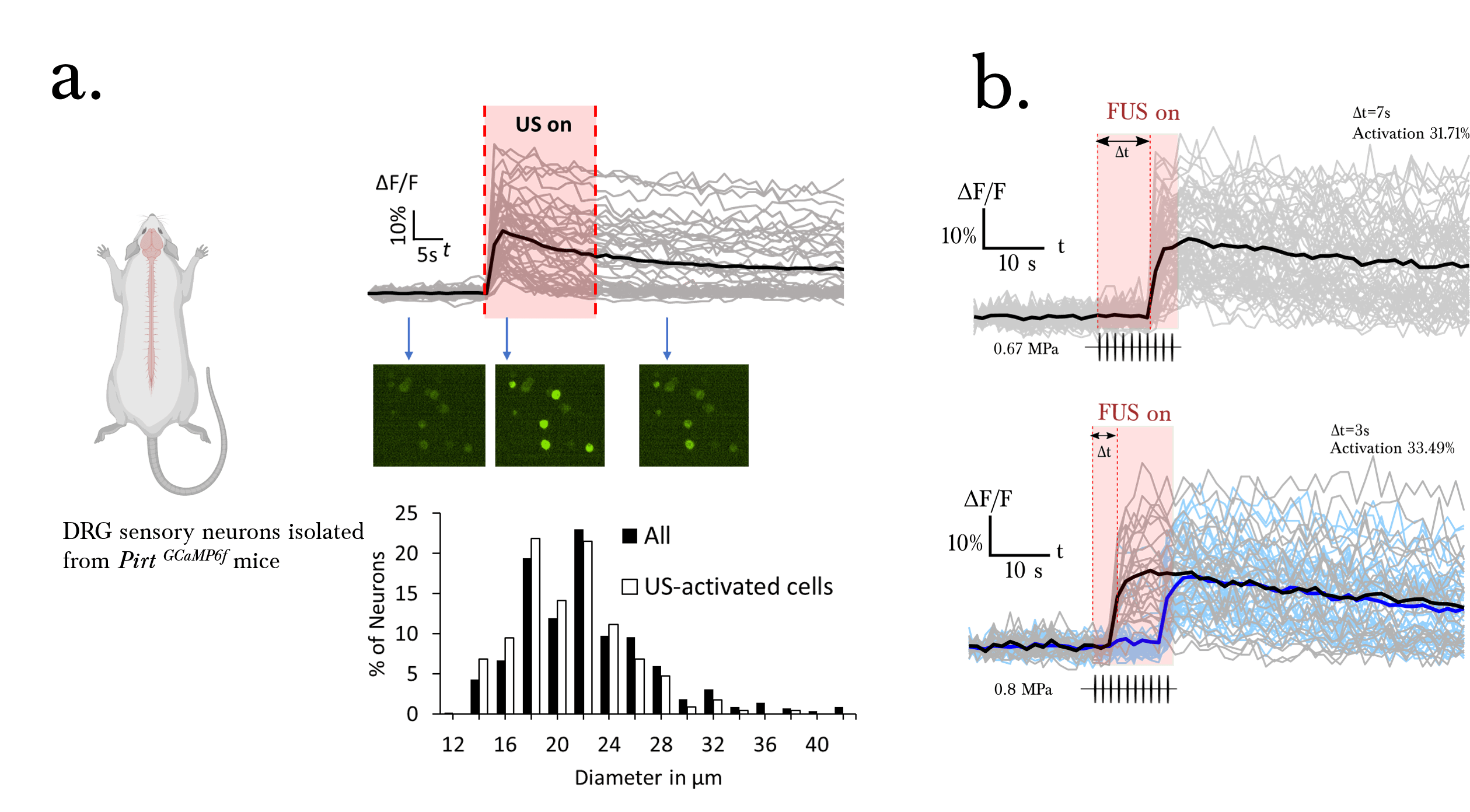}
    \caption[DRG neurons expressing GCaMP6f stimulated by ultrasound]
    { DRG neurons expressing GCaMP6f stimulated by ultrasound. a) Top: Several cells show increased fluorescence levels. This demonstrates preliminary US-mediated neuronal activation. Bottom: Quantification of the increase in relative fluorescence levels during US application. Bottom: Histogram showing that the US activated a heterogeneous group of neurons, including small-, medium-, and large-diameter neurons. b) Long pulse durations (100 msec) at different pressures (top: 0.67 MPa and bottom:0.80 MPa) lead to delayed onset of DRG neuron activation. }
    \label{fig:DRGNprelimResults}
\end{figure*}

We identified exploratory US settings (excitation frequency: 0.5 MHz; pressure: 0.67 MPa; pulse duration: 1 msec; pulse repetition frequency: 1 Hz; Number of pulses: 10) where more than 20\% of dissociated DRG sensory neurons indicated qualitative activation (Figure~\ref{fig:DRGNprelimResults}a). Upon using a long pulse duration (100 ms) with a pulse repetition frequency of 1 Hz and 10 pulses, we observed activation of 31.7\% at 0.67 MPa and 33.49\% at 0.8 MPa (Figure~\ref{fig:DRGNprelimResults}b). Interestingly, we observed a delay $\Delta t$ of $\SI{7} {sec}$ in the trigger of an action potential at 0.67 MPa, which reduced to $\SI{3}{sec}$ at 0.8 MPa. Long pulse durations are associated with streaming-induced shear stresses and heating, which may play a role in the delayed onset of neuronal activation.

\begin{figure*}[!htb]
    \centering
    \includegraphics[width=0.8\textwidth,trim= 8 8 8 8,clip]{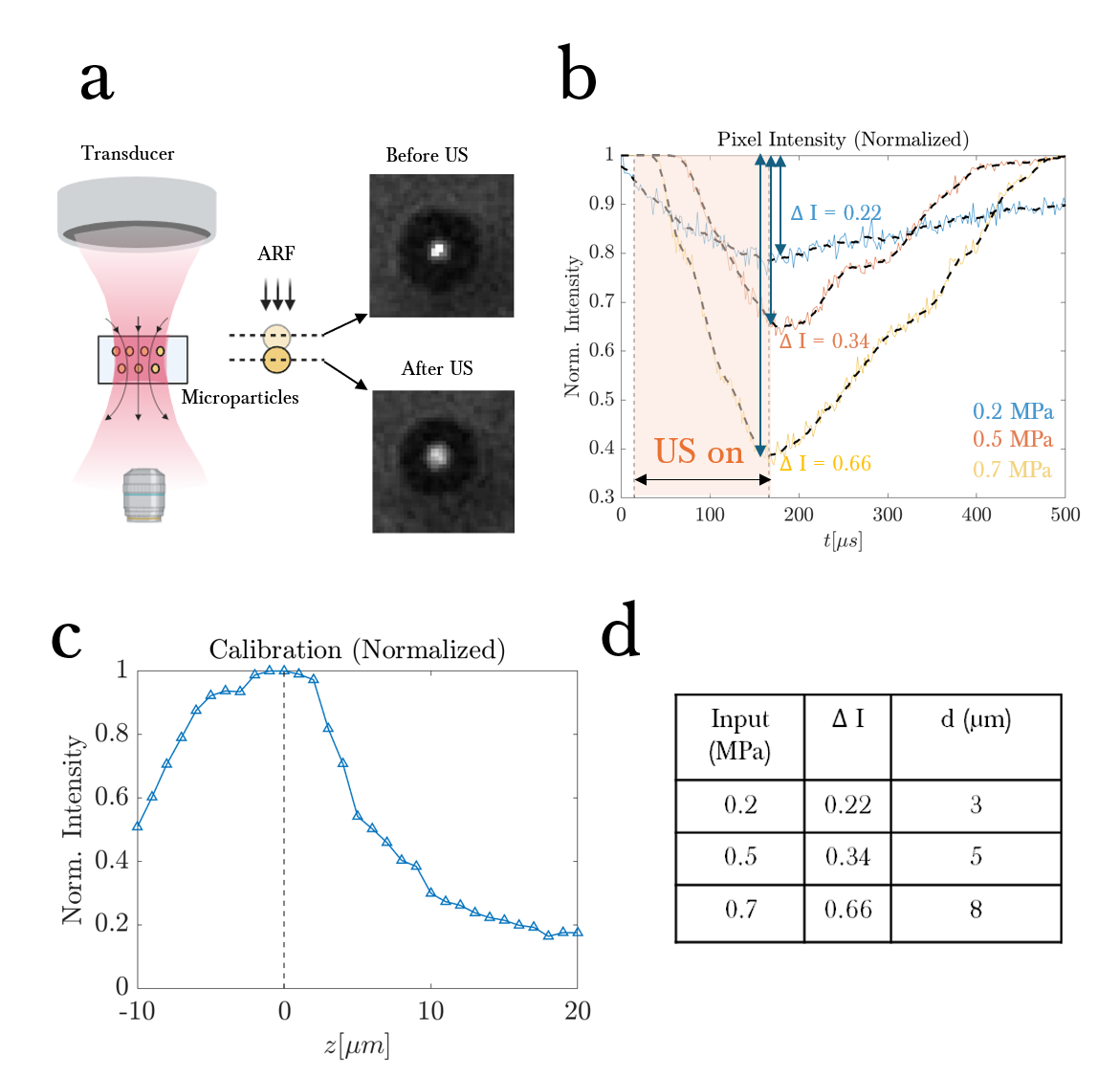}
    \caption [Quantification of ARF-induced Bio-sphere Displacement] {a) Schematic of the experimental setup indicating micro-spheres being pushed down by Ultrasound ARF leading to blurring of the image once in focus, b) Normalized intensity drop of the central pixel, c) Calibration curve obtained by z-stack images of the same sphere, d) quantification of the ARF induced displacement. }
    \label{fig:biosphereDisplacement}
\end{figure*}

\subsection{Physical Validation of ARF-induced Bio-sphere Displacement}
An experimental demonstration of the optical tracking technique is shown in ~\ref{fig:biosphereDisplacement}. $\SI{6}{\mu m}$ biospheres (Banglabs, USA) were embedded in an agarose gel matrix ($\SI{1}{\%} wt/vol$ ). A 500-cycle ultrasound pulse train at 3.3 MHz was applied at varying pressures (0.2, 0.5, and 0.7 MPa). This displaced the spheres out of the microscope focal plane and resulted in a blurred image (Figure~\ref{fig:biosphereDisplacement}a).

We quantified the drop in the central pixel's intensity by tracking it during sonication at 0.5 million frames per second using high-speed videography with a Shimadzu HPV-X2 camera (Kyoto, Japan). A moving average was applied to the normalized traces to remove the high-frequency measurement noise. The drop in the mid-pixel intensity $\Delta I$ was then quantified (Figure~\ref{fig:biosphereDisplacement}b). Additionally, a set of z-stack images of the sphere was obtained to provide a mapping between the z-position and central pixel intensity. This calibration trace served as a look-up table to translate the drop in intensity $\Delta I$ to displacement $d$ in microns (Figure~\ref{fig:biosphereDisplacement}c). The sphere moved by $\SI{8}{\mu m}$ at an input of 0.7 MPa (Figure~\ref{fig:biosphereDisplacement}d). We suggest that this magnitude of displacement, when localized to a cellular membrane, could generate sufficient local mechanical stress to elicit a functional response.

\subsection{ Observation of ARF on Bio-sphere-labeled 4T1 Cells}
\begin{figure*}[!htb]
    \centering
    \includegraphics[width=1.0\textwidth,trim= 20 10 10 20,clip]{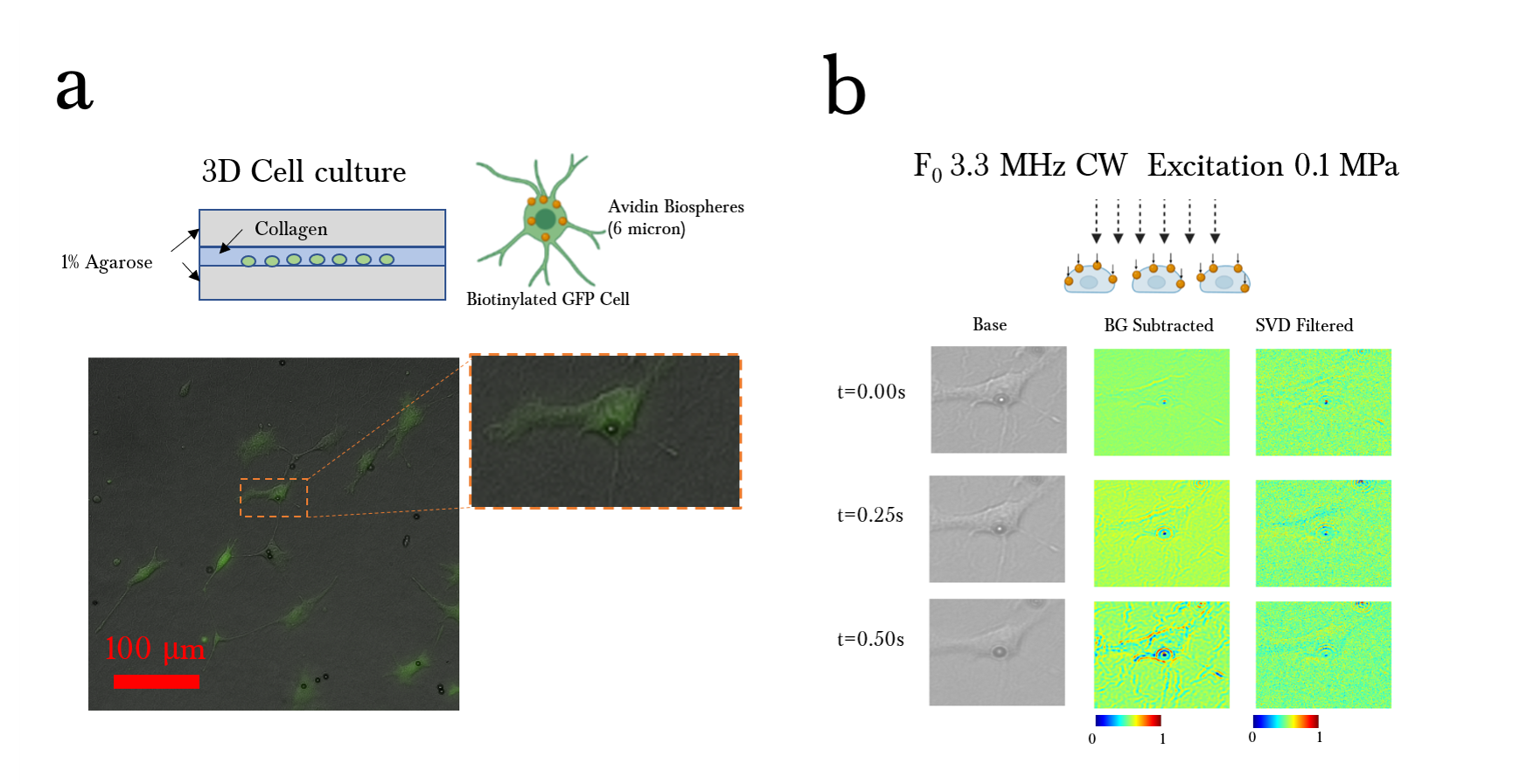}
    \caption [ARF-induced Bio-sphere labeled GFP 4T1 Cells] {a) Schematic showing 3D cell culture where  GFP 4T1 cells cultured on collagen are sandwiched between two layers of agarose. Fluorescent microscopy in the bottom confirms cell viability and successful attachment of biospheres. b) Ultrasound sonication trial with 3.3 MHz CW excitation at three different time points showing the base, background subtracted and SVD filtered images to quantify the ARF-induced bead displacement qualitatively. }
    \label{fig:4T1BiosphereSonication}
\end{figure*}

We prepared biosphere-labeled GFP 4T1 cells using the previously outlined 3D culture procedure.  We verified their viability and confirmed successful physical attachment to the biospheres (Figure~\ref{fig:4T1BiosphereSonication}a) using fluorescent imaging.

We then applied continuous wave (CW) excitation at 3.3 MHz at three different pressure levels (0.05, 0.075, and 0.1 MPa) to facilitate optical tracking. While short-pulse excitations are ideal for studying purely ARF-induced displacement, they require ultra-high frame rate acquisition to discern sphere motion. The sphere was demonstrably pushed downward by ultrasound. This is evident from the blurred base image corresponding to a sonication pressure of 0.1 MPa (Figure~\ref{fig:4T1BiosphereSonication}b).

Background subtraction and SVD filtering showed the differential motion of the sphere relative to the surrounding cell membrane. We believe that this observed motion arises from the radiation force exerted on the sphere owing to its contrasting acoustic properties. These results provide an initial proof-of-concept for targeted cellular mechanostimulation using our in-vitro setup.

\section{Discussion}

The aim of this chapter was to develop and validate an \textsl{in-vitro} platform designed to explore the poly-modal mechanisms underlying ultrasonic neuromodulation. By creating an apparatus that allows for simultaneous ultrasound exposure and high-throughput calcium imaging, we established a framework to evaluate specific acoustic parameters while minimizing macroscopic environmental confounders.

Although the biological investigations presented here are still in their early stages, they offer valuable initial observations that confirm the apparatus's utility. For example, the preliminary data show a different temporal response of DRG neurons to varying pulse durations. Longer pulses (\SI{100}{msec}) led to a delayed onset ($\Delta t \approx \SI{3}{sec}-\SI{7}{sec}$). This suggests the potential involvement of separate transduction pathways and slow-acting mechanisms. This is likely due to residual thermal accumulation or shear stress as opposed to the immediate effect of acoustic radiation force on ion channels. This shows the necessity of our agarose overlay and degassing protocols for isolating mechanotransduction events from bulk fluid dynamics.

Moreover, the introduction and quantification of biosphere labeling represent a promising step toward bridging the gap between applying forces on the cell membrane, similar to precise microindentation using AFM, and non-invasive, low-spatial-resolution therapeutic ultrasound.

The selection of 4T1 cells for the bio-sphere experiments indicates the preliminary nature of this research. Demonstrating the actuation of biospheres on easily manipulated 4T1 cells was a pragmatic step before adapting the protocol to more sensitive primary neuronal populations. Our data suggest that this contrast-enhanced approach can generate localized displacements of up to \SI{8}{\mu m} at clinically relevant frequencies. Thus, localized membrane deformation can be achieved without sonicating the entire tissue volume by using the acoustic impedance mismatch of the bio-spheres. Essentially, we devised a way to convert a macroscopic acoustic field into a microscopic, targeted mechanical stimulus.

Significant future work is required to translate these preliminary findings into firm conclusions. Through biological controls, larger sample sizes, and rigorous statistical evaluations, we must confirm mechanosensitive channel gating. The specificity of the ligand-receptor binding used to attach the bio-spheres also requires further optimization to ensure the targeted excitation of specific neuronal subtypes. Lastly, evaluating the theoretical Dynamic Acoustic Radiation Force (DRF) formulations proposed here will be important to determine if oscillating mechanostimuli can better match the time constants of specific neural circuits.

\section{Conclusions}

 The primary contribution of this work are laying the engineering groundwork for a targeted \textsl{in-vitro} ultrasound neuromodulation platform.

    \begin{itemize}    \item \textbf{Design of an Artifact-Mitigated \textsl{In-Vitro} Apparatus:} We designed a high-throughput platform pairing focused ultrasound with fluorescence imaging. This configuration ensures accurate acoustic targeting while suppressing bulk fluid streaming.

    \item \textbf{Quantification of Acoustic Radiation Force:} We developed an optical tracking method capable of measuring micro-scale displacements (~8 $\mu m$) of targets subjected to localized acoustic radiation forces. This technique provides a reliable calibration metric for future mechano-stimulation research.

    \item \textbf{Framework for Contrast-Enhanced Mechanostimulation:} We introduced a stimulation approach that uses biocompatible metallic microspheres as localized acoustic stress concentrators. We validated the labeling protocols in 4T1 cell lines and confirmed the actuation mechanisms; we then showed that macroscopic acoustic fields can be translated into targeted, subcellular mechanical stimuli at clinically viable frequencies.
\end{itemize}

In conclusion, although future controlled studies are needed to fully determine the biological mechanisms underlying the efficacy of ultrasound-sonicated neuronal activation, the in vitro platform developed here may serve as an essential foundation for exploring the molecular mechanisms underlying targeted, non-invasive neuromodulation.

\chapter{Summary and Conclusions}
\label{chap:Conclusions}

\section{Thesis Objectives and Context}

This thesis bridges the gap between the potential of acoustic holography and its clinical realization in transcranial ultrasound (TUS) therapies. Focused ultrasound has become an effective tool for non-invasive neuro-intervention; however, its widespread adoption faces key challenges, including the prohibitive cost of phased arrays, electronic packaging constraints, and the correction of skull aberrations at high frequencies \cite{hynynen1998demonstration, jolesz2014intraoperative}.

Acoustic holography, although, offers an accessible alternative by using passive, patient-specific lenses to correct skull-induced distortions \cite{melde2016holograms}, it has two bottlenecks toward clinical translation: (1) low-fidelity design approximations that fail at clinical sub-megahertz frequencies, (2) reliance on imaging modalities such as magnetic resonance imaging (MRI) for precise lens–skull registration \cite{kyriakou2014review}. We addressed these limitations by modeling the holographic lens as an acoustically thick volume. Furthermore, we modeled the human skull as a nonlinear medium to use the parametric array effect for skull-lens registration. In addition, we addressed the challenge of noninvasive monitoring of the intracranial environment during treatment and designed an artifact-mitigated \textsl{in vitro} platform that isolates specific ultrasonic mechanotransduction mechanisms for targeted neuromodulation.

\section{Thesis Contributions}

The chapter-wise key contributions of the thesis are listed  below in juxtaposition with the current state-of-the-art.

\subsection{Hologram Topology Optimization (HASA-ADAM)}

\textbf{State of the Art:} Current rapid design methods for acoustic holograms, such as the Iterative Angular Spectrum Approach (IASA), rely on the Thin-Element Approximation (TEA) to generate the lens post optimization \cite{melde2016holograms}. These models treat the lens as a simple 2D phase screen, ignoring internal wave dynamics. While full-wave time-domain solvers can capture these physics, they remain computationally prohibitive for iterative clinical-scale lens design \cite{jimenez2019holograms}.

\textbf{Contribution:} In \textbf{Chapter~\ref{ch:hologram_design}}, we showed that TEA breaks down at the sub megahertz frequencies required for transcranial targeting. At frequencies near \SI{1}{MHz} and below, the lens enters an acoustically thick regime ($L \gg \lambda$), where the high aspect ratio of lens features induces \textit{refractive walk-off} and diffraction spreading the energy laterally leading to pixel migration and volumetric cross talk.

To overcome this limitation, we introduced \textbf{HASA-ADAM}. HASA-ADAM shifts from phase-screen estimation to true volumetric topology optimization. By optimizing the 3D topology rather than a 2D phase profile, this framework accounts for internal wave propagation, amplitude modulation, and edge diffraction without the computational burden of time-domain solvers due to frequency domain propagation. This approach achieved a PSNR improvement of approximately \SI{7}{dB} over conventional methods, enabling the rapid generation ($<20$ min) of large-aperture lenses ($\approx \SI{60}{mm}$) for transcranial focusing.

\subsection{Nonlinear Acoustic Registration}

\textbf{State of the Art:} The effectiveness of hologram lenses to correct for skull-aberration depends on accurate registration with the skull anatomy. Currently, sub-millimeter registration relies on MR-guided TUS (MRgFUS) \cite{elias2016thalamotomy}, which monopolizes expensive imaging infrastructure, or optical neuronavigation \cite{chen2020neuronavigation}, which suffers from error margins ($\sim$1.5--3.5~mm) insufficient for high-frequency targeting. These limitations make passive transcranial holography inaccessible and inaccurate.

\textbf{Contribution:} Historically, acoustic models have treated the human skull purely as a passive, linear aberrator. In \textbf{Chapter~\ref{ch:hologram_registration}}, we reconceptualized the skull as an \textit{active nonlinear emitter}. By exploiting the extreme contrast in the nonlinearity parameter ($\beta$) between cortical bone and soft tissue \cite{duck2013physical}, we introduced the Parametric Array (PA) signal as a direct, real-time feedback mechanism for active alignment of the lens with skull geometry.

We showed the mechanisms underlying this sensitivity: angular misregistration triggers \textbf{Volumetric energy trapping } and \textbf{ shear mode conversion } through which misaligned primary acoustic energy is trapped within the diplo\"{e} layer and may be converted into slow-moving shear waves. We also showed that parametric generation efficiency scales inversely with the cube of wave velocity ($c^{-3}$). This trapped energy may act as a volumetric pump, amplifying the difference-frequency ($\Delta f$) signal. Our approach thus provides an alternative hologram-skull registration method by correlating optimal sub-millimeter alignment ($\theta = 0^\circ$) with a drop in the PA signal ($\geq 20\%$), which circumvents the spatial limitations of optical tracking and the infrastructure burden of MRI.

\subsection{Non-invasive ICP Monitoring in Hydrocephalus}

\textbf{State of the Art:} The clinical management of hydrocephalus relies on invasive External Ventricular Drains (EVDs) \cite{zhang2017invasive}, which carry risks of infection and hemorrhage, whereas Non-invasive surrogates, such as Transcranial Doppler \cite{qiu2025noninvasive, jiang2025advancements}, are indirect measurements often confounded by systemic hemodynamics and operator variability.

\textbf{Contribution:} Building upon our nonlinear acoustic findings, in \textbf{Chapter~\ref{chap:ICPMonitoring}} we extended the parametric array effect to introduce a novel diagnostic paradigm based on \textbf{volumetric suppression}. We established that the volumetric replacement of higher-nonlinearity brain tissue ($\beta \approx 4.7$) with lower-nonlinearity cerebrospinal fluid ($\beta \approx 3.6$) in the focal zone causes a reduction in the nonlinear amplifier gain of the intracranial medium.

We showed that clinically relevant ventricular expansion yields a detectable $\sim$10\% signal attenuation in transmission mode which is above the detection threshold of a hydrophone. By proving that this diagnostic margin remains stable across a wide operational frequency band (75--125kHz) and is highly tolerant to lateral probe misalignment ($\pm 5$mm), this contribution provides groundwork for wearable, operator-independent intracranial monitoring of ventricle expansion in hydrocephalus.

\subsection{In-Vitro Neuromodulation Platform Development}

\textbf{State of the Art:} \textsl{In-vivo} and \textsl{in-vitro} ultrasonic neuromodulation studies are frequently confounded by bulk fluid streaming, thermal accumulation, and off-target auditory artifacts \cite{sato_ultrasonic_2017}. Tools like atomic force microscopy (AFM) provide high-resolution mechanical stimulation but lack therapeutic penetration depth. Thus, to target the subcellular membrane, we need to use frequencies above 10 MHz, which restricts the modality to superficial targets \cite{cadoni2023ectopic}.

\textbf{Contribution:} In \textbf{Chapter~\ref{chap:inVitroNeuroModulation}}, we developed an artifact-mitigated \textsl{in-vitro} high-throughput hardware platform. We isolated the Acoustic Radiation Force (ARF) from thermal and streaming confounders using specific geometric spacing, degassing protocols, and agarose tissue-mimicking overlays.

We also developed a proof-of-concept prototype for contrast-enhanced mechanostimulation. We showed that the acoustic impedance mismatch acts as a local stress concentrator by attaching bio-spheres to cellular membranes. We quantified that this approach converted a macroscopic acoustic field (0.5– 1 Hz) into a targeted subcellular mechanical displacement ($\sim\SI{8}{\micro\metre}$) using optical tracking.  Thus, using our approach, localized mechanical stimuli can be achieved without using high-frequency, high-intensity fields.

\section{Significance}

The four contributions define a framework for \textbf{Precision Passive Acoustics}. By treating the holographic lens as a volumetric refractive element and the skull as a nonlinear medium, we converted obstacles into engineering tools.

\begin{enumerate}
    \item \textbf{Design:} A patient-specific lens is generated using topology optimization (HASA-ADAM) to ensure optimal diffraction-limited focal quality for both complex and point targeting. We did this by overcoming the limitations of the thin-film approximation in phase-based hologram optimization.
    \item \textbf{Register:} During the procedure, the lens is aligned using the nonlinear parametric acoustic signature of the skull itself. This eliminates the need for continuous MRI monitoring, making transcranial ultrasound therapy more accessible and portable.
    \item \textbf{Monitor:} The parametric array principles can be used diagnostically to track intracranial volumetric expansion of the ventricles. This approach ensures patient safety during hydrocephalus monitoring or treatment without the need for invasive probes.
    \item \textbf{Modulate:} The engineered \textsl{in-vitro} platform provides a standardized apparatus to isolate and study the mechanotransduction pathways required for future targeted non-invasive neuromodulation therapies.
\end{enumerate}

\section{Limitations}

\noindent\textbf{Biological Validation of Mechanostimulation.}
The neuromodulation platform (Chapter~\ref{chap:inVitroNeuroModulation}) represents exploratory proof-of-concept work. Although the actuation of the biospheres was validated, the preliminary biological observations lacked the exhaustive controls, such as sham sonications, precise continuous temperature monitoring, and pharmacological blockers required to confirm mechanosensitive channel gating.

\noindent\textbf{Material Characterization.}
The fidelity of topology optimization is highly sensitive to the exact material properties of the 3D-printed lens. Sub-wavelength discrepancies in the speed of sound owing to manufacturing and curing variations can lead to axial focal shifts, thereby requiring post-print calibration.

\noindent\textbf{Elastic Modeling Cost.}
The HASA propagator currently used for iterative lens design relies on a fluid model. Although our registration investigations highlighted the critical role of shear modes in the skull and lens, incorporating full elastic wave propagation into a topology optimization loop remains computationally expensive for rapid clinical optimization and prototyping.

\noindent\textbf{Skull Variability and Trapped Gas.}
Nonlinear sensing strategies rely on the cumulative generation of parametric waves. Clinical scenarios involving trapped postoperative gas (pneumocephalus) or large variations in bone porosity induce severe acoustic impedance mismatches ($Z_\text{skull}/Z_\text{air} \approx 7500$) that scatter primary energy and destroy phase coherence, which can extinguish the diagnostic feedback loop. This though represents a small population of patients.

\section{Future Directions}

\noindent\textbf{Sonogenetics and Targeted Mechanopharmacology.}
Future studies should apply biological controls to validate targeted neuromodulation, building on the in vitro biosphere platform. Researchers could functionalize these spheres to target specific ion channels, enabling precise, low-frequency ultrasonic modulation of neural circuits.

\noindent\textbf{Elastic Topology Optimization.}
Future iterations of HASA-ADAM should integrate computationally efficient elastic wave approximations. This could pioneer shear-mode holography, in which passive lenses are deliberately crafted to exploit mode conversion to minimize off-target skull heating or enhance transcranial energy transfer.

\noindent\textbf{Autonomous Closed-Loop Alignment.}
Future work could realize autonomous, closed-loop lens registration in outpatient clinical environments by coupling the real-time nonlinear feedback signal with six-degree-of-freedom robotic positioning systems and eliminating operator dependency.

\noindent\textbf{\textsl{In-Vivo} Validation of Diagnostic Sensing.}

The hydrocephalus monitoring work has so far been validated only computationally. Translation to a large animal model — non-human primates being the most relevant anatomically — is the necessary next step. Physiological confounders that are absent from phantoms, including pulsatile cerebral blood flow, respiratory motion, and the variable acoustic attenuation of scalp and subcutaneous fat, will all need to be characterized before the approach can be taken into a clinical trial.

\section{Conclusion}

This thesis shows that the key obstacles to holography-assisted high-precision transcranial ultrasound therapy, specifically aberration correction and MRI-free registration, can be solved. The solution lies in modeling wave propagation in acoustically thick diffractive elements and exploiting the intrinsic nonlinear acoustic signatures of cranial tissues. We have developed a novel framework for acoustic hologram design and registration, and also demonstrated its use for monitoring ventricular expansion in hydrocephalus. These contributions make transcranial neuro-therapeutic and diagnostic tools more accessible to everyone, providing the engineering foundation needed to translate them into routine clinical practice.

\begin{appendices}

\addtocontents{toc}{\protect\renewcommand{\protect\cftchappresnum}{\appendixname\space}}
\addtocontents{toc}{\protect\renewcommand{\protect\cftchapnumwidth}{6em}}

\chapter{ HASA Wave Propagation Assumptions}
\section{The Parabolic (One-Way) Wave Equation}
\label{app:parabolic_HASA_equation}

The core of our HASA-ADAM hologram optimization is the Heterogeneous Angular Spectrum Approach (HASA), which uses a frequency-domain forward-marching scheme to model wave propagation through an aberrating skull layer. It relies on the parabolic (also known as one-way) approximation of the wave equation. In this section, we will derive the paraxial wave equation without backward-propagating reflections.

Wave propagation in a heterogeneous, non-absorbing medium is governed by the Helmholtz equation. For a time-harmonic acoustic pressure field $P(\mathbf{r})$ operating at an angular frequency $\omega$, the governing equation is~\cite{schoen2019heterogeneous}
\begin{equation}
\nabla^2 P(\mathbf{r}) + k^2(\mathbf{r}) P(\mathbf{r}) = 0
\label{eq:app_helmholtz}
\end{equation}
where $k(\mathbf{r}) = \omega / c(\mathbf{r})$ is the spatially varying wavenumber and $c(\mathbf{r})$ represents the local speed of sound.

To isolate the forward-propagating behavior, the total pressure field is factored into a slowly varying complex envelope $U(\mathbf{r})$ \cite{morse1946methods}that modulates a fast-oscillating carrier wave traveling along the primary propagation axis $z$:
\begin{equation}
P(x,y,z) = U(x,y,z) e^{i k_0 z}
\label{eq:app_ansatz}
\end{equation}
Here, $k_0 = \omega/c_0$ serves as a constant reference background wavenumber. Substituting this assumed solution into the Helmholtz equation requires evaluating the spatial derivatives with respect to the axial coordinate $z$. Applying the product rule yields the first and second derivatives:
\begin{equation}
\frac{\partial P}{\partial z} = \left( \frac{\partial U}{\partial z} + i k_0 U \right) e^{i k_0 z}
\end{equation}
\begin{equation}
\frac{\partial^2 P}{\partial z^2} = \left( \frac{\partial^2 U}{\partial z^2} + 2 i k_0 \frac{\partial U}{\partial z} - k_0^2 U \right) e^{i k_0 z}
\end{equation}

Inserting these expansions back into Equation~\ref{eq:app_helmholtz} and dividing out the common exponential phase term $e^{i k_0 z}$ results in the exact envelope equation:
\begin{equation}
\nabla_\perp^2 U + \frac{\partial^2 U}{\partial z^2} + 2 i k_0 \frac{\partial U}{\partial z} + \left( k^2(\mathbf{r}) - k_0^2 \right) U = 0
\label{eq:app_exact_envelope}
\end{equation}
where the transverse Laplacian operator is defined as $\nabla_\perp^2 = \frac{\partial^2}{\partial x^2} + \frac{\partial^2}{\partial y^2}$.

The parabolic approximation imposes a  restriction: the envelope $U$ must evolve slowly along the propagation axis relative to the acoustic wavelength making the second-order axial variation is vanishingly small compared to the first-order spatial variation:
\begin{equation}
\left| \frac{\partial^2 U}{\partial z^2} \right| \ll \left| 2 k_0 \frac{\partial U}{\partial z} \right|
\end{equation}

Applying this condition and neglecting the second-order axial derivative ($\frac{\partial^2 U}{\partial z^2} \approx 0$), Equation~\ref{eq:app_exact_envelope} simplifies into the paraxial wave equation:
\begin{equation}
2 i k_0 \frac{\partial U}{\partial z} = -\nabla_\perp^2 U - \left( k^2(\mathbf{r}) - k_0^2 \right) U
\label{eq:app_paraxial}
\end{equation}

This approximation carries a real world implication. A second-order spatial differential equation supports two independent solutions. They represent forward- and backward-traveling waves. Dropping the $\frac{\partial^2 U}{\partial z^2}$ term reduces the system to a first-order differential equation in $z$ for forward only propagation.

Thus, our formulation assumes 100\% forward energy transmission across all interfaces. When this model encounters high-contrast boundaries (such as the water-to-skull interface, where the reflection coefficient is large ($R \approx 0.57$)), the algorithm enforces $R = 0$. This overestimates transcranial transmission.

The optimizer as a result is blind to the phase decorrelation caused by internal standing waves and backscattering within the diplo\"{e} layer of the skull.

\section{WKB (Slowly Varying Envelope) Approximation}
\label{app:wkb_derivation}

Spectral convolution methods for heterogeneous media rely on the Wentzel–Kramers–Brillouin (WKB) approximation \cite{morse1946methods} to get analytical solutions for wave propagation through media with spatially varying (gradually varying compared to wavelength) properties. We derive the continuity constraint using the WKB approximation to show why sharp acoustic boundaries can cause spectral leakage in our HASA model.

Consider the simplified one-dimensional Helmholtz equation for a wave propagating through a heterogeneous medium:
\begin{equation}
\frac{d^2 P}{dz^2} + k^2(z) P = 0
\label{eq:app_1d_helmholtz}
\end{equation}
where $k(z)$ is the spatially dependent wavenumber. The WKB method gives us a solution where the amplitude $A(z)$ and the accumulated phase $\phi(z)$ are decoupled:
\begin{equation}
P(z) = A(z) e^{i \phi(z)}
\end{equation}

Let's substitute this decoupled expression for $P(z)$  into Equation~\ref{eq:app_1d_helmholtz}:
\begin{equation}
\frac{d^2}{dz^2} \left( A(z) e^{i \phi(z)} \right) + k^2(z) A(z) e^{i \phi(z)} = 0
\end{equation}
Now let's expand the second derivative through the product rule, get both real and imaginary terms:
\begin{equation}
\left( \frac{d^2 A}{dz^2} - A \left( \frac{d\phi}{dz} \right)^2 + i \left( 2 \frac{dA}{dz} \frac{d\phi}{dz} + A \frac{d^2\phi}{dz^2} \right) \right) e^{i \phi(z)} + k^2(z) A e^{i \phi(z)} = 0
\end{equation}

Isolating the real components provides the governing relationship for the phase accumulation:
\begin{equation}
\frac{d^2 A}{dz^2} - A \left( \frac{d\phi}{dz} \right)^2 + k^2(z) A = 0
\end{equation}
Dividing by the amplitude $A(z)$ gives us a expression for the square of the local phase gradient:
\begin{equation}
\left( \frac{d\phi}{dz} \right)^2 = k^2(z) + \frac{1}{A} \frac{d^2 A}{dz^2}
\label{eq:app_wkb_phase}
\end{equation}

 The WKB approximation requires the amplitude profile to vary so gradually that the wave amplitude $A(z)$ remains decoupled from rapid phase fluctuations. Thus, its second spatial derivative is negligible compared to the square of the local wavenumber:
\begin{equation}
\left| \frac{1}{A} \frac{d^2 A}{dz^2} \right| \ll k^2(z)
\end{equation}

Under this condition, Equation~\ref{eq:app_wkb_phase} allows the phase gradient to be approximated by the local wavenumber ($\frac{d\phi}{dz} \approx \pm k(z)$). However, this truncation requires the medium's properties to remain stable over a single acoustic wavelength. In other words, the fractional change in the local wavenumber must be small over one wavelength:
\begin{equation}
\left| \frac{1}{k^2(z)} \frac{dk}{dz} \right| \ll 1
\label{eq:app_continuity_k}
\end{equation}

Now let's relate this wavenumber constraint to material properties. We substitute the definition of the wavenumber $k(z) = \omega / c(z)$q and apply the chain rule:
\begin{equation}
\frac{dk}{dz} = \frac{d}{dz} \left( \frac{\omega}{c(z)} \right) = -\frac{\omega}{c^2(z)} \frac{dc}{dz}
\end{equation}

Substituting this derivative back into the continuity constraint (Equation~\ref{eq:app_continuity_k}) and expanding this into three dimensions and normalizing against the reference background wavenumber $k_0 = \omega / c_0$, the requirement becomes:
\begin{equation}
\left| \frac{c^2}{\omega^2} \left( -\frac{\omega}{c^2} \nabla c \right) \right| = \frac{1}{\omega} |\nabla c| = \frac{c}{\omega} \left| \frac{\nabla c}{c} \right| \approx \frac{1}{k_0} \left| \frac{\nabla c}{c} \right| \ll 1
\label{eq:app_wkb_final}
\end{equation}

Equation~\ref{eq:app_wkb_final} defines the limit of the slowly varying envelope approximation, which requires the local speed of sound to vary continuously. At the water-skull interface, however, the speed of sound jumps abruptly from $\sim 1480$ m/s to over $2500$ m/s across a fraction of the acoustic wavelength. This results in a step-function discontinuity that violates the $\ll 1$  inequality and breaks the decoupled phase assumption. Propagating waves thus produce artificial phase accumulation and spectral leakage in the Fourier transforms.

\chapter{Shear Wave Propagation}
\section{Shear Mode Conversion at the Lens Interface}
\label{app:shear_conversion_temporal_evolution}

We analyzed the normal and shear stress temporal evolution using at the lens interface k-Wave elastic solvers assuming a shear attenuation of 10 dB/MHz within the lens. The numerical solver models the propagation of elastic waves within an isotropic solid medium by integrating the coupled first-order velocity-stress equations: $\rho \frac{\partial v_i}{\partial t} = \frac{\partial \sigma_{ij}}{\partial x_j}$ and $\frac{\partial \sigma_{ij}}{\partial t} = \lambda \delta_{ij} \frac{\partial v_k}{\partial x_k} + \mu \left( \frac{\partial v_i}{\partial x_j} + \frac{\partial v_j}{\partial x_i} \right)$ where $v_i$ is the particle velocity vector, $\sigma_{ij}$ is the stress tensor, $\rho$ is the mass density, $\delta_{ij}$ is the Kronecker delta, $\lambda$ and $\mu$ are the Lam\'e parameters of the lens material. Here, the Einstein summation convention is used for repeated indices.

\begin{figure}[htbp]
    \centering

    \includegraphics[width=0.8\textwidth]{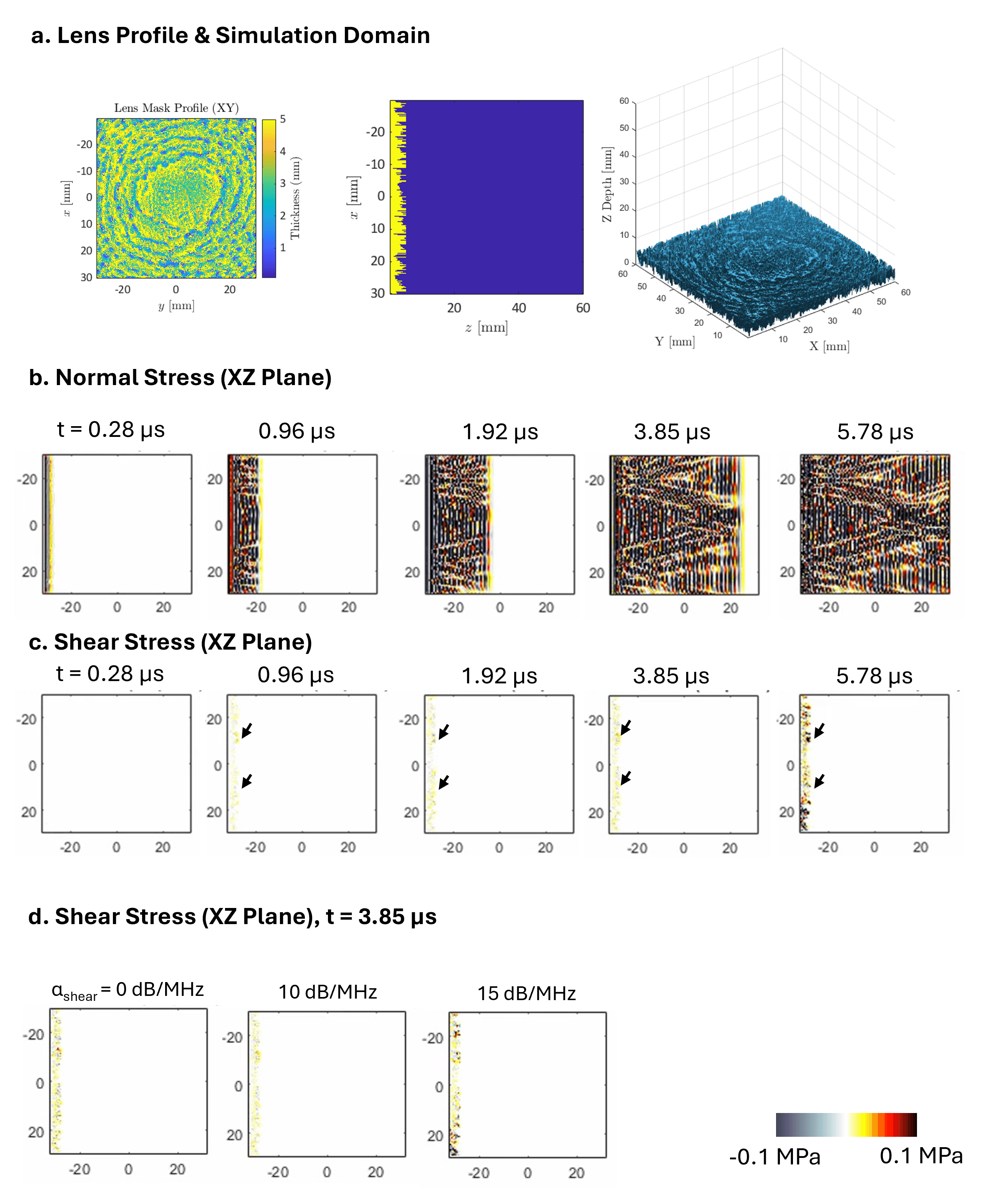}

    \caption[Simulation of elastic wave propagation through lens topology.]{\textbf{Simulation of elastic wave propagation through lens topology.} \textbf{(a)} Acoustic lens topology and the simulation domain. (Left) A 2D heat map shows the Lens Mask Profile (XY plane) with thickness ranging from 0 to 5 mm. (Middle) A 2D visualization of the computational domain in the XZ plane. (Right) A 3D view illustrating the simulation domain. \textbf{(b)} Temporal evolution of Normal Stress within the XZ plane. A sequence of five time snapshots ($t = 0.28$ \textmu s, $t = 0.96$ \textmu s, $t = 1.92$ \textmu s, $t = 3.85$ \textmu s, and $t = 5.78$ \textmu s) shows the progression of normal stress waves. \textbf{(c)} Temporal evolution of Shear Stress in the XZ plane for the same five time points as in (b). Shear waves are generated at the solid-fluid interface within the domain. Black arrows indicate the origin points at the lens interface. \textbf{(d)} A parametric study showing Shear Stress distribution in the XZ plane at a fixed time, $t = 3.85$ \textmu s, for three different attenuation coefficients ($\alpha_{\text{shear}}$): 0 dB/MHz, 10 dB/MHz, and 15 dB/MHz. The higher the shear attenuation, the larger the reduction in the amplitude of shear stress .}
    \label{fig:lens_shear_simulation}
\end{figure}

First, we looked at the shear and compressional wave propagation effects at different time points. The compressional (longitudinal) wave speed $c_p$ and shear (transverse) wave speed $c_s$ are governed by the following elastic material properties: $c_p = \sqrt{\frac{\lambda + 2\mu}{\rho}}, \quad c_s = \sqrt{\frac{\mu}{\rho}}$. Because both $\lambda$ and $\mu$ are positive for solid media, $c_p > c_s$. As expected the normal stress propagated rapidly through the lens (Figure~\ref{fig:lens_shear_simulation}b), the shear wave was noticeably delayed owing to its relatively lower speed ($c_s \approx 1300$~m/s), as shown in Figure~\ref{fig:lens_shear_simulation}c. Moreover, the effect of shear was only accentuated at the steep lens-water interface. Shear waves vanished beyond the lens interface as water cannot support shear propagation.

In a lossy medium, the decay of the propagating shear stress $\sigma_{\text{shear}}(x)$ can be described by: $\sigma_{\text{shear}}(x) = \sigma_0 e^{-\alpha_{\text{shear}} x}$ where $\sigma_0$ is the initial amplitude, $x$ is the propagation distance, and $\alpha_{\text{shear}}$ is the attenuation coefficient. Without attenuation ($\alpha_{\text{shear}} = 0$~dB/MHz), the shear stress was more prominent, which reduced in amplitude as we increased the attenuation to 10 and 15 dB/MHz. It is to be noted that these attenuation values are much higher than that for compressional waves for a resin-based 3D-printed lens (approx. 3 dB/MHz \cite{bakaric2021measurement}) and are only used here as worst-case scenarios.

\chapter{Experimental Equipment}
\section{Acoustic Hologram Experiment}
\begin{figure}[h]
    \centering
    \includegraphics[width=0.8\linewidth]{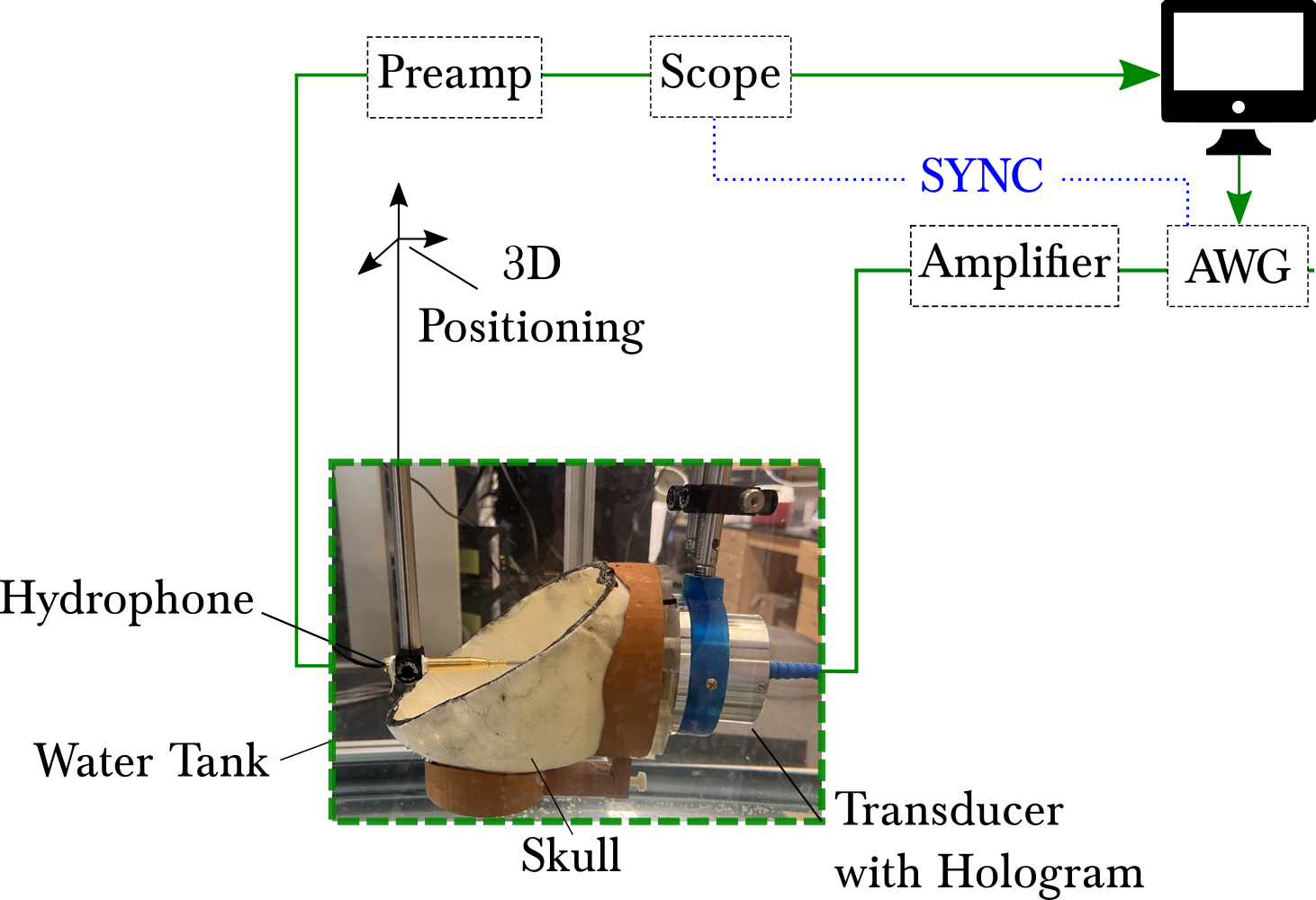}
    \caption[Experimental Setup for Hologram Validation.] {An arbitrary waveform generator (AWG) generates a signal, amplified to drive a transducer with an acoustic hologram. Ultrasound waves propagate through a skull in a water tank. A hydrophone on a 3D positioning system measures the transmitted acoustic pressure field. The signal is conditioned by a preamplifier (Preamp) and digitized by an oscilloscope (Scope). A computer controls data acquisition and synchronizes (SYNC) the AWG and oscilloscope.}
    \label{fig:hologramDesignSetup}
\end{figure}
\clearpage
\section{Acoustic Holography Registration}
\begin{figure}[h]
    \centering
    \includegraphics[width=0.8\linewidth]{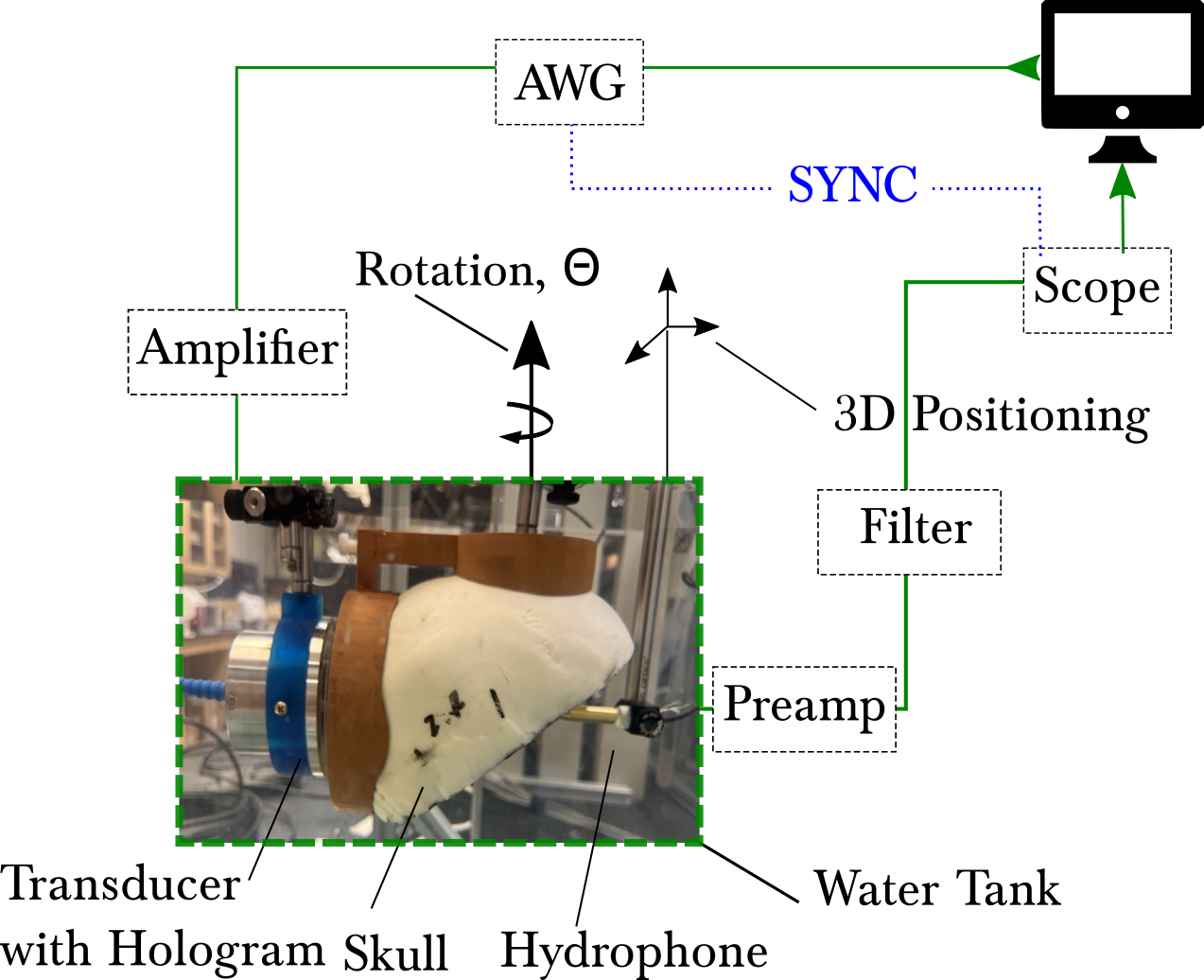}
    \caption [We affixed the transducer, hologram, and skull assembly to a rotational axis (Rotation, $\Theta$) ]{We affixed the transducer, hologram, and skull assembly to a rotational axis (Rotation, $\Theta$) to cause misregistration. A 3D positioning system moves the hydrophone along the z-axis.  We apply an additional 40 dB analog low-pass gain between the preamplifier (Preamp) and the oscilloscope (Scope).}
    \label{fig:hologramRegdSetup}
\end{figure}

\chapter{Literature Review}

\section{Acoustic Holography}

In Table~\ref{tab:lit_review} below, we provide a summary of various design methodologies, operational parameters, and specific insights and limitations found in recent research related to acoustic holography.
{\footnotesize
\renewcommand{\arraystretch}{1.3}
\begin{longtable}{
  >{\raggedright\arraybackslash}p{0.18\textwidth}
  >{\raggedright\arraybackslash}p{0.30\textwidth}
  >{\raggedright\arraybackslash}p{0.20\textwidth}
  >{\raggedright\arraybackslash}p{0.25\textwidth}
}
\caption{Study of Acoustic Holography Literature} \label{tab:lit_review} \\
\toprule
\textbf{Citation} & \textbf{Methodology \& Parameters} & \textbf{Application} & \textbf{Key Findings \& Limitations} \\
\midrule
\endfirsthead

\multicolumn{4}{c}{{\bfseries \tablename\ \thetable{} -- continued from previous page}} \\
\toprule
\textbf{Citation} & \textbf{Methodology \& Parameters} & \textbf{Application} & \textbf{Key Findings \& Limitations} \\
\midrule
\endhead

\midrule \multicolumn{4}{r}{{Continued on next page}} \\
\endfoot

\bottomrule
\endlastfoot

Melde et al. (2016)~\cite{melde2016holograms} & \textbf{Passive Lens (IASA Optimization)} \newline 2.0 MHz \newline $50\,\mu$m (SLA printing) \newline Aperture 50mm; $Z=20$mm & Pushing PDMS particles, 2D free-field & Reconstructed diffraction-limited beams. Limit: Assumes thin-element approximation. \\ \addlinespace
Brown et al. (2017)~\cite{Brown2017} & \textbf{Kinoforms (Binary Search)} \newline 1.9, 2.5, 3.1 MHz \newline Target planes & Multi-frequency field generation & Encoded different patterns on different frequencies. Limit: Crosstalk, lack of full-wave modeling. \\ \addlinespace
Brown et al. (2019)~\cite {brown2019} & \textbf{Composite Lens (Phase \& Amp)} \newline 2.7 MHz \newline Far-field & Independent Phase/Amp modulation & Limit: Loss of amplitude due to multiple plates; narrowband. \\ \addlinespace

Brown et al. (2020)~\cite{Brown2020} & \textbf{Stackable Holograms} \newline 3.0 MHz \newline Far-field & Reconfigurable combined fields & Holograms can be translated relative to each other to shift target patterns. \\ \addlinespace
Fushimi et al. (2021)~\cite{fushimi2021acoustic} & \textbf{Diff-PAT (Automatic Diff.)} \newline 40 kHz \newline PAT array / $150\,\mu$m pixels \newline Near-field / 3D Volumetric & Complex image reconstruction & Achieved much higher PSNR than IASA (by $\sim$8~dB). Limit: Hyperparameter tuning needed. \\ \addlinespace
Li et al. (2021)~\cite{Li2021Comparison} & \textbf{Direct Search vs IASA}  \newline Near-field & General Holography & Direct search balanced target and whole-region metrics better than IASA. \\ \addlinespace
Lee et al. (2022)~\cite{lee2022deep} & \textbf{Deep Learning Framework} \newline 2.0 MHz \newline Near-field & High-res image generation & Extremely fast generation times. Limit: Generalizability to novel constraints. \\ \addlinespace
Maimbourg et al. (2018)~\cite{maimbourg2018printed} & \textbf{Adaptive Lens (FDTD)} \newline 0.914 MHz \newline Curved (D=67, R=59) \newline Trans-skull & Single focus (TUS) & Restored focus through \textit{ex-vivo} skull. Limit: Susceptible to physical registration errors. \\ \addlinespace
Jim\'{e}nez-Gamb\'{i}n et al. (2019)~\cite{jimenez2019holograms} & \textbf{Time-Reversal / Phase Conj.} \newline 1 MHz \newline Flat (D=50) \newline Trans-skull & Single, double, volumetric & Produced complex fields compensating for aberration. Limit: Computationally heavy. \\ \addlinespace
Acquaticci et al. (2019)~\cite{Acquaticci2019} & \textbf{Axicon Lenses (k-Wave)} \newline 0.445 MHz \newline Flat (D=28) \newline Trans-skull (5mm) & Deep focal depth & Improved spatial resolution. Limit: Phase aberration not fully accounted for. \\ \addlinespace
Ferri et al. (2019)~\cite{ferri2019enhanced} & \textbf{Enhanced FDTD (Shear/Absorp)} \newline 0.760 MHz \newline Curved (D=67, R=59) \newline Trans-skull & Single (BBB levels, 100~kPa) & Proved inclusion of shear waves improves focus quality. Limit: Simulation only. \\ \addlinespace

Hu et al. (2022)~\cite{hu2022binary} & \textbf{Binary Metasurfaces (BAM)} \newline 0.45--0.55 MHz \newline Flat (D=120) \newline Trans-skull & Dynamic multi-point & Corrected aberrations and dynamically steered focus by changing frequency. \\ \addlinespace
Stanziola et al. (2023)~\cite{stanziola2023physics} & \textbf{Physics-Based Deep Learning} \newline Hologram property map \newline Water / Skull & High-fidelity beam shaping & Differentiable model improved field fidelity over thin-element methods. \\ \addlinespace
Jim\'{e}nez-Gamb\'{i}n et al. (2022)~\cite{JimenezGambin2022} & \textbf{Acoustic Holograms} \newline 1.68 MHz \newline Flat/Curved \newline Bilateral mouse skull & BBB Opening in mice & Enabled simultaneous multi-target bilateral BBB opening \textit{in vivo}. \\ \addlinespace
He et al. (2022)~\cite{he2022multitarget} & \textbf{Phase-only Hologram} \newline Mouse skull & Small animal tFUS & Multi-target neuromodulation without phased arrays. \\ \addlinespace
Marzo \& Drinkwater (2018)~\cite{Marzo2018} & \textbf{Holographic Acoustic Tweezers} \newline 0.040 MHz \newline  two opposing $16\times16$ PAT elements \newline Mid-air levitation & Particle manipulation & Demonstrated independent 3D manipulation of multiple particles using HATs. \\ \addlinespace

Jim\'{e}nez et al. (2021)~\cite{Jimenez2021} & \textbf{Self-Demodulation} \newline Far-field & Acoustic Vortex Beams & Explored subwavelength acoustic vortex beams using self-demodulation. \\ \addlinespace
Sallam et al. (2024)~\cite{sallam2024gradient} & \textbf{Gradient Descent Opt.} \newline Trans-skull & Nonlinear Holography & Gradient descent optimization of acoustic holograms for transcranial FUS. \\ \addlinespace
Kruizinga et al. (2017)~\cite{kruizinga2017compressive} & \textbf{Compressive Holography} \newline Single Sensor \newline 3D Imaging & 3D Ultrasound Imaging & Utilized a single sensor and compressive sensing for 3D ultrasound imaging. \\ \addlinespace
Zhong et al. (2024)~\cite{zhong2023} & \textbf{Physics-Based Deep Learning} \newline PAT Array \newline Micro-nano & Robotic Manipulation & Real-time calculation of Phase-Only Holograms (POH) for dexterous manipulation. \\ \addlinespace
Wang et al. (2025)~\cite{wang2025} & \textbf{Semi-Supervised Neural Net} \newline Holographic field & Real-Time Reconstruction & Knowledge-driven method for real-time acoustic holographic field reconstruction. \\ \addlinespace
Khan \& Kim (2025)~\cite{khan2025} & \textbf{IASA Simulation Study} \newline 0.75--4.0 MHz \newline Fresnel zone & Fidelity Analysis & Higher frequencies improve resolution but introduce edge ringing and high attenuation above 4~MHz. \\ \addlinespace
Baresch \& Garbin (2020)~\cite{baresch2020} & \textbf{Holographic Trapping} \newline Complex environments & Payload release & Acoustic trapping of microbubbles in complex environments. \\ \addlinespace
Ma et al. (2020)~\cite{ma2020} & \textbf{Holographic Patterning} \newline 5.0 MHz \newline Biocompatible hydrogel & Cell patterning & Acoustic holographic cell patterning in a biocompatible hydrogel. \\ \addlinespace
Pinton et al. (2011)~\cite{pinton2011effects} & \textbf{Full-Wave Nonlinear Sim.} \newline Transcranial & High-intensity brain therapy & Effects of nonlinear ultrasound propagation on high-intensity brain therapy. \\ \addlinespace
Kook et al. (2023)~\cite{kook2023multifocal} & \textbf{Skull-Compensated TUS} \newline Trans-skull & Neuromodulation & Multifocal skull-compensated system for targeted neuromodulation applications. \\ \addlinespace
Estrada et al. (2021)~\cite{estrada2021} & \textbf{Spherical Array System} \newline Multi-element Array \newline Rodent skull & TUS and Optoacoustics & High-precision transcranial ultrasound stimulation and optoacoustic imaging in rodents. \\

\end{longtable}
}
\end{appendices}

{ \singlespacing
\clearpage
\phantomsection
\addcontentsline{toc}{chapter}{References}
\bibliographystyle{bibliography/jasanum}
\bibliography{bibliography/allReferences}

\begin{thebibliography}{100}
\newcommand{\enquote}[1]{``#1''}
\expandafter\ifx\csname url\endcsname\relax
  \def\url#1{\texttt{#1}}\fi
\expandafter\ifx\csname urlprefix\endcsname\relax\def\urlprefix{URL }\fi
\providecommand{\bibinfo}[2]{#2}
\providecommand{\noopsort}[1]{}
\providecommand{\switchargs}[2]{#2#1}

\bibitem{szabo_diagnostic_2013}
\bibinfo{author}{T.~L. Szabo}, \emph{\bibinfo{title}{Diagnostic {Ultrasound} {Imaging}: {Inside} {Out}, {Second} {Edition}}}, \bibinfo{edition}{2 edition} edition (\bibinfo{publisher}{Academic Press}, \bibinfo{address}{Amsterdam ; Boston}) (\bibinfo{year}{2013}).

\bibitem{miller_overview_2012}
\bibinfo{author}{D.~L. Miller}, \bibinfo{author}{N.~B. Smith}, \bibinfo{author}{M.~R. Bailey}, \bibinfo{author}{G.~J. Czarnota}, \bibinfo{author}{K.~Hynynen}, and \bibinfo{author}{I.~R.~S. Makin}, \enquote{\bibinfo{title}{Overview of {Therapeutic} {Ultrasound} {Applications} and {Safety} {Considerations}}}, \bibinfo{journal}{Journal of Ultrasound in Medicine} \textbf{\bibinfo{volume}{31}}, \bibinfo{pages}{623--634} (\bibinfo{year}{2012}).

\bibitem{chaussy_extracorporeal_1984}
\bibinfo{author}{C.~Chaussy}, \bibinfo{author}{E.~Schmiedt}, \bibinfo{author}{D.~Jocham}, \bibinfo{author}{J.~Sch{\"u}ller}, \bibinfo{author}{H.~Brandl}, and \bibinfo{author}{B.~Liedl}, \enquote{\bibinfo{title}{Extracorporeal shock-wave lithotripsy ({ESWL}) for treatment of urolithiasis}}, \bibinfo{journal}{Urology} \textbf{\bibinfo{volume}{23}}, \bibinfo{pages}{59--66} (\bibinfo{year}{1984}).

\bibitem{skolarikos_extracorporeal_2006}
\bibinfo{author}{A.~Skolarikos}, \bibinfo{author}{G.~Alivizatos}, and \bibinfo{author}{J.~de~la Rosette}, \enquote{\bibinfo{title}{Extracorporeal {Shock} {Wave} {Lithotripsy} 25 {Years} {Later}: {Complications} and {Their} {Prevention}}}, \bibinfo{journal}{European Urology} \textbf{\bibinfo{volume}{50}}, \bibinfo{pages}{981--990} (\bibinfo{year}{2006}).

\bibitem{bachu2021high}
\bibinfo{author}{V.~S. Bachu}, \bibinfo{author}{J.~Kedda}, \bibinfo{author}{I.~Suk}, \bibinfo{author}{J.~J. Green}, and \bibinfo{author}{B.~Tyler}, \enquote{\bibinfo{title}{High-intensity focused ultrasound: A review of mechanisms and clinical applications}}, \bibinfo{journal}{Annals of Biomedical Engineering} \textbf{\bibinfo{volume}{49}}, \bibinfo{pages}{1975--1991} (\bibinfo{year}{2021}).

\bibitem{chaussy2005technology}
\bibinfo{author}{C.~Chaussy}, \bibinfo{author}{S.~Th{\"u}roff}, \bibinfo{author}{X.~Rebillard}, and \bibinfo{author}{A.~Gelet}, \enquote{\bibinfo{title}{Technology insight: high-intensity focused ultrasound for urologic cancers}}, \bibinfo{journal}{Nature Clinical Practice Urology} \textbf{\bibinfo{volume}{2}}, \bibinfo{pages}{191--198} (\bibinfo{year}{2005}).

\bibitem{izadifar2020introduction}
\bibinfo{author}{Z.~Izadifar}, \bibinfo{author}{Z.~Izadifar}, \bibinfo{author}{D.~Chapman}, and \bibinfo{author}{P.~Babyn}, \enquote{\bibinfo{title}{An introduction to high intensity focused ultrasound: systematic review on principles, devices, and clinical applications}}, \bibinfo{journal}{Journal of Clinical Medicine} \textbf{\bibinfo{volume}{9}}, \bibinfo{pages}{460} (\bibinfo{year}{2020}).

\bibitem{cline1992mr}
\bibinfo{author}{H.~E. Cline}, \bibinfo{author}{J.~F. Schenck}, \bibinfo{author}{K.~Hynynen}, \bibinfo{author}{R.~D. Watkins}, \bibinfo{author}{S.~P. Souza}, and \bibinfo{author}{F.~A. Jolesz}, \enquote{\bibinfo{title}{Mr-guided focused ultrasound surgery}}, \bibinfo{journal}{Journal of computer assisted tomography} \textbf{\bibinfo{volume}{16}}, \bibinfo{pages}{956--965} (\bibinfo{year}{1992}).

\bibitem{tanter1998focusing}
\bibinfo{author}{M.~Tanter}, \bibinfo{author}{J.-L. Thomas}, and \bibinfo{author}{M.~Fink}, \enquote{\bibinfo{title}{Focusing and steering through absorbing and aberrating layers: Application to ultrasonic propagation through the skull}}, \bibinfo{journal}{The Journal of the Acoustical Society of America} \textbf{\bibinfo{volume}{103}}, \bibinfo{pages}{2403--2410} (\bibinfo{year}{1998}).

\bibitem{hynynen1998demonstration}
\bibinfo{author}{K.~Hynynen} and \bibinfo{author}{F.~A. Jolesz}, \enquote{\bibinfo{title}{Demonstration of potential noninvasive ultrasound brain therapy through an intact skull}}, \bibinfo{journal}{Ultrasound in medicine \& biology} \textbf{\bibinfo{volume}{24}}, \bibinfo{pages}{275--283} (\bibinfo{year}{1998}).

\bibitem{pernot2003high}
\bibinfo{author}{M.~Pernot}, \bibinfo{author}{J.-F. Aubry}, \bibinfo{author}{M.~Tanter}, \bibinfo{author}{J.-L. Thomas}, and \bibinfo{author}{M.~Fink}, \enquote{\bibinfo{title}{High power transcranial beam steering for ultrasonic brain therapy}}, \bibinfo{journal}{Physics in Medicine \& Biology} \textbf{\bibinfo{volume}{48}}, \bibinfo{pages}{2577} (\bibinfo{year}{2003}).

\bibitem{aubry2003experimental}
\bibinfo{author}{J.-F. Aubry}, \bibinfo{author}{M.~Tanter}, \bibinfo{author}{M.~Pernot}, \bibinfo{author}{J.-L. Thomas}, and \bibinfo{author}{M.~Fink}, \enquote{\bibinfo{title}{Experimental demonstration of noninvasive transskull adaptive focusing based on prior computed tomography scans}}, \bibinfo{journal}{The Journal of the Acoustical Society of America} \textbf{\bibinfo{volume}{113}}, \bibinfo{pages}{84--93} (\bibinfo{year}{2003}).

\bibitem{mcdannold_transcranial_2010}
\bibinfo{author}{N.~McDannold}, \bibinfo{author}{G.~T. Clement}, \bibinfo{author}{P.~Black}, \bibinfo{author}{F.~Jolesz}, and \bibinfo{author}{K.~Hynynen}, \enquote{\bibinfo{title}{Transcranial {Magnetic} {Resonance} {Imaging}--{Guided} {Focused} {Ultrasound} {Surgery} of {Brain} {Tumors}: {Initial} {Findings} in 3 {Patients}}}, \bibinfo{journal}{Neurosurgery} \textbf{\bibinfo{volume}{66}}, \bibinfo{pages}{323--332} (\bibinfo{year}{2010}).

\bibitem{jeanmonod_transcranial_2012}
\bibinfo{author}{D.~Jeanmonod}, \bibinfo{author}{B.~Werner}, \bibinfo{author}{A.~Morel}, \bibinfo{author}{L.~Michels}, \bibinfo{author}{E.~Zadicario}, \bibinfo{author}{G.~Schiff}, and \bibinfo{author}{E.~Martin}, \enquote{\bibinfo{title}{Transcranial magnetic resonance imaging-guided focused ultrasound: noninvasive central lateral thalamotomy for chronic neuropathic pain}}, \bibinfo{journal}{Neurosurgical Focus} \textbf{\bibinfo{volume}{32}}, \bibinfo{pages}{1--11} (\bibinfo{year}{2012}).

\bibitem{elias_pilot_2013}
\bibinfo{author}{W.~J. Elias}, \bibinfo{author}{D.~Huss}, \bibinfo{author}{T.~Voss}, \bibinfo{author}{J.~Loomba}, \bibinfo{author}{M.~Khaled}, \bibinfo{author}{E.~Zadicario}, \bibinfo{author}{R.~C. Frysinger}, \bibinfo{author}{S.~A. Sperling}, \bibinfo{author}{S.~Wylie}, \bibinfo{author}{S.~J. Monteith}, \bibinfo{author}{J.~Druzgal}, \bibinfo{author}{B.~B. Shah}, \bibinfo{author}{M.~Harrison}, and \bibinfo{author}{M.~Wintermark}, \enquote{\bibinfo{title}{A pilot study of focused ultrasound thalamotomy for essential tremor}}, \bibinfo{journal}{The New England Journal of Medicine} \textbf{\bibinfo{volume}{369}}, \bibinfo{pages}{640--648} (\bibinfo{year}{2013}).

\bibitem{jung_bilateral_2015}
\bibinfo{author}{H.~H. Jung}, \bibinfo{author}{S.~J. Kim}, \bibinfo{author}{D.~Roh}, \bibinfo{author}{J.~G. Chang}, \bibinfo{author}{W.~S. Chang}, \bibinfo{author}{E.~J. Kweon}, \bibinfo{author}{C.-H. Kim}, and \bibinfo{author}{J.~W. Chang}, \enquote{\bibinfo{title}{Bilateral thermal capsulotomy with {MR}-guided focused ultrasound for patients with treatment-refractory obsessive-compulsive disorder: a proof-of-concept study}}, \bibinfo{journal}{Molecular Psychiatry} \textbf{\bibinfo{volume}{20}}, \bibinfo{pages}{1205--1211} (\bibinfo{year}{2015}).

\bibitem{fasano_mri-guided_2017}
\bibinfo{author}{A.~Fasano}, \bibinfo{author}{M.~Llinas}, \bibinfo{author}{R.~P. Munoz}, \bibinfo{author}{E.~Hlasny}, \bibinfo{author}{W.~Kucharczyk}, and \bibinfo{author}{A.~M. Lozano}, \enquote{\bibinfo{title}{{MRI}-guided focused ultrasound thalamotomy in non-{ET} tremor syndromes}}, \bibinfo{journal}{Neurology} \textbf{\bibinfo{volume}{89}}, \bibinfo{pages}{771--775} (\bibinfo{year}{2017}).

\bibitem{carpentier_clinical_2016}
\bibinfo{author}{A.~Carpentier}, \bibinfo{author}{M.~Canney}, \bibinfo{author}{A.~Vignot}, \bibinfo{author}{V.~Reina}, \bibinfo{author}{K.~Beccaria}, \bibinfo{author}{C.~Horodyckid}, \bibinfo{author}{C.~Karachi}, \bibinfo{author}{D.~Leclercq}, \bibinfo{author}{C.~Lafon}, \bibinfo{author}{J.-Y. Chapelon}, \bibinfo{author}{L.~Capelle}, \bibinfo{author}{P.~Cornu}, \bibinfo{author}{M.~Sanson}, \bibinfo{author}{K.~Hoang-Xuan}, \bibinfo{author}{J.-Y. Delattre}, and \bibinfo{author}{A.~Idbaih}, \enquote{\bibinfo{title}{Clinical trial of blood-brain barrier disruption by pulsed ultrasound}}, \bibinfo{journal}{Science Translational Medicine} \textbf{\bibinfo{volume}{8}}, \bibinfo{pages}{343re2--343re2} (\bibinfo{year}{2016}).

\bibitem{abrahao_first--human_2019}
\bibinfo{author}{A.~Abrahao}, \bibinfo{author}{Y.~Meng}, \bibinfo{author}{M.~Llinas}, \bibinfo{author}{Y.~Huang}, \bibinfo{author}{C.~Hamani}, \bibinfo{author}{T.~Mainprize}, \bibinfo{author}{I.~Aubert}, \bibinfo{author}{C.~Heyn}, \bibinfo{author}{S.~E. Black}, \bibinfo{author}{K.~Hynynen}, \bibinfo{author}{N.~Lipsman}, and \bibinfo{author}{L.~Zinman}, \enquote{\bibinfo{title}{First-in-human trial of blood--brain barrier opening in amyotrophic lateral sclerosis using {MR}-guided focused ultrasound}}, \bibinfo{journal}{Nature Communications} \textbf{\bibinfo{volume}{10}}, \bibinfo{pages}{1--9} (\bibinfo{year}{2019}).

\bibitem{lipsman2018blood}
\bibinfo{author}{N.~Lipsman} \emph{et~al.}, \enquote{\bibinfo{title}{Blood-brain barrier opening in alzheimer's disease using mr-guided focused ultrasound}}, \bibinfo{journal}{Nature Communications}  (\bibinfo{year}{2018}).

\bibitem{jolesz2014intraoperative}
\bibinfo{author}{F.~A. Jolesz}, \emph{\bibinfo{title}{Intraoperative imaging and image-guided therapy}} (\bibinfo{publisher}{Springer Science \& Business Media}) (\bibinfo{year}{2014}).

\bibitem{hertzberg2010ultrasound}
\bibinfo{author}{Y.~Hertzberg}, \bibinfo{author}{A.~Volovick}, \bibinfo{author}{Y.~Zur}, \bibinfo{author}{Y.~Medan}, \bibinfo{author}{S.~Vitek}, and \bibinfo{author}{G.~Navon}, \enquote{\bibinfo{title}{Ultrasound focusing using magnetic resonance acoustic radiation force imaging: application to ultrasound transcranial therapy}}, \bibinfo{journal}{Medical physics} \textbf{\bibinfo{volume}{37}}, \bibinfo{pages}{2934--2942} (\bibinfo{year}{2010}).

\bibitem{melde2016holograms}
\bibinfo{author}{K.~Melde}, \bibinfo{author}{A.~G. Mark}, \bibinfo{author}{T.~Qiu}, and \bibinfo{author}{P.~Fischer}, \enquote{\bibinfo{title}{Holograms for acoustics}}, \bibinfo{journal}{Nature} \textbf{\bibinfo{volume}{537}}, \bibinfo{pages}{518--522} (\bibinfo{year}{2016}).

\bibitem{shen2014anisotropic}
\bibinfo{author}{C.~Shen}, \bibinfo{author}{J.~Xu}, \bibinfo{author}{N.~X. Fang}, and \bibinfo{author}{Y.~Jing}, \enquote{\bibinfo{title}{Anisotropic complementary acoustic metamaterial for canceling out aberrating layers}}, \bibinfo{journal}{Physical Review X} \textbf{\bibinfo{volume}{4}}, \bibinfo{pages}{041033} (\bibinfo{year}{2014}).

\bibitem{maimbourg20183d}
\bibinfo{author}{G.~Maimbourg}, \bibinfo{author}{A.~Houdouin}, \bibinfo{author}{T.~Deffieux}, \bibinfo{author}{M.~Tanter}, and \bibinfo{author}{J.-F. Aubry}, \enquote{\bibinfo{title}{3d-printed adaptive acoustic lens as a disruptive technology for transcranial ultrasound therapy using single-element transducers}}, \bibinfo{journal}{Physics in Medicine \& Biology} \textbf{\bibinfo{volume}{63}}, \bibinfo{pages}{025026} (\bibinfo{year}{2018}).

\bibitem{ferri2019enhanced}
\bibinfo{author}{M.~Ferri}, \bibinfo{author}{J.~M. Bravo}, \bibinfo{author}{J.~Redondo}, and \bibinfo{author}{J.~V. S{\'a}nchez-P{\'e}rez}, \enquote{\bibinfo{title}{Enhanced numerical method for the design of 3-d-printed holographic acoustic lenses for aberration correction of single-element transcranial focused ultrasound}}, \bibinfo{journal}{Ultrasound in medicine \& biology} \textbf{\bibinfo{volume}{45}}, \bibinfo{pages}{867--884} (\bibinfo{year}{2019}).

\bibitem{jimenez2023primate}
\bibinfo{author}{S.~Jim{\'{e}}nez-Gamb{\'{i}}n}, \bibinfo{author}{S.~Bae}, \bibinfo{author}{R.~Ji}, \bibinfo{author}{F.~Tsitsos}, and \bibinfo{author}{E.~E. Konofagou}, \enquote{\bibinfo{title}{Feasibility of hologram-assisted bilateral blood-brain barrier opening in non-human primates}}, \bibinfo{journal}{IEEE Transactions on Ultrasonics, Ferroelectrics, and Frequency Control} \textbf{\bibinfo{volume}{71}}, \bibinfo{pages}{164--173} (\bibinfo{year}{2024}).

\bibitem{jimenez2021mouse}
\bibinfo{author}{S.~Jim{\'{e}}nez-Gamb{\'{i}}n} \emph{et~al.}, \enquote{\bibinfo{title}{Acoustic holograms for bilateral blood-brain barrier opening in a mouse model}}, \bibinfo{journal}{IEEE Transactions on Biomedical Engineering} \textbf{\bibinfo{volume}{69}}, \bibinfo{pages}{1359--1368} (\bibinfo{year}{2021}).

\bibitem{andres2022numerical}
\bibinfo{author}{D.~Andr{\'e}s}, \bibinfo{author}{N.~Jim{\'e}nez}, \bibinfo{author}{J.~M. Benlloch}, and \bibinfo{author}{F.~Camarena}, \enquote{\bibinfo{title}{Numerical study of acoustic holograms for deep-brain targeting through the temporal bone window}}, \bibinfo{journal}{Ultrasound in Medicine \& Biology} \textbf{\bibinfo{volume}{48}}, \bibinfo{pages}{872--886} (\bibinfo{year}{2022}).

\bibitem{glickstein2024}
\bibinfo{author}{B.~Glickstein} \emph{et~al.}, \enquote{\bibinfo{title}{Rationally designed acoustic holograms for uniform nanodroplet-mediated tissue ablation}}, \bibinfo{journal}{IEEE Transactions on Ultrasonics, Ferroelectrics, and Frequency Control} \textbf{\bibinfo{volume}{71}} (\bibinfo{year}{2024}).

\bibitem{he2022multitarget}
\bibinfo{author}{J.~He}, \bibinfo{author}{J.~Wu}, \bibinfo{author}{Y.~Zhu}, \bibinfo{author}{Y.~Chen}, \bibinfo{author}{M.~Yuan}, \bibinfo{author}{L.~Zeng}, and \bibinfo{author}{X.~Ji}, \enquote{\bibinfo{title}{Multitarget transcranial ultrasound therapy in small animals based on phase-only acoustic holographic lens}}, \bibinfo{journal}{IEEE Transactions on Ultrasonics, Ferroelectrics, and Frequency Control} \textbf{\bibinfo{volume}{69}}, \bibinfo{pages}{662--671} (\bibinfo{year}{2021}).

\bibitem{jimenez2019generating}
\bibinfo{author}{S.~Jim{\'e}nez-Gamb{\'\i}n}, \bibinfo{author}{N.~Jimenez}, \bibinfo{author}{J.~M. Benlloch}, and \bibinfo{author}{F.~Camarena}, \enquote{\bibinfo{title}{Generating bessel beams with broad depth-of-field by using phase-only acoustic holograms}}, \bibinfo{journal}{Scientific reports} \textbf{\bibinfo{volume}{9}}, \bibinfo{pages}{20104} (\bibinfo{year}{2019}).

\bibitem{kruizinga2017compressive}
\bibinfo{author}{P.~Kruizinga}, \bibinfo{author}{P.~van~der Meulen}, \bibinfo{author}{A.~Fedjajevs}, \bibinfo{author}{F.~Mastik}, \bibinfo{author}{G.~Springeling}, \bibinfo{author}{N.~de~Jong}, \bibinfo{author}{J.~G. Bosch}, and \bibinfo{author}{G.~Leus}, \enquote{\bibinfo{title}{Compressive 3d ultrasound imaging using a single sensor}}, \bibinfo{journal}{Science advances} \textbf{\bibinfo{volume}{3}}, \bibinfo{pages}{e1701423} (\bibinfo{year}{2017}).

\bibitem{Li2021Comparison}
\bibinfo{author}{J.~Li}, \bibinfo{author}{Z.~Lv}, \bibinfo{author}{Z.~Hou}, and \bibinfo{author}{Y.~Pei}, \enquote{\bibinfo{title}{Comparison of balanced direct search and iterative angular spectrum approaches for designing acoustic holography structure}}, \bibinfo{journal}{Applied Acoustics} \textbf{\bibinfo{volume}{175}}, \bibinfo{pages}{107848} (\bibinfo{year}{2021}), \urlprefix\url{http://dx.doi.org/10.1016/j.apacoust.2020.107848}.

\bibitem{lee2022}
\bibinfo{author}{M.~H. Lee}, \bibinfo{author}{H.~M. Lew}, \bibinfo{author}{S.~Youn}, \bibinfo{author}{T.~Kim}, and \bibinfo{author}{J.~Y. Hwang}, \enquote{\bibinfo{title}{Deep learning-based framework for fast and accurate acoustic hologram generation}}, \bibinfo{journal}{IEEE Transactions on Ultrasonics, Ferroelectrics, and Frequency Control} \textbf{\bibinfo{volume}{69}}, \bibinfo{pages}{3353--3366} (\bibinfo{year}{2022}).

\bibitem{fushimi2021acoustic}
\bibinfo{author}{T.~Fushimi}, \bibinfo{author}{K.~Yamamoto}, and \bibinfo{author}{Y.~Ochiai}, \enquote{\bibinfo{title}{Acoustic hologram optimisation using automatic differentiation}}, \bibinfo{journal}{Scientific reports} \textbf{\bibinfo{volume}{11}}, \bibinfo{pages}{12678} (\bibinfo{year}{2021}).

\bibitem{jimenez2019holograms}
\bibinfo{author}{S.~Jim{\'e}nez-Gamb{\'i}n}, \bibinfo{author}{N.~Jim{\'e}nez}, \bibinfo{author}{J.~M. Benlloch}, and \bibinfo{author}{F.~Camarena}, \enquote{\bibinfo{title}{Holograms to focus arbitrary ultrasonic fields through the skull}}, \bibinfo{journal}{Physical Review Applied} \textbf{\bibinfo{volume}{12}}, \bibinfo{pages}{014016} (\bibinfo{year}{2019}).

\bibitem{schoen2021experimental}
\bibinfo{author}{S.~Schoen}, \bibinfo{author}{P.~Dash}, and \bibinfo{author}{C.~D. Arvanitis}, \enquote{\bibinfo{title}{Experimental demonstration of trans-skull volumetric passive acoustic mapping with the heterogeneous angular spectrum approach}}, \bibinfo{journal}{IEEE transactions on ultrasonics, ferroelectrics, and frequency control} \textbf{\bibinfo{volume}{69}}, \bibinfo{pages}{534--542} (\bibinfo{year}{2021}).

\bibitem{o2016registration}
\bibinfo{author}{M.~A. O'Reilly}, \bibinfo{author}{R.~M. Jones}, \bibinfo{author}{G.~Birman}, and \bibinfo{author}{K.~Hynynen}, \enquote{\bibinfo{title}{Registration of human skull computed tomography data to an ultrasound treatment space using a sparse high frequency ultrasound hemispherical array}}, \bibinfo{journal}{Medical Physics} \textbf{\bibinfo{volume}{43}}, \bibinfo{pages}{5063--5071} (\bibinfo{year}{2016}).

\bibitem{de2007mr}
\bibinfo{author}{B.~D. de~Senneville}, \bibinfo{author}{C.~Mougenot}, \bibinfo{author}{B.~Quesson}, \bibinfo{author}{I.~Dragonu}, \bibinfo{author}{N.~Grenier}, and \bibinfo{author}{C.~T. Moonen}, \enquote{\bibinfo{title}{Mr thermometry for monitoring tumor ablation}}, \bibinfo{journal}{European radiology} \textbf{\bibinfo{volume}{17}}, \bibinfo{pages}{2401--2410} (\bibinfo{year}{2007}).

\bibitem{kyriakou2014review}
\bibinfo{author}{A.~Kyriakou}, \bibinfo{author}{E.~Neufeld}, \bibinfo{author}{B.~Werner}, \bibinfo{author}{M.~M. Paulides}, \bibinfo{author}{G.~Szekely}, and \bibinfo{author}{N.~Kuster}, \enquote{\bibinfo{title}{A review of numerical and experimental compensation techniques for skull-induced phase aberrations in transcranial focused ultrasound}}, \bibinfo{journal}{International journal of hyperthermia} \textbf{\bibinfo{volume}{30}}, \bibinfo{pages}{36--46} (\bibinfo{year}{2014}).

\bibitem{chen2020neuronavigation}
\bibinfo{author}{K.-T. Chen}, \bibinfo{author}{Y.-J. Lin}, \bibinfo{author}{W.-Y. Chai}, \bibinfo{author}{C.-J. Lin}, \bibinfo{author}{P.-Y. Chen}, \bibinfo{author}{C.-Y. Huang}, \bibinfo{author}{J.~S. Kuo}, \bibinfo{author}{H.-L. Liu}, and \bibinfo{author}{K.-C. Wei}, \enquote{\bibinfo{title}{Neuronavigation-guided focused ultrasound (navifus) for transcranial blood-brain barrier opening in recurrent glioblastoma patients: Clinical trial protocol}}, \bibinfo{journal}{Annals of Translational Medicine} \textbf{\bibinfo{volume}{8}} (\bibinfo{year}{2020}).

\bibitem{vandoormaal2019clinical}
\bibinfo{author}{J.~A. van Doormaal} \emph{et~al.}, \enquote{\bibinfo{title}{Clinical accuracy of holographic navigation using a head-mounted display}}, \bibinfo{journal}{Operative Neurosurgery}  (\bibinfo{year}{2019}).

\bibitem{elias2016thalamotomy}
\bibinfo{author}{W.~J. Elias} \emph{et~al.}, \enquote{\bibinfo{title}{A randomized trial of focused ultrasound thalamotomy for essential tremor}}, \bibinfo{journal}{New England Journal of Medicine}  (\bibinfo{year}{2016}).

\bibitem{wei2013neuronavigation}
\bibinfo{author}{K.~C. Wei} \emph{et~al.}, \enquote{\bibinfo{title}{Neuronavigation-guided focused ultrasound-induced blood-brain barrier opening}}, \bibinfo{journal}{American Journal of Neuroradiology}  (\bibinfo{year}{2013}).

\bibitem{sallam2023nonlinear}
\bibinfo{author}{A.~Sallam} and \bibinfo{author}{S.~Shahab}, \enquote{\bibinfo{title}{Nonlinear acoustic holography with adaptive sampling}}, \bibinfo{journal}{IEEE Transactions on Ultrasonics, Ferroelectrics, and Frequency Control} \textbf{\bibinfo{volume}{70}}, \bibinfo{pages}{1516--1526} (\bibinfo{year}{2023}).

\bibitem{duck2013physical}
\bibinfo{author}{F.~A. Duck}, \emph{\bibinfo{title}{Physical properties of tissue: A comprehensive reference book}} (\bibinfo{publisher}{Academic press}) (\bibinfo{year}{2013}).

\bibitem{jiang2022invention}
\bibinfo{author}{Y.~Jiang}, \bibinfo{author}{W.~Huang}, \bibinfo{author}{X.-J. Wu}, \bibinfo{author}{X.-L. Shi}, \bibinfo{author}{R.-R. Hu}, \bibinfo{author}{W.~Chen}, \bibinfo{author}{T.-F. Zhang}, \bibinfo{author}{X.-L. Xu}, \bibinfo{author}{C.-G. Huang}, and \bibinfo{author}{L.-J. Hou}, \enquote{\bibinfo{title}{Invention of a non-invasive intracranial pressure (icp) monitoring system--an enlightenment from a hydrocephalus study}}, \bibinfo{journal}{British Journal of Neurosurgery} \textbf{\bibinfo{volume}{36}}, \bibinfo{pages}{693--698} (\bibinfo{year}{2022}).

\bibitem{nitsche2008transcranial}
\bibinfo{author}{M.~A. Nitsche}, \bibinfo{author}{L.~G. Cohen}, \bibinfo{author}{E.~M. Wassermann}, \bibinfo{author}{A.~Priori}, \bibinfo{author}{N.~Lang}, \bibinfo{author}{A.~Antal}, \bibinfo{author}{W.~Paulus}, \bibinfo{author}{F.~Hummel}, \bibinfo{author}{P.~S. Boggio}, \bibinfo{author}{F.~Fregni}, \emph{et~al.}, \enquote{\bibinfo{title}{Transcranial direct current stimulation: state of the art 2008}}, \bibinfo{journal}{Brain stimulation} \textbf{\bibinfo{volume}{1}}, \bibinfo{pages}{206--223} (\bibinfo{year}{2008}).

\bibitem{walsh2000transcranial}
\bibinfo{author}{V.~Walsh} and \bibinfo{author}{A.~Cowey}, \enquote{\bibinfo{title}{Transcranial magnetic stimulation and cognitive neuroscience}}, \bibinfo{journal}{Nature Reviews Neuroscience} \textbf{\bibinfo{volume}{1}}, \bibinfo{pages}{73--80} (\bibinfo{year}{2000}).

\bibitem{faria2011finite}
\bibinfo{author}{P.~Faria}, \bibinfo{author}{M.~Hallett}, and \bibinfo{author}{P.~C. Miranda}, \enquote{\bibinfo{title}{A finite element analysis of the effect of electrode area and inter-electrode distance on the spatial distribution of the current density in tdcs}}, \bibinfo{journal}{Journal of neural engineering} \textbf{\bibinfo{volume}{8}}, \bibinfo{pages}{066017} (\bibinfo{year}{2011}).

\bibitem{deng2013electric}
\bibinfo{author}{Z.-D. Deng}, \bibinfo{author}{S.~H. Lisanby}, and \bibinfo{author}{A.~V. Peterchev}, \enquote{\bibinfo{title}{Electric field depth--focality tradeoff in transcranial magnetic stimulation: simulation comparison of 50 coil designs}}, \bibinfo{journal}{Brain stimulation} \textbf{\bibinfo{volume}{6}}, \bibinfo{pages}{1--13} (\bibinfo{year}{2013}).

\bibitem{haar2010ultrasound}
\bibinfo{author}{G.~T. Haar}, \enquote{\bibinfo{title}{Ultrasound bioeffects and safety}}, \bibinfo{journal}{Proceedings of the Institution of Mechanical Engineers, Part H: Journal of Engineering in Medicine} \textbf{\bibinfo{volume}{224}}, \bibinfo{pages}{363--373} (\bibinfo{year}{2010}).

\bibitem{cadoni2023ectopic}
\bibinfo{author}{S.~Cadoni}, \bibinfo{author}{C.~Demen{\'e}}, \bibinfo{author}{I.~Alcala}, \bibinfo{author}{M.~Provansal}, \bibinfo{author}{D.~Nguyen}, \bibinfo{author}{D.~Nelidova}, \bibinfo{author}{G.~Labern{\`e}de}, \bibinfo{author}{J.~Lubetzki}, \bibinfo{author}{R.~Goulet}, \bibinfo{author}{E.~Burban}, \emph{et~al.}, \enquote{\bibinfo{title}{Ectopic expression of a mechanosensitive channel confers spatiotemporal resolution to ultrasound stimulations of neurons for visual restoration}}, \bibinfo{journal}{Nature Nanotechnology} \textbf{\bibinfo{volume}{18}}, \bibinfo{pages}{667--676} (\bibinfo{year}{2023}).

\bibitem{sato_ultrasonic_2017}
\bibinfo{author}{T.~Sato}, \bibinfo{author}{M.~Shapiro}, and \bibinfo{author}{D.~Tsao}, \enquote{\bibinfo{title}{Ultrasonic {Neuromodulation} {Causes} {Widespread} {Cortical} {Activation} via an {Indirect} {Auditory} {Mechanism}}},  (\bibinfo{year}{2018}).

\bibitem{guo_ultrasound_2018}
\bibinfo{author}{H.~Guo}, \bibinfo{author}{M.~Hamilton}, \bibinfo{author}{S.~Offutt}, \bibinfo{author}{C.~Gloeckner}, \bibinfo{author}{T.~Li}, \bibinfo{author}{Y.~Kim}, \bibinfo{author}{W.~Legon}, \bibinfo{author}{J.~Alford}, and \bibinfo{author}{H.~Lim}, \enquote{\bibinfo{title}{Ultrasound {Produces} {Extensive} {Brain} {Activation} via a {Cochlear} {Pathway}}}, \bibinfo{journal}{Neuron} \textbf{\bibinfo{volume}{98}}, \bibinfo{pages}{1020--1030} (\bibinfo{year}{2018}).

\bibitem{Dash2025}
\bibinfo{author}{P.~P. Dash} and \bibinfo{author}{C.~D. Arvanitis}, \enquote{\bibinfo{title}{Acoustic {Holography} in the {Megahertz} {Frequency} {Range} with {Optimal} {Lens} {Topologies} and {Nonlinear} {Acoustic} {Feedback}}},   (\bibinfo{year}{2025}), \urlprefix\url{https://arxiv.org/abs/2508.07103}.

\bibitem{melde2023compact}
\bibinfo{author}{K.~Melde}, \bibinfo{author}{H.~Kremer}, \bibinfo{author}{M.~Shi}, \bibinfo{author}{S.~Seneca}, \bibinfo{author}{C.~Frey}, \bibinfo{author}{I.~Platzman}, \bibinfo{author}{C.~Degel}, \bibinfo{author}{D.~Schmitt}, \bibinfo{author}{B.~Sch{\"o}lkopf}, and \bibinfo{author}{P.~Fischer}, \enquote{\bibinfo{title}{Compact holographic sound fields enable rapid one-step assembly of matter in {3D}}}, \bibinfo{journal}{Science Advances} \textbf{\bibinfo{volume}{9}}, \bibinfo{pages}{eadf6182} (\bibinfo{year}{2023}).

\bibitem{hirayama2019volumetric}
\bibinfo{author}{R.~Hirayama}, \bibinfo{author}{D.~M. Plasencia}, \bibinfo{author}{N.~Masuda}, and \bibinfo{author}{S.~Subramanian}, \enquote{\bibinfo{title}{A volumetric display for visual, tactile and audio presentation using acoustic trapping}}, \bibinfo{journal}{Nature} \textbf{\bibinfo{volume}{575}}, \bibinfo{pages}{320--323} (\bibinfo{year}{2019}).

\bibitem{xie2016acoustic}
\bibinfo{author}{Y.~Xie}, \bibinfo{author}{C.~Shen}, \bibinfo{author}{W.~Wang}, \bibinfo{author}{J.~Li}, \bibinfo{author}{D.~Suo}, \bibinfo{author}{B.-I. Popa}, \bibinfo{author}{Y.~Jing}, and \bibinfo{author}{S.~A. Cummer}, \enquote{\bibinfo{title}{Acoustic holographic rendering with two-dimensional metamaterial-based passive phased array}}, \bibinfo{journal}{Scientific Reports} \textbf{\bibinfo{volume}{6}}, \bibinfo{pages}{35437} (\bibinfo{year}{2016}).

\bibitem{hu2022airy}
\bibinfo{author}{Z.~Hu}, \bibinfo{author}{Y.~Yang}, \bibinfo{author}{L.~Yang}, \bibinfo{author}{Y.~Gong}, \bibinfo{author}{C.~Chukwu}, \bibinfo{author}{D.~Ye}, \bibinfo{author}{Y.~Yue}, \bibinfo{author}{J.~Yuan}, \bibinfo{author}{A.~V. Kravitz}, and \bibinfo{author}{H.~Chen}, \enquote{\bibinfo{title}{Airy-beam holographic sonogenetics for advancing neuromodulation precision and flexibility}}, \bibinfo{journal}{Proceedings of the National Academy of Sciences} \textbf{\bibinfo{volume}{121}}, \bibinfo{pages}{e2402200121} (\bibinfo{year}{2024}).

\bibitem{andres2023holographic}
\bibinfo{author}{D.~Andr{\'e}s}, \bibinfo{author}{I.~Rivens}, \bibinfo{author}{P.~Mouratidis}, \bibinfo{author}{N.~Jim{\'e}nez}, \bibinfo{author}{F.~Camarena}, and \bibinfo{author}{G.~ter Haar}, \enquote{\bibinfo{title}{Holographic focused ultrasound hyperthermia system for uniform simultaneous thermal exposure of multiple tumor spheroids}}, \bibinfo{journal}{Cancers} \textbf{\bibinfo{volume}{15}}, \bibinfo{pages}{2540} (\bibinfo{year}{2023}).

\bibitem{daniel2024multifrequency}
\bibinfo{author}{M.~Daniel}, \bibinfo{author}{D.~Attali}, \bibinfo{author}{T.~Tiennot}, \bibinfo{author}{M.~Tanter}, and \bibinfo{author}{J.-F. Aubry}, \enquote{\bibinfo{title}{Multifrequency transcranial ultrasound holography with acoustic lenses}}, \bibinfo{journal}{Physical Review Applied} \textbf{\bibinfo{volume}{21}}, \bibinfo{pages}{014011} (\bibinfo{year}{2024}).

\bibitem{jimenez2024feasibility}
\bibinfo{author}{S.~Jim{\'e}nez-Gamb{\'i}n}, \bibinfo{author}{S.~Bae}, \bibinfo{author}{R.~Ji}, \bibinfo{author}{F.~Tsitsos}, and \bibinfo{author}{E.~E. Konofagou}, \enquote{\bibinfo{title}{Feasibility of hologram-assisted bilateral blood--brain barrier opening in non-human primates}}, \bibinfo{journal}{IEEE Transactions on Ultrasonics, Ferroelectrics, and Frequency Control} \textbf{\bibinfo{volume}{71}}, \bibinfo{pages}{1172--1185} (\bibinfo{year}{2024}).

\bibitem{kook2023multifocal}
\bibinfo{author}{G.~Kook}, \bibinfo{author}{Y.~Jo}, \bibinfo{author}{C.~Oh}, \bibinfo{author}{X.~Liang}, \bibinfo{author}{J.~Kim}, \bibinfo{author}{S.-M. Lee}, \bibinfo{author}{S.~Kim}, \bibinfo{author}{J.-W. Choi}, and \bibinfo{author}{H.~J. Lee}, \enquote{\bibinfo{title}{Multifocal skull-compensated transcranial focused ultrasound system for neuromodulation applications based on acoustic holography}}, \bibinfo{journal}{Microsystems \& Nanoengineering} \textbf{\bibinfo{volume}{9}}, \bibinfo{pages}{45} (\bibinfo{year}{2023}).

\bibitem{yao2025acoustic}
\bibinfo{author}{X.~Yao}, \bibinfo{author}{X.~Piao}, \bibinfo{author}{S.~Hong}, \bibinfo{author}{C.~Ji}, \bibinfo{author}{M.~Wang}, \bibinfo{author}{Y.~Wei}, \bibinfo{author}{Z.~Xu}, \bibinfo{author}{J.-J. Pan}, \bibinfo{author}{Y.~Pei}, and \bibinfo{author}{B.~Cheng}, \enquote{\bibinfo{title}{Acoustic hologram-enabled simultaneous multi-target blood-brain barrier opening ({AH}-{SiMBO})}}, \bibinfo{journal}{Communications Engineering} \textbf{\bibinfo{volume}{4}}, \bibinfo{pages}{99} (\bibinfo{year}{2025}).

\bibitem{meng2021applications}
\bibinfo{author}{Y.~Meng}, \bibinfo{author}{K.~Hynynen}, and \bibinfo{author}{N.~Lipsman}, \enquote{\bibinfo{title}{Applications of focused ultrasound in the brain: from thermoablation to drug delivery}}, \bibinfo{journal}{Nature Reviews Neurology} \textbf{\bibinfo{volume}{17}}, \bibinfo{pages}{7--22} (\bibinfo{year}{2021}).

\bibitem{schoen2022towards}
\bibinfo{author}{S.~Schoen}, \bibinfo{author}{M.~S. Kilinc}, \bibinfo{author}{H.~Lee}, \bibinfo{author}{Y.~Guo}, \bibinfo{author}{F.~L. Degertekin}, \bibinfo{author}{G.~F. Woodworth}, and \bibinfo{author}{C.~Arvanitis}, \enquote{\bibinfo{title}{Towards controlled drug delivery in brain tumors with microbubble-enhanced focused ultrasound}}, \bibinfo{journal}{Advanced Drug Delivery Reviews} \textbf{\bibinfo{volume}{180}}, \bibinfo{pages}{114043} (\bibinfo{year}{2022}).

\bibitem{choi2024neuronavigation}
\bibinfo{author}{S.~W. Choi}, \bibinfo{author}{M.~Komaiha}, \bibinfo{author}{D.~Choi}, \bibinfo{author}{N.~Lu}, \bibinfo{author}{T.~I. Gerhardson}, \bibinfo{author}{A.~Fox}, \bibinfo{author}{N.~Chaudhary}, \bibinfo{author}{S.~Camelo-Piragua}, \bibinfo{author}{T.~L. Hall}, \bibinfo{author}{A.~S. Pandey}, \bibinfo{author}{Z.~Xu}, and \bibinfo{author}{J.~R. Sukovich}, \enquote{\bibinfo{title}{Neuronavigation-guided transcranial histotripsy ({NaviTH}) system}}, \bibinfo{journal}{Ultrasound in Medicine \& Biology} \textbf{\bibinfo{volume}{50}}, \bibinfo{pages}{1155--1166} (\bibinfo{year}{2024}).

\bibitem{gu2020modified}
\bibinfo{author}{J.~Gu} and \bibinfo{author}{Y.~Jing}, \enquote{\bibinfo{title}{A modified mixed domain method for modeling acoustic wave propagation in strongly heterogeneous media}}, \bibinfo{journal}{The Journal of the Acoustical Society of America} \textbf{\bibinfo{volume}{147}}, \bibinfo{pages}{4055--4068} (\bibinfo{year}{2020}).

\bibitem{sallam2024gradient}
\bibinfo{author}{A.~Sallam}, \bibinfo{author}{C.~Cengiz}, \bibinfo{author}{M.~Pewekar}, \bibinfo{author}{E.~Hoffmann}, \bibinfo{author}{W.~Legon}, \bibinfo{author}{E.~Vlaisavljevich}, and \bibinfo{author}{S.~Shahab}, \enquote{\bibinfo{title}{Gradient descent optimization of acoustic holograms for transcranial focused ultrasound}}, \bibinfo{journal}{Journal of Applied Physics} \textbf{\bibinfo{volume}{136}}, \bibinfo{pages}{144901} (\bibinfo{year}{2024}).

\bibitem{schoen2020heterogeneous}
\bibinfo{author}{S.~Schoen} and \bibinfo{author}{C.~D. Arvanitis}, \enquote{\bibinfo{title}{Heterogeneous angular spectrum method for trans-skull imaging and focusing}}, \bibinfo{journal}{IEEE Transactions on Medical Imaging} \textbf{\bibinfo{volume}{39}}, \bibinfo{pages}{1605--1614} (\bibinfo{year}{2019}).

\bibitem{pichardo_multi-frequency_2011}
\bibinfo{author}{S.~Pichardo}, \bibinfo{author}{V.~W. Sin}, and \bibinfo{author}{K.~Hynynen}, \enquote{\bibinfo{title}{Multi-frequency characterization of the speed of sound and attenuation coefficient for longitudinal transmission of freshly excised human skulls}}, \bibinfo{journal}{Physics in Medicine and Biology} \textbf{\bibinfo{volume}{56}}, \bibinfo{pages}{219--250} (\bibinfo{year}{2011}).

\bibitem{morse1946methods}
\bibinfo{author}{P.~M. Morse} and \bibinfo{author}{H.~Feshbach}, \emph{\bibinfo{title}{Methods of Theoretical Physics, Part {I}}} (\bibinfo{publisher}{McGraw-Hill Book Company}, \bibinfo{address}{New York}) (\bibinfo{year}{1946}).

\bibitem{Dash2023Heterogenous}
\bibinfo{author}{P.~P. Dash} and \bibinfo{author}{C.~Arvanitis}, \enquote{\bibinfo{title}{Heterogenous angular spectrum approach based holograms for trans-skull focused ultrasound therapy}}, \bibinfo{journal}{The Journal of the Acoustical Society of America} \textbf{\bibinfo{volume}{153}}, \bibinfo{pages}{A103--A103} (\bibinfo{year}{2023}), \urlprefix\url{http://dx.doi.org/10.1121/10.0018312}.

\bibitem{arvanitis2025trans}
\bibinfo{author}{C.~Arvanitis} and \bibinfo{author}{P.~P. Dash}, \enquote{\bibinfo{title}{Trans-skull focused ultrasound using acoustic hologram and heterogenous angular spectrum approach, and hologram registration}},  (\bibinfo{year}{2025}), \bibinfo{note}{uS Patent 12,502,558}.

\bibitem{bakaric2021measurement}
\bibinfo{author}{M.~Bakaric}, \bibinfo{author}{P.~Miloro}, \bibinfo{author}{A.~Javaherian}, \bibinfo{author}{B.~T. Cox}, \bibinfo{author}{B.~E. Treeby}, and \bibinfo{author}{M.~D. Brown}, \enquote{\bibinfo{title}{Measurement of the ultrasound attenuation and dispersion in 3d-printed photopolymer materials from 1 to 3.5 mhz}}, \bibinfo{journal}{The Journal of the Acoustical Society of America} \textbf{\bibinfo{volume}{150}}, \bibinfo{pages}{2798--2805} (\bibinfo{year}{2021}).

\bibitem{treeby2010k}
\bibinfo{author}{B.~E. Treeby} and \bibinfo{author}{B.~T. Cox}, \enquote{\bibinfo{title}{k-wave: {MATLAB} toolbox for the simulation and reconstruction of photoacoustic wave fields}}, \bibinfo{journal}{Journal of Biomedical Optics} \textbf{\bibinfo{volume}{15}}, \bibinfo{pages}{021314} (\bibinfo{year}{2010}).

\bibitem{blackstock_fundamentals_2000}
\bibinfo{author}{D.~T. Blackstock}, \emph{\bibinfo{title}{Fundamentals of {Physical} {Acoustics}}} (\bibinfo{publisher}{John Wiley \& Sons}) (\bibinfo{year}{2000}).

\bibitem{schoen2019heterogeneous}
\bibinfo{author}{S.~Schoen} and \bibinfo{author}{C.~D. Arvanitis}, \enquote{\bibinfo{title}{Heterogeneous angular spectrum method for trans-skull imaging and focusing}}, \bibinfo{journal}{IEEE transactions on medical imaging} \textbf{\bibinfo{volume}{39}}, \bibinfo{pages}{1605--1614} (\bibinfo{year}{2019}).

\bibitem{Kilinc2025Piezo}
\bibinfo{author}{M.~S. Kilinc}, \bibinfo{author}{H.~Lee}, \bibinfo{author}{Y.~R. Ferry}, \bibinfo{author}{B.~Ingram}, \bibinfo{author}{B.~Skowronski}, \bibinfo{author}{P.~P. Dash}, \bibinfo{author}{R.~P. Zangabad}, \bibinfo{author}{C.~Arvanitis}, and \bibinfo{author}{F.~L. Degertekin}, \enquote{\bibinfo{title}{A {Piezo}-{Cmut} {Hybrid} {Hemispherical} {Transmit} {Array} for {Passive} {Acoustic} {Mapping} of {Microbubble} {Activity}}}, in \emph{\bibinfo{booktitle}{2025 {IEEE} {International} {Ultrasonics} {Symposium} ({IUS})}}, \bibinfo{pages}{1--5} (\bibinfo{organization}{IEEE}) (\bibinfo{year}{2025}), \urlprefix\url{http://dx.doi.org/10.1109/IUS62464.2025.11201469}.

\bibitem{angla2023transcranial}
\bibinfo{author}{C.~Angla}, \bibinfo{author}{B.~Larrat}, \bibinfo{author}{J.-L. Gennisson}, and \bibinfo{author}{S.~Chatillon}, \enquote{\bibinfo{title}{Transcranial ultrasound simulations: A review}}, \bibinfo{journal}{Medical Physics} \textbf{\bibinfo{volume}{50}}, \bibinfo{pages}{1051--1072} (\bibinfo{year}{2023}).

\bibitem{bu2024deep}
\bibinfo{author}{M.~Bu}, \bibinfo{author}{W.~Gu}, \bibinfo{author}{B.~Li}, \bibinfo{author}{Q.~Zhu}, \bibinfo{author}{X.~Jiang}, \bibinfo{author}{D.~Ta}, and \bibinfo{author}{X.~Liu}, \enquote{\bibinfo{title}{A deep learning-based method of acoustic holographic lens generation for transcranial focused ultrasound}}, \bibinfo{journal}{AIP Advances} \textbf{\bibinfo{volume}{14}}, \bibinfo{pages}{125026} (\bibinfo{year}{2024}).

\bibitem{li2022acoustic}
\bibinfo{author}{B.~Li}, \bibinfo{author}{M.~Lu}, \bibinfo{author}{C.~Liu}, \bibinfo{author}{X.~Liu}, and \bibinfo{author}{D.~Ta}, \enquote{\bibinfo{title}{Acoustic hologram reconstruction with unsupervised neural network}}, \bibinfo{journal}{Frontiers in Materials} \textbf{\bibinfo{volume}{9}} (\bibinfo{year}{2022}).

\bibitem{lee2022deep}
\bibinfo{author}{M.~H. Lee}, \bibinfo{author}{H.~M. Lew}, \bibinfo{author}{S.~Youn}, \bibinfo{author}{T.~Kim}, and \bibinfo{author}{J.~Y. Hwang}, \enquote{\bibinfo{title}{Deep learning-based framework for fast and accurate acoustic hologram generation}}, \bibinfo{journal}{IEEE Transactions on Ultrasonics, Ferroelectrics, and Frequency Control} \textbf{\bibinfo{volume}{69}}, \bibinfo{pages}{3353--3366} (\bibinfo{year}{2022}).

\bibitem{jiang2022flexible}
\bibinfo{author}{L.~Jiang}, \bibinfo{author}{G.~Lu}, \bibinfo{author}{Y.~Zeng}, \bibinfo{author}{Y.~Sun}, \bibinfo{author}{H.~Kang}, \bibinfo{author}{J.~Burford}, \bibinfo{author}{C.~Gong}, \bibinfo{author}{M.~S. Humayun}, \bibinfo{author}{Y.~Chen}, and \bibinfo{author}{Q.~Zhou}, \enquote{\bibinfo{title}{Flexible ultrasound-induced retinal stimulating piezo-arrays for biomimetic visual prostheses}}, \bibinfo{journal}{Nature Communications} \textbf{\bibinfo{volume}{13}}, \bibinfo{pages}{3853} (\bibinfo{year}{2022}).

\bibitem{naor2012towards}
\bibinfo{author}{O.~Naor}, \bibinfo{author}{Y.~Hertzberg}, \bibinfo{author}{E.~Zemel}, \bibinfo{author}{E.~Kimmel}, and \bibinfo{author}{S.~Shoham}, \enquote{\bibinfo{title}{Towards multifocal ultrasonic neural stimulation {II}: design considerations for an acoustic retinal prosthesis}}, \bibinfo{journal}{Journal of Neural Engineering} \textbf{\bibinfo{volume}{9}}, \bibinfo{pages}{026006} (\bibinfo{year}{2012}).

\bibitem{lagerburg2025dimensional}
\bibinfo{author}{V.~Lagerburg}, \bibinfo{author}{A.~Vrancken}, \bibinfo{author}{S.~Bergsma}, \bibinfo{author}{J.~Dekker}, \bibinfo{author}{W.~Diemer}, \bibinfo{author}{J.~Waldner-Troost}, and \bibinfo{author}{M.~Koenrades}, \enquote{\bibinfo{title}{Dimensional accuracy and resolution assessment of the formlabs form {3B} {3D} printer for medical applications}}, \bibinfo{journal}{Annals of 3D Printed Medicine} \textbf{\bibinfo{volume}{19}}, \bibinfo{pages}{100204} (\bibinfo{year}{2025}).

\bibitem{stanziola2023physics}
\bibinfo{author}{A.~Stanziola}, \bibinfo{author}{B.~T. Cox}, \bibinfo{author}{B.~E. Treeby}, and \bibinfo{author}{M.~D. Brown}, \enquote{\bibinfo{title}{Physics-based acoustic holograms}}, \bibinfo{howpublished}{arXiv preprint arXiv:2305.03625} (\bibinfo{year}{2023}), \urlprefix\url{https://arxiv.org/abs/2305.03625}.

\bibitem{maimbourg2018printed}
\bibinfo{author}{G.~Maimbourg}, \bibinfo{author}{A.~Houdouin}, \bibinfo{author}{T.~Deffieux}, \bibinfo{author}{M.~Tanter}, and \bibinfo{author}{J.-F. Aubry}, \enquote{\bibinfo{title}{{3D}-printed adaptive acoustic lens as a disruptive technology for transcranial ultrasound therapy using single-element transducers}}, \bibinfo{journal}{Physics in Medicine \& Biology} \textbf{\bibinfo{volume}{63}}, \bibinfo{pages}{025026} (\bibinfo{year}{2018}).

\bibitem{chen2021neuronavigation}
\bibinfo{author}{K.-T. Chen}, \bibinfo{author}{W.-Y. Chai}, \bibinfo{author}{Y.-J. Lin}, \bibinfo{author}{C.-J. Lin}, \bibinfo{author}{P.-Y. Chen}, \bibinfo{author}{H.-C. Tsai}, \bibinfo{author}{C.-Y. Huang}, \bibinfo{author}{J.~S. Kuo}, \bibinfo{author}{H.-L. Liu}, and \bibinfo{author}{K.-C. Wei}, \enquote{\bibinfo{title}{Neuronavigation-guided focused ultrasound for transcranial blood-brain barrier opening and immunostimulation in brain tumors}}, \bibinfo{journal}{Science Advances} \textbf{\bibinfo{volume}{7}}, \bibinfo{pages}{eabd0772} (\bibinfo{year}{2021}).

\bibitem{pouliopoulos2020clinical}
\bibinfo{author}{A.~N. Pouliopoulos}, \bibinfo{author}{S.-Y. Wu}, \bibinfo{author}{M.~T. Burgess}, \bibinfo{author}{M.~E. Karakatsani}, \bibinfo{author}{H.~A.~S. Kamimura}, and \bibinfo{author}{E.~E. Konofagou}, \enquote{\bibinfo{title}{A clinical system for non-invasive blood--brain barrier opening using a neuronavigation-guided single-element focused ultrasound transducer}}, \bibinfo{journal}{Ultrasound in Medicine \& Biology} \textbf{\bibinfo{volume}{46}}, \bibinfo{pages}{73--89} (\bibinfo{year}{2020}).

\bibitem{Dash2025Leveraging}
\bibinfo{author}{P.~P. Dash} and \bibinfo{author}{C.~Arvanitis}, \enquote{\bibinfo{title}{Leveraging the parametric array effect for transcranial focused ultrasound interventions}}, \bibinfo{journal}{The Journal of the Acoustical Society of America} \textbf{\bibinfo{volume}{157}}, \bibinfo{pages}{A307--A307} (\bibinfo{year}{2025}), \urlprefix\url{http://dx.doi.org/10.1121/10.0038401}.

\bibitem{westervelt1963parametric}
\bibinfo{author}{P.~J. Westervelt}, \enquote{\bibinfo{title}{Parametric acoustic array}}, \bibinfo{journal}{The Journal of the acoustical society of America} \textbf{\bibinfo{volume}{35}}, \bibinfo{pages}{535--537} (\bibinfo{year}{1963}).

\bibitem{Dash2021Non}
\bibinfo{author}{P.~P. Dash} and \bibinfo{author}{C.~Arvanitis}, \enquote{\bibinfo{title}{Non-linearities under highly focused high-intensity ultrasound fields}}, \bibinfo{journal}{The Journal of the Acoustical Society of America} \textbf{\bibinfo{volume}{150}}, \bibinfo{pages}{A125--A125} (\bibinfo{year}{2021}), \urlprefix\url{http://dx.doi.org/10.1121/10.0007852}.

\bibitem{hamilton_nonlinear_2008}
\bibinfo{author}{M.~F. Hamilton} and \bibinfo{author}{D.~T. Blackstock}, \emph{\bibinfo{title}{Nonlinear Acoustics}} (\bibinfo{publisher}{Acoustical Society of America}) (\bibinfo{year}{2008}).

\bibitem{clement_enhanced_2004}
\bibinfo{author}{G.~T. Clement}, \bibinfo{author}{P.~J. White}, and \bibinfo{author}{K.~Hynynen}, \enquote{\bibinfo{title}{Enhanced ultrasound transmission through the human skull using shear mode conversion}}, \bibinfo{journal}{The Journal of the Acoustical Society of America} \textbf{\bibinfo{volume}{115}}, \bibinfo{pages}{1356--1364} (\bibinfo{year}{2004}).

\bibitem{renaud2008exploration}
\bibinfo{author}{G.~Renaud}, \bibinfo{author}{S.~Calle}, \bibinfo{author}{J.~P. Remenieras}, and \bibinfo{author}{M.~Defontaine}, \enquote{\bibinfo{title}{Exploration of trabecular bone nonlinear elasticity using time-of-flight modulation}}, \bibinfo{journal}{IEEE Transactions on Ultrasonics, Ferroelectrics, and Frequency Control} \textbf{\bibinfo{volume}{55}}, \bibinfo{pages}{1497--1507} (\bibinfo{year}{2008}).

\bibitem{pinton2011effects}
\bibinfo{author}{G.~Pinton}, \bibinfo{author}{J.-F. Aubry}, \bibinfo{author}{M.~Fink}, and \bibinfo{author}{M.~Tanter}, \enquote{\bibinfo{title}{Effects of nonlinear ultrasound propagation on high intensity brain therapy}}, \bibinfo{journal}{Medical Physics} \textbf{\bibinfo{volume}{38}}, \bibinfo{pages}{1207--1216} (\bibinfo{year}{2011}).

\bibitem{baron2009simulation}
\bibinfo{author}{C.~Baron}, \bibinfo{author}{J.-F. Aubry}, \bibinfo{author}{M.~Tanter}, \bibinfo{author}{S.~Meairs}, and \bibinfo{author}{M.~Fink}, \enquote{\bibinfo{title}{Simulation of intracranial acoustic fields in clinical trials of sonothrombolysis}}, \bibinfo{journal}{Ultrasound in Medicine \& Biology} \textbf{\bibinfo{volume}{35}}, \bibinfo{pages}{1148--1158} (\bibinfo{year}{2009}).

\bibitem{song2021experimental}
\bibinfo{author}{J.~Song}, \bibinfo{author}{D.~Jung}, \bibinfo{author}{J.~S. Kim}, and \bibinfo{author}{J.~Lee}, \enquote{\bibinfo{title}{Experimental evaluation of pseudo-sound in a parametric array}}, \bibinfo{journal}{Journal of the Acoustical Society of America} \textbf{\bibinfo{volume}{150}}, \bibinfo{pages}{3787--3796} (\bibinfo{year}{2021}).

\bibitem{karpov2003}
\bibinfo{author}{S.~Karpov}, \bibinfo{author}{A.~Prosperetti}, and \bibinfo{author}{L.~Ostrovsky}, \enquote{\bibinfo{title}{Nonlinear wave interactions in bubble layers}}, \bibinfo{journal}{The Journal of the Acoustical Society of America} \textbf{\bibinfo{volume}{113}}, \bibinfo{pages}{1304--1316} (\bibinfo{year}{2003}), \urlprefix\url{http://dx.doi.org/10.1121/1.1539519}.

\bibitem{airan2017neuromodulation}
\bibinfo{author}{R.~Airan}, \enquote{\bibinfo{title}{Neuromodulation with nanoparticles}}, \bibinfo{journal}{Science} \textbf{\bibinfo{volume}{357}}, \bibinfo{pages}{465} (\bibinfo{year}{2017}).

\bibitem{rincon2022biomarkers}
\bibinfo{author}{J.~Rincon-Torroella}, \bibinfo{author}{H.~Khela}, \bibinfo{author}{A.~Bettegowda}, and \bibinfo{author}{C.~Bettegowda}, \enquote{\bibinfo{title}{Biomarkers and focused ultrasound: the future of liquid biopsy for brain tumor patients}}, \bibinfo{journal}{Journal of Neuro-Oncology} \textbf{\bibinfo{volume}{156}}, \bibinfo{pages}{33--48} (\bibinfo{year}{2022}).

\bibitem{hofmann2023variations}
\bibinfo{author}{B.~Hofmann}, \bibinfo{author}{I.~{\O}. Brandsaeter}, and \bibinfo{author}{E.~Kjelle}, \enquote{\bibinfo{title}{Variations in wait times for imaging services: a register-based study of self-reported wait times for specific examinations in {Norway}}}, \bibinfo{journal}{BMC Health Services Research} \textbf{\bibinfo{volume}{23}}, \bibinfo{pages}{1287} (\bibinfo{year}{2023}).

\bibitem{panfilova2021review}
\bibinfo{author}{A.~Panfilova}, \bibinfo{author}{R.~J.~G. van Sloun}, \bibinfo{author}{H.~Wijkstra}, \bibinfo{author}{O.~A. Sapozhnikov}, and \bibinfo{author}{M.~Mischi}, \enquote{\bibinfo{title}{A review on {B/A} measurement methods with a clinical perspective}}, \bibinfo{journal}{Journal of the Acoustical Society of America} \textbf{\bibinfo{volume}{149}}, \bibinfo{pages}{2200--2237} (\bibinfo{year}{2021}).

\bibitem{zhang2001experimental}
\bibinfo{author}{D.~Zhang}, \bibinfo{author}{X.~Gong}, and \bibinfo{author}{X.~Chen}, \enquote{\bibinfo{title}{Experimental imaging of the acoustic nonlinearity parameter {B/A} for biological tissues via a parametric array}}, \bibinfo{journal}{Ultrasound in Medicine \& Biology} \textbf{\bibinfo{volume}{27}}, \bibinfo{pages}{1359--1365} (\bibinfo{year}{2001}).

\bibitem{tang2011effect}
\bibinfo{author}{M.-X. Tang}, \bibinfo{author}{J.~Loughran}, \bibinfo{author}{E.~Stride}, \bibinfo{author}{D.~Zhang}, and \bibinfo{author}{R.~J. Eckersley}, \enquote{\bibinfo{title}{Effect of bubble shell nonlinearity on ultrasound nonlinear propagation through microbubble populations}}, \bibinfo{journal}{Journal of the Acoustical Society of America} \textbf{\bibinfo{volume}{129}}, \bibinfo{pages}{EL76--82} (\bibinfo{year}{2011}).

\bibitem{cavaro2011microbubble}
\bibinfo{author}{M.~Cavaro}, \bibinfo{author}{C.~Payan}, \bibinfo{author}{J.~Moysan}, and \bibinfo{author}{F.~Baqu{\'e}}, \enquote{\bibinfo{title}{Microbubble cloud characterization by nonlinear frequency mixing}}, \bibinfo{journal}{Journal of the Acoustical Society of America} \textbf{\bibinfo{volume}{129}}, \bibinfo{pages}{EL179--183} (\bibinfo{year}{2011}).

\bibitem{overvelde2010nonlinear}
\bibinfo{author}{M.~Overvelde}, \bibinfo{author}{V.~Garbin}, \bibinfo{author}{J.~Sijl}, \bibinfo{author}{B.~Dollet}, \bibinfo{author}{N.~de~Jong}, \bibinfo{author}{D.~Lohse}, and \bibinfo{author}{M.~Versluis}, \enquote{\bibinfo{title}{Nonlinear shell behavior of phospholipid-coated microbubbles}}, \bibinfo{journal}{Ultrasound in Medicine \& Biology} \textbf{\bibinfo{volume}{36}}, \bibinfo{pages}{2080--2092} (\bibinfo{year}{2010}).

\bibitem{pinton2012attenuation}
\bibinfo{author}{G.~Pinton}, \bibinfo{author}{J.-F. Aubry}, \bibinfo{author}{E.~Bossy}, \bibinfo{author}{M.~Muller}, \bibinfo{author}{M.~Pernot}, and \bibinfo{author}{M.~Tanter}, \enquote{\bibinfo{title}{Attenuation, scattering, and absorption of ultrasound in the skull bone}}, \bibinfo{journal}{Medical Physics} \textbf{\bibinfo{volume}{39}}, \bibinfo{pages}{299--307} (\bibinfo{year}{2012}).

\bibitem{zhang2017invasive}
\bibinfo{author}{X.~Zhang}, \bibinfo{author}{J.~E. Medow}, \bibinfo{author}{B.~J. Iskandar}, \bibinfo{author}{F.~Wang}, \bibinfo{author}{M.~Shokoueinejad}, \bibinfo{author}{J.~Koueik}, and \bibinfo{author}{J.~G. Webster}, \enquote{\bibinfo{title}{Invasive and noninvasive means of measuring intracranial pressure: a review}}, \bibinfo{journal}{Physiological measurement} \textbf{\bibinfo{volume}{38}}, \bibinfo{pages}{R143--R182} (\bibinfo{year}{2017}).

\bibitem{fischer2020non}
\bibinfo{author}{J.~B. Fischer}, \bibinfo{author}{A.~Ghouse}, \bibinfo{author}{S.~Tagliabue}, \bibinfo{author}{F.~Maruccia}, \bibinfo{author}{A.~Rey-Perez}, \bibinfo{author}{M.~B{\'a}guena}, \bibinfo{author}{P.~Cano}, \bibinfo{author}{R.~Zucca}, \bibinfo{author}{U.~M. Weigel}, \bibinfo{author}{J.~Sahuquillo}, \bibinfo{author}{M.~A. Poca}, and \bibinfo{author}{T.~Durduran}, \enquote{\bibinfo{title}{Non-invasive estimation of intracranial pressure by diffuse optics: A proof-of-concept study}}, \bibinfo{journal}{Journal of Neurotrauma} \textbf{\bibinfo{volume}{37}}, \bibinfo{pages}{2569--2579} (\bibinfo{year}{2020}).

\bibitem{qiu2025noninvasive}
\bibinfo{author}{J.~Qiu}, \bibinfo{author}{T.-J. Zou}, \bibinfo{author}{D.-M. Wang}, \bibinfo{author}{H.-R. Luo}, \bibinfo{author}{H.-T. Yu}, \bibinfo{author}{L.~Lei}, and \bibinfo{author}{W.-H. Yin}, \enquote{\bibinfo{title}{Noninvasive intracranial pressure prediction using a multimodal ultrasound-based hemispheric modeling strategy: A prospective dual-center study}}, \bibinfo{journal}{Neurocritical Care} \textbf{\bibinfo{volume}{43}}, \bibinfo{pages}{911--926} (\bibinfo{year}{2025}).

\bibitem{jiang2025advancements}
\bibinfo{author}{X.~Jiang}, \bibinfo{author}{H.~Guo}, \bibinfo{author}{W.~Xiao}, \bibinfo{author}{L.~Wang}, \bibinfo{author}{D.~Wu}, \bibinfo{author}{J.~Liu}, \bibinfo{author}{Q.~Zhao}, and \bibinfo{author}{Y.~Shao}, \enquote{\bibinfo{title}{Advancements in non-invasive intracranial pressure monitoring via optic nerve sheath diameter measurement}}, \bibinfo{journal}{Medical Science Monitor: International Medical Journal of Experimental and Clinical Research} \textbf{\bibinfo{volume}{31}}, \bibinfo{pages}{e947237} (\bibinfo{year}{2025}).

\bibitem{czosnyka2004monitoring}
\bibinfo{author}{M.~Czosnyka} and \bibinfo{author}{J.~D. Pickard}, \enquote{\bibinfo{title}{Monitoring and interpretation of intracranial pressure}}, \bibinfo{journal}{Journal of Neurology, Neurosurgery \& Psychiatry} \textbf{\bibinfo{volume}{75}}, \bibinfo{pages}{813--821} (\bibinfo{year}{2004}).

\bibitem{blomqvist2021sulfatide}
\bibinfo{author}{M.~Blomqvist}, \bibinfo{author}{H.~Zetterberg}, \bibinfo{author}{K.~Blennow}, and \bibinfo{author}{J.-E. M{\aa}nsson}, \enquote{\bibinfo{title}{Sulfatide in health and disease. the evaluation of sulfatide in cerebrospinal fluid as a possible biomarker for neurodegeneration}}, \bibinfo{journal}{Molecular and Cellular Neuroscience} \textbf{\bibinfo{volume}{116}}, \bibinfo{pages}{103670} (\bibinfo{year}{2021}).

\bibitem{kwave_toolbox}
\bibinfo{author}{B.~E. Treeby} and \bibinfo{author}{B.~T. Cox}, \enquote{\bibinfo{title}{k-wave: Matlab toolbox for the simulation and reconstruction of photoacoustic wave fields}},  (\bibinfo{year}{2010}).

\bibitem{jaraj2017estimated}
\bibinfo{author}{D.~Jaraj}, \bibinfo{author}{K.~Rabiei}, \bibinfo{author}{T.~Marlow}, \bibinfo{author}{C.~Jensen}, \bibinfo{author}{I.~Skoog}, and \bibinfo{author}{C.~Wikkels{\o}}, \enquote{\bibinfo{title}{Estimated ventricle size using evans index: reference values from a population-based sample}}, \bibinfo{journal}{European journal of neurology} \textbf{\bibinfo{volume}{24}}, \bibinfo{pages}{468--474} (\bibinfo{year}{2017}).

\bibitem{bendella2024brain}
\bibinfo{author}{Z.~Bendella}, \bibinfo{author}{V.~Purrer}, \bibinfo{author}{R.~Haase}, \bibinfo{author}{S.~Z{\"u}low}, \bibinfo{author}{C.~Kindler}, \bibinfo{author}{V.~Borger}, \bibinfo{author}{M.~Banat}, \bibinfo{author}{F.~Dorn}, \bibinfo{author}{U.~W{\"u}llner}, \bibinfo{author}{A.~Radbruch}, and \bibinfo{author}{F.~C. Schmeel}, \enquote{\bibinfo{title}{Brain and ventricle volume alterations in idiopathic normal pressure hydrocephalus determined by artificial intelligence-based {MRI} volumetry}}, \bibinfo{journal}{Diagnostics} \textbf{\bibinfo{volume}{14}}, \bibinfo{pages}{1422} (\bibinfo{year}{2024}).

\bibitem{westervelt_parametric_1963}
\bibinfo{author}{P.~J. Westervelt}, \enquote{\bibinfo{title}{Parametric acoustic array}}, \bibinfo{journal}{Journal of the Acoustical Society of America} \textbf{\bibinfo{volume}{35}}, \bibinfo{pages}{535--537} (\bibinfo{year}{1963}).

\bibitem{geraldini2022transcranial}
\bibinfo{author}{F.~Geraldini}, \bibinfo{author}{A.~D. Cassai}, and \bibinfo{author}{M.~Munari}, \enquote{\bibinfo{title}{Transcranial ultrasound as a useful tool in early detection and follow-up of hydrocephalus in acute subarachnoid hemorrhage}}, \bibinfo{journal}{Journal of Neurosurgical Anesthesiology} \textbf{\bibinfo{volume}{34}}, \bibinfo{pages}{e75} (\bibinfo{year}{2022}).

\bibitem{caricato2014echography}
\bibinfo{author}{A.~Caricato}, \bibinfo{author}{S.~Pitoni}, \bibinfo{author}{L.~Montini}, \bibinfo{author}{M.~G. Bocci}, \bibinfo{author}{P.~Annetta}, and \bibinfo{author}{M.~Antonelli}, \enquote{\bibinfo{title}{Echography in brain imaging in intensive care unit: State of the art}}, \bibinfo{journal}{World Journal of Radiology} \textbf{\bibinfo{volume}{6}}, \bibinfo{pages}{636--642} (\bibinfo{year}{2014}).

\bibitem{zhang2021wearable}
\bibinfo{author}{B.~Zhang}, \bibinfo{author}{Z.~Huang}, \bibinfo{author}{H.~Song}, \bibinfo{author}{H.~S. Kim}, and \bibinfo{author}{J.~Park}, \enquote{\bibinfo{title}{Wearable intracranial pressure monitoring sensor for infants}}, \bibinfo{journal}{Biosensors} \textbf{\bibinfo{volume}{11}}, \bibinfo{pages}{213} (\bibinfo{year}{2021}).

\bibitem{moskowitz2010cumulative}
\bibinfo{author}{S.~I. Moskowitz}, \bibinfo{author}{W.~J. Davros}, \bibinfo{author}{M.~E. Kelly}, \bibinfo{author}{D.~Fiorella}, \bibinfo{author}{P.~A. Rasmussen}, and \bibinfo{author}{T.~J. Masaryk}, \enquote{\bibinfo{title}{Cumulative radiation dose during hospitalization for aneurysmal subarachnoid hemorrhage}}, \bibinfo{journal}{American Journal of Neuroradiology} \textbf{\bibinfo{volume}{31}}, \bibinfo{pages}{1377--1382} (\bibinfo{year}{2010}).

\bibitem{filippou2018recent}
\bibinfo{author}{V.~Filippou} and \bibinfo{author}{C.~Tsoumpas}, \enquote{\bibinfo{title}{Recent advances on the development of phantoms using {3D} printing for imaging with {CT}, {MRI}, {PET}, {SPECT}, and ultrasound}}, \bibinfo{journal}{Medical Physics} \textbf{\bibinfo{volume}{45}}, \bibinfo{pages}{e740--e760} (\bibinfo{year}{2018}).

\bibitem{krasovitski_intramembrane_2011}
\bibinfo{author}{B.~Krasovitski}, \bibinfo{author}{V.~Frenkel}, \bibinfo{author}{S.~Shoham}, and \bibinfo{author}{E.~Kimmel}, \enquote{\bibinfo{title}{Intramembrane cavitation as a unifying mechanism for ultrasound-induced bioeffects}}, \bibinfo{journal}{Proc. Natl. Acad. Sci} \textbf{\bibinfo{volume}{108}}, \bibinfo{pages}{3258--3263} (\bibinfo{year}{2011}).

\bibitem{king_effective_2013}
\bibinfo{author}{R.~King}, \bibinfo{author}{J.~Brown}, \bibinfo{author}{W.~Newsome}, and \bibinfo{author}{K.~Pauly}, \enquote{\bibinfo{title}{Effective {Parameters} for {Ultrasound}-{Induced} {In} {Vivo} {Neurostimulation}}}, \bibinfo{journal}{Ultrasound Med. Biol} \textbf{\bibinfo{volume}{39}}, \bibinfo{pages}{312--331} (\bibinfo{year}{2013}).

\bibitem{plaksin2014intramembrane}
\bibinfo{author}{M.~Plaksin}, \bibinfo{author}{S.~Shoham}, and \bibinfo{author}{E.~Kimmel}, \enquote{\bibinfo{title}{Intramembrane cavitation as a predictive bio-piezoelectric mechanism for ultrasonic brain stimulation}}, \bibinfo{journal}{Physical review X} \textbf{\bibinfo{volume}{4}}, \bibinfo{pages}{011004} (\bibinfo{year}{2014}).

\bibitem{prieto_activation_2018}
\bibinfo{author}{M.~Prieto}, \bibinfo{author}{K.~Firouzi}, \bibinfo{author}{B.~Khuri-Yakub}, and \bibinfo{author}{M.~Maduke}, \enquote{\bibinfo{title}{Activation of {Piezo1} but {Not} {NaV1}.2 {Channels} by {Ultrasound} at 43 {MHz}}}, \bibinfo{journal}{Ultrasound Med. Biol} \textbf{\bibinfo{volume}{44}}, \bibinfo{pages}{1217--1232} (\bibinfo{year}{2018}).

\bibitem{liao_optimal_2021}
\bibinfo{author}{D.~Liao}, \bibinfo{author}{M.-Y. Hsiao}, \bibinfo{author}{G.~Xiang}, and \bibinfo{author}{P.~Zhong}, \enquote{\bibinfo{title}{Optimal pulse length of insonification for {Piezo1} activation and intracellular calcium response}}, \bibinfo{journal}{Sci. Rep} \textbf{\bibinfo{volume}{11}}, \bibinfo{pages}{709} (\bibinfo{year}{2021}).

\bibitem{hoffman_focused_2022}
\bibinfo{author}{B.~Hoffman}, \bibinfo{author}{Y.~Baba}, \bibinfo{author}{S.~Lee}, \bibinfo{author}{C.-K. Tong}, \bibinfo{author}{E.~Konofagou}, and \bibinfo{author}{E.~Lumpkin}, \enquote{\bibinfo{title}{Focused ultrasound excites action potentials in mammalian peripheral neurons in part through the mechanically gated ion channel {PIEZO2}}}, \bibinfo{journal}{Proc. Natl. Acad. Sci} \textbf{\bibinfo{volume}{119, e2115821119}} (\bibinfo{year}{2022}).

\bibitem{kubanek_ultrasound_2018}
\bibinfo{author}{J.~Kubanek}, \bibinfo{author}{P.~Shukla}, \bibinfo{author}{A.~Das}, \bibinfo{author}{S.~A. Baccus}, and \bibinfo{author}{M.~B. Goodman}, \enquote{\bibinfo{title}{Ultrasound {Elicits} {Behavioral} {Responses} through {Mechanical} {Effects} on {Neurons} and {Ion} {Channels} in a {Simple} {Nervous} {System}}}, \bibinfo{journal}{Journal of Neuroscience} \textbf{\bibinfo{volume}{38}}, \bibinfo{pages}{3081--3091} (\bibinfo{year}{2018}).

\bibitem{yoo_focused_2022}
\bibinfo{author}{S.~Yoo}, \bibinfo{author}{D.~Mittelstein}, \bibinfo{author}{R.~Hurt}, \bibinfo{author}{J.~Lacroix}, and \bibinfo{author}{M.~Shapiro}, \enquote{\bibinfo{title}{Focused ultrasound excites cortical neurons via mechanosensitive calcium accumulation and ion channel amplification}}, \bibinfo{journal}{Nat. Commun} \textbf{\bibinfo{volume}{13}}, \bibinfo{pages}{493} (\bibinfo{year}{2022}).

\bibitem{cotero_noninvasive_2019}
\bibinfo{author}{V.~Cotero}, \bibinfo{author}{Y.~Fan}, \bibinfo{author}{T.~Tsaava}, \bibinfo{author}{A.~Kressel}, \bibinfo{author}{I.~Hancu}, \bibinfo{author}{P.~Fitzgerald}, \bibinfo{author}{K.~Wallace}, \bibinfo{author}{S.~Kaanumalle}, \bibinfo{author}{J.~Graf}, \bibinfo{author}{W.~Rigby}, \bibinfo{author}{T.-J. Kao}, \bibinfo{author}{J.~Roberts}, \bibinfo{author}{C.~Bhushan}, \bibinfo{author}{S.~Joel}, \bibinfo{author}{T.~Coleman}, \bibinfo{author}{S.~Zanos}, \bibinfo{author}{K.~Tracey}, \bibinfo{author}{J.~Ashe}, \bibinfo{author}{S.~Chavan}, and \bibinfo{author}{C.~Puleo}, \enquote{\bibinfo{title}{Noninvasive sub-organ ultrasound stimulation for targeted neuromodulation}}, \bibinfo{journal}{Nat. Commun} \textbf{\bibinfo{volume}{10}}, \bibinfo{pages}{952} (\bibinfo{year}{2019}).

\bibitem{folloni_manipulation_2019}
\bibinfo{author}{D.~Folloni}, \bibinfo{author}{L.~Verhagen}, \bibinfo{author}{R.~Mars}, \bibinfo{author}{E.~Fouragnan}, \bibinfo{author}{C.~Constans}, \bibinfo{author}{J.-F. Aubry}, \bibinfo{author}{M.~Rushworth}, and \bibinfo{author}{J.~Sallet}, \enquote{\bibinfo{title}{Manipulation of {Subcortical} and {Deep} {Cortical} {Activity} in the {Primate} {Brain} {Using} {Transcranial} {Focused} {Ultrasound} {Stimulation}}}, \bibinfo{journal}{Neuron} \textbf{\bibinfo{volume}{101}}, \bibinfo{pages}{1109--1116} (\bibinfo{year}{2019}).

\bibitem{tufail_transcranial_2010}
\bibinfo{author}{Y.~Tufail}, \bibinfo{author}{A.~Matyushov}, \bibinfo{author}{N.~Baldwin}, \bibinfo{author}{M.~L. Tauchmann}, \bibinfo{author}{J.~Georges}, \bibinfo{author}{A.~Yoshihiro}, \bibinfo{author}{S.~I.~H. Tillery}, and \bibinfo{author}{W.~J. Tyler}, \enquote{\bibinfo{title}{Transcranial {Pulsed} {Ultrasound} {Stimulates} {Intact} {Brain} {Circuits}}}, \bibinfo{journal}{Neuron} \textbf{\bibinfo{volume}{66}}, \bibinfo{pages}{681--694} (\bibinfo{year}{2010}).

\bibitem{kubanek_remote_2020}
\bibinfo{author}{J.~Kubanek}, \bibinfo{author}{J.~Brown}, \bibinfo{author}{P.~Ye}, \bibinfo{author}{K.~Pauly}, \bibinfo{author}{T.~Moore}, and \bibinfo{author}{W.~Newsome}, \enquote{\bibinfo{title}{Remote, brain region--specific control of choice behavior with ultrasonic waves}}, \bibinfo{journal}{Sci. Adv} \textbf{\bibinfo{volume}{6, eaaz4193}} (\bibinfo{year}{2020}).

\bibitem{blackmore_ultrasound_2019}
\bibinfo{author}{J.~Blackmore}, \bibinfo{author}{S.~Shrivastava}, \bibinfo{author}{J.~Sallet}, \bibinfo{author}{C.~Butler}, and \bibinfo{author}{R.~Cleveland}, \enquote{\bibinfo{title}{Ultrasound {Neuromodulation}: {A} {Review} of {Results}, {Mechanisms} and {Safety}}}, \bibinfo{journal}{Ultrasound Med. Biol} \textbf{\bibinfo{volume}{45}}, \bibinfo{pages}{1509--1536} (\bibinfo{year}{2019}).

\bibitem{darmani_non-invasive_2022}
\bibinfo{author}{G.~Darmani}, \bibinfo{author}{T.~O. Bergmann}, \bibinfo{author}{K.~Butts~Pauly}, \bibinfo{author}{C.~F. Caskey}, \bibinfo{author}{L.~de~Lecea}, \bibinfo{author}{A.~Fomenko}, \bibinfo{author}{E.~Fouragnan}, \bibinfo{author}{W.~Legon}, \bibinfo{author}{K.~R. Murphy}, \bibinfo{author}{T.~Nandi}, \bibinfo{author}{M.~A. Phipps}, \bibinfo{author}{G.~Pinton}, \bibinfo{author}{H.~Ramezanpour}, \bibinfo{author}{J.~Sallet}, \bibinfo{author}{S.~N. Yaakub}, \bibinfo{author}{S.~S. Yoo}, and \bibinfo{author}{R.~Chen}, \enquote{\bibinfo{title}{Non-invasive transcranial ultrasound stimulation for neuromodulation}}, \bibinfo{journal}{Clinical Neurophysiology} \textbf{\bibinfo{volume}{135}}, \bibinfo{pages}{51--73} (\bibinfo{year}{2022}).

\bibitem{gaub2020neurons}
\bibinfo{author}{B.~M. Gaub}, \bibinfo{author}{K.~C. Kasuba}, \bibinfo{author}{E.~Mace}, \bibinfo{author}{T.~Strittmatter}, \bibinfo{author}{P.~R. Laskowski}, \bibinfo{author}{S.~A. Geissler}, \bibinfo{author}{A.~Hierlemann}, \bibinfo{author}{M.~Fussenegger}, \bibinfo{author}{B.~Roska}, and \bibinfo{author}{D.~J. M{\"u}ller}, \enquote{\bibinfo{title}{Neurons differentiate magnitude and location of mechanical stimuli}}, \bibinfo{journal}{Proceedings of the National Academy of Sciences} \textbf{\bibinfo{volume}{117}}, \bibinfo{pages}{848--856} (\bibinfo{year}{2020}).

\bibitem{alsteens2017atomic}
\bibinfo{author}{D.~Alsteens}, \bibinfo{author}{H.~E. Gaub}, \bibinfo{author}{R.~Newton}, \bibinfo{author}{M.~Pfreundschuh}, \bibinfo{author}{C.~Gerber}, and \bibinfo{author}{D.~J. M{\"u}ller}, \enquote{\bibinfo{title}{Atomic force microscopy-based characterization and design of biointerfaces}}, \bibinfo{journal}{Nature Reviews Materials} \textbf{\bibinfo{volume}{2}}, \bibinfo{pages}{1--16} (\bibinfo{year}{2017}).

\bibitem{krieg2019atomic}
\bibinfo{author}{M.~Krieg}, \bibinfo{author}{G.~Fl{\"a}schner}, \bibinfo{author}{D.~Alsteens}, \bibinfo{author}{B.~M. Gaub}, \bibinfo{author}{W.~H. Roos}, \bibinfo{author}{G.~J. Wuite}, \bibinfo{author}{H.~E. Gaub}, \bibinfo{author}{C.~Gerber}, \bibinfo{author}{Y.~F. Dufr{\^e}ne}, and \bibinfo{author}{D.~J. M{\"u}ller}, \enquote{\bibinfo{title}{Atomic force microscopy-based mechanobiology}}, \bibinfo{journal}{Nature Reviews Physics} \textbf{\bibinfo{volume}{1}}, \bibinfo{pages}{41--57} (\bibinfo{year}{2019}).

\bibitem{rabut2020ultrasound}
\bibinfo{author}{C.~Rabut}, \bibinfo{author}{S.~Yoo}, \bibinfo{author}{R.~C. Hurt}, \bibinfo{author}{Z.~Jin}, \bibinfo{author}{H.~Li}, \bibinfo{author}{H.~Guo}, \bibinfo{author}{B.~Ling}, and \bibinfo{author}{M.~G. Shapiro}, \enquote{\bibinfo{title}{Ultrasound technologies for imaging and modulating neural activity}}, \bibinfo{journal}{Neuron} \textbf{\bibinfo{volume}{108}}, \bibinfo{pages}{93--110} (\bibinfo{year}{2020}).

\bibitem{mohammadjavadi2022transcranial}
\bibinfo{author}{M.~Mohammadjavadi}, \bibinfo{author}{R.~T. Ash}, \bibinfo{author}{N.~Li}, \bibinfo{author}{P.~Gaur}, \bibinfo{author}{J.~Kubanek}, \bibinfo{author}{Y.~Saenz}, \bibinfo{author}{G.~H. Glover}, \bibinfo{author}{G.~R. Popelka}, \bibinfo{author}{A.~M. Norcia}, and \bibinfo{author}{K.~B. Pauly}, \enquote{\bibinfo{title}{Transcranial ultrasound neuromodulation of the thalamic visual pathway in a large animal model and the dose-response relationship with mr-arfi}}, \bibinfo{journal}{Scientific Reports} \textbf{\bibinfo{volume}{12}}, \bibinfo{pages}{19588} (\bibinfo{year}{2022}).

\bibitem{Arvanitis2023Ultrasound}
\bibinfo{author}{C.~Arvanitis}, \bibinfo{author}{P.~P. Dash}, and \bibinfo{author}{C.~Kim}, \enquote{\bibinfo{title}{Ultrasound mediated control of neurons and immune cells}}, \bibinfo{journal}{The Journal of the Acoustical Society of America} \textbf{\bibinfo{volume}{153}}, \bibinfo{pages}{A31--A31} (\bibinfo{year}{2023}), \urlprefix\url{http://dx.doi.org/10.1121/10.0018046}.

\bibitem{sarvazyan2010biomedical}
\bibinfo{author}{A.~P. Sarvazyan}, \bibinfo{author}{O.~V. Rudenko}, and \bibinfo{author}{W.~L. Nyborg}, \enquote{\bibinfo{title}{Biomedical applications of radiation force of ultrasound: historical roots and physical basis}}, \bibinfo{journal}{Ultrasound in medicine \& biology} \textbf{\bibinfo{volume}{36}}, \bibinfo{pages}{1379--1394} (\bibinfo{year}{2010}).

\bibitem{sapozhnikov2013radiation}
\bibinfo{author}{O.~A. Sapozhnikov} and \bibinfo{author}{M.~R. Bailey}, \enquote{\bibinfo{title}{Radiation force of an arbitrary acoustic beam on an elastic sphere in a fluid}}, \bibinfo{journal}{The Journal of the Acoustical Society of America} \textbf{\bibinfo{volume}{133}}, \bibinfo{pages}{661--676} (\bibinfo{year}{2013}).

\bibitem{gor2014forces}
\bibinfo{author}{L.~P. Gor'kov}, \enquote{\bibinfo{title}{On the forces acting on a small particle in an acoustical field in an ideal fluid}}, in \emph{\bibinfo{booktitle}{Selected Papers of Lev P. Gor'kov}}, \bibinfo{pages}{315--317} (\bibinfo{year}{2014}).

\bibitem{yosioka1955acoustic}
\bibinfo{author}{K.~Yosioka} and \bibinfo{author}{Y.~Kawasima}, \enquote{\bibinfo{title}{Acoustic radiation pressure on a compressible sphere}}, \bibinfo{journal}{Acta Acustica united with Acustica} \textbf{\bibinfo{volume}{5}}, \bibinfo{pages}{167--173} (\bibinfo{year}{1955}).

\bibitem{glynne2013efficient}
\bibinfo{author}{P.~Glynne-Jones}, \bibinfo{author}{P.~P. Mishra}, \bibinfo{author}{R.~J. Boltryk}, and \bibinfo{author}{M.~Hill}, \enquote{\bibinfo{title}{Efficient finite element modeling of radiation forces on elastic particles of arbitrary size and geometry}}, \bibinfo{journal}{The Journal of the Acoustical Society of America} \textbf{\bibinfo{volume}{133}}, \bibinfo{pages}{1885--1893} (\bibinfo{year}{2013}).

\bibitem{sapozhnikov2006radiation}
\bibinfo{author}{O.~A. Sapozhnikov}, \bibinfo{author}{L.~A. Trusov}, \bibinfo{author}{A.~I. Gromov}, \bibinfo{author}{N.~R. Owen}, \bibinfo{author}{M.~R. Bailey}, and \bibinfo{author}{L.~A. Crum}, \enquote{\bibinfo{title}{Radiation force imparted on a kidney stone by a doppler-mode diagnostic pulse}}, \bibinfo{journal}{The Journal of the Acoustical Society of America} \textbf{\bibinfo{volume}{120}}, \bibinfo{pages}{3109--3109} (\bibinfo{year}{2006}).

\bibitem{bruus2007theoretical}
\bibinfo{author}{H.~Bruus}, \emph{\bibinfo{title}{Theoretical microfluidics}}, volume~\bibinfo{volume}{18} (\bibinfo{publisher}{Oxford university press}) (\bibinfo{year}{2007}).

\bibitem{settnes2012forces}
\bibinfo{author}{M.~Settnes} and \bibinfo{author}{H.~Bruus}, \enquote{\bibinfo{title}{Forces acting on a small particle in an acoustical field in a viscous fluid}}, \bibinfo{journal}{Physical Review E} \textbf{\bibinfo{volume}{85}}, \bibinfo{pages}{016327} (\bibinfo{year}{2012}).

\bibitem{gavrilov1984use}
\bibinfo{author}{L.~R. Gavrilov}, \enquote{\bibinfo{title}{Use of focused ultrasound for stimulation of nerve structures}}, \bibinfo{journal}{Ultrasonics} \textbf{\bibinfo{volume}{22}}, \bibinfo{pages}{132--138} (\bibinfo{year}{1984}).

\bibitem{gavrilov2012focused}
\bibinfo{author}{L.~Gavrilov} and \bibinfo{author}{E.~Tsirulnikov}, \enquote{\bibinfo{title}{Focused ultrasound as a tool to input sensory information to humans}}, \bibinfo{journal}{Acoustical Physics} \textbf{\bibinfo{volume}{58}}, \bibinfo{pages}{1--21} (\bibinfo{year}{2012}).

\bibitem{fatemi2000probing}
\bibinfo{author}{M.~Fatemi} and \bibinfo{author}{J.~F. Greenleaf}, \enquote{\bibinfo{title}{Probing the dynamics of tissue at low frequencies with the radiation force of ultrasound}}, \bibinfo{journal}{Physics in Medicine \& Biology} \textbf{\bibinfo{volume}{45}}, \bibinfo{pages}{1449} (\bibinfo{year}{2000}).

\bibitem{fatemi1999vibro}
\bibinfo{author}{M.~Fatemi} and \bibinfo{author}{J.~F. Greenleaf}, \enquote{\bibinfo{title}{Vibro-acoustography: An imaging modality based on ultrasound-stimulated acoustic emission}}, \bibinfo{journal}{Proceedings of the National Academy of Sciences} \textbf{\bibinfo{volume}{96}}, \bibinfo{pages}{6603--6608} (\bibinfo{year}{1999}).

\bibitem{silva2006parametric}
\bibinfo{author}{G.~T. Silva}, \bibinfo{author}{S.~Chen}, and \bibinfo{author}{L.~P. Viana}, \enquote{\bibinfo{title}{Parametric amplification of the dynamic radiation force of acoustic waves in fluids}}, \bibinfo{journal}{Physical review letters} \textbf{\bibinfo{volume}{96}}, \bibinfo{pages}{234301} (\bibinfo{year}{2006}).

\bibitem{Brown2017}
\bibinfo{author}{M.~D. Brown}, \bibinfo{author}{B.~T. Cox}, and \bibinfo{author}{B.~E. Treeby}, \enquote{\bibinfo{title}{Design of multi-frequency acoustic kinoforms}}, \bibinfo{journal}{Applied Physics Letters} \textbf{\bibinfo{volume}{111}}, \bibinfo{pages}{244101} (\bibinfo{year}{2017}).

\bibitem{Brown2020}
\bibinfo{author}{M.~D. Brown}, \bibinfo{author}{B.~T. Cox}, and \bibinfo{author}{B.~E. Treeby}, \enquote{\bibinfo{title}{Stackable acoustic holograms}}, \bibinfo{journal}{Applied Physics Letters} \textbf{\bibinfo{volume}{116}}, \bibinfo{pages}{261901} (\bibinfo{year}{2020}).

\bibitem{Acquaticci2019}
\bibinfo{author}{F.~Acquaticci} \emph{et~al.}, \enquote{\bibinfo{title}{Ultrasound axicon: Systematic approach to optimize focusing resolution through human skull bone}}, \bibinfo{journal}{Materials} \textbf{\bibinfo{volume}{12}}, \bibinfo{pages}{3433} (\bibinfo{year}{2019}).

\bibitem{hu2022binary}
\bibinfo{author}{Z.~Hu}, \bibinfo{author}{Y.~Yang}, \bibinfo{author}{L.~Xu}, \bibinfo{author}{Y.~Hao}, and \bibinfo{author}{H.~Chen}, \enquote{\bibinfo{title}{Binary acoustic metasurfaces for dynamic focusing of transcranial ultrasound}}, \bibinfo{journal}{Frontiers in Neuroscience} \textbf{\bibinfo{volume}{16}}, \bibinfo{pages}{984953} (\bibinfo{year}{2022}).

\bibitem{JimenezGambin2022}
\bibinfo{author}{S.~Jim\'{e}nez-Gamb\'{i}n} \emph{et~al.}, \enquote{\bibinfo{title}{Acoustic holograms for bilateral blood-brain barrier opening in a mouse model}}, \bibinfo{journal}{IEEE Transactions on Biomedical Engineering} \textbf{\bibinfo{volume}{69}}, \bibinfo{pages}{1359--1368} (\bibinfo{year}{2021}).

\bibitem{Marzo2018}
\bibinfo{author}{A.~Marzo} and \bibinfo{author}{B.~W. Drinkwater}, \enquote{\bibinfo{title}{Holographic acoustic tweezers}}, \bibinfo{journal}{Proceedings of the National Academy of Sciences} \textbf{\bibinfo{volume}{116}}, \bibinfo{pages}{84--89} (\bibinfo{year}{2019}), \urlprefix\url{http://dx.doi.org/10.1073/pnas.1813047115}.

\bibitem{Jimenez2021}
\bibinfo{author}{N.~Jim{\'e}nez}, \bibinfo{author}{J.~Ealo}, \bibinfo{author}{R.~D. Muelas-Hurtado}, \bibinfo{author}{A.~Duclos}, and \bibinfo{author}{V.~Romero-Garc{\'i}a}, \enquote{\bibinfo{title}{Subwavelength {Acoustic} {Vortex} {Beams} {Using} {Self}-{Demodulation}}}, \bibinfo{journal}{Physical Review Applied} \textbf{\bibinfo{volume}{15}} (\bibinfo{year}{2021}), \urlprefix\url{http://dx.doi.org/10.1103/PhysRevApplied.15.054027}.

\bibitem{zhong2023}
\bibinfo{author}{C.~Zhong} \emph{et~al.}, \enquote{\bibinfo{title}{Real-time acoustic holography with physics-based deep learning for robotic manipulation}}, \bibinfo{journal}{IEEE Transactions on Automation Science and Engineering} \textbf{\bibinfo{volume}{21}}, \bibinfo{pages}{2951--2962} (\bibinfo{year}{2023}).

\bibitem{wang2025}
\bibinfo{author}{S.~Wang}, \bibinfo{author}{F.~You}, \bibinfo{author}{X.~Wang}, and \bibinfo{author}{H.~Xiao}, \enquote{\bibinfo{title}{A knowledge-driven method for real-time acoustic holographic field reconstruction using physical modeling and semi-supervised neural networks}}, \bibinfo{journal}{Knowledge-Based Systems} \textbf{\bibinfo{volume}{311}}, \bibinfo{pages}{113044} (\bibinfo{year}{2025}), \urlprefix\url{http://dx.doi.org/10.1016/j.knosys.2025.113044}.

\bibitem{khan2025}
\bibinfo{author}{H.~Khan} and \bibinfo{author}{J.~Kim}, \enquote{\bibinfo{title}{Reconstruction {Fidelity} of {Acoustic} {Holograms} {Across} 0.75--4.0 {MHz} {Excitation} {Frequencies}: A {Simulation} {Study}}}, \bibinfo{journal}{Applied Sciences} \textbf{\bibinfo{volume}{15}}, \bibinfo{pages}{10991} (\bibinfo{year}{2025}), \urlprefix\url{http://dx.doi.org/10.3390/app152010991}.

\bibitem{baresch2020}
\bibinfo{author}{D.~Baresch} and \bibinfo{author}{V.~Garbin}, \enquote{\bibinfo{title}{Acoustic trapping of microbubbles in complex environments}}, \bibinfo{journal}{Proceedings of the National Academy of Sciences} \textbf{\bibinfo{volume}{117}}, \bibinfo{pages}{15490--15496} (\bibinfo{year}{2020}).

\bibitem{ma2020}
\bibinfo{author}{Z.~Ma} \emph{et~al.}, \enquote{\bibinfo{title}{Acoustic holographic cell patterning in a biocompatible hydrogel}}, \bibinfo{journal}{Advanced Materials} \textbf{\bibinfo{volume}{32}}, \bibinfo{pages}{1904181} (\bibinfo{year}{2020}).

\bibitem{dash2020operational}
\bibinfo{author}{P.~P. Dash}, \enquote{\bibinfo{title}{Operational modal analysis of rolling tire: A tire cavity accelerometer mediated approach}}, Ph.D. thesis, \bibinfo{school}{Virginia Tech} (\bibinfo{year}{2020}).

\end{thebibliography}
}

\chapter*{Vita}
\addcontentsline{toc}{chapter}{Vita}

Pradosh P. Dash earned his Bachelor of Technology in Mechanical Engineering from the National Institute of Technology in Rourkela, India, in 2015 and started his career in Noise, Vibration, and Harshness (NVH) engineering for powertrain systems in the R\&D division at Bajaj Auto Ltd. Pradosh then attended Virginia Tech to pursue his interest in vibrations and acoustics. He received a Master of Science in 2020 for his research on operational modal analysis\cite{dash2020operational} to predict structure-borne noise.

In 2020, he began his Ph.D. at the Georgia Institute of Technology, working with Prof. Costas D. Arvanitis in the Ultrasound Biophysics and Bioengineering Laboratory. His doctoral research focused on holography-based transcranial focused ultrasound therapy and monitoring using nonlinear acoustics. While he was a graduate student, he also completed a Research Scientist Internship at Meta Reality Labs, where he worked on ultrasound-based sensors for robotics applications. Pradosh has actively engaged with the academic community at Georgia Tech during his PhD, serving as president of the IEEE UFFC Society's Georgia Tech Chapter and as the national representative for the Acoustical Society of America. Post graduation, he looks forward to continuing his work on connecting wave physics and acoustics.

\vspace{1cm}

\begin{center}
    \faGlobe \ \href{https://pradosh-dash.github.io}{pradosh-dash.github.io} \hspace{1cm}
    \faLinkedin \ \href{https://linkedin.com/in/ppdash}{linkedin.com/in/ppdash}
\end{center}

\end{document}